\documentclass[a4paper,12pt]{book}

\usepackage[latin1]{inputenc}

\usepackage[usenames]{color}
\definecolor{webblue}{rgb}{0,0,.5}
\definecolor{webgreen}{rgb}{0,.5,0}

\usepackage[colorlinks=true,linkcolor=webblue,citecolor=webblue,bookmarks=true,pdftitle={Vientos, Turbulencia y Ondas en las Nubes de la Atmósfera de Venus},pdfauthor={Javier Peralta Calvillo}]{hyperref}

\usepackage{amsmath,amsthm,amssymb}

\usepackage{graphics,graphicx,float}
\usepackage{epsfig}

\usepackage{multirow}
\usepackage{array}
\usepackage{xtab}

\usepackage{setspace}

\usepackage{natbib}

\usepackage{pdfpages}

\usepackage{makeidx}
\makeindex 

\begin{document}



\includepdf[pages=1-2]{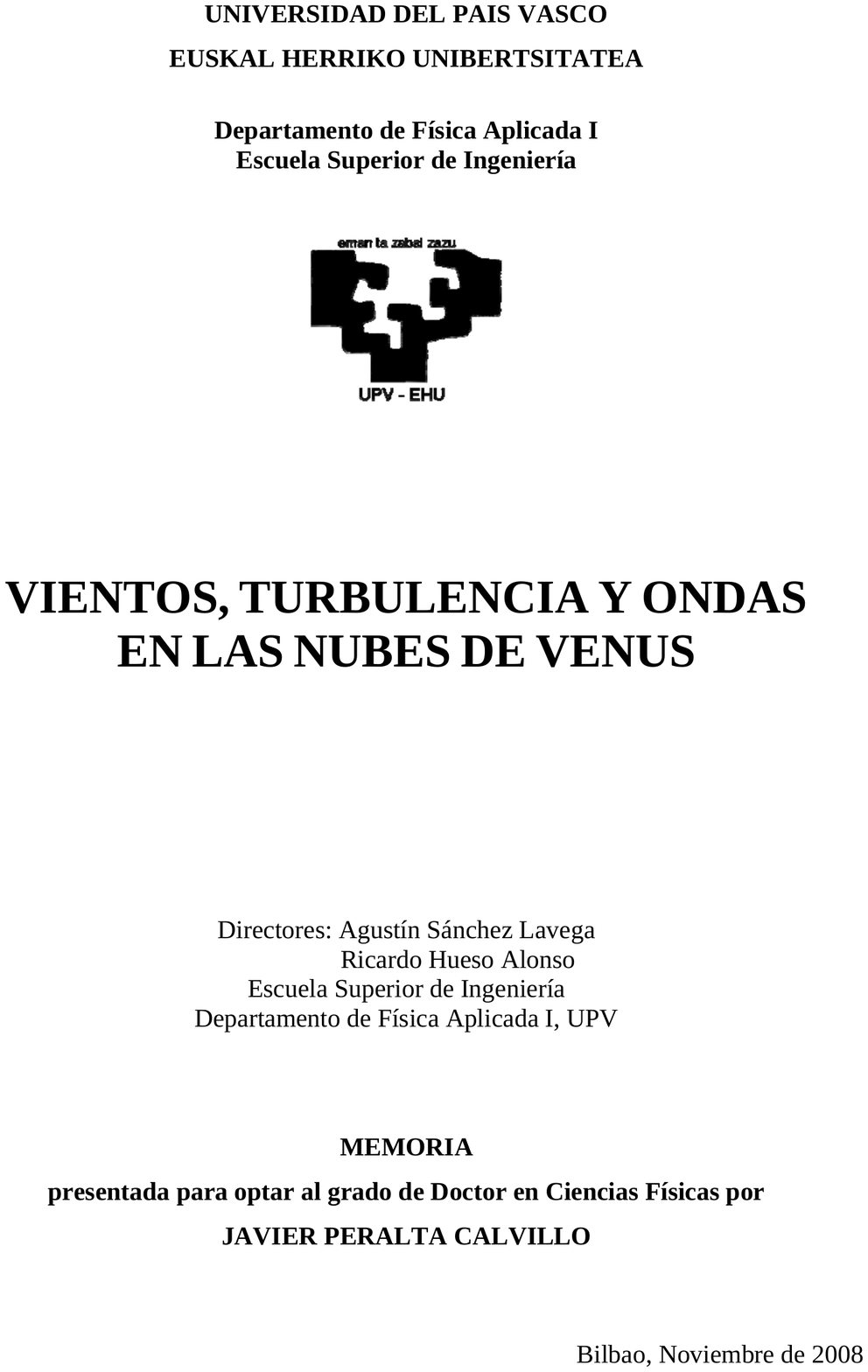}

\newpage

\chapter{Agradecimientos}\indent

Parece que fuera ayer cuando, calado hasta los huesos de fantasías de la bendita ni\~{n}ez, so\~{n}aba con ser astronauta y viajar a otros planetas. Dicha fantasía no quedó en el pasado, sino que decidió evolucionar y crecer hacia un destino aún más formidable cuando decidí que la ciencia era la due\~{n}a de las huellas que debía seguir. Tras meses de escritura, al fin me encuentro ante este apartado de la tesis, y me doy cuenta de lo afortunado que debo sentirme por haber cumplido uno de los sue\~{n}os de mi vida. No dudo un sólo instante que mi aportación a la ciencia y los conocimientos que he llegado adquirir en estos cuatro a\~{n}os son sólo los pasos finales de un camino que conducía a la consecución de un sue\~{n}o.\\

Son mis padres, Rafael y Teresa, las primeras personas a las que quiero dar las gracias. Ellos han supuesto mi principal apoyo y aliento durante toda mi vida, incluida la carrera de Física y estos cuatro a\~{n}os de doctorado. Sé que si no fuera por ellos, jamás habría llegado tan lejos. Y por supuesto a mis hermanas María y Teresa, que me han hecho recordar continuamente lo que significa sentirse querido incondicionalmente, María por sus ánimos sin descanso, y Teresa por enfrascarse en que no olvide nunca lo mucho que me quiere.\\

Sin duda que nada de esto hubiera sido posible sin mis dos directores, Agustín Sánchez Lavega y Ricardo Hueso, quienes no sólo han resultado ser dos investigadores modélicos por lo que siento una gran admiración, sino también dos personas que me han acogido desde el principio con amabilidad y con cari\~{n}o. Y si bien es conocida (y fácilmente demostrable) su fama de bri\-llan\-tes y apasionados investigadores, más importante es aún su valía como personas. A Agustín le doy gracias por contagiarme de su pasión por la ciencia, por ser tan paciente conmigo y por confiar más en mi de lo que yo mismo hago. A Ricardo quiero agradecerle la cantidad de horas y sacrificios que ha realizado para formarme como científico, su incansable paciencia en su tarea como director de tesis (sé que no soy una persona fácil) y sobre todo el haber sido la persona que más se ha esforzado para que no me sienta solo en un lugar donde cuesta mucho hacer amistades y donde la gente suele ser reacia a permitirte entrar en sus grupos de amistades fuera del trabajo.\\

No hay sustento sin un tercer pilar, y mi compa\~{n}ero de despacho Santiago ha sido para mi una fuente de ánimos y apoyo en todos los sentidos, además de maestro, amigo y confidente. Tanto él como su esposa Yolanda se han portado conmigo de forma excelente. Quiero dar también las gracias a Naiara Barrado, Marta Massot, Estíbaliz Api\~{n}ániz, Sara García y Asier Agos, además de los que ya no están como son Aleksander Morgado y Nora Madariaga. Todos han sido excelentes compa\~{n}eros en el Departamento, en especial Marta Massot (con quien tanto he compartido) y Naiara Barrado, quien tanto me ha ayudado y animado. Mención especial merecen José Félix Rojas, Jon Legarreta y Enrique García, quienes me han aportado tanto apoyo como agradables tertulias. También expreso mis más sinceros agradecimientos a toda la gente del Departamento de Física Aplicada I de la Escuela de Ingenieros de Bilbao, en cuanto que me han tratado con una hospitalidad de la que merecidamente tiene fama Euskadi. En especial, destacaré a Teresa del Río, Carmen Orde\~{n}ana, Asun Illaramendi, Ibón Aramburu y Agustín Salazar, por la atención especial que siempre me han brindado. No puedo olvidar tampoco a todos mis compa\~{n}eros italianos, quienes me hicieron sentir como en casa durante mi estancia en Roma. Quiero dar las gracias a Giuseppe Piccioni, Alejandro Cardesín, Orietta Lanciano, Romolo Politi y Alessandra Migliorini, y con especial cari\~{n}o a Alejandro, Valeria Cottini y Teresa Mastroianni, sin faltar la hospitalidad y cari\~{n}o de Elisabetta Visalberghi y por supuesto de Goccia (os echo mucho de menos).\\

Quiero también expresar mi más profundo agradecimiento a todos los profesores que me han acompa\~{n}ado en este camino y me inspiraron el amor a las matemáticas y las ciencias, desde mi maestra Inés, pasando por Antonio Trujillo, Amelia Esteve, Ana Villaexcusa, Manuel Jiménez, Paco Valdivia y Fernando Torres, sin olvidar a Inés Rodríguez Hidalgo, Fernando Moreno Insertis y Jon Sáenz. Y no olvidaré mencionar a Manuel López Puertas, quien me ayudó cuando nadie más lo hacía y me abrió las puertas a la investigación y a participar en lo que sería mi primera publicación.\\

Aunque no estén directamente relacionadas con mi vida laboral, hay amigos y amigas muy importantes a las que quiero mencionar. Para empezar mi mejor amiga, Marta Puerta, que me ha demostrado con creces en estos cuatro a\~{n}os que siempre estará a mi lado. También a Fátima Rubio, por su valiosísima ayuda y apoyo en los momentos que más lo necesitaba. Y por supuesto, no pueden faltar Ana Ursúa, Coral Sáez, Cristina Merchán y Aitziber Otaola, que me han prestado un apoyo moral innegable. A todos mis amigos de Algeciras, Sevilla, Canarias y Granada. A toda mi familia al completo. Y aunque sé que no es muy común en un tesis científica, dado que la ausencia de prueba no es la prueba de la ausencia, doy las gracias a los que están arriba y siempre han velado por mi.\\

Finalmente, queda agradecer a la Universidad del País Vasco su financiación durante estos cuatro a\~{n}os como becario pre-doctoral. Con este trabajo creo poder decir con orgullo que he logrado rentabilizar una ayuda sin la cual nada de esto hubiera sido posible.\\

Y agradezco al lector que se haya molestado en leer esta larga serie de agradecimientos, previo a la lectura de una tesis que ha sido el producto de mucho esfuerzo, crecimiento y superación. Y no sólo es mi humilde aportación a la ciencia, sino la prueba de lo que el trabajo y la constancia pueden llegar a lograr, tal y como tantas veces me ha repetido mi estimado director y amigo Agustín Sánchez Lavega.

\tableofcontents

\addcontentsline{toc}{chapter}{índice de tablas}
\listoftables
\addcontentsline{toc}{chapter}{índice de figuras}
\listoffigures

\chapter{Thesis Summary}\indent

Venus constitutes a unique framework for meteorological research. Its slow rotation places its atmospheric dynamics outside of the familiar regime of quasi-geostrophic motions, which have been the focus of most atmospheric studies. The principal mode of atmospheric circulation on Venus is a zonal retrograde superrotation of the entire atmosphere, from just above the lo\-west scale height to $\approx100$ km. Approximately cyclostrophic balance prevails throughout the nonequatorial lower atmosphere of Venus wherever the westward winds are large. Eddies, the mean meridional circulation, and planetary-scale waves may all be involved in the upward transport of retrograde angular momemtum to maintain the atmospheric superrotation. Many spacecraft observations have tried to describe the properties of the Venus atmosphere and its circulation, but up to now these data hardly serve to define the basic circulation of the bulk of the atmosphere. Nevertheless, most recent mission Venus Express not only has supposed a reborn of the spatial exploration of Venus after a significative gap of time, but also has allowed the usage of modern instrumentation capable of getting over the limitations from previous missions. This thesis focuses on the experimental characterization of three important aspects of the atmospheric dynamics in Venus at the level of the clouds: global winds, atmospheric turbulence and mesoscale gravity waves.\\

Characterization of the winds is crucial for an adequate determination of the dynamic regime that rules the circulation. The results that I present here provide for the first time a three-dimensional view of the winds in Venus thanks to the images obtained with the imaging spectrometer VIRTIS onboard Venus Express, capable of sounding different levels of the atmosphere (top of the clouds and lower clouds) simultaneously. Moreover, the inferred meridional circulation has brought strong new evidences of the existence of a Hadley circulation at the top of the clouds. A thermal tide and 5-day global scale oscillation of the mean wind have been detected too, and long and short-term temporal variations are apparent from our wind measurements.\\

The study of the turbulence at the cloud level can provide some highlights on the role played by eddies in the general circulation. Unfortunately, the \mbox{error} in wind speed measurements inhibits the usage of the traditional tools to study turbulence through the kinetic energy power spectra and an alternative approach through the cloud brightness distribution from images obtained by the Galileo spacecraft is employed. Temporal and spatial changes are detected through this analysis at the top of the clouds. On the one hand, the cloud brightness power spectra display noticiable alterations in the slope values when comparing results belonging to the periods of Mariner 10, Pio\-neer Venus and the Galileo flyby. On the other hand, the equatorial and midlatitudes regions at the top of the clouds exhibit brightness power spectra with a slope $\approx-5/3$ (similar to Kolmogorov's law and what's expected for the three-dimensional turbulence), whereas in the subpolar regions the brightness power spectra display a slope $\approx-3$ for the highest spatial scales and $\approx-5/3$ for the lowest ones (a result close to what has been obtained for kinetic energy power spectrum in the case of the Earth). It is also shown that this behaviour seems highly related to the meridional distribution of the zonal winds, with nearly constant values for the equator and midlatitudes, and a steepen decrease of the wind from the subpolar regions.\\

The third aspect in this thesis is about gravity waves. Images obtained by Venus Express have allowed us to study mesoscale waves at different vertical levels of the Venusian atmosphere. In particular, and for the first time, in the lower clouds. We have characterized the morphology and dynamics of these waves, identifying them as internal gravity waves type. These waves are probably convectively induced, with no apparent relation to solar-locked or surface elevations sources. And examination of their properties in terms of a linear model was also carried out, resulting in vertical wavelengths in accordance with previous results from different observations. Gravity waves are important to the characterization of the atmosphere as their mere pre\-sen\-ce serves as an indicator of the atmospheric stability and because they are probable means of transport for the linear momemtum in the atmospheric general circulation.\\

This thesis starts with a brief introduction to Venus and the composition, vertical structure and dynamics of its atmosphere (chapter \ref{chapter-intro}), followed by a description of the data employed, instrumentation and the software used (chapter \ref{chapter-observs}). I will present the results obtained from our measurements of winds in Venus at different altitudes of the cloud level from images belonging to the mentioned spatial missions (chapter \ref{chapter-winds}), and also introduce the study about the cloud brightness distribution at the top of the clouds (chapter \ref{chapter-turbulence}). Finally, I will end with the study about the mesoscale gravity waves observed at the top of the clouds and lower clouds (chapter \ref{chapter-gravitywaves}).

\mainmatter


\chapter{Introducción}\label{chapter-intro}

\section{La atmósfera de Venus}\label{chapter-intro-VenusAtmosphere}\indent

Venus es el segundo planeta del sistema solar en cercanía al Sol. Describe una órbita en torno al Sol prácticamente circular, a una distancia promedio de 0.72 UA y con un periodo orbital de 224.70 días. Al igual que Mercurio, Venus no tiene ningún satélite. Su eje de rotación tiene una inclinación de 177.4$^{\circ}$, por lo que es prácticamente perpendicular a la eclíptica (por tanto, a diferencia de la Tierra, es de esperar que Venus carezca de estaciones) y es tal que gira en sentido opuesto al resto de planetas del sistema solar. Su periodo de rotación es de 243.02 días, con lo que el día sidéreo es más largo que el a\~{n}o venusiano. El día solar (el periodo de tiempo entre dos pasos sucesivos del Sol por el mismo meridiano) es significativamente más corto, de unos 117 días terrestres. Su tama\~{n}o, masa y densidad son comparables a las de la Tierra, por lo que en el pasado se pensaba en Venus como un planeta gemelo de ésta, aunque una comparación más exhaustiva revela que las diferencias entre ambos son mucho más que notables (ver Tabla \ref{tab:tabla-VenusData}).\\

\begin{table}[h!]
  \caption{Venus comparado con la Tierra.}
	\label{tab:tabla-VenusData}
	\centering
  \begin{spacing}{0.6}
		\begin{tabular}{*{3}{>{\scriptsize}c}}
			& & \\
			\hline\hline
			& & \\
			\textit{Propiedad} & \textit{\textbf{Venus}} & \textit{\textbf{La Tierra}} \\
			& & \\
			\hline
			& & \\
			Masa                 &  $4.87\cdot10^{24}~kg$   &  $5.97\cdot10^{24}~kg$   \\
			Diámetro             &       $12,104~km$        &  $12,756~km$             \\
			Densidad             &  $5,250~km\cdot m^{-3}$  &  $5,515~km\cdot m^{-3}$  \\
			Gravedad Superf.     &   $8.87~m\cdot s^{-2}$   &  $9.78~m\cdot s^{-2}$    \\
			Temperatura Superf.  &          730$^{\circ}$K          &          288$^{\circ}$K          \\
			& & \\
			Periodo de Rotación  &       243.02 días        &       23.93 horas        \\
			Periodo Orbital      &       224.70 días        &      356.26 días         \\
			Distancia al Sol     &   $1.082\cdot10^{8}~km$  &   $1.496\cdot10^{8}~km$  \\
			Excentricidad        &           0.007          &           0.017          \\
			Oblicuidad           &            177$^{\circ}$          &           23.45$^{\circ}$         \\
			& & \\
			\hline
			& & \\
		\end{tabular}
  \end{spacing}
\end{table}

Generalmente las atmósferas planetarias no colapsan formando una fina capa sobre la superficie debido a su propio peso, sino que se encuentran en una situación de equilibrio en la que su peso es contrarrestado por la presión. Decimos en este caso que la atmósfera está en \textit{equilibrio hidrostático}\index{Equilibrio!hidrostático}. Dicho equilibrio ($\partial p/\partial z=-\rho\cdot g$) conlleva una estratificación vertical de la atmósfera con la densidad y presión disminuyendo de forma exponencial con la altura ($p=p_{0}\cdot e^{-z/H}$, donde $H$ es la escala vertical de alturas). Por otro lado, la atmósfera de un planeta queda bien caracterizada por la manera en que su temperatura varía con la altura (ver Figura 1.1). En Venus el carácter de esta variación divide a la atmósfera en tres regiones. La \textit{troposfera}\index{Troposfera} (0-60 km) abarca desde la superficie hasta la altura en que finalizan las nubes y en ella la temperatura disminuye con la altura. Debido a la capa nubosa la troposfera es una región oculta a la observación directa con luz visible. Sin embargo, numerosas medidas ``in situ'' fueron obtenidas por las 16 sondas (Pioneer Venus de NASA, Venera de la extinta URSS y los globos franco-rusos VEGA) que descendieron en su atmósfera en las décadas de los 70 y 80, en su mayoría por latitudes cercanas al ecuador \citep{Seiff1983,Seiff1985}. La troposfera posee un gradiente térmico vertical\footnote{El gradiente térmico vertical\index{Gradiente!térmico vertical} es el ritmo con el que la temperatura varía con la altura ($\Gamma=\partial T/\partial z$).} de aproximadamente 9$^{\circ}$K/km, valor muy cercano al gradiente adiabático\index{Gradiente!adiabático}\footnote{Variación de temperatura que experimenta una parcela de aire al moverse verticalmente sin intercambiar calor con el medio que lo rodea, es decir, de forma adiabática ($\Gamma_{d}=g/c_{p}=7.39~K\cdot km^{-1}$).} por lo que es probable que la convección domine esta región. La \textit{mesosfera}\index{Mesosfera} (60-100 km) es una región de la atmósfera más accesible a las medidas de forma remota \citep{Taylor1980,Zasova1999}, en la que la temperatura sigue descendiendo con la altura aunque de manera menos acusada. La particu\-la\-ri\-dad de esta región reside en las variaciones horizontales de la temperatura con la latitud, creciendo ésta del ecuador hacia los polos \citep{Taylor1980} y evidenciando la posible existencia de una circulación de Hadley. Por último, la \textit{termosfera}\index{Termosfera} (100-200 km) es una región controlada por el equilibrio entre la radiación solar UV incidente y la conductividad de las moléculas presentes, y donde la temperatura tiende a aumentar con la altura en el lado diurno del planeta y se mantiene constante en el lado nocturno.\\

\begin{figure}[h!]
	\centering
		\includegraphics[width=0.8\textwidth]{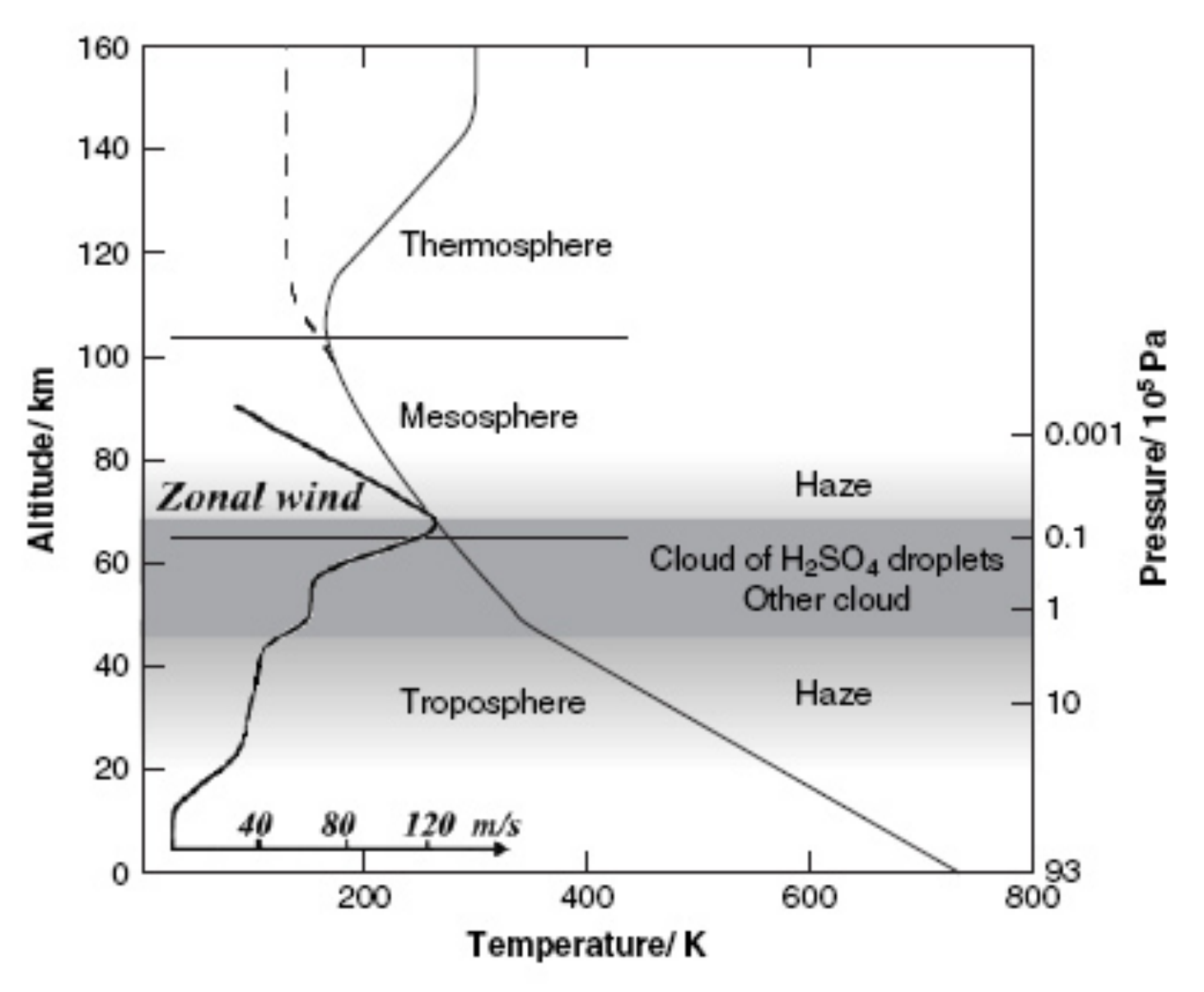}
	\label{fig:Venus-Thermal-Profile}
	\caption[Perfil térmico vertical de la atmósfera de Venus.]{\scriptsize{Perfil térmico vertical de la atmósfera de Venus, donde se distinguen principalmente tres capas: la troposfera (0-60 km), la mesosfera (60-100 km) y la termosfera (100-200 km). La línea a trazos en la termosfera muestra el perfil térmico vertical en la parte nocturna de Venus. La capa de nubes se localiza aproximadamente entre 48 y 70 km de altura. Se visualiza, además, el perfil vertical de los vientos zonales.}}
\end{figure}

La atmósfera de Venus está compuesta principalmente por $CO_{2}$ (96.5\%) y $N_{2}$ (3.5\%), tal como puede observarse en la Tabla \ref{tab:tabla-VenusComposition}. Aunque en cantidades muy peque\~{n}as, también están presentes compuestos de sulfuro, carbono y cloro, así como vapor de agua, en concentraciones que van de unos pocos a centenares de partes por millón \citep{Esposito1997,deBergh2006}. Venus está completamente cubierto por una capa de nubes de unos 22 km de extensión vertical, localizada entre 48 y 70 km de altura (ver Figura 1.1). Aunque las nubes no son muy densas y la visibilidad dentro de ellas es de unos centenares de metros, la opacidad total de la capa nubosa sí que resulta significativa. Las nubes apenas muestran rasgos en el rango del visible, si bien no ocurre lo mismo para la región del ultravioleta en el que una gran variedad de detalles nubosos son visibles cuando observamos la luz reflejada del lado diurno. El descubrimiento de \citet{Allen1984} de una ventana transparente\footnote{A través de las ``ventanas'' en $\lambda=$ 1, 1.27, 1.74 y $2.3~\mu m$ la radiación térmica de la atmósfera inferior puede escapar al espacio.} en el IR cercano proporcionó una nueva y poderosa herramienta para investigar las nubes más profundas y las variaciones que experimenta la opacidad total de las nubes. En estas longitudes de onda las nubes también muestran detalles, si bien el origen de éstos es distinto de los observados en el lado diurno: los contrastes de las nubes en las imágenes UV se deben a cambios en la reflectividad de las nubes a la radiación UV, mientras que en el lado nocturno son variaciones de la opacidad a la radiación térmica emitida desde los niveles inferiores y representan nubes más densas y profundas.\\

\begin{table}[h!]
  \caption{Composición de la atmósfera de Venus.}
	\label{tab:tabla-VenusComposition}
	\centering
  \begin{spacing}{0.6}
		\begin{tabular}{*{5}{>{\scriptsize}c}}
			& & \\
			\hline\hline
			& & \\
			\multirow{2}{*}{\textit{Compuesto}} & \textit{Abundancia} &  &  \multirow{2}{*}{\textit{Compuesto}} & \textit{Abundancia}  \\
			  & \textit{(fracción molar)} &  &  &  \textit{(fracción molar)}  \\
			& & & & \\
			\hline
			& & & & \\
			$CO_{2}$  &  0.965   &   &  $He$      &   12 ppm  \\
			$N_{2}$   &  0.035   &   &  $Ne$      &    7 ppm  \\
			$SO_{2}$  & 150 ppm  &   &  $H_{2}S$  &    3 ppm  \\
			$Ar$      &  70 ppm  &   &  $HCl$     &  400 ppb  \\
			$CO$      &  30 ppm  &   &  $Kr$      &   30 ppb  \\
			$H_{2}O$  &  20 ppm  &   &  $HF$      &    5 ppb  \\
			& & & & \\
			\hline
			& & & & \\
		\end{tabular}
  \end{spacing}
\end{table}

Las observaciones espectroscópicas y de polarimetría en telescopios de tierra nos muestran que las nubes superiores están formadas por gotas de $1~\mu m$ ricas en $H_{2}SO_{4}$ (75\%). Las sondas Venera y Pioneer proporcionaron datos sobre la estructura vertical de las nubes y las propiedades físicas de los aerosoles, descubriendo tres modos distintos de tama\~{n}o de partículas \citep{Ragent1985}:
\begin{itemize}
  \item Modo 1 (diámetro promedio de $\sim0.5~\mu m$): la composición de esta partícula es desconocida y su concentración varía con la altura. Sus variaciones horizontales y verticales son las que producen los contrastes que exhiben las nubes en las imágenes en UV.
  \item Modo 2 (diámetro promedio de $\sim2-3~\mu m$): casi con total seguridad se trata de ácido sulfúrico altamente concentrado (75\%).
  \item Modo 3 (diámetro promedio $>3~\mu m$): son las partículas de mayor tama\~{n}o y se desconoce su composición. Se especula que muchas de ellas podrían tratarse de cristales, carentes de ácido sulfúrico en caso de ser esféricas y homogéneas.
\end{itemize}

\section{Dinámica atmosférica y estudios previos}\label{chapter-intro-PrevStudies}\indent

La atmósfera de Venus presenta a escala global dos regímenes dinámicos claros variables en altura (ver Figura 1.1): la \textit{superrotación zonal retrógrada} de este a oeste en la troposfera y mesosfera \citep{Gierasch1997} y la \textit{circulación solar-antisolar} a través del amanecer y el anochecer en la termosfera \citep{Bougher1997}. El fenómeno de la superrotación debe su nombre al hecho de que en una región bastante amplia de la atmósfera de Venus los vientos tienen velocidades muy superiores a la rotación del pla\-ne\-ta (en la cima de las nubes la atmósfera se mueve 60 veces más rápido que el planeta y supera los $120~m\cdot s^{-1}$), hecho que contrasta con el caso de la Tierra con una atmósfera que gira sincrónicamente con el planeta sólido situado en su interior. Tal como puede observarse en la Figura 1.1, la superrotación aumenta con la altura, mostrando su máxima magnitud en la cima de las nubes y decreciendo de forma acusada a partir de esa altura \citep{Gierasch1997}. Este comportamiento resulta intrigante en cuanto a que los vientos zonales vuelven a ser importantes en la baja termosfera \citep{Bougher1997}, conformando el segundo régimen dinámico destacable mencionado anteriormente. Superpuesto a la superrotación de la alta troposfera parece existir una circulación de la atmósfera mucho más lenta del ecuador hacia los polos, con velocidades meridionales que apenas alcanzan los $10~m\cdot s^{-1}$ y gigantescos vórtices en los polos (ver Figura 1.2) que sugieren fuertes movimientos de subsidencia \citep{Taylor1980,Piccioni2007}.\\

\begin{figure}[h!]
	\centering
		\includegraphics[width=0.8\textwidth]{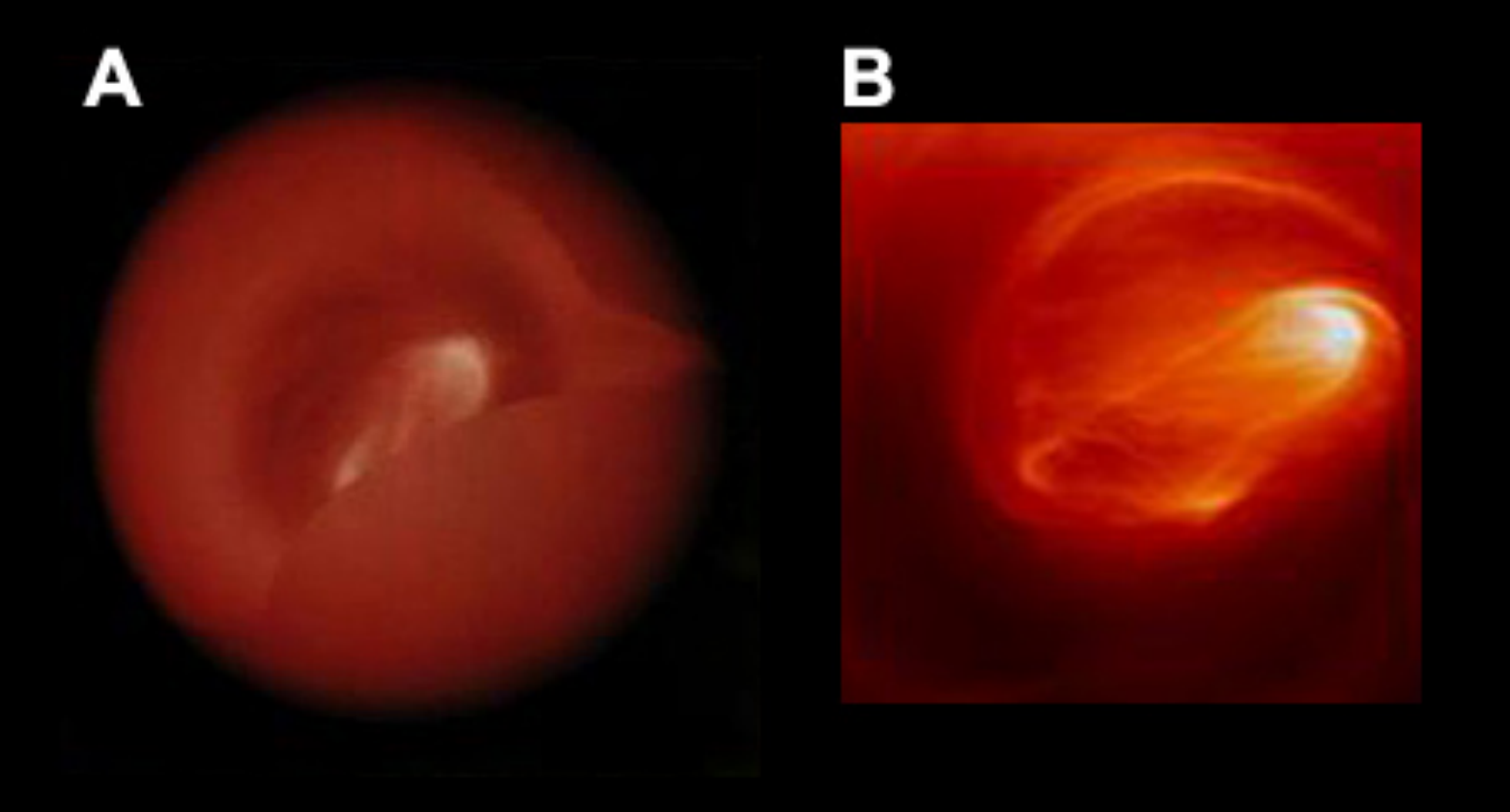}
	\label{fig:Venus-Polar-Vortex}
	\caption[Vortices polares de Venus.]{\scriptsize{Vortices polares de Venus captados gracias a la emisión térmica del planeta. (A) es una imagen del vórtice del polo norte de Venus tomada por el instrumento OIR de Pioneer Venus \citep{Taylor1980} a una longitud de onda de $11.5~\mu m$. (B) es una imagen del vórtice del polo sur de Venus tomada por el instrumento VIRTIS de Venus Express \citep{Piccioni2007} a una longitud de onda de $5~\mu m$ ($\sim59$ km de altura). Las regiones más brillantes se corresponden con aquellas en las que llega más radiación de las capas calientes inferiores.}}
\end{figure}

Hasta la fecha todos los intentos de modelizar la superrotación zonal han tenido un éxito limitado, por lo que los mecanismos básicos que rigen este fenómeno siguen sin estar del todo claros. Otro de los aspectos más intri\-gantes tiene que ver con el modo en que un planeta débilmente rotante es capaz de acelerar tanto su atmósfera. No sabemos si la circulación meridional consiste en una única célula de Hadley que se extiende desde la superficie hasta la parte superior de la atmósfera, o si por el contrario tenemos una serie de células apiladas verticalmente (ver Figura 1.3). Por otro lado, las ondas de gravedad o los fenómenos turbillonarios todavía sin caracterizar podrían jugar un papel esencial en el transporte de momento angular y alimentación de la superrotación. También sigue siendo una incógnita el modo en que los vórtices polares acoplan las dos regímenes dinámicos mencionados anteriormente, así como el comportamiento complejo y variable de dichos vórtices.

\begin{figure}[h!]
	\centering
		\includegraphics[width=0.6\textwidth]{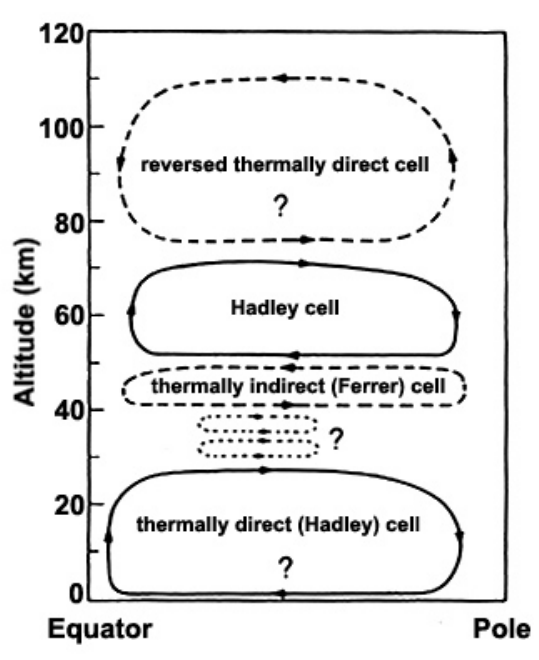}
	\label{fig:Venus-meridional-circulation}
	\caption[Esquema vertical de la circulación meridional.]{\scriptsize{Posible esquema vertical de la circulación meridional en la atmósfera de Venus \citep{Prinn1985}. En caso de que exista, se espera que la circulación de Hadley sea un mecanismo eficiente de transporte de aire caliente hacia los polos y aire frío hacia el ecuador. Discontinuidades en el gradiente vertical de temperatura detectadas por las sondas Venera, Pionee Venus y los globos Vega sugieren un esquema como el de la figura.}}
\end{figure}

La dinámica atmosférica de Venus ha sido estudiada mediante el uso de muy diversas técnicas. Muchas de ellas consisten en mediciones directas como las realizadas ``in situ'' por las sondas espaciales Venera \citep{Schubert1977,Marov1978}, Pioneer Venus \citep{Schubert1983,Gierasch1997} y los globos VEGA \citep{Preston1986}, que penetraron en la atmósfera y midieron los vientos en determinadas localizaciones y momentos concretos. También se han realizado medidas indirectas del viento mediante espectroscopía Doppler de las líneas de emisión de diversos componentes atmosféricos \citep{Widemann2007,Lellouch2008,Gaulme2008} o trabajando bajo la hipótesis de que la atmósfera de Venus está en equilibrio ciclostrófico y estimando la velocidad del viento con la ecuación del viento térmico y medidas de temperatura \citep{Newman1984,Roos-Serote1995,Piccialli2008}.\\

La nave Galileo (NASA) llevó a cabo en 1990 un sobrevuelo de Venus, momento que se aprovechó para estudiar la atmósfera de éste. Desde entonces hasta el a\~{n}o 2006 no ha habido más misiones espaciales que visitaran este mundo\footnote{Excepto la nave Magallanes, que estudió la superficie de Venus entre 1990 y 1994 usando imágenes de radar.} y observaran su atmósfera. La nave Venus Express llegó a Venus en abril de 2006 y puso fin a esta sequía observacional. Desde entonces orbita el planeta tomando imágenes de las nubes con dos instrumentos: VIRTIS y VMC. La cámara VMC capta imágenes en 4 filtros distintos y obtiene imágenes de más alta resolución \citep{Markiewicz2007a}, mientras que VIRTIS\index{VIRTIS} es un espectrómetro visual capaz de obtener simultáneamente imágenes de menor resolución espacial pero en un amplio rango de longitudes de onda \citep{Drossart2007}. En ambas misiones se han obtenido perfiles de viento mediante el seguimiento de las estructuras nubosas observadas, ya sea haciendo uso de técnicas automáticas \citep{Belton1991,Toigo1994,Markiewicz2007b} o utilizando la inspección visual \citep{Peralta2007b,Sanchez-Lavega2008}.\\

\section{Objetivos de esta tesis doctoral}\indent

Esta tesis tiene como objeto profundizar en tres aspectos importantes de la dinámica atmosférica de Venus al nivel de las nubes: \textbf{(1)} la circulación general atmosférica, \textbf{(2)} la turbulencia y \textbf{(3)} los fenómenos ondulatorios. Todos estos elementos han sido estudiados a través de imágenes de la atmósfera tomadas por las misiones espaciales Galileo y Venus Express en diferentes longitudes de onda. Estas imágenes proporcionan información de distintos niveles de altura. Los detalles sobre los instrumentos (cámaras y espectrómetros visuales) y la metodología de estudio seguida se explican en el capítulo \ref{chapter-observs}.\\

El estudio de los vientos en Venus (ver capítulo \ref{chapter-winds}) supone un paso primordial a la hora de caracterizar la circulación general. Con este trabajo se pretende construir un cuadro tridimensional de los vientos a la altura de las nubes en Venus a partir de las imágenes en diferentes longitudes de onda que proporciona el instrumento VIRTIS a bordo de la nave Venus Express (ver capítulo \ref{chapter-observs}). Tres son las seleccionadas en este caso: $380~nm$ (ultravioleta) que en el lado diurno permite captar la cima de las nubes a $\sim66$ km de altura, $980~nm$ (infrarrojo cercano) que visualiza en el lado diurno también la base de la cima de las nubes a $\sim61$ km, y $1.74~\mu m$ (infrarrojo) que permite visualizar en el lado nocturno las nubes inferiores a $\sim47$ km. También se incluye un estudio de la variabilidad temporal de los vientos en Venus en distintos intervalos de tiempo: Por un lado la insolación periódica que el Sol ejerce sobre la atmósfera (marea térmica) así como la presencia de diferentes ondas deberían producir alteraciones temporales de corto plazo detectables en el seno del campo de velocidades, hecho que tratará de detectarse usan\-do las imágenes de VIRTIS. Por otro lado, la amplia cobertura temporal de datos que proporciona Venus Express nos permitirá estudiar también las variaciones del viento a medio plazo (semanas a meses) y la comparación con las medidas de la sonda Galileo permitirá asímismo investigar las de largo plazo (a\~{n}os).\\

Además de la circulación general, en esta tesis se investiga la turbulencia presente en la atmósfera de Venus a través de las características de sus nubes superiores. La estimación directa de las características de la turbulencia en la atmósfera es un objetivo científico difícil de alcanzar porque implicaría rea\-li\-zar medidas de los movimientos con una precisión más allá de lo que los datos espaciales permiten en la actualidad. Por ello, hemos analizado la distribución horizontal de tama\~{n}os de las nubes y los espectros espaciales de variación de brillo de éstas como un indicador del régimen turbulento de la atmósfera. Los resultados se comparan con resultados de diferentes modelos de turbulencia atmosférica (ver capítulo \ref{chapter-turbulence}).\\

Finalmente, se presenta un estudio exhaustivo de ondas atmosféricas de meso\-escala presentes en las observaciones de las nubes efectuadas por el ins\-tru\-men\-to VIRTIS. Estas ondas han sido encontradas en los tres niveles de altura que pueden detectarse con este instrumento y sus características han llevado a identificar dichas ondas como ondas internas de gravedad. Tras una caracterización de la morfología y dinámica de estas ondas se determinan algunas de sus características no directamente observadas (como la longitud de onda vertical) comparando las medidas efectuadas con un modelo teórico sencillo. Las ondas de gravedad son importantes ya que su presencia nos indica la existencia de regiones de estabilidad estática atmosférica positiva permitiendo deducir este  parámetro. Por otro lado las ondas constituyen probablemente un importante medio de transporte de momento lineal y ener\-gía en la circulación atmosférica global del planeta (ver capítulo \ref{chapter-gravitywaves}).

\chapter{Observaciones y Métodos de Análisis}\label{chapter-observs}\indent

A lo largo de las últimas décadas Venus ha sido observado por multitud de sondas espaciales y por telescopios terrestres. En esta tesis he utilizado datos provenientes de las naves Galileo y Venus Express. La primera realizó un sobrevuelo de Venus de unos pocos días en febrero de 1990 en su camino hacia su objetivo final: Júpiter. La segunda es una misión completamente dedicada al estudio de Venus desde abril del 2006, adquiriendo datos sobre su atmósfera, superficie y entorno espacial. La nave Galileo fue el último ingenio espacial que estudió la atmósfera de Venus, hasta que la llegada de Venus Express puso fin a un abandono de décadas en la exploración científica de este mundo.\\

En este capítulo presentaré los datos con los que he trabajado en esta tesis, incluyendo no sólo las imágenes empleadas y sus características sino también los intrumentos con los que fueron tomadas y el software con el que se analizaron. Finalizaré el capítulo describiendo las implicaciones que tiene en Venus estudiar la atmósfera a diferentes longitudes de onda así como la técnica empleada en el estudio del movimiento de las estructuras nubosas.\\

\section{La cámara SSI de Galileo}\label{chapter-observs-GAL}\indent

La misión espacial Galileo al planeta Júpiter fue llevada a cabo por la agencia espacial estadounidense NASA. Fue lanzada el 18 de octubre de 1989 y constaba de un orbitador y una sonda. El orbitador realizaría numerosos sobrevuelos de Júpiter y sus lunas, mientras que la sonda estaba destinada a sumergirse en la atmósfera superior de Júpiter. Para alcanzar este destino la trayectoria de Galileo incluía un sobrevuelo cercano de Venus.\\

El orbitador Galileo contaba con un amplio conjunto de instrumentos científicos, entre ellos la cámara \textit{SSI} (Solid State Imaging)\index{Cámara!SSI}, consistente en un telescopio Cassegrain de 1.5 m de focal, acoplado a un detector CCD (charge-coupled device) de $800\times800$ píxeles (ver Figura 2.1). La cámara incluía hasta ocho filtros que permitían obtener imágenes en distintas longitudes de onda (ver Tabla \ref{tab:tabla-filtersGalileo}). Los filtros violeta, verde y rojo eran filtros de banda ancha, mientras que los de metano eran de banda estrecha. El obturador de la cámara permitía dos modos de exposición: el modo ``normal'' que incluía un número discreto de tiempos de exposición entre 4.167 ms y 0.8 s, y el modo ``extendido'' con tiempos de exposición desde 1.067 s hasta 51.2 s. \citet{Belton1992} y \citet{Klaasen1997} proporcionan una descripición más detallada de las especificaciones técnicas de la cámara SSI.\\

\begin{table}[h!]
  \caption{Filtros de la cámara SSI.}
	\label{tab:tabla-filtersGalileo}
	\centering
  \begin{spacing}{0.6}
		\begin{tabular}{*{3}{>{\scriptsize}c}}
			& & \\
			\hline\hline
			& & \\
			\textit{Filtro} & \textit{$\lambda_{eff}$ (nm)} & \textit{$\Delta\lambda$ (nm)} \\
			& & \\
			\hline
			& & \\
			Violeta      & 413 &  45  \\
			Verde        & 559 &  65  \\
			Transparente & 654 & 440  \\
			Rojo         & 665 &  60  \\
			Metano I     & 731 &  10  \\
			Continuo     & 757 &  19  \\
			Metano II    & 887 &  16  \\
			Infrarrojo   & 991 &  50  \\
			& & \\
			\hline
		\end{tabular}
    \begin{center}
\scriptsize{\textit{Nota:} $\lambda_{eff}$ es la longitud de onda efectiva y\\$\Delta\lambda$ es la anchura del filtro a mitad de altura.}
    \end{center}
  \end{spacing}
\end{table}

\begin{figure}[h!]
	\centering
		\includegraphics[width=0.7\textwidth]{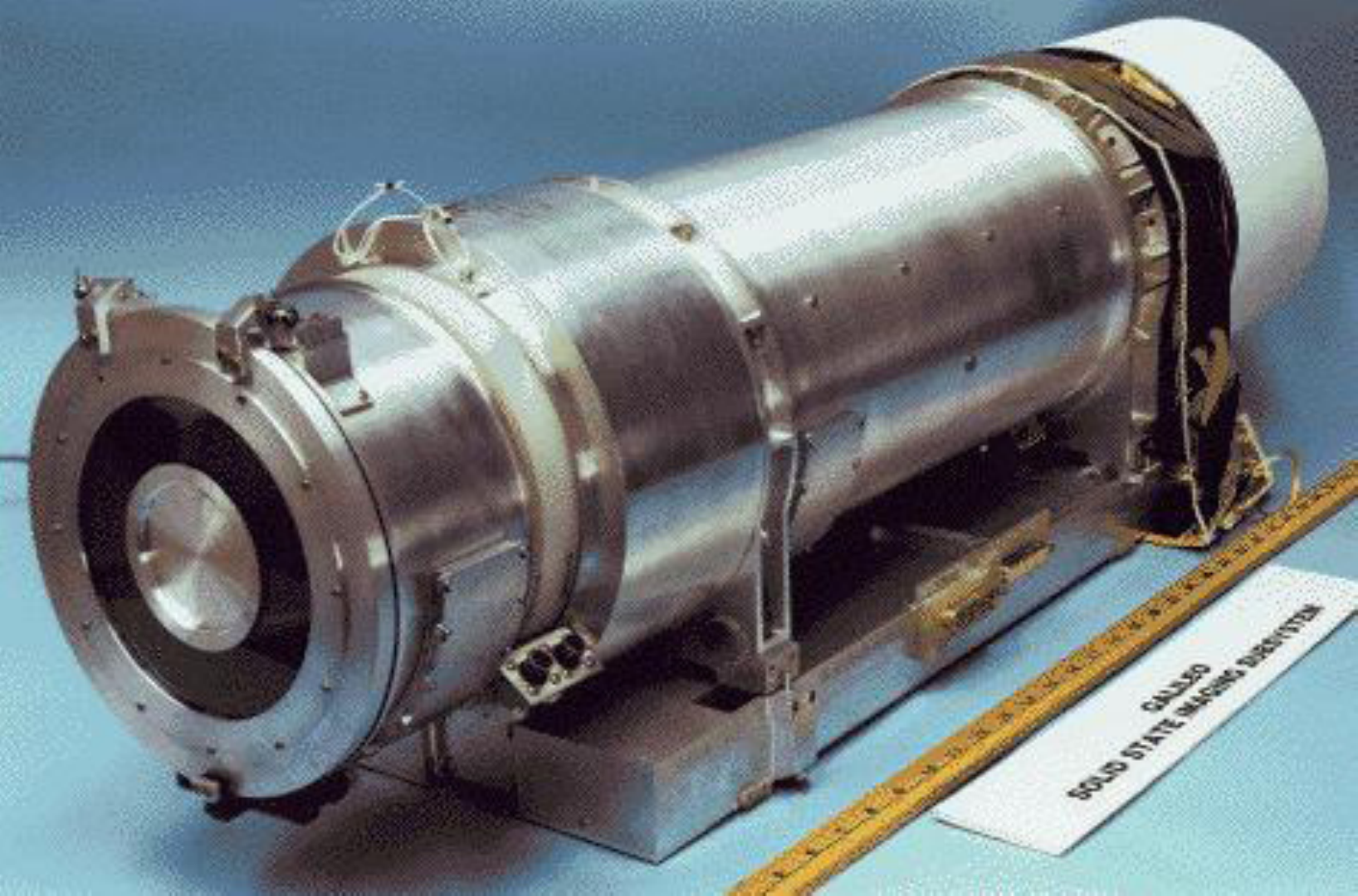}
	\label{fig:Galileo-SSI}
	\caption[Cámara SSI de la nave Galileo.]{\scriptsize{Cámara SSI de la nave Galileo. A la izquierda se muestra la apertura de entrada de la cámara con la cubierta protectora puesta. La pieza cilíndrica blanca a la derecha es el refrigerador para el detector. La electrónica de la cámara está ubicada en la caja rectangular situada bajo el telescopio.}}
\end{figure}

Las imágenes que obtuvo la cámara SSI no estaban exentas de defectos\index{Defectos!de cámara SSI}. La corrección de ``campo de respuesta plano'' o ``flat field''\index{Flat Field} se usa para mejorar la calidad de las imágenes digitales y para eliminar defectos causados por distorsiones ópticas y la variación pixel a pixel de la sensibilidad del detector. Consiste en tomar una imagen de una superficie uniformemente iluminada con el mismo sistema óptico con el que se ha realizado la imagen, a la misma distancia focal y con la misma apertura. De ese modo, puesto que partimos de una superficie uniforme, ese flat field representará fielmente la respuesta de todo el sistema óptico. Los variaciones pixel a pixel más destacables de la cámara SSI pueden visualizarse en la versión contrastada del ``flat field'' mostrado en la Figura 2.2A. Los anillos oscuros son sombras de las peque\~{n}as partículas de polvo ($\sim10~\mu m$ de diámetro) depositadas en la ventana del detector. El oscurecimiento en las esquinas de la imagen es causado por el ligero vi\~{n}eteo que provoca el diámetro de apertura del filtro. Las columnas verticales brillantes que se suceden cada 33 píxeles son debidas a que el área de los píxeles es mayor en estas columnas, consecuencia de la técnica usada en la fabricación del detector CCD \citep{Belton1992}. Por otro lado, en la Figura 2.2B podemos ver una imagen de Venus tomada por la cámara SSI durante el sobrevuelo de Galileo. En ella podemos ver defectos adicionales: píxeles, filas y columnas muertas (oscuras) y sobresaturadas (brillantes). Estos defectos son consecuencia del ruido de lectura debido a las ``trampas de carga'' en el CCD \citep{Belton1992} o a da\~{n}os en el detector debido a partículas de alta energía. Todos estos defectos son descritos de manera más amplia por \citet{Klaasen1997}.\\

\begin{figure}[h!]
	\centering
		\includegraphics[width=1.0\textwidth]{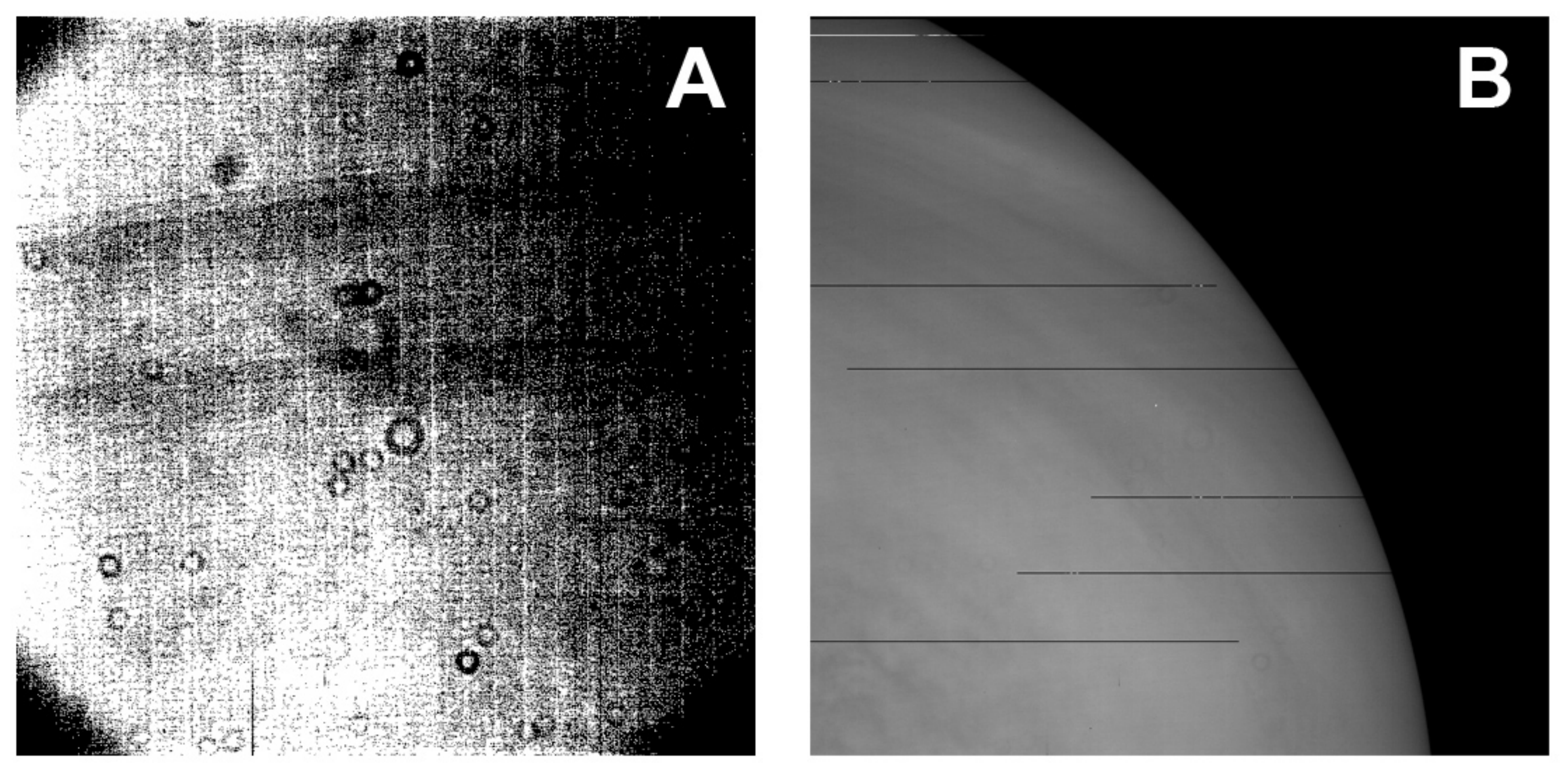}
	\label{fig:Galileo-SSI-defects}
	\caption[Defectos de la cámara SSI de Galileo.]{\scriptsize{La imagen \textbf{A} corresponde a un ``flat field'' de la cámara SSI tomado sin la cubierta óptica y cuyo contraste ha sido aumentado. En él se muestran los defectos más comunes de la cámara, destacando los anillos (sombras de las partículas de polvo depositadas), el vi\~{n}eteo en las esquinas de la imagen y el patrón repetitivo de columnas verticales brillantes. La imagen \textbf{B} muestra una imagen de Venus captada con la cámara SSI, con un vi\~{n}eteo inapreciable por tomarse con la cubierta óptica, y con los típicos píxeles y líneas defectuosas.}}
\end{figure}

En febrero de 1990 la nave Galileo sobrevoló el planeta Venus, momento que supuso la primera oportunidad de probar la instrumentación a bordo de la nave. Tras su máximo acercamiento al planeta el 10 de febrero, la nave comenzó a alejarse observándolo con un ángulo de fase de 45$^{\circ}$ y con una mejor visualización del hemisferio norte que del hemisferio sur. Entre el 10 y el 17 de febrero la cámara SSI tomó un total de 67 imágenes en el filtro de violeta ($418~nm$) y 10 en el filtro de $1~\mu m$ ($986~nm$, a partir de ahora NIR o infrarrojo cercano), imágenes que son de acceso público a través del Planetary Data System de NASA (http://pds.jpl.nasa.gov). La resolución espacial de estas imágenes varía con el momento en que fueron tomadas, teniendo para el máximo acercamiento al planeta ($\sim 29,900$ km) una re\-so\-lu\-ción de $\sim 246$ m/pixel, y para la distancia máxima ($\sim 3,774,000$ km) una resolución de $\sim 38.26$ km/pixel. Además, las imágenes de este sobrevuelo presentan ciertas peculiaridades debido a que fueron tomadas en un momento de la misión en que la cámara SSI tenía puesta la cubierta óptica. Dicha cubierta (cuya misión era preservar el sistema óptico en las mejores condiciones posibles hasta la llegada a Júpiter) no sólo afecta al proceso de calibración sino también a la respuesta del sistema óptico de la cámara. Además, la cubierta óptica reduce el diámetro de apertura del telescopio y el cono de luz entrante, con lo que: (1) el diámetro de los anillos de polvo también se reducen de forma proporcional, y (2) el vi\~{n}eteo de las esquinas prácticamente desaparece. No fue posible crear ``flat fields'' durante el pe\-rio\-do en que estuvo puesta la cubierta óptica debido a la ausencia de objetivos uniformemente iluminados que visualizar.\\

Por otro lado, también se produjo una caída de aproximadamente un 30\% en la sensibilidad de la cámara entre diciembre de 1989 y diciembre de 1990, debido posiblemente a la contaminación de la cubierta óptica por deposición de partículas\footnote{La nave quemó una cantidad importante de combustible ($\approx40$ kg) para correcciones de la trayectoria antes de llegar a Venus.}, resultando en imágenes de Venus ligeramente subespuestas y con un rango dinámico\index{Imágenes!rango dinámico de} de $6-7$ bits en vez de los 8 bits que teóricamente se debían alcanzar. Así, para el filtro violeta las estructuras nubosas alcanzan contrastes entre el 2\% para altas latitudes, y el 11\% para latitudes ecuatoriales. Sin embargo, en el filtro del NIR sólo se llega al 5\% de contraste \citep{Peralta2007b,Peralta2007a}.\\

\section{VIRTIS-M en Venus Express}\label{chapter-observs-VEX}\indent

Venus Express ha sido la primera misión de la Agencia Espacial Europea (ESA) al planeta Venus. Entre los objetivos científicos de esta misión se encuentra estudiar la atmósfera y alta atmósfera en detalle, la superficie del planera así como también las interacciones superficie-atmósfera \citep{Svedhem2007}. Si bien la misión comenzó en noviembre de 2005 e inicialmente estaba previsto que durase unos 500 días (aproximadamente dos días de Venus), en febrero de 2007 se decidió extenderla hasta principios de mayo de 2009. Actualmente ESA debate una posible segunda extensión de la misión.\\

Venus Express lleva a bordo siete instrumentos de los cuales sólo los datos de VIRTIS (Visible and Infrared Thermal Imaging Spectrometer) han sido utilizados para esta tesis\footnote{Los directores de esta tesis fueron seleccionados por ESA como co-investigador (Agustín Sánchez Lavega) e investigador asociado (Ricardo Hueso Alonso) en este ins\-tru\-men\-to, teniendo por tanto acceso directo a los datos.}. Tal como puede observarse en la Figura 2.3, VIRTIS\index{VIRTIS} es un instrumento dual con telescopios separados que proporcionan dos canales: (1) VIRTIS-M, un espectrómetro de imágenes capaz de trabajar en el visible ($0.3-1~\mu m$) y en el infrarrojo ($1-5~\mu m$), y (2) VIRTIS-H, un espectrómetro de alta resolución que cubre el rango espectral infrarrojo $2-5~\mu m$. La rendija de VIRTIS-M tiene un campo de visión de $0.25\times64~mrad$, lo que abarca un tercio del diámetro de Venus cuando la nave está en el apocentro de la órbita ($\sim60,000$ km del planeta). Sin embargo, el espejo secundario de VIRTIS-M tiene un total de 256 posiciones, por lo que tras un escaneado completo se cubre un campo de visión más amplio (para un completo resumen de las especificaciones técnicas de VIRTIS consultar la Tabla 1 de \citealt*{Drossart2007}).\\

\begin{figure}[h!]
	\centering
		\includegraphics[width=0.7\textwidth]{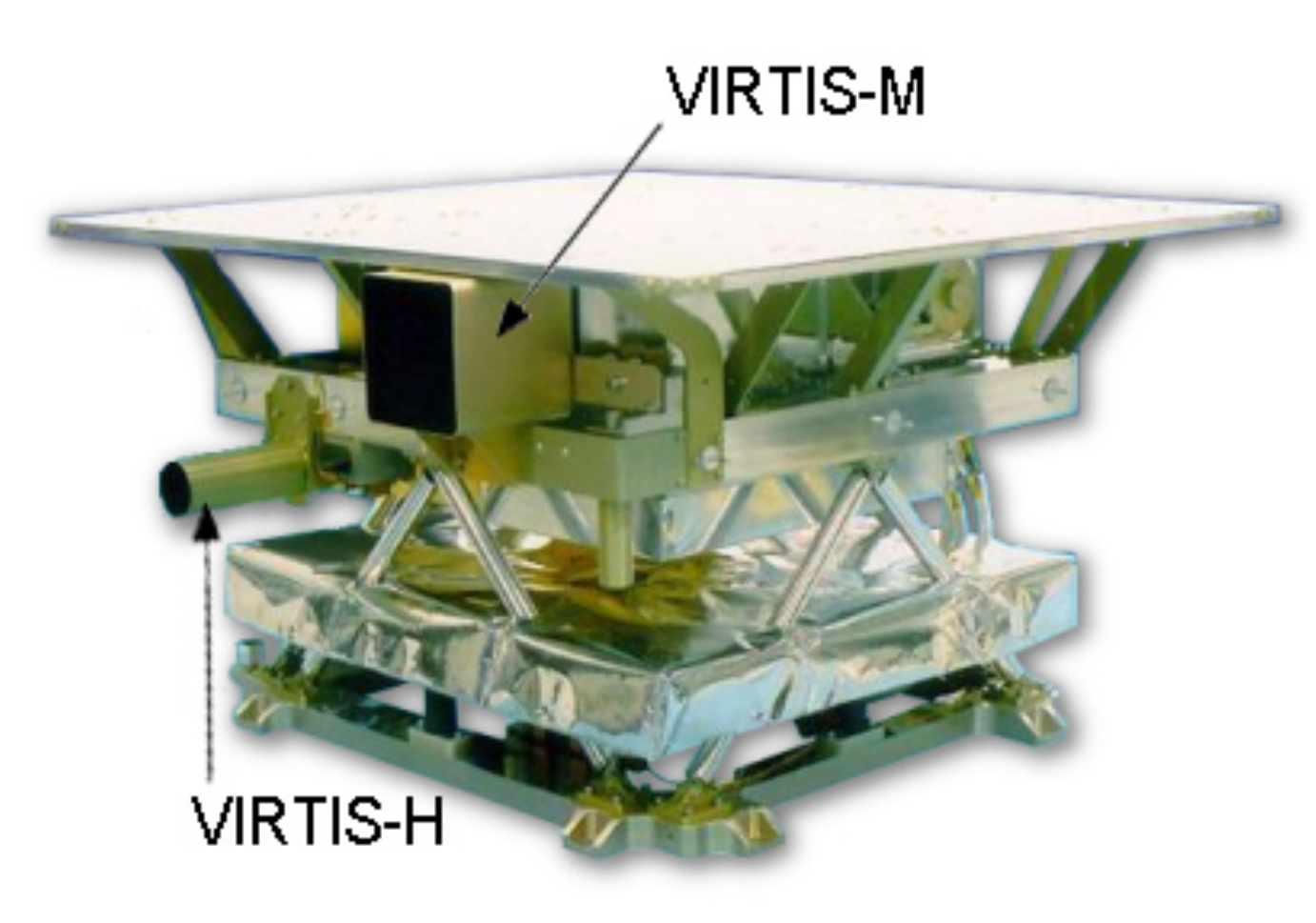}
	\label{fig:Vex-VIRTIS}
	\caption[Instrumento VIRTIS de Venus Express.]{\scriptsize{Instrumento VIRTIS a bordo de la nave Venus Express. VIRTIS se compone de dos telescopios independientes: VIRTIS-M, un espectrómetro de imágenes capaz de trabajar en el visible ($0.3-1~\mu m$) y en el infrarrojo ($1-5~\mu m$), y VIRTIS-H, un espectrómetro de alta resolución que cubre el rango espectral infrarrojo $2-5~\mu m$. Ambos telescopios están situados dentro de una caja enfríada por un radiador (placa blanca superior). El módulo electrónico está situado justamente debajo de la caja fría, constituyendo la base inferior del instrumento que se conecta a la nave.}}
\end{figure}

El 11 de abril de 2006 la nave Venus Express llegó a Venus realizando su inserción orbital. Desde entonces describe órbitas altamente elípticas cada 24 horas (con el apocentro a 60,000 km sobre el polo sur de Venus, y el pericentro a 250 km sobre los 80$^{\circ}$N). Debido a la excentricidad de la órbita el instrumento VIRTIS sólo es capaz de tomar imágenes del hemisferio sur de Venus y con resoluciones espaciales que varían entre los 10 km/pixel hasta los 30 km/pixel en latitudes subpolares y 100 km/pixel en latitudes cercanas al ecuador. Inicialmente se definieron 11 casos científicos de observaciones de Venus con los diferentes instrumentos. De interés particular para la caracterización de la atmósfera con VIRTIS son los tres primeros casos \citep{Titov2006}:
\begin{itemize}
  \item Caso 1: Observaciones desde el pericentro de la órbita. En este caso la nave está a sólo 350 km sobre la cima de las nubes de Venus, con lo que el campo de visión es muy peque\~{n}o y la obtención de imágenes se complica al ser mayor el tiempo que se tarda en formar una imagen que el tiempo que se mantiene el objetivo a captar en el campo de visión. Sin embargo en este caso pueden obtenerse espectros de alta calidad y excelente relación se\~{n}al/ruido.
  \item Caso 2: Fuera de la región del pericentro la nave está mucho más alejada de las nubes (más de $\sim10,000$ km) por lo que se pueden obtener imágenes de alta resolución espacial. En este modo VIRTIS trabaja conjuntamente con la cámara VMC realizando tomas de imágenes con objeto de abarcar una región de interés en Venus (ver Figura 2.4A).
  \item Caso 3: Observaciones desde el apocentro de la órbita. En este caso el campo de visión de VIRTIS no es lo suficientemente amplio como para cubrir todo el disco del planeta, por lo que se toman varias imágenes con objeto de formar un ``mosaico'' del disco del planeta  (ver Figura 2.4B). Estas operaciones pueden complementarse con la cámara VMC, que sí posee una vista completa del disco. El elevado volumen de datos que implica este caso ha reducido necesariamente el número global de mosaicos finalmente adquiridos.
\end{itemize}

\begin{figure}[h!]
	\centering
		\includegraphics[width=1.0\textwidth]{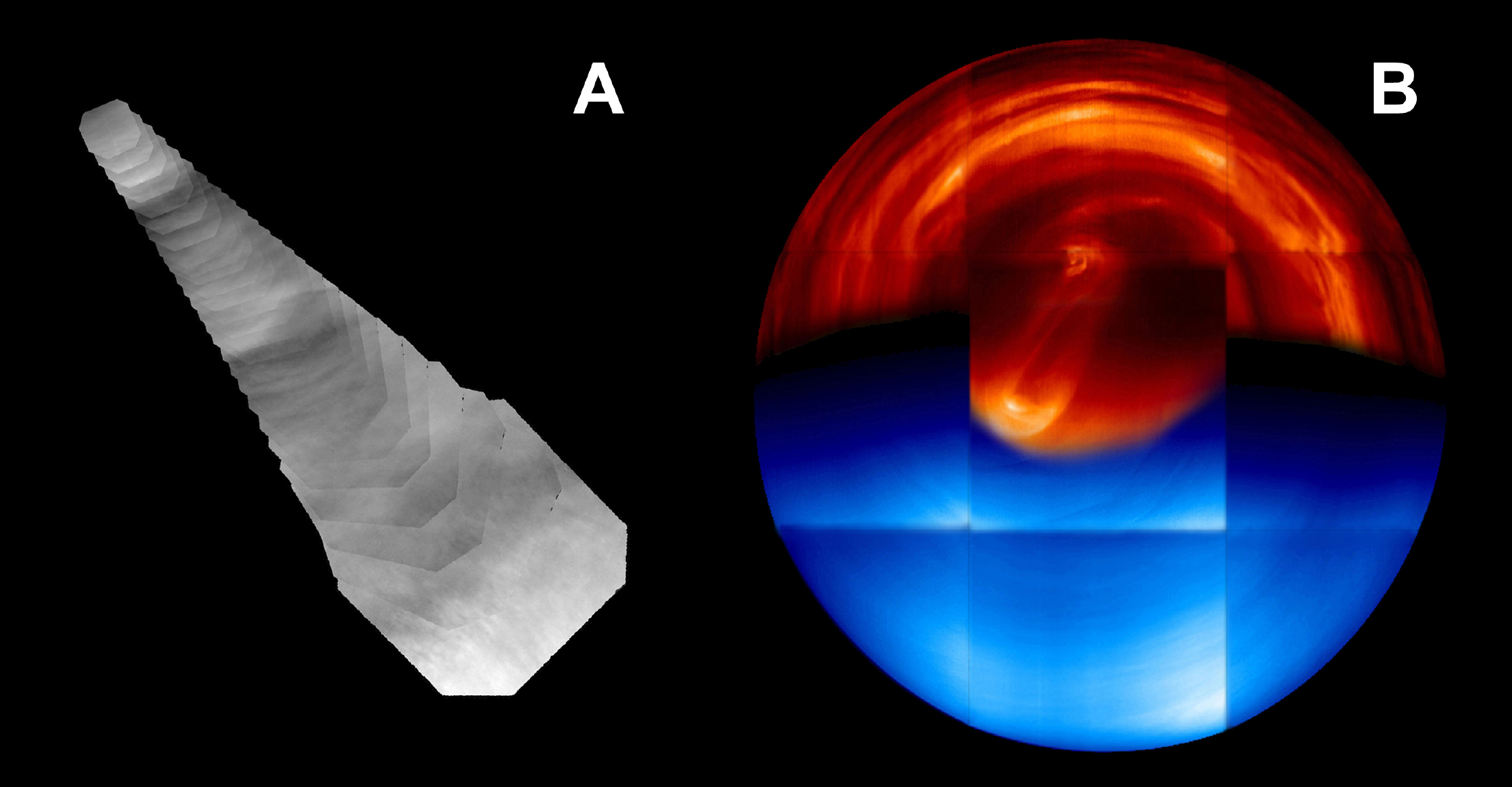}
	\label{fig:VEX-ObservationCases}
	\caption[Casos de observación con VIRTIS-M.]{\scriptsize{\textbf{(A)} Imágenes en UV de Venus tomadas por la cámara VMC en las que se ilustra el Caso 2 de las observaciones (imágenes de alta resolución enfocando una región de interés del hemisferio norte). \textbf{(B)} Imágenes en UV del lado diurno (color azul) e IR del lado nocturno (color rojo) de Venus tomadas por VIRTIS-M e ilustrando el Caso 3 de las observaciones (imágenes orientadas a formar un mosaido para cubrir todo el disco del planeta).}}
\end{figure}

VIRTIS-M proporciona los datos en forma de ``cubos de imágenes'', cada imagen del cubo correspondiente a una determinada longitud de onda (ver Figura 2.5), y cada cubo conteniendo imágenes dentro de uno de los dos rangos espectrales que cubre VIRTIS-M: visible ($0.25-1~\mu m$) e infrarrojo ($1-5~\mu m$). Los cubos de imágenes de VIRTIS-M están disponibles en la página web del archivo de ciencias planetarias de la Agencia Espacial Europea (ESA). Ambos tipos de cubos (visible e infarrojo) suelen tomarse en cada observación de manera sucesiva, de tal forma que por cada cubo del visible siempre hay un cubo de infarrojo. Los cubos de VIRTIS-M, además, se presentan de manera muy diversa. Si bien en la mayoría de los casos poseen una sóla imagen para cada longitud de onda, en ocasiones el espejo secundario realiza más de una pasada y encontramos varias imágenes montadas unas sobre otras. El tama\~{n}o de las imágenes también es variable, aunque en la mayoría de los casos presentan únicamente dos tipos de tama\~{n}o: peque\~{n}o ($64\times64$ píxeles) y grande ($256\times256$ píxeles). Generalmente se almacenan imágenes correspondientes a un total de 432 longitudes de onda. Los tiempos de exposición suelen ajustarse a las estrategias de observación, en la mayoría de los casos favoreciendo las observaciones del lado nocturno del planeta (cubos del infrarrojo con buena calidad) en detrimento del lado diurno (cubos del visible de baja calidad), si bien en algunas ocasiones se escogen tiempos de exposición que favorecen las observaciones del visible e incluso valores intermedios para favorecer ambas.\\

\begin{figure}[h!]
	\centering
		\includegraphics[width=1.0\textwidth]{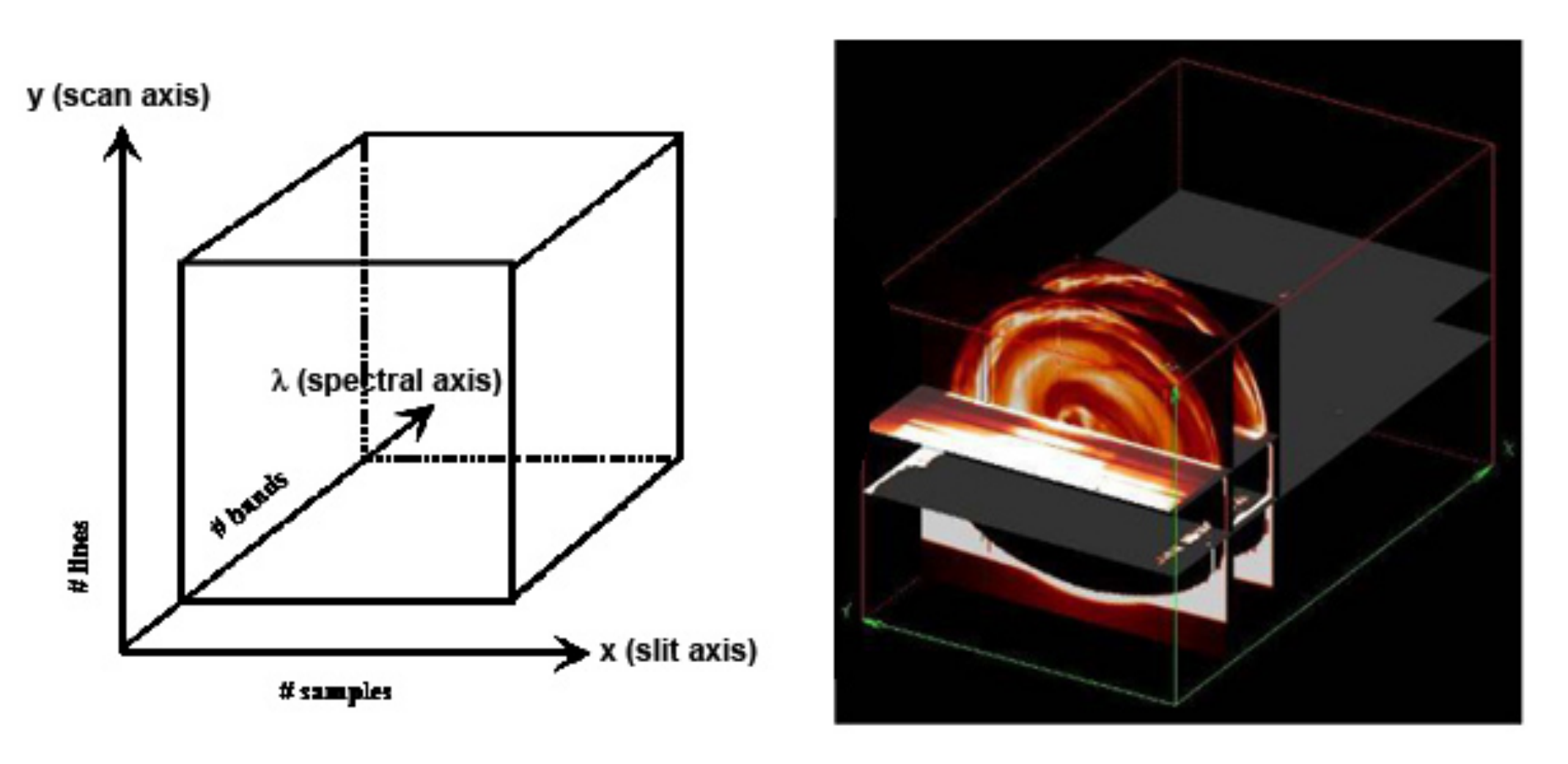}
	\label{fig:Vex-VIRTIS-Cube}
	\caption[Esquema 3D de un cubo de VIRTIS-M.]{\scriptsize{Esquema 3D de un cubo de VIRTIS-M. Los ejes del cubo se denominan \textit{banda} (banda espectral, corresponde a una longitud de onda), \textit{muestra} (dirección espacial a lo largo de la rendija) y \textit{línea} (datos adquiridos en pasos de tiempo sucesivos). El orden de adquisición durante una observación estándar de VIRTIS-M es siempre banda-muestra-línea. Para cada valor de \textit{banda} VIRTIS-M conforma una imagen, mientras que si fijamos \textit{muestra} y \textit{línea} tenemos un espectro.}}
\end{figure}

Entre los defectos\index{Defectos!de VIRTIS-M} que suelen mostrar las imágenes de los cubos de VIRTIS-M (ver Figura 2.6) cabe destacar los siguientes:
\begin{itemize}
  \item Durante el proceso de adquisición de un cubo por lo general se cierra el obturador en varias ocasiones con objeto de enfriar el detector, pues éste se calienta y pierde sensibilidad. Como consecuencia se crean en las imágenes un número determinado de líneas horizontales o ``dark frames'' sin información útil.
  \item Las imágenes de los cubos también presentan el ``odd-even defect'' consistente en un efecto de bandeado en las líneas verticales de las imágenes que provoca variaciones bruscas en el valor medio del brillo de líneas verticales consecutivas. Este defecto tiene su origen en el proceso de lectura del CCD.
  \item Fallos en los sensores electrónicos (principalmente en al amplificador de la se\~{n}al) así como el bombardeo de partículas de alta energía ge\-ne\-ran píxeles muertos y sobresaturados, además de líneas espúreas tanto vertical como horizontalmente.
\end{itemize}

\begin{figure}[h!]
	\centering
		\includegraphics[width=0.7\textwidth]{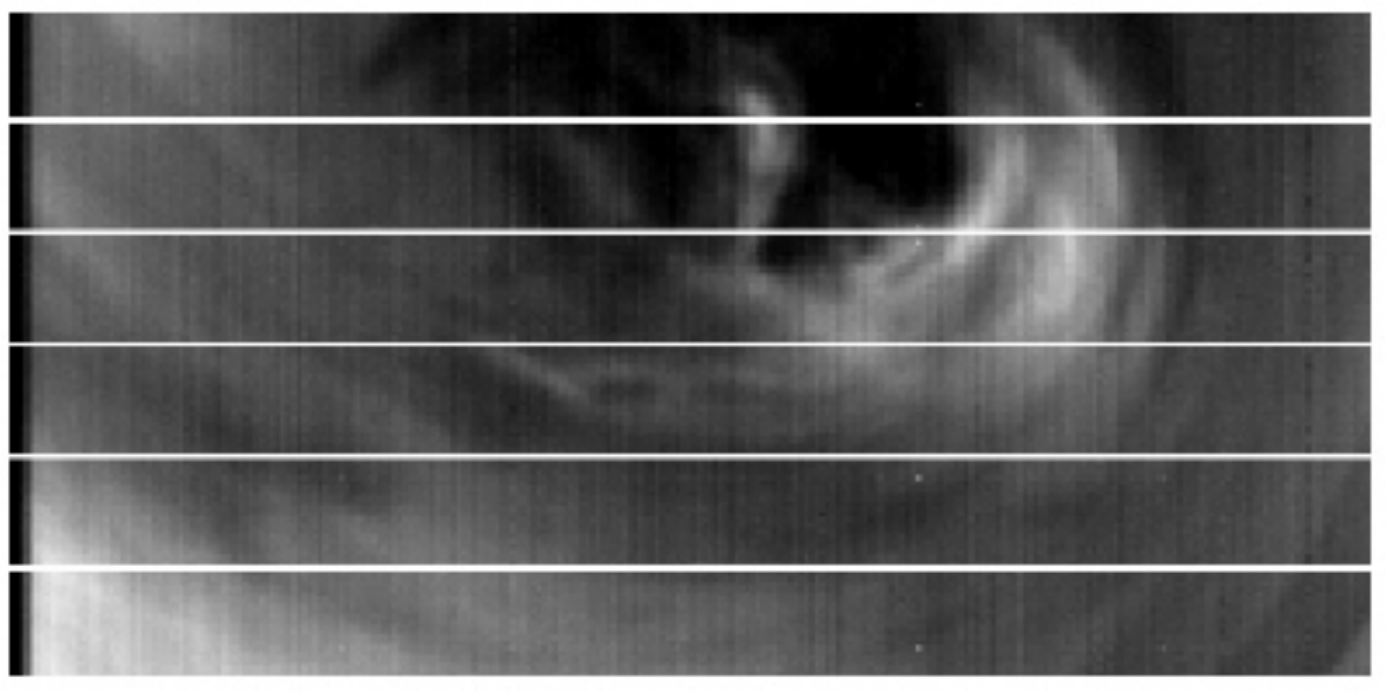}
	\label{fig:Vex-VIRTIS-Defects}
	\caption[Defectos de las imágenes de VIRTIS-M.]{\scriptsize{Defectos característicos de una imagen de VIRTIS-M. En este ejemplo podemos ver claramente las líneas horizontales brillantes o ``dark frames'', el bandeado de las líneas verticales u ``odd-even defect'', y algunos píxeles sobresaturados.}}
\end{figure}

\section{El software de trabajo: PLIA}\label{chapter-observs-PLIA}\indent

Los datos planetarios utilizados en esta tesis se redujeron y analizaron utilizando el software PLIA (Planetary Laboratory for Image Analysis; \citealt*{Peralta2005,Hueso2008b}). PLIA es un conjunto integrado de programas escritos en el lenguaje de programación IDL que posee una interfaz gráfica que facilita su uso (ver Figura 2.7) y está dise\~{n}ado para el estudio de imágenes planetarias provenientes de diferentes misiones espaciales. El software se utilizó para navegar las imágenes (asignar coordenadas de longitud y latitud a cada píxel), corregir del oscurecimiento hacia el limbo, obtener cortes fotométricos (útiles en el estudio de la estructura de las nubes y la turbulencia), y para realizar proyecciones geométricas tanto cilíndricas como polares. Durante esta tesis participé activamente en el desarrollo de algunos de estas herramientas y su implementación en PLIA.\\

\begin{figure}[h!]
	\centering
		\includegraphics[width=1.0\textwidth]{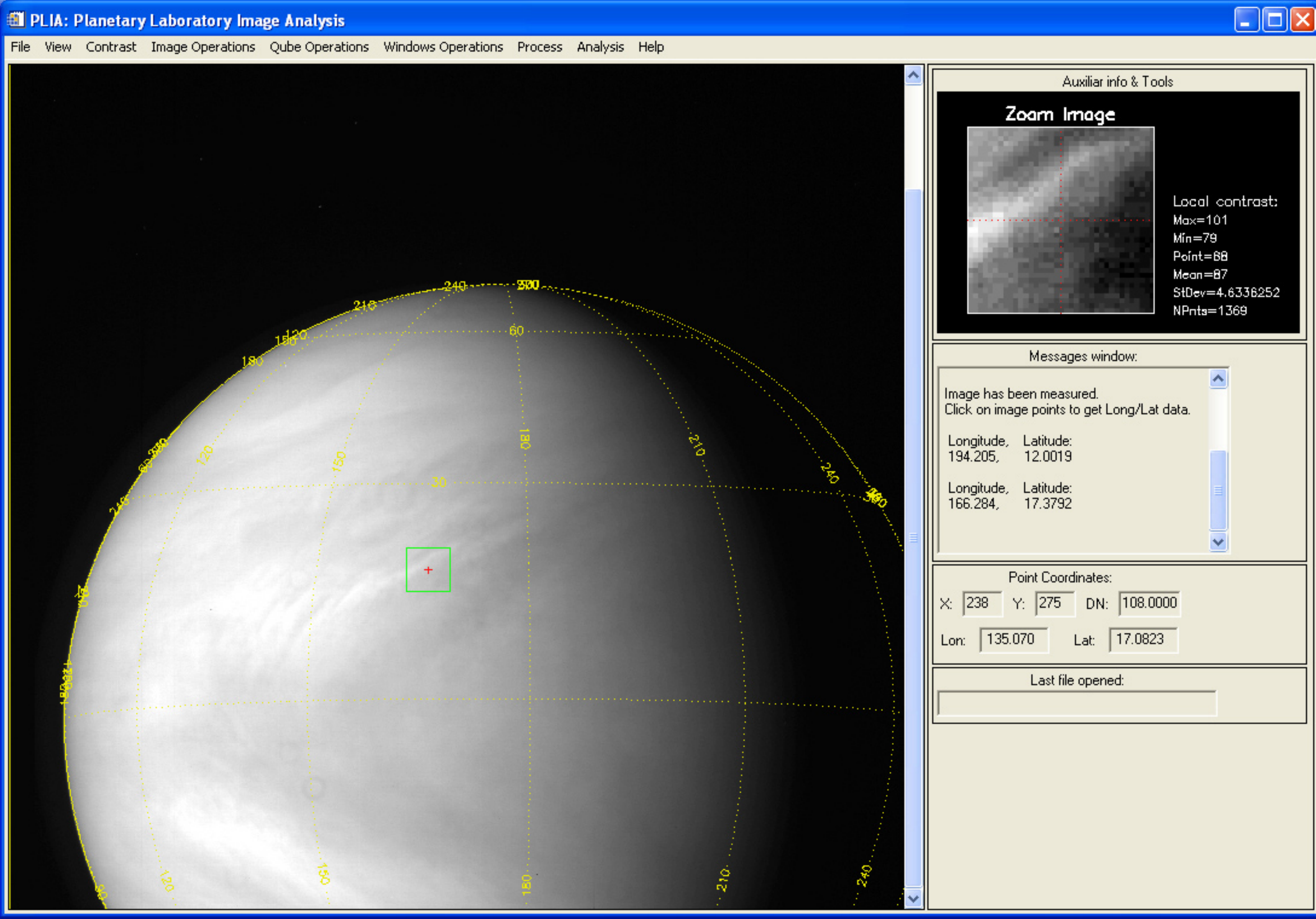}
	\label{fig:PLIA}
	\caption[Interfaz gráfica de PLIA.]{\scriptsize{Interfaz gráfica de PLIA. En la parte superior de la ventana están horizontalmente distribuidos los botones que dan acceso a las funciones del programa. La ventana principal nos muestra la imagen con la que estemos trabajando, en este caso una imagen de Venus tomada por la cámara SSI de Galileo. La columna a la derecha nos aporta información adicional sobre la imagen: visión ampliada de una región de la imagen, coordenadas geográficas, información fotométrica y otros datos.}}
\end{figure}

PLIA es capaz de trabajar con imágenes obtenidas por las cámaras de la nave Mariner 10, la OCPP de Pioneer Venus, la SSI de Galileo, la ISS de Cassini y el instrumento VIRTIS-M de Venus Express. También puede leer datos astronómicos en formato FITS y posee algunas utilidades para trabajar con las cámaras WFPC2 y ACS del telescopio espacial Hubble. Los datos de interés para nuestra tesis (los obtenidos por las misiones Galileo y Venus Express) están disponibles en el formato estándar PDS \citep{McMahon1996} y pueden ser leídas en IDL usando rutinas de carácter público que están incorporadas en PLIA. En el caso de las imágenes de Venus tomadas por la cámara SSI, cada imagen consta de un par de ficheros, un fichero *.IMG que contiene la imagen en sí, y un fichero *.LBL que proporciona la cabecera de la imagen (datos temporales, de navegación, etc...) en formato PDS. Por otro lado, los cubos del instrumento VIRTIS-M de Venus Express están en formato PDS y contienen dos direcciones espaciales y una espectral. En este caso PLIA es capaz de extraer las imágenes de las longitudes de onda que se deseen, realizar operaciones con ellas y compararlas. Los cubos suelen constar de tres ficheros distintos: un fichero *.QUB con las imágenes sin calibrar, un fichero *.CAL con las imágenes calibradas y un fichero *.GEO con los datos de navegación de los cubos. Los cubos del visible no están calibrados radiométricamente y no existen las versiones *.CAL correspondientes. Todos los ficheros incluyen una cabecera con la información pertinente en formato PDS.\\

\textit{Corrección de defectos}: PLIA posee rutinas para corregir defectos\index{Imágenes!corrección de defectos} tanto generales como específicos de cada instrumento (ver Figura 2.8). Durante el sobrevuelo de Venus la nave Galileo tomó imágenes con la cubierta óptica de la cámara SSI puesta, por lo que las imágenes que usamos carecían de vi\~{n}eteo. Primeramente corregimos las líneas verticales brillantes de posición fija (ver Figura 2.2A). A continuación eliminamos las filas y columnas defectuosas. Estos defectos son identificados de forma automática escaneando las imágenes con una caja de $1\times5$ pixeles y buscando los valores locales más elevados de la desviación estándar del brillo. Las líneas y píxeles defectuosos se corrigen entonces promediándolos con vecinos próximos.\\

La corrección de los cubos de VIRTIS-M se llevó a cabo de la siguiente manera:
\begin{itemize}
  \item Las posiciones de los ``dark frames'' están especificadas en la cabecera de los cubos, por lo que se usó esta información para eliminarlos.
  \item El efecto de ``bandeado'' fue corregido con filtros direccionales que estimaban el brillo promedio de cada línea vertical, sua\-vi\-zan\-do la variancia de la línea de brillo entre vecinos próximos.
  \item Las líneas espúreas y los píxeles muertos y sobresaturados fueron identificados y corregidos de la misma manera que en las imágenes de SSI de Galileo.
\end{itemize}

\begin{figure}[h!]
	\centering
		\includegraphics[width=1.0\textwidth]{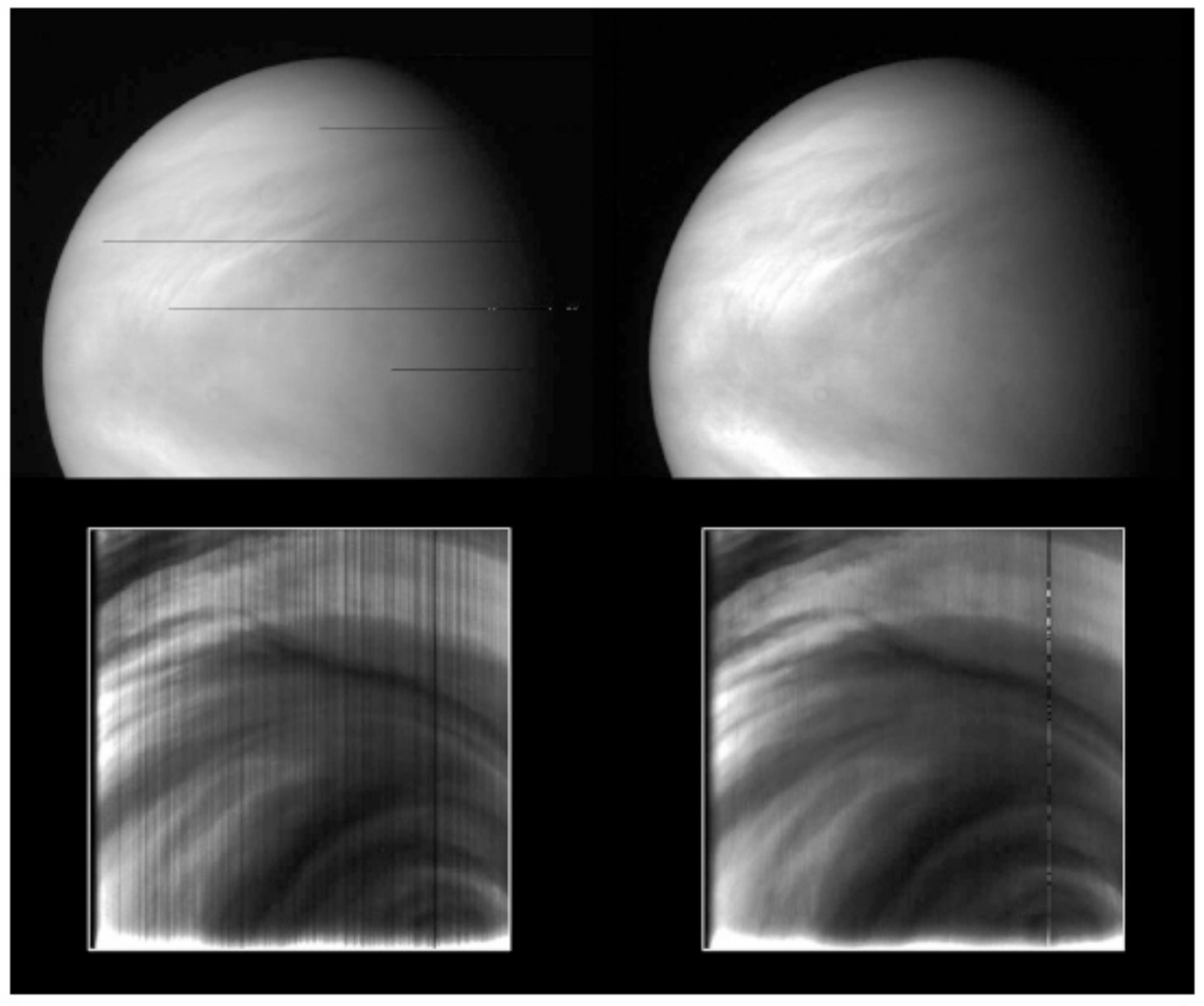}
	\label{fig:Venus-ImagesCorrected}
	\caption[Corrección de defectos en imágenes SSI y VIRTIS-M.]{\scriptsize{Corrección de defectos en imágenes SSI y VIRTIS-M de Venus. Arriba tenemos original (izquierda) y versión corregida y procesada (derecha) de una imagen violeta tomada por la cámara SSI de Galileo. Nótese que no se pudieron corregir los anillos oscuros. Abajo tenemos imagen original (izquierda) y corregida (derecha) de un cubo infrarrojo tomado por VIRTIS-M. Vemos que la corrección del bandeado vertical no llega a ser del todo perfecta pero es satisfactoria en la mayor parte de la imagen.}}
\end{figure}

\textit{Navegación de las imágenes}: Una vez eliminados los defectos nave\-gamos las imágenes. La navegación\index{Imágenes!navegación} de una imagen planetaria es un proceso ma\-te\-má\-ti\-co por el cual a cada píxel de una imagen se le asocia una coordenada de latitud y longitud. Aunque las imágenes de algunas misiones nos proporcionan directamente la navegación de cada píxel así como información geométrica importante (así sucede con VIRTIS-M de Venus Express), no es lo que ocurre con la mayoría de las misiones. PLIA puede navegar la imagen si se dispone de cierta información sobre la geometría de obsrvación y características del instrumento: longitud y latitud del punto subnave\index{Punto!subnave}\footnote{El punto subnave se corresponde con perpendicular del planeta que intersecciona con la posición donde está la nave.}, píxel que se corresponde con el punto subnave, distancia de la nave al centro del planeta y ángulo azimutal norte\index{ángulo azimutal norte} (el que forma la vertical de la imagen y el eje de rotación del planeta) \citep{Hueso2008b}. En el caso de las imágenes de la cámara SSI de Galileo, estos datos están contenidos en la cabecera PDS de cada imagen contenida en los ficheros *.LBL. En el caso de los cubos de VIRTIS la navegación está contenida en los ficheros *.GEO que PLIA lee automáticamente al acceder al cubo de datos.\\

\textit{Corrección del oscurecimiento hacia el limbo}: Se define la reflectividad\index{Reflectividad} como la fracción de radiación solar incidente que refleja la atmósfera. Ésta varía de un punto a otro de la atmósfera de un planeta (ya que también varían el ángulo de incidencia solar $i$ y el ángulo emergente $e$) dando lugar al efecto conocido como oscurecimiento hacia el limbo\index{Oscurecimiento hacia el limbo}, característico de las imágenes de la atmósfera en el lado diurno del planeta. PLIA posee una rutina que puede corregir este efecto usando para ello la corrección de primer orden de Minnaert\index{Minnaert!corrección de} \citep{Minnaert1941}. El factor de corrección para la reflectividad se define como:
\begin{equation}
	\frac{(I/F)}{(I/F)_{0}}=(\mu_{0})^{k}\cdot(\mu)^{k-1},
	\label{Minnaert-Law}
\end{equation}
donde $(I/F)$ es la reflectividad observada, $(I/F)_{0}$ es la reflectividad normalizada para un ángulo de incidencia $i=0$ y ángulo de emergencia $e=0$. Ambos ángulos se han expresado aquí con los términos $\mu=\cos e$ y $\mu_{0}=\cos i$. El parámetro $k$ es el coeficiente de oscurecimiento hacia el limbo que depende de las propiedades ópticas de la atmósfera (que pueden variar con la latitud y dependiendo de la longitud de onda de la imagen). En el caso $k=1$ tenemos la corrección de Lambert\index{Lambert!corrección de}, equivalente a suponer que la cantidad de radiación que se refleja no depende del ángulo de reflexión. Veremos que la corrección del efecto de oscurecimiento hacia el limbo resulta importante a la hora de extraer cortes fotométricos de la atmósfera planetaria (ver capítulo \ref{chapter-turbulence}) y medir el movimiento de detalles nubosos cerca próximos a la región del limbo (ver capítulo \ref{chapter-winds}). PLIA puede calcular el valor de los factores $\mu$ y $\mu_{0}$ en cada píxel de la imagen siempre y cuando la imagen proporcione el ángulo de fase y la longitud y latitud subsolar (tanto las imágenes de Galileo como las de Venus Express disponen de estos datos).\\

\textit{Procesado de imágenes}: Con objeto de una más fácil indentificación de los detalles nubosos de la atmósfera, PLIA también dispone de herramientas de procesado\index{Imágenes!procesado} de la imagen que ayudan a mejorar la visualización. Entre ellas cabe destacar la posibilidad de especificar el umbral de histograma o ``thres\-hol\-ding'', y una herramienta de autoescalado capaz de sugerir en la mayoría de las imágenes cuáles son los niveles más adecuados para una visualización óptima. También posee un amplio conjunto de herramientas que modifican el brillo de los píxeles para mejorar tanto la visualización global como la de detalles específicos \citep{Gonzalez1992}. En este grupo merece la pena mencionar la máscara de enfoque, filtros Butter\-worth, ecua\-li\-za\-ción de histogramas, realzado adaptativo del contraste (local histogram equalization) y operadores de primera y segunda derivada (Sobel y Roberts) para extraer bordes de la imagen \citep{Pajares2001}. Todas estas herramientas pueden combinarse en el orden que queramos hasta conseguir el resultado deseado.\\

\textit{Proyecciones geométricas}: PLIA es capaz de llevar a cabo proyecciones geométricas\index{Imágenes!proyección geométrica} de las imágenes planetarias, tanto de forma cilíndrica (coordenadas cartesianas) como polar (coordenadas polares). Las proyecciones cilíndricas han mostrado ser de mucha utilidad para la medición de vientos en latitudes bajas (ver capítulo \ref{chapter-winds}) y para la extracción de cortes fotométricos con valores equiespaciados (ver capítulo \ref{chapter-turbulence}), mientras que las proyecciones polares han resultado cruciales para una buena visualización e identificación de detalles nubosos a la hora de medir vientos en latitudes altas con las imágenes de Galileo (ver capítulo \ref{chapter-winds}). Una vez que el usuario especifica los parámetros de la proyección (resolución espacial y cotas mínima y máxima de latitud y longitud) PLIA trata las imágenes como un conjunto de datos distribuidos espacialmente de forma irregular (triangulación de Delaunay; \citealp{Akima1978}) y los proyecta en en el tipo de coordenadas deseado.\\

\section{Niveles de altura de las nubes}

\subsection{Imágenes en diferentes longitudes de onda}\label{chapter-observs-WavelenImag}\indent

Tanto en las imágenes obtenidas por la cámara SSI en Galileo como en las observaciones realizadas por el instrumento VIRTIS en Venus Express en cada longitud de onda se aprecian estructuras nubosas de diferente morfología y propiedades (ver Figura 2.9). Así, las observaciones de Galileo de Venus efectuadas en los filtros de $418~nm$ y $986~nm$ no sólo muestran morfologías nubosas diferentes sino también movimientos distintos (más rápidos en las imágenes violetas) que fueron interpretadas como cizalla vertical del viento en la cima de las nubes \citep{Belton1991}. Es de destacar también las fuertes diferencias existentes entre las imágenes en UV y NIR, obtenidas en el lado diurno del planeta en luz solar reflejada por las nubes, y las imágenes en IR observadas en el lado nocturno del planeta. Esta última corresponde a radiación térmica emitida por la cálida atmósfera inferior y retenida parcialmente por las nubes profundas que aparecen como fuentes de opacidad en las imágenes IR. Diferentes estudios de transporte radiativo permiten estimar la altura de las nubes sobre la superficie que se observan en las imágenes en cada longitud de onda (ver Figura 2.10).\\

\begin{figure}[h!]
	\centering
		\includegraphics[width=1.0\textwidth]{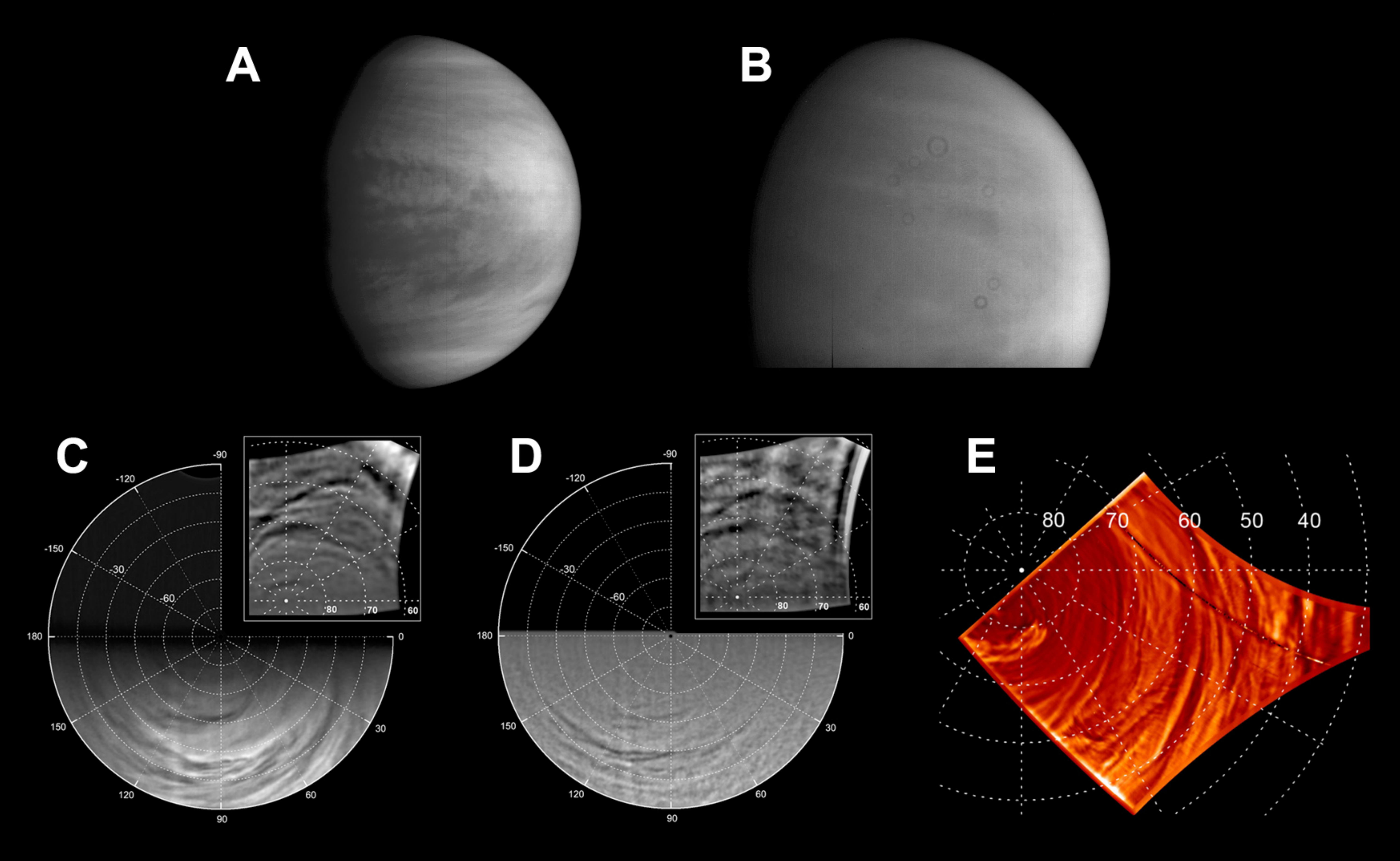}
	\label{fig:VenusAtmos-diff-filters}
	\caption[Venus observado con diferentes filtros.]{\scriptsize{Arriba: imágenes de Venus tomadas por la cámara SSI de Galileo de la cima de las nubes a $418~nm$ (A) y de la base de la cima de las nubes a $986~nm$ (B). Abajo: proyecciones polares de imágenes captadas por VIRTIS-M de la cima de las nubes a $380~nm$ (C), de la base de la cima de las nubes a $980~nm$ (D) y de las nubes inferiores a $1.74~\mu m$ (E). Las imágenes A-D fueron tomadas a partir de la luz reflejada (lado diurno) mientras que la imagen E se captó a partir de la emisión térmica emitida (lado nocturno). Claramente se observa que las nubes exhiben distinta morfología, correspondiente a diferentes niveles verticales de la atmósfera de Venus.}}
\end{figure}

\begin{figure}[h!]
	\centering
		\includegraphics[width=0.8\textwidth]{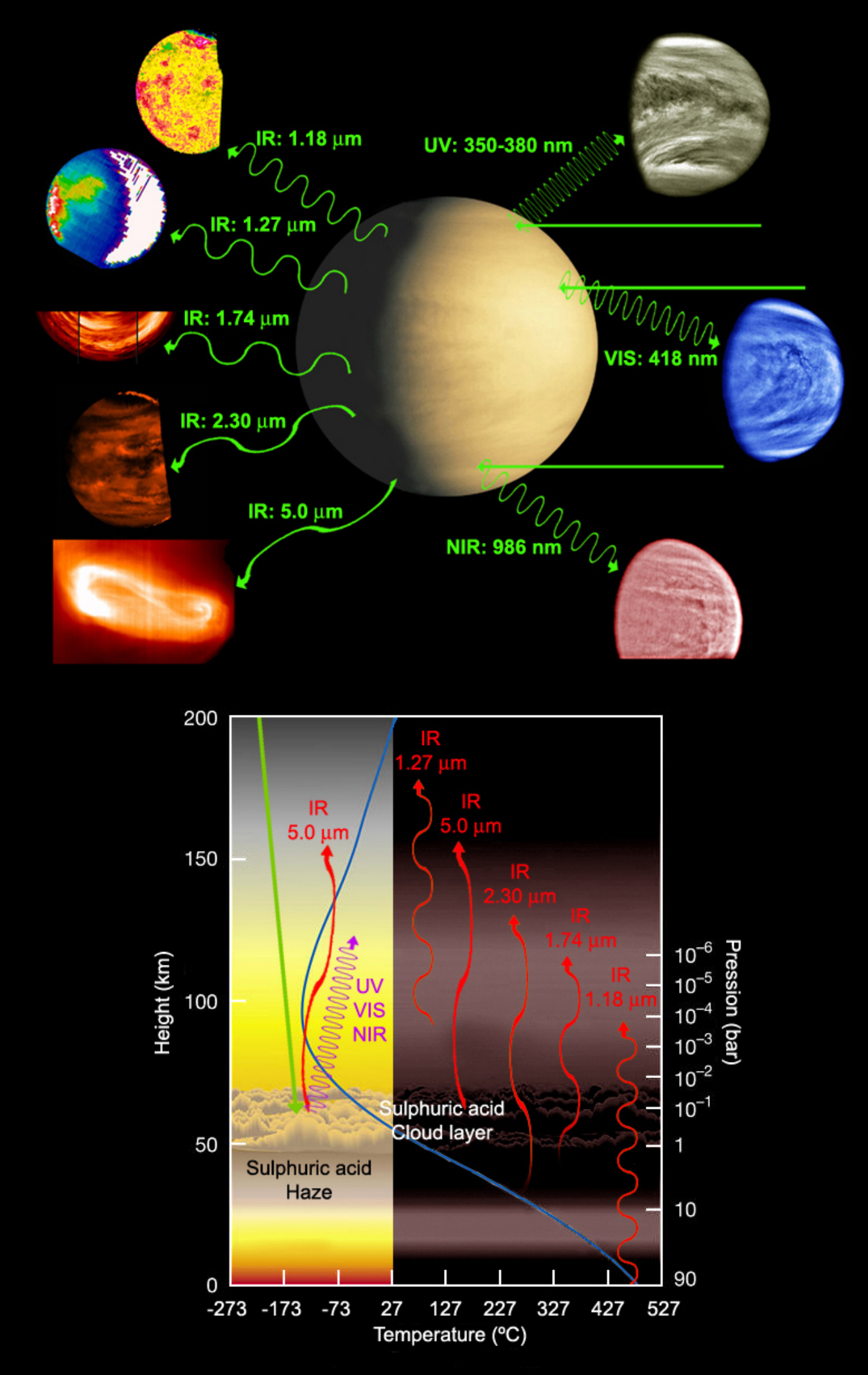}
	\label{fig:Venus-AtmosphereStructure}
	\caption[Correspondencia entre filtros y altura atmosférica.]{\scriptsize{Rango espectral de mayor relevancia en imágenes y su equivalencia en niveles de altura dentro de la atmósfera: $350-380~nm$ (UV), $980~nm$ (NIR) y $5.0~\mu m$ (IR) se corresponden en el lado diurno con luz solar reflejada en las nubes superiores, mientras que a $1.74~\mu m$, $2.30~\mu m$ y $5.0~\mu m$ tenemos emisión térmica opacitada por las nubes, a $1.27~\mu m$ contribuye esencialmente la luminiscencia del oxígeno molecular, y en $1.18~\mu m$ tenemos mayoritariamente emisión térmica de la superficie modulada por los accidentes geográficos.}}
\end{figure}

En el \textit{lado diurno} las imágenes violetas de $418~nm$ (Galileo) y ultravioletas de $380~nm$ (Venus Express) permiten visualizar estructuras nubosas de la cima de las nubes, nivel que se corresponde con un rango de alturas de $62-70$ km (a partir de ahora referido como el nivel a 66 km). En el caso del infrarrojo cercano de $980~nm$ (Galileo y Venus Express) los fotones son más penetrantes y llegan hasta la base de la cima de las nubes, localizada en un intervalo de altitud de $58-64$ km (a partir de ahora 61 km) \citep{Belton1991,Peralta2007b,Sanchez-Lavega2008}. En el lado nocturno las imágenes infrarrojas de 1.74 y $2.3~\mu m$ miden la radiancia que proviene de altitudes por debajo de las nubes principales y es atenuada al atravesar la capa de las nubes inferiores a $44-48$ km \citep{Carlson1991,Crisp1991}. Debido a que desconocemos el modo en que varía la estructura vertical de las nubes, existe una incertidumbre asociada a los valores de altitud de las estructuras nubosas observadas en cada longitud de onda. Sin embargo, los valores aquí presentados en base a cálculos detallados sobre las observaciones de VIRTIS-M (ver Auxiliary Material en \citealp{Sanchez-Lavega2008}) son consistentes con estimaciones publicadas previamente \citep{Esposito1997}.\\

\chapter{Medidas de Vientos en Venus}\label{chapter-winds}\indent

En este capítulo presentaré nuestras medidas de vientos obtenidas a partir del seguimiento de detalles nubosos en imágenes de las misiones espaciales Galileo y Venus Express. Incluiré también resultados sobre su distribución meridional, variabilidad temporal y el modo en que son afectados por la marea térmica y otros factores de naturaleza periódica. En el caso de las observaciones de Venus Express, es la primera vez que ha sido posible usar imágenes de un mismo instrumento para obtener medidas de viento simultáneamente en tres niveles de altura de la región de nubes, representando una medida en tres dimensiones (longitudinal, latitudinal y vertical) de los vientos en Venus.\\

\section{Observaciones}\label{chapter-winds-observs}\indent

Históricamente, para medir las velocidades de las formaciones nubosas se han usado tanto imágenes tomadas desde observatorios terrestres \citep{Crisp1991,Carlson1991} como imágenes de alta resolución tomadas desde diferentes naves: Mariner 10 en 1974 \citep{Limaye1977,Schubert1977}, Pioneer Venus durante los a\~{n}os 1979-1985 \citep{Rossow1980,Rossow1990,Limaye1981,Limaye1982,Limaye1988a,Limaye2007}, Galileo en 1990 \citep{Belton1991,Toigo1994,Peralta2007b} y desde el a\~{n}o 2006 Venus Express \citep{Markiewicz2007b,Sanchez-Lavega2008}.\\

La cámara SSI a bordo de la nave Galileo obtuvo imágenes en dos longitudes de onda: violeta ($418~nm$, $\sim66$ km) y NIR ($418~nm$, $\sim61$ km). \citet{Belton1991} midieron los vientos de Venus en estas dos longitudes de onda empleando un conjunto de imágenes que abarcaba todo el sobrevuelo. Por otro lado, \citet{Toigo1994} también realizaron mediciones con estas imágenes, aunque utilizando únicamente las de más alta resolución corres\-pon\-dien\-tes al 11 de febrero. El conjunto de imágenes de Galileo que usamos en este trabajo viene descrito en la Tabla \ref{tab:tabla-imagsGalileo}. Las imágenes en violeta de los días 12-14 de febrero y las tomadas en NIR el día 13 de febrero fueron usadas para medir los vientos del hemisferio norte, debido a que éstas sólo visualizaban dicho hemisferio. Por otro lado, las imágenes en violeta correspondientes a los días 15-17 de febrero, si bien tenían peor resolución espacial, abarcaban ambos hemisferios por lo que se emplearon para medir también los vientos del hemisferio sur.\\

\begin{table}[h!]
  \caption{Imágenes de Galileo usadas para calcular los vientos.}
	\label{tab:tabla-imagsGalileo}
	\centering
  \begin{spacing}{0.6}

		\begin{tabular}{*{7}{>{\scriptsize}c}}
			& & & & & & \\
			\hline\hline
			& & & & & & \\
			\textit{Número} & \textit{Fecha} & \textit{Hora} & \multirow{2}{*}{\textit{Filtro}} & \textit{Resolución en} & \textit{Resolución en} \\
      \textit{de Imagen} & \textit{(dd/mm/aa)} & \textit{(hh:mm:ss)} & & \textit{60$^{\circ}$N (km/pix)} & \textit{05$^{\circ}$N (km/pix)}  \\
			& & & & & & \\
			\hline
			& & & & & & \\
			18222000 & 12/02/1990 & 05:58:27 &  Violeta   & 24 & 11  \\
			18223900 & 12/02/1990 & 07:58:46 &  Violeta   & 26 & 12  \\
			18225700 & 12/02/1990 & 09:58:06 &  Violeta   & 26 & 12  \\
			18227600 & 12/02/1990 & 11:58:24 &  Violeta   & 27 & 13  \\
			18229400 & 12/02/1990 & 13:57:43 &  Violeta   & 28 & 13  \\
			18224400 & 13/02/1990 & 05:58:16 &  Violeta   & 33 & 16  \\
			18224445 & 13/02/1990 & 05:58:47 & IR cercano & 33 & 16  \\
			18226300 & 13/02/1990 & 07:58:36 &  Violeta   & 36 & 19  \\
			18226345 & 13/02/1990 & 07:59:07 & IR cercano & 36 & 19  \\
			18228400 & 13/02/1990 & 10:00:56 &  Violeta   & 36 & 19  \\
			18228445 & 13/02/1990 & 10:01:27 & IR cercano & 36 & 19  \\
			18220200 & 13/02/1990 & 12:00:15 &  Violeta   & 37 & 20  \\
			18220245 & 13/02/1990 & 12:00:46 & IR cercano & 37 & 20  \\
			18223900 & 14/02/1990 & 05:28:46 &  Violeta   & 48 & 23  \\
			18225700 & 14/02/1990 & 07:28:05 &  Violeta   & 48 & 24  \\
			18223300 & 15/02/1990 & 04:58:16 &  Violeta   & 54 & 27  \\
			18225200 & 15/02/1990 & 06:58:35 &  Violeta   & 59 & 29  \\
			18222800 & 16/02/1990 & 04:28:46 &  Violeta   & 68 & 33  \\
			18224600 & 16/02/1990 & 06:28:04 &  Violeta   & 69 & 34  \\
			18222200 & 17/02/1990 & 03:58:15 &  Violeta   & 80 & 38  \\
			18224100 & 17/02/1990 & 05:58:34 &  Violeta   & 82 & 40  \\
			& & & & & & \\
			\hline
			& & & & & & \\
			
		\end{tabular}

  \end{spacing}
  
\end{table}

Las imágenes de VIRTIS empleadas para las mediciones de viento fueron obtenidas entre abril de 2006 y junio de 2007 y se presentan en la Tabla \ref{tab:tabla-imagsVEX}, donde para cada longitud de onda se detallan el número de trazadores medidos y el rango de latitudes y horas locales abarcados por cada par de imágenes. Estas imágenes sólo abarcan el hemisferio sur y las tres longitudes de onda utilizadas son: UV ($380~nm$, $\sim66$ km), NIR ($980~nm$, $\sim61$ km) e IR ($1.74~\mu m$, $\sim47$ km). El número de trazadores totales usados en cada longitud de onda son mostrados en la parte inferior de la Tabla junto con el rango total de latitudes y horas locales.\\

\begin{table}[h!]
  \caption{Imágenes de Venus Express usadas para calcular los vientos.}
	\label{tab:tabla-imagsVEX}
	\centering
  \begin{spacing}{0.6}
		\begin{tabular}{*{11}{>{\tiny}c}}
			& & & & & & & & & & \\
			\hline\hline
			& & & & & & & & & & \\
			\multirow{2}{*}{\textit{órbita}} & \textit{Fecha} &  & \textit{UV} &  &  & \textit{NIR} &  &  & \textit{IR} &  \\
			  & \textit{(dd/mm/aa)} & \textit{N$^{\circ}$} & \textit{Latitudes} & \textit{Hora Local} & \textit{N$^{\circ}$} & \textit{Latitudes} & \textit{Hora Local} & \textit{N$^{\circ}$} & \textit{Latitudes} & \textit{Hora Local}  \\
			& & & & & & & & & &  \\
			\hline
			& & & & & & & & & &  \\
			VOI00 &  12/04/2006 & \textemdash & \textemdash & \textemdash & \textemdash & \textemdash & \textemdash &      86     &  10$^{\circ}$S-86$^{\circ}$S  &  19.6-05.5   \\
			VOI01 &  13/04/2006 & \textemdash & \textemdash & \textemdash & \textemdash & \textemdash & \textemdash &      14     &  32$^{\circ}$S-81$^{\circ}$S  &  19.2-03.7   \\
			VOI03 &  14/04/2006 & \textemdash & \textemdash & \textemdash & \textemdash & \textemdash & \textemdash &       5     &  44$^{\circ}$S-84$^{\circ}$S  &  21.2-03.0   \\
			VOI04 &  17/04/2006 & \textemdash & \textemdash & \textemdash & \textemdash & \textemdash & \textemdash &       8     &  45$^{\circ}$S-84$^{\circ}$S  &  22.1-04.1   \\
			VOI05 &  19/04/2006 &     145     &  03$^{\circ}$S-71$^{\circ}$S  &  08.8-16.1  &     106     &  04$^{\circ}$S-63$^{\circ}$S   &  08.1-14.3  & \textemdash & \textemdash & \textemdash  \\
			  34  &  24/05/2006 &      25     &  17$^{\circ}$S-40$^{\circ}$S  &  11.2-13.5  &      32     &   00$^{\circ}$-37$^{\circ}$S    &  11.2-13.3  & \textemdash & \textemdash & \textemdash  \\
			  69  &  28/06/2006 &      98     &  07$^{\circ}$S-60$^{\circ}$S  &  09.2-16.1  &      20     &  19$^{\circ}$S-73$^{\circ}$S   &  11.7-13.8  & \textemdash & \textemdash & \textemdash  \\
			  70  &  29/06/2006 &      90     &  34$^{\circ}$S-66$^{\circ}$S  &  11.3-15.3  &      52     &  19$^{\circ}$S-59$^{\circ}$S   &  11.3-14.5  & \textemdash & \textemdash & \textemdash  \\
			  72  &  01/07/2006 & \textemdash & \textemdash & \textemdash & \textemdash & \textemdash & \textemdash &     128     &  01$^{\circ}$S-85$^{\circ}$S  &  18.4-04.6   \\
			  73  &  02/07/2006 &       8     &  50$^{\circ}$S-80$^{\circ}$S  &  09.6-14.3  &       3     &  51$^{\circ}$S-76$^{\circ}$S   &  09.5-13.4  &      84     &  40$^{\circ}$S-84$^{\circ}$S  &  19.1-02.9   \\
			  74  &  03/07/2006 &      10     &  52$^{\circ}$S-74$^{\circ}$S  &  09.2-12.8  &      10     &  52$^{\circ}$S-74$^{\circ}$S   &  09.2-12.8  &      10     &  52$^{\circ}$S-86$^{\circ}$S  &  20.6-03.4   \\
			  75  &  04/07/2006 &      16     &  53$^{\circ}$S-76$^{\circ}$S  &  09.0-14.4  &      14     &  51$^{\circ}$S-72$^{\circ}$S   &  07.0-12.9  &      32     &  40$^{\circ}$S-84$^{\circ}$S  &  19.6-04.4   \\
			  76  &  05/07/2006 &      18     &  45$^{\circ}$S-73$^{\circ}$S  &  12.3-16.1  &      30     &  45$^{\circ}$S-75$^{\circ}$S   &  11.9-15.2  & \textemdash & \textemdash & \textemdash  \\
			  77  &  06/07/2006 &       7     &  52$^{\circ}$S-80$^{\circ}$S  &  13.2-16.3  &      24     &  49$^{\circ}$S-81$^{\circ}$S   &  12.2-15.8  &      65     &  38$^{\circ}$S-69$^{\circ}$S  &  18.9-00.1   \\
			  78  &  07/07/2006 &      13     &  45$^{\circ}$S-65$^{\circ}$S  &  12.1-16.3  &      29     &  54$^{\circ}$S-80$^{\circ}$S   &  12.4-17.0  &       5     &  80$^{\circ}$S-84$^{\circ}$S  &  20.4-23.6   \\
			  79  &  08/07/2006 &       8     &  50$^{\circ}$S-65$^{\circ}$S  &  13.5-16.2  & \textemdash & \textemdash & \textemdash &       5     &  79$^{\circ}$S-82$^{\circ}$S  &  22.1-23.8   \\
			  80  &  09/07/2006 &      15     &  50$^{\circ}$S-70$^{\circ}$S  &  11.9-14.6  &      42     &  51$^{\circ}$S-81$^{\circ}$S   &  12.1-17.7  &      20     &  44$^{\circ}$S-81$^{\circ}$S  &  19.9-23.9   \\
			  81  &  10/07/2006 &       9     &  53$^{\circ}$S-62$^{\circ}$S  &  13.2-15.6  &      29     &  49$^{\circ}$S-73$^{\circ}$S   &  12.7-16.2  & \textemdash & \textemdash & \textemdash  \\
			  82  &  11/07/2006 &      10     &  59$^{\circ}$S-78$^{\circ}$S  &  12.1-15.1  &      20     &  50$^{\circ}$S-74$^{\circ}$S   &  12.4-15.3  & \textemdash & \textemdash & \textemdash  \\
			  84  &  13/07/2006 &       8     &  59$^{\circ}$S-81$^{\circ}$S  &  09.5-12.9  &      17     &  61$^{\circ}$S-79$^{\circ}$S   &  08.6-13.3  &      98     &  36$^{\circ}$S-82$^{\circ}$S  &  20.9-02.5   \\
			  85  &  14/07/2006 &      32     &  54$^{\circ}$S-85$^{\circ}$S  &  07.9-13.4  &      30     &  52$^{\circ}$S-76$^{\circ}$S   &  08.5-14.2  & \textemdash & \textemdash & \textemdash  \\
			  86  &  15/07/2006 &      10     &  57$^{\circ}$S-82$^{\circ}$S  &  08.5-12.9  &      23     &  56$^{\circ}$S-83$^{\circ}$S   &  07.7-13.1  & \textemdash & \textemdash & \textemdash  \\
			  88  &  17/07/2006 & \textemdash & \textemdash & \textemdash & \textemdash & \textemdash & \textemdash &      34     &  08$^{\circ}$S-49$^{\circ}$S  &  21.8-01.2   \\
			  94  &  23/07/2006 &       8     &  65$^{\circ}$S-76$^{\circ}$S  &  07.1-12.5  &      22     &  59$^{\circ}$S-87$^{\circ}$S   &  06.9-13.2  & \textemdash & \textemdash & \textemdash  \\
			  95  &  24/07/2006 &      29     &  56$^{\circ}$S-83$^{\circ}$S  &  06.6-12.6  &      23     &  60$^{\circ}$S-82$^{\circ}$S   &  06.6-14.1  & \textemdash & \textemdash & \textemdash  \\
			  96  &  25/07/2006 &       7     &  65$^{\circ}$S-82$^{\circ}$S  &  07.0-12.8  &      19     &  60$^{\circ}$S-85$^{\circ}$S   &  06.3-15.5  & \textemdash & \textemdash & \textemdash  \\
			  97  &  26/07/2006 &      16     &  55$^{\circ}$S-87$^{\circ}$S  &  10.8-16.2  &      24     &  57$^{\circ}$S-85$^{\circ}$S   &  11.9-17.5  & \textemdash & \textemdash & \textemdash  \\
			  98  &  27/07/2006 &       4     &  68$^{\circ}$S-78$^{\circ}$S  &  12.5-13.7  &      25     &  63$^{\circ}$S-83$^{\circ}$S   &  10.6-16.4  & \textemdash & \textemdash & \textemdash  \\
			 220  &  26/11/2006 & \textemdash & \textemdash & \textemdash &      26     &  65$^{\circ}$S-84$^{\circ}$S   &  07.6-15.1  & \textemdash & \textemdash & \textemdash  \\
			 244  &  20/12/2006 &      15     &  67$^{\circ}$S-84$^{\circ}$S  &  08.6-15.3  &      42     &  69$^{\circ}$S-86$^{\circ}$S   &  07.5-15.9  & \textemdash & \textemdash & \textemdash  \\
			 283  &  28/01/2007 &      24     &  55$^{\circ}$S-78$^{\circ}$S  &  10.6-16.1  & \textemdash & \textemdash & \textemdash & \textemdash & \textemdash & \textemdash  \\
			 299  &  13/02/2007 & \textemdash & \textemdash & \textemdash & \textemdash & \textemdash & \textemdash &     126     &  14$^{\circ}$S-64$^{\circ}$S  &  18.5-21.2   \\
			 300  &  14/02/2007 & \textemdash & \textemdash & \textemdash & \textemdash & \textemdash & \textemdash &      84     &  14$^{\circ}$S-64$^{\circ}$S  &  18.4-21.1   \\
			 410  &  05/06/2007 & \textemdash & \textemdash & \textemdash & \textemdash & \textemdash & \textemdash &      44     &  02$^{\circ}$S-26$^{\circ}$S  &  20.2-23.0   \\
			 411  &  06/06/2007 & \textemdash & \textemdash & \textemdash & \textemdash & \textemdash & \textemdash &      84     &  04$^{\circ}$S-38$^{\circ}$S  &  19.2-23.4   \\
			& & & & & & & & & &  \\
			\multicolumn{2}{c}{\scriptsize{TOTAL}} & 625 & 03$^{\circ}$S-87$^{\circ}$S & 06.6-16.3 & 662 & 00$^{\circ}$-87$^{\circ}$S & 06.3-17.7 & 932 & 04$^{\circ}$S-86$^{\circ}$S & 18.4-05.5  \\
			& & & & & & & & & &  \\
			\hline
			& & & & & & & & & &  \\
		\end{tabular}
  \end{spacing}
\end{table}

Todas las imágenes usadas para las mediciones de vientos fueron na\-ve\-ga\-das, corregidas de defectos (tanto los propios de la cámara como los arbitrarios de la imagen), procesadas y proyectadas de forma cilíndrica y polar usando el software PLIA \citep{Hueso2008b}. En el caso de las imágenes de Galileo corregimos la variación centro-limbo de la reflectividad aplicando la ley de dispersión de Lambert \citep{Rossow1980}. Aumentamos el contraste combinando máscaras de enfoque\footnote{Estas herramientas resultaron esenciales para buscar detalles tanto a altas latitudes en las imágenes de Galileo como para las imágenes UV y NIR de VIRTIS.} (``unsharp mask'') y filtros Butterworth \citep{Gonzalez1992}. Una vez procesamos una imagen, llevamos a cabo una proyección cilíndrica cuando pretendimos medir detalles en bajas latitudes, o polar cuando lo hacíamos a altas latitudes. En muchas ocasiones las imágenes proyectadas eran procesadas de nuevo para mejorar aún más la visibilidad de las estructuras nubosas.\\

El estudio del movimiento de las formaciones nubosas observadas en imágenes separadas en el tiempo permite llevar a cabo mediciones del viento en cada uno de estos niveles. Para ello hemos de suponer que las nubes se mueven conforme al viento y las desviaciones locales que éste presente, por lo que éstas pueden usarse como trazadores de los movimientos atmosféricos\footnote{En 1985 los globos VEGA confirmaron esta hipótesis, proporcionando información detallada sobre la dinámica en las nubes intermedias del ecuador.}. Las componentes zonal y meridional de las velocidades del viento son calculadas de acuerdo con las siguientes expresiones:
\begin{align}
	u&=(R+H)\cdot\cos\phi\cdot\frac{\Delta\lambda}{\Delta t}\cdot\frac{\pi}{180},                  \label{vel-zonal}\\
	v&=(R+H)\cdot\frac{\Delta\phi}{\Delta t}\cdot\frac{\pi}{180}.  \label{vel-merid}
\end{align}
donde $\Delta\lambda$ y $\Delta\phi$ son los desplazamientos del detalle nuboso (medidos en grados) en longitud y latitud respectivamente, $\Delta t$ es el intervalo de tiempo entre el par de imágenes, $\phi$ es la latitud promedio del detalle nuboso, $R$ es el radio planetario y $H$ es la altura correspondiente al nivel de nubes que estemos observando.\\

Para cada par de imágenes identificamos los detalles nubososos en común y medimos sus posiciones en ambas usando las ecuaciones (\ref{vel-zonal}) y (\ref{vel-merid}). Los pares examinados están separados entre sí 2 horas en el caso de las imágenes de la cámara SSI de Galileo, y de 20-74 minutos en el caso de la cámara espectral VIRTIS. Si bien la utilización de pares separados por intervalos de tiempo mayores ayuda a incrementar la precisión de las medidas, la mayoría de las estructuras y detalles finos resultan no ser identificables para espacios de tiempo mayores que 2 horas, especialmente a altas latitudes.\\

La calidad de las imágenes juega un papel esencial a la hora de estimar la incertidumbre de nuestras medidas (ver Apéndice \ref{appendix-errors}). En el caso de las imágenes SSI de Galileo, al considerar una resolución espacial en latitudes medias de 0.15$^{\circ}$ e imágenes separadas 2 horas entre sí, el error de medida en las imágenes del violeta resulta ser del orden de $1~m\cdot s^{-1}$. A causa del bajo contraste de las imágenes en NIR tuvimos que emplear detalles de mayor tama\~{n}o con errores de medida de $\sim5~m\cdot s^{-1}$. Para latitudes mayores la incertidumbre de las medidas se incrementaba en ambos filtros hasta los $10~m\cdot s^{-1}$ en $\pm$70$^{\circ}$ debido a la convergencia de los meridianos al aproximarnos a los polos. Para las imágenes VIRTIS de Venus Express, teniendo en cuenta tanto las diferentes resoluciones espaciales como la precisión en la navegación de las imágenes, entre latitudes medias y polares (de 30$^{\circ}$S a 90$^{\circ}$S) el error de medida oscilaba entre $4-12~m\cdot s^{-1}$, mientras que en latitudes tropicales y ecuatoriales (de 0$^{\circ}$ a 30$^{\circ}$S) tenemos $13-20~m\cdot s^{-1}$ \citep{Sanchez-Lavega2008}.\\

\section{Morfología de las nubes}\indent

Mientras que la órbita polar de Venus Express permite obtener una buena perspectiva de las regiones polares de Venus, el sobrevuelo ecuatorial que llevó a cabo la nave Galileo impidió tener una buena visión de éstas, por lo que optamos por realizar proyecciones polares de las imágenes originales de SSI usando el software PLIA. Las proyecciones polares tienen la ventaja de que permiten visualizar con mayor claridad los patrones nubosos de altas latitudes que la vista oblicua. En 1978 Suomi y Limaye obtuvieron la primera proyección estereográfica polar (ver Fig.1 en \citealp{Suomi1978}) usan\-do 22 imágenes UV tomadas por la nave Mariner 10, simulando así lo que se hubiera visto en caso de haber volado sobre el polo sur de Venus. En nues\-tro caso realizamos una composición del hemisferio norte de Venus usando proyecciones polares de hasta 12 imágenes del violeta tomadas por la cámara SSI y asumiendo un periodo de rotación fijo de 4.4 días para todas las latitudes. Aunque veremos más adelante que en altas latitudes la atmósfera de Venus está lejos de rotar como un sólido rígido, el valor promedio de rotación usado para la composición proporciona una visión aceptable de la región polar norte de Venus, tal y como puede observarse en la Figura 3.1A.\\

\begin{figure}[h!]
	\centering
		\includegraphics[width=1.0\textwidth]{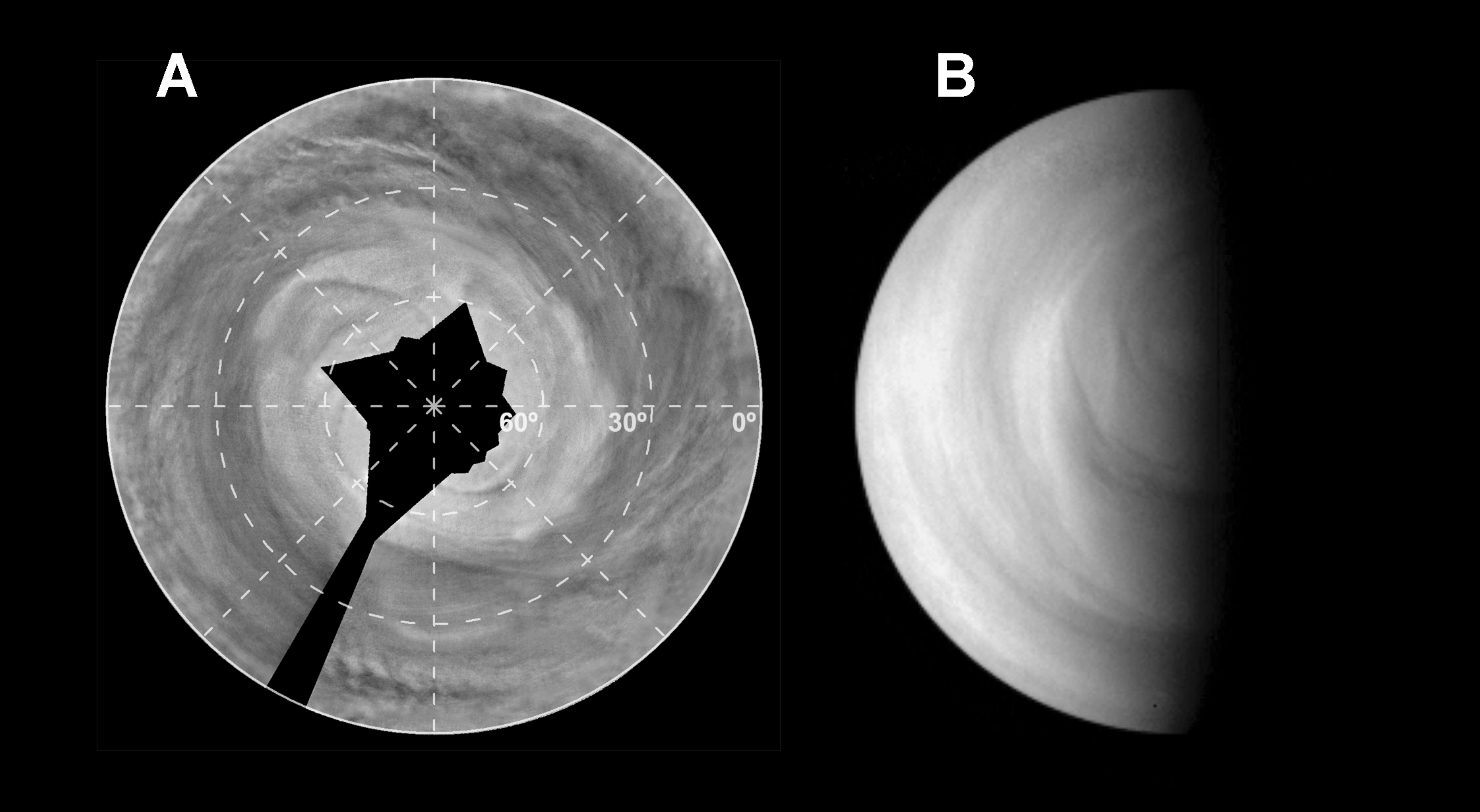}
	\label{fig:GalileoVEX-Venus-UV}
	\caption[La cima de las nubes en Venus.]{\scriptsize{Composición del hemisferio norte en violeta(\textbf{A}, imágenes de Galileo-SSI) e imagen real del hemisferio sur en ultravioleta (\textbf{B}, imágenes de Vex-VIRTIS) de Venus, ambas correspondientes a la cima de las nubes.}}
\end{figure}

En ambas imágenes de la Figura 3.1 vemos que la estructura más llamativa de la cima de las nubes es sin duda la brillante banda polar con forma espiral, conocida como el \textit{collar polar}\index{Collar polar}. Dicho collar rodea al polo a partir de latitudes superiores a los 45$^{\circ}$N, y se rompe en estrías espirales tanto dentro de la brillante \textit{capa polar}\index{Capa polar} como en su límite más externo. Al igual que puede observarse en las imágenes UV de Mariner 10 \citep{Suomi1978}, para la cima de las nubes tenemos que la región polar es mucho más brillante que el resto del hemisferio. Debido a la trayectoria ecuatorial del sobrevuelo de Galileo, vemos que a partir de $\sim$60$^{\circ}$N empiezan a faltar datos del polo, si bien en limitados casos tenemos información nubosa hasta 70$^{\circ}$N, lo que resultó útil para estudiar el movimiento de las nubes. Esta estructura de collar de las imágenes UV no sólo se restringe a la cima de las nubes \citep{Rossow1980,Peralta2007b,Piccioni2006} sino que se manifiesta en diversos niveles de altura y parece estar asociado al dipolo polar\footnote{Si bien en la cima de las nubes no parece ser una estructura permanente en el tiempo \citep{Taylor1980} y desaparece por encima de los 75 km de altura \citep{Piccioni2007}.}. Es razonable, por tanto, pensar que la persistencia del collar polar es una prueba indirecta de la estabilidad del vórtice polar. De acuerdo con los modelos estándar de nubes y de transferencia radiativa, el nivel de altura de las imágenes en UV se corresponde con el intervalo $z\sim64-68$ km (ver Supplementary Material en \citealt*{Sanchez-Lavega2008}).\\

El nivel de las nubes que captan las imágenes en NIR ($980~nm$) fue es\-tu\-dia\-do por vez primera en 1990 con la cámara SSI de Galileo \citep{Belton1991,Peralta2007b}, y posteriormente gracias a las imágenes de la cámara espectral VIRTIS de Venus Express \citep{Sanchez-Lavega2008}. \citet{Belton1991} situaron este nivel a pocos kilómetros por debajo de la cima de las nubes e hicieron hincapié en que los patrones nubosos eran similares a los de las imágenes UV aunque exhibiendo un contraste ``anticorrelacionado'', detalle que corroboramos en nuestro propio análisis de estas imágenes (ver Figura 3.2). Además de esto, las estructuras alargadas típicas de latitudes medias muestran en las imágenes NIR una disposición más zonal que las de las imágenes UV, y aunque a bajas latitudes las estructuras se corresponden bastante bien entre ambos niveles, las velocidades zonales a las que se mueven son distintas \citep{Belton1991,Peralta2007b,Sanchez-Lavega2008}. Más recientemente, este nivel de nubes se ha ubicado a una altura $z\sim60$ km variable en latitud (ver Supplementary Material en \citealt*{Sanchez-Lavega2008}).\\

\begin{figure}[h!]
	\centering
		\includegraphics[width=0.60\textwidth]{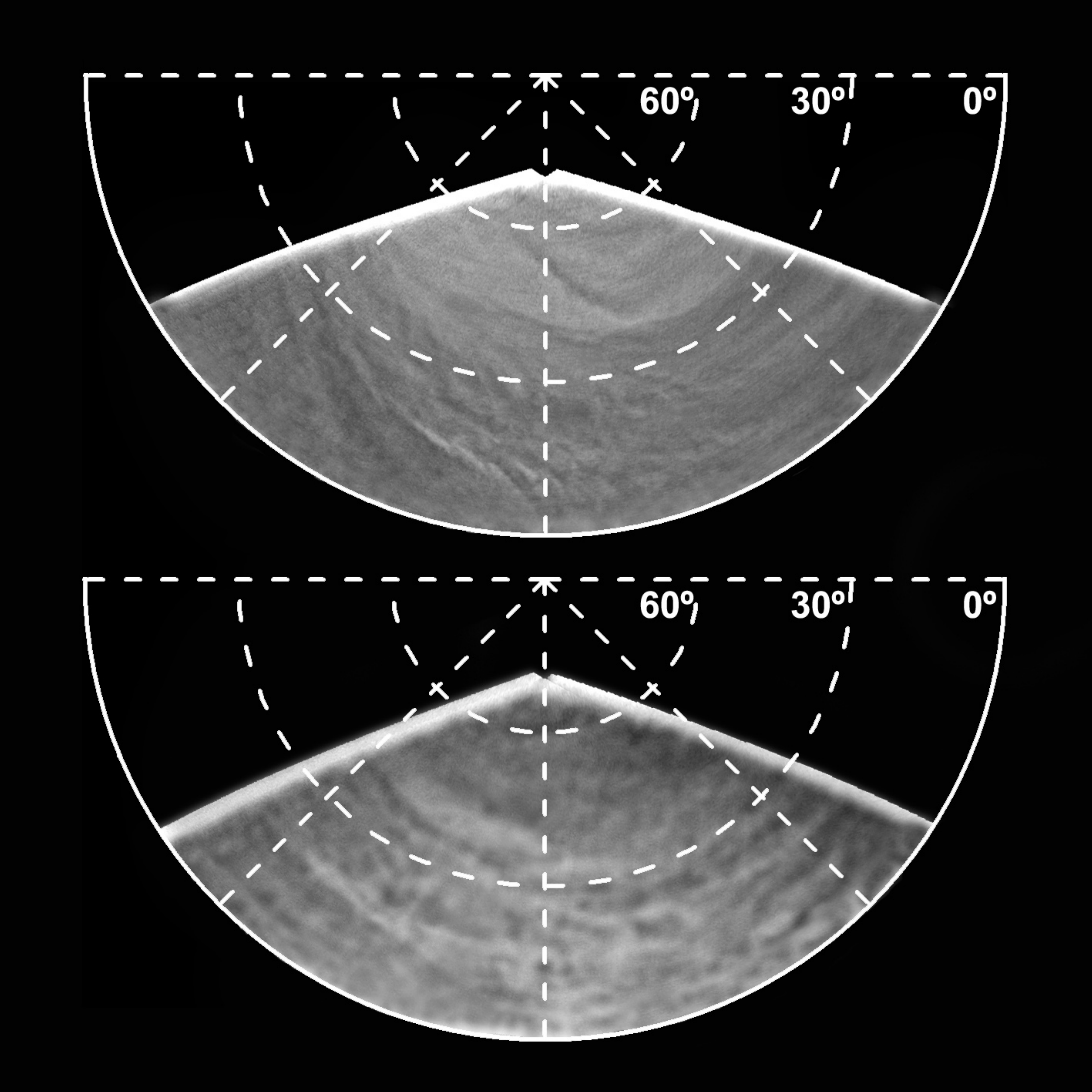}
	\label{fig:GalileoVenus-VIO-NIR}
	\caption[Comparación de la morfología nubosa en violeta y NIR.]{\scriptsize{Proyecciones polares del mismo sector del hemisferio norte a partir de imágenes casi simultáneas en violeta ($418~nm$, arriba) y NIR ($986~nm$, abajo).}}
\end{figure}

Respecto a las nubes profundas estudiadas en imágenes nocturnas en la ventana de observación de $1.74~\mu m$ con VIRTIS, podemos distinguir en la mayoría de los casos tres regiones claramente diferenciadas. La zona ecua\-to\-rial hasta los 40$^{\circ}$ de latitud muestra por lo general un patrón nuboso desorganizado (ver Fig. 2.9E en la página \pageref{fig:VenusAtmos-diff-filters} y Fig. 5.2D en la página \pageref{fig:VEXVenus-WavesExamples}), que podría ser el producto de actividad turbulenta y convectiva \citep{Carlson1991,Peralta2008}. La región de latitudes entre 40$^{\circ}$ y 70$^{\circ}$ de latitud muestra nubes de textura ``suave'' y forma alargada, exhibiendo cierta inclinación respecto a los paralelos y una alternancia de estructuras con fuerte y baja opacidad. Dicha alternancia de opacidad es más intensa en la zona que se corresponde con el límite externo de la \textit{capa polar} (donde el collar polar se rompe en estrías en la cima de las nubes), y suele ser menos acusada a medida que nos acercamos a las latitudes polares. Por encima de los 70$^{\circ}$ las nubes se tornan más oscuras y opacas, y comienza de forma clara la región de influencia del dipolo, tal como puede observarse en la Fig. 2.9E (página \pageref{fig:VenusAtmos-diff-filters}).\\

\section{Vientos Zonales en Venus}\label{chapter-winds-zonalwind}\indent

El seguimiento de las estructuras nubosas fue llevado a cabo usando un amplio número de imágenes tomadas por misiones espaciales de épocas dife\-rentes: Galileo en 1990 y Venus Express entre los a\~{n}os 2006 y 2007. En la parte diurna del planeta las imágenes de ambas misiones nos permiten medir los movimientos en dos niveles de altura: las imágenes UV la cima de las nubes a $\sim66~km$ y las imágenes NIR la base de la cima de las nubes a $\sim61~km$. En la parte nocturna la cámara VIRTIS de Venus Express nos proporcionó imágenes IR de la parte baja de las nubes a $\sim48~km$. Para estas medidas se usaron pares de imágenes pertenecientes al mismo día y con un intervalo temporal que oscilaba entre 20 minutos y 2 horas, dependiendo de la disponibilidad de imágenes útiles en nuestro estudio. Mientras que las imágenes UV de SSI en Galileo cubren ambos hemisferios, las imágenes NIR sólo abarcan el hemisferio norte, y se obtuvieron un total de 421 mediciones en UV y 108 en NIR. En el caso de Venus Express las imágenes de VIRTIS proporcionan datos exclusivamente del hemisferio sur, y realizamos 625 mediciones en UV, 662 en NIR y 932 en IR. Por convenio, el error que asociamos a los valores de velocidad promediados\footnote{La dispersión de los datos suele reflejar la magnitud del error de medida, si bien en muchos casos también puede deberse a la presencia de ondas y turbulencia en la atmósfera \citep{Smith1996}.} será la desviación estándar de las medidas, excepto cuando tengamos un número peque\~{n}o de ellas (momento en el que emplearemos el error típico de medida).\\

\subsection{Venus con Galileo (datos de 1990)}\indent

Un perfil promediado así como las velocidades individuales están representados en la Figura 3.3. El perfil promediado se obtuvo tomando bandas de latitud de 2$^{\circ}$ de ancho y calculando el valor medio de las medidas de velocidad contenidas en cada banda ($<u>(\bar{\varphi})=\left|\frac{\sum u_{i}}{N}\right|_{\varphi_{1}}^{\varphi_{2}}$, donde $N$ es el número de medidas de velocidad contenidas en el rango de latitud $\varphi_{1}-\varphi_{2}$). Se toma como error de los valores promedio la desviación estándar de las medidas contenidas en cada banda de latitud, excepto cuando ésta sea inferior al error de medida o cuando tengamos muy pocas mediciones (en cuyo caso el error de medida sustituye a la desviación estándar).\\

\begin{figure}[!h]
	\centering
		\includegraphics[width=0.60\textwidth]{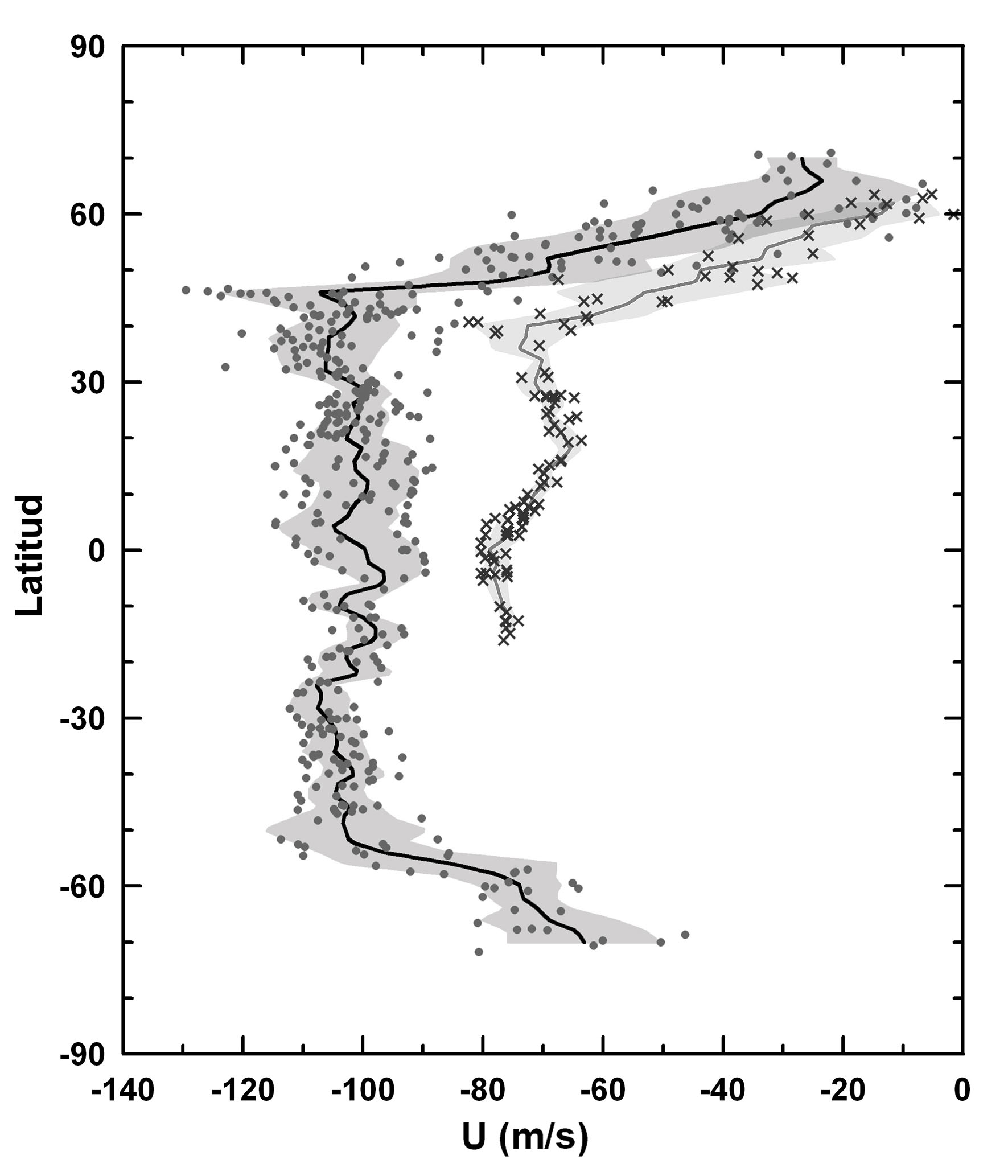}
	\label{fig:GalileoVenus-ZonalWind}
	\caption[Viento zonal a distintas alturas con imágenes de Galileo.]{\scriptsize{Componente zonal del viento medida a partir del seguimiento de nubes en imágenes de Galileo. Las medidas en UV y NIR aparecen como círculos negros y cruces respectivamente. Las líneas continuas oscura y clara hacen alusión a los perfiles promediados, y las regiones sombreadas definen el error.}}
\end{figure}
 
En ambos niveles de altura se puede comprobar que la componente zonal del viento a altas latitudes experimenta una acusada disminución que se intensifica a medida que nos aproximamos a los polos. En la cima de las nubes (UV) este decrecimiento no parece estrictamente simétrico respecto del ecuador, comenzando aproximadamente a 45$^{\circ}$N en el hemisferio norte y a 50$^{\circ}$S en el hemisferio sur, y con la caída en el hemisferio norte más acentuada que en el sur. En la base de la cima de las nubes (NIR) observamos también esta caída de los vientos, produciéndose en este caso a partir de los 40$^{\circ}$N. Estimamos para el decaimiento del viento zonal en ambos niveles una cizalla meridional\index{Cizalla!meridional} de $\frac{\partial<u>}{\partial y}=0.03\pm0.06~m\cdot s^{-1}\cdot km^{-1}$ en el hemisferio norte, y una cizalla de $0.02\pm0.06~m\cdot s^{-1}\cdot km^{-1}$ para la cima de las nubes en el hemisferio sur, donde $<u>$ es la componente zonal del viento promediada e $y$ es la coordenada meridional.\\

\subsection{Venus con Venus Express (datos de 2006-2007)}\indent

Los valores promedio del viento zonal en los tres niveles verticales mencionados a partir de las imágenes de la cámara espectral VIRTIS aparecen en la Figura 3.4. Dichos perfiles están promediados en bandas de latitud de 2$^{\circ}$ y temporalmente, correspondiendo las barras de error a la desviación estándar de todas las medidas contenidas en cada banda. Observamos que los perfiles presentan un comportamiento muy similar en las tres longitudes de onda (es decir, en los tres niveles de altura), si bien a continuación concretamos algunas propiedades específicas de cada caso.\\

En la \textit{cima de las nubes} (lado diurno y altitud de $\sim66~km$) la componente zonal del viento tiene un valor $<u>=-102\pm 10~m\cdot s^{-1}$ constante desde el ecuador hasta 50$^{\circ}$S, momento en el cual empieza a decrecer con una cizalla meridional de $\frac{\partial<u>}{\partial y}=0.026\pm0.06~m\cdot s^{-1}\cdot km^{-1}$ hasta anularse en el polo.\\

En la \textit{base de la cima de las nubes} (lado diurno y altura de $\sim61~km$) la componente zonal del viento se comporta de modo casi idéntico al nivel inmediatamente superior, aunque los vientos zonales tienen menor magnitud ($<u>=-60\pm 10~m\cdot s^{-1}$) y el decaimiento se produce un poco más hacia el polo, a partir de los 55$^{\circ}$S.\\

En las \textit{nubes inferiores} (lado nocturno y altura de $\sim47~km$) el perfil meridional también es parecido a los anteriores, con un valor de velocidad zonal aproximadamente constante ($<u>=-60\pm 10~m\cdot s^{-1}$) desde el ecuador hasta los 65$^{\circ}$S, y desde aquí experimenta un decrecimiento (cizalla de $\frac{\partial<u>}{\partial y}=0.021\pm0.06~m\cdot s^{-1}\cdot km^{-1}$) para finalmente anularse en el polo.\\

\begin{figure}[h!]
	\centering
		\includegraphics[width=0.60\textwidth]{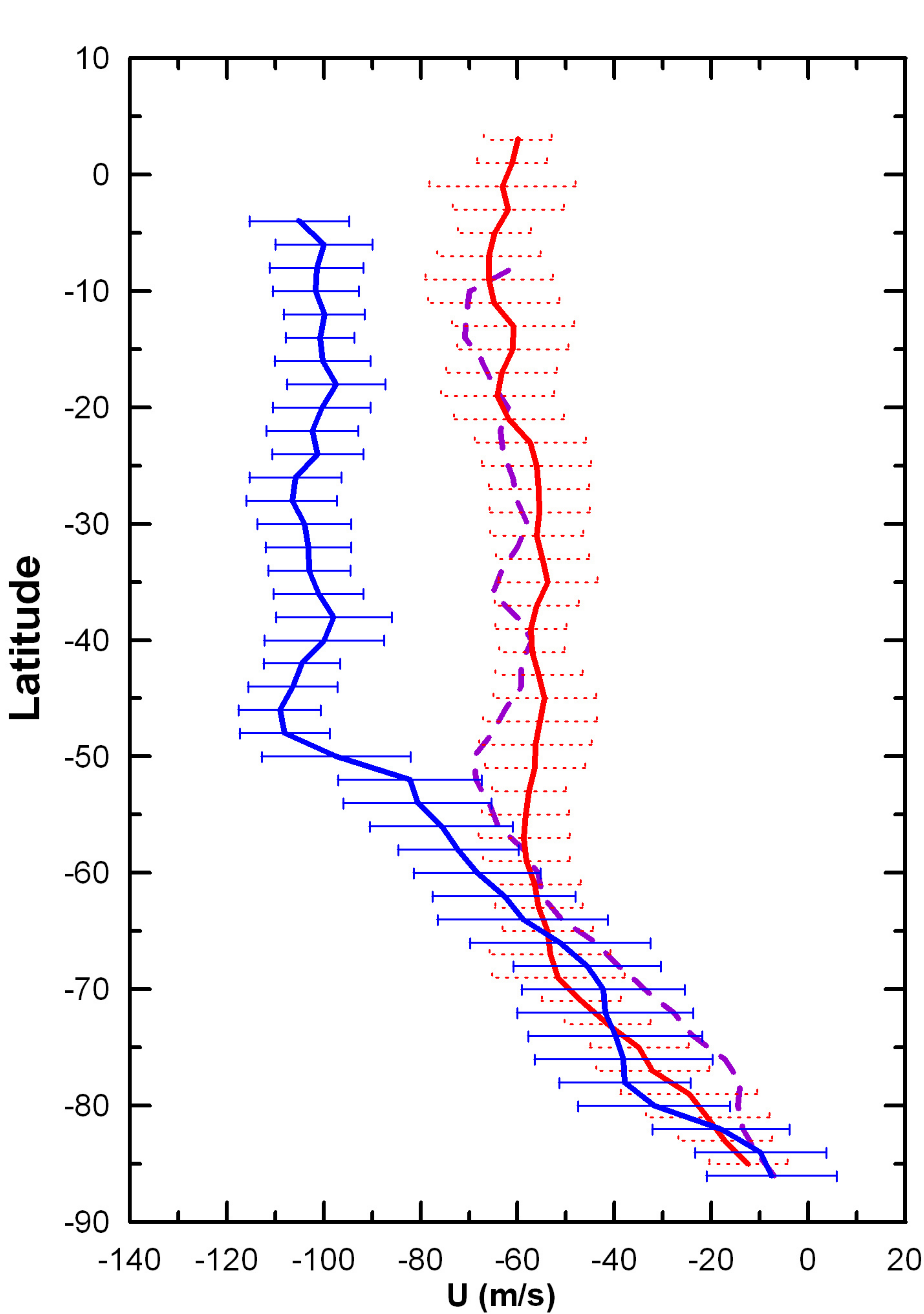}
	\label{fig:VirtisVenus-ZonalWind}
	\caption[Viento zonal a distintas alturas con imágenes de VIRTIS-M.]{\scriptsize{Componente zonal del viento en el hemisferio sur medida a partir del seguimiento de nubes en imágenes de VIRTIS de Venus Express (abril 2006-julio 2007). Los perfiles están promediados temporalmente y en bandas de latitud de 2$^{\circ}$ para tres longitudes de onda: ultravioleta (línea azul, $380~nm$; cima de las nubes, altura $\sim66~km$; lado diurno), infrarrojo cercano (línea violeta, $980~nm$; base de la cima de las nubes, altura $\sim61~km$; lado diurno) e infrarrojo (línea roja, $980~nm$; nubes profundas, altura $\sim61~km$; lado nocturno).}}
\end{figure}

\subsection{Cizalla Vertical del viento}\indent

Al comparar sus resultados con los perfiles verticales de viento obtenidos por los globos VEGA \citep{Linkin1986}, \citet{Belton1991} sugirieron que los niveles de las imágenes violeta y NIR captadas por la cámara SSI de Galileo no debían estar separados entre sí más de 15 km. Las sondas Pioneer Venus también proporcionaron datos sobre la estructura vertical de los vientos en dos latitudes que resultan de interés para nuestro trabajo: la sonda ``north'' transmitió datos a una latitud de 5.7$^{\circ}$N, mientras que la sonda ``large'' lo hizo a 59.7$^{\circ}$N, ambas en el lado diurno del planeta \citep{Colin1979,Schubert1983}. Entre 50 y 60 km de altura, el valor medio de la cizalla vertical\index{Cizalla!vertical} elevada al cuadrado $\left(\frac{\partial<u>}{\partial z}\right)^{2}$ era de $7\cdot10^{-6}$ y $11\cdot10^{-6}~s^{-2}$ para las sondas ``north'' y ``large'' respectivamente (ver Figura 6 en \citealp{Gierasch1997}). Si comparamos estos valores con los deducidos a partir de nuestros perfiles de Galileo se puede comprobar que el mejor ajuste se produce cuando suponemos que el filtro violeta ($418~nm$) y NIR ($986~nm$) están separados verticalmente unos 8 km.\\

\begin{figure}[h!]
	\centering
		\includegraphics[width=0.7\textwidth]{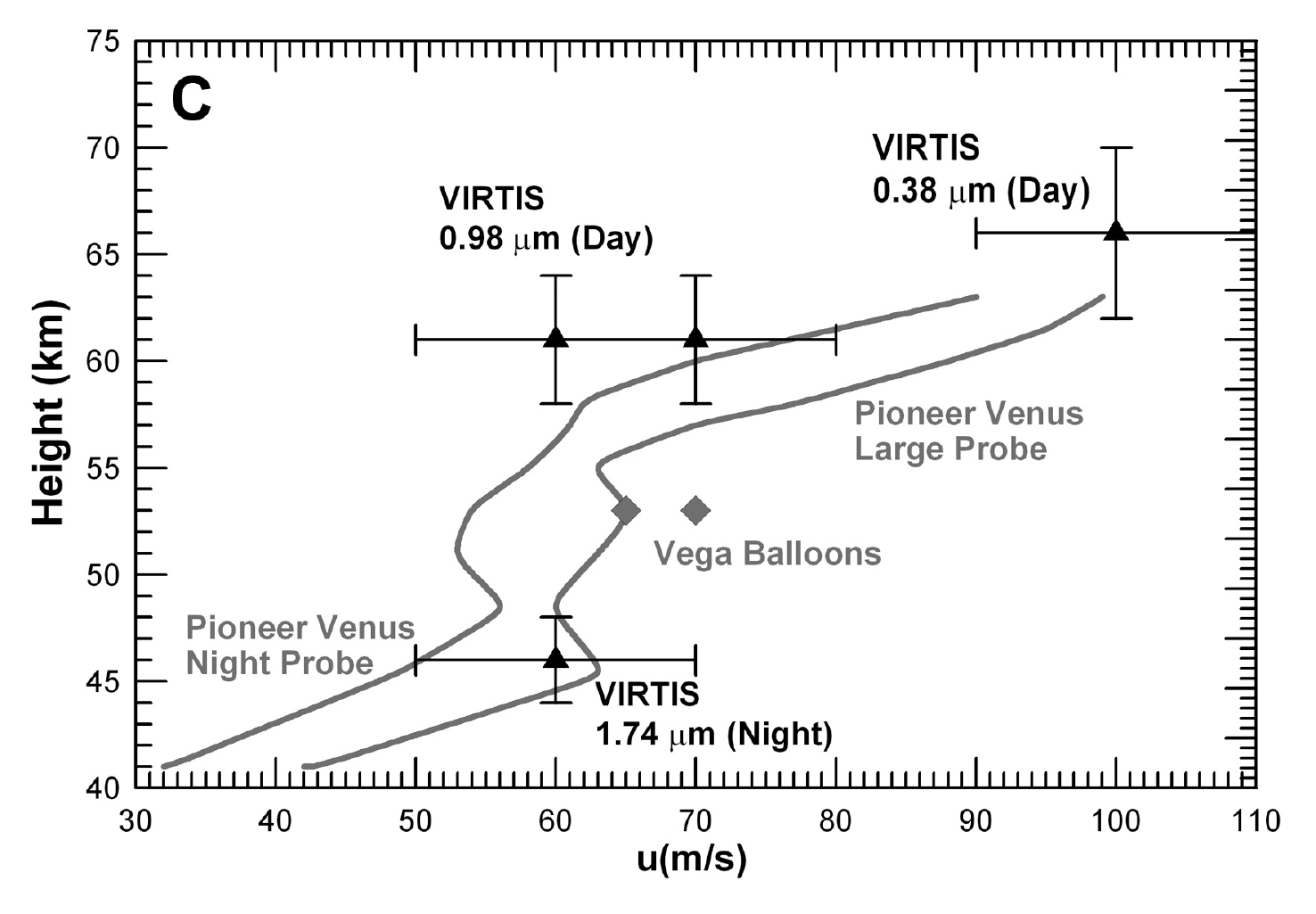}
	\label{fig:Winds-Vertical-Profile}
	\caption[Perfiles verticales del viento zonal en Venus.]{\scriptsize{Perfiles verticales del viento zonal obtenidos por las sondas de Pioneer Venus Large y Night (línea continua), los globos VEGA (diamantes) y con las imágenes de VIRTIS de Venus Express (triángulos). En general existe un buen acuerdo entre las medidas realizadas por las misiones en diferentes épocas, indicativo de que el perfil vertical del viento es bastante estable.}}
\end{figure}

Nuestras medidas de vientos con las imágenes de VIRTIS y las alturas propuestas anteriormente para las diferentes longitudes de onda también son consistentes con la cizalla vertical medida tanto por las sondas Pioneer Venus como por los globos VEGA en diferentes localizaciones de la atmósfera \citep{Schubert1983,Gierasch1997} (ver Figura 3.5). En el ecuador y a latitudes medias (0$^{\circ}$ a 55$^{\circ}$S) la cizalla vertical del viento es de $\frac{\partial<u>}{\partial z}=8\pm2~m\cdot s^{-1}\cdot km^{-1}$ entre 61 y 66 km de altura (NIR y UV), mientras que $\frac{\partial<u>}{\partial z}<1~m\cdot s^{-1}\cdot km^{-1}$ entre 47 y 61 km de altura (IR y NIR). En la región subpolar (de 50$^{\circ}$-60$^{\circ}$S hasta el polo) la cizalla vertical dentro de las nubes es débil, con valores $\frac{\partial<u>}{\partial z}\leq2~m\cdot s^{-1}\cdot km^{-1}$.\\

\begin{figure}[h!]
	\centering
		\includegraphics[width=1.0\textwidth]{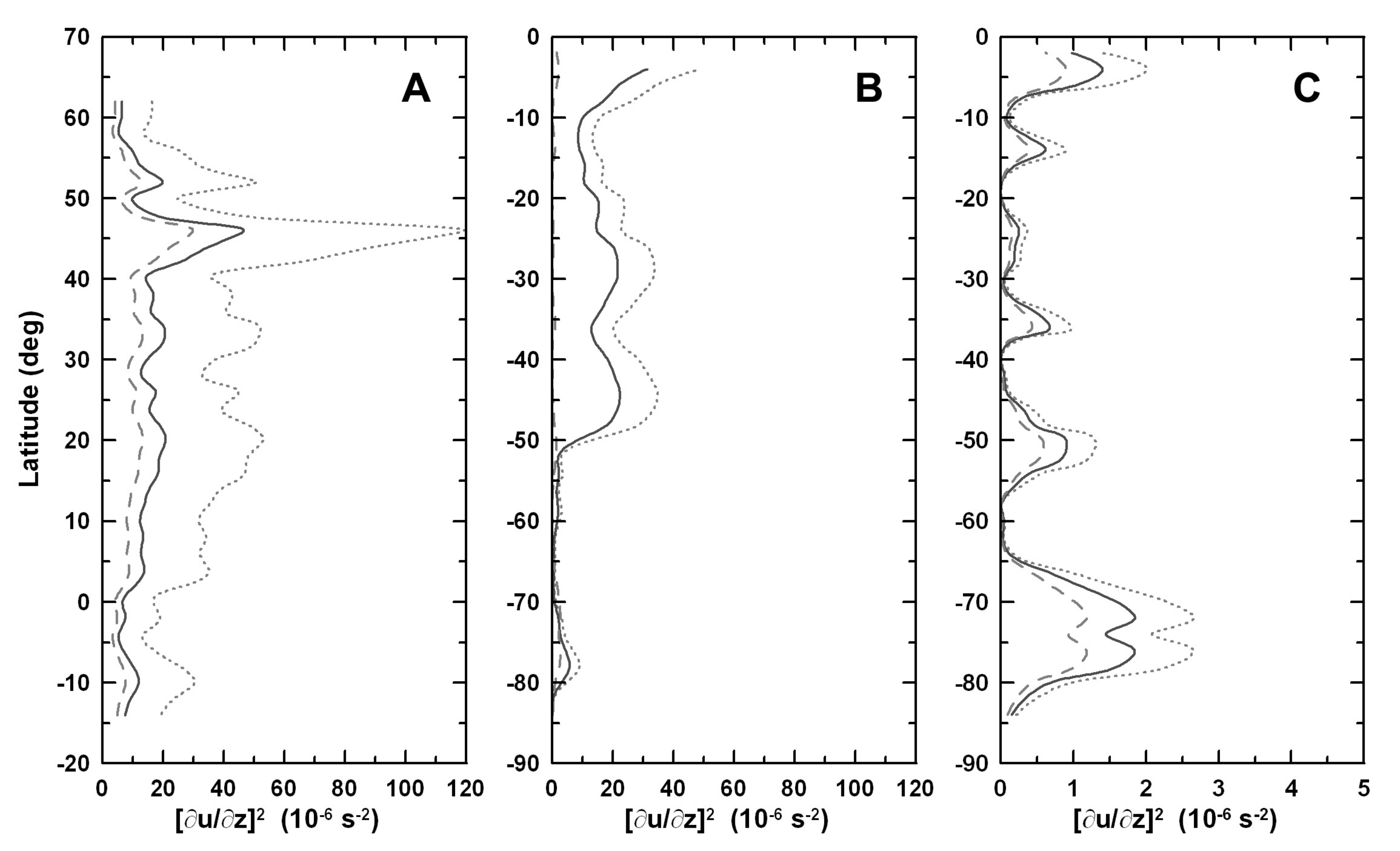}
	\label{fig:Venus-VertShear}
	\caption[Cizalla vertical del viento zonal en Venus.]{\scriptsize{Cizalla Vertical al cuadrado del viento zonal obtenido con los datos de Galileo y Venus Expresss. (A) y (B) nos muestran la cizalla entre los niveles UV y NIR con datos de Galileo y Venus Express respectivamente. Ya que la diferencia de altura no está determinada con exactitud, se sugieren varios valores: 5 km (línea a trazos), 8 km (línea continua) y 10 km (línea en puntos suspensivos). (C) nos muestra la cizalla entre los niveles NIR e IR con datos de Venus Express. Los valores de diferencia de altura sugeridos en este caso son: 10 km (línea a trazos), 12 km (línea continua) y 15 km (línea en puntos suspensivos).}}
\end{figure}

Usando los perfiles de la componente zonal del viento extraída a partir de las imágenes de Galileo y Venus Express, la Figura 3.6 muestra el perfil meridional de la cizalla vertical del viento zonal elevada al cuadrado $\left(\frac{\partial<u>}{\partial z}\right)^{2}$ para varios valores de separación vertical entre el nivel UV y NIR (5, 8 y 10 km) y entre el nivel NIR e IR (10, 12 y 15 km). En este análisis se supone que la distancia $\Delta z$ entre los niveles de altura no cambia con la latitud\footnote{Algunos análisis sugieren que $\Delta z$ entre los niveles estudiados difiere entre el ecuador y el polo.} ni con el tiempo al comparar datos de Galileo y Venus Express. El valor más elevado de cizalla vertical se da para 45$^{\circ}$N en el caso de las imágenes de Galileo (Figura 3.6A), si bien el perfil es suave en general con valores de cizalla entre $\frac{\partial<u>}{\partial z}=1.4-4.4\cdot10^{-3}~s^{-1}$, dependiendo la separación vertical que hayamos considerado entre ambos niveles de la cima de las nubes. Las Figuras 3.6B y 3.6C nos muestran la cizalla entre UV-NIR y NIR-IR respectivamente, a partir de las medidas de viento con imágenes de Venus Express. Vemos que los valores de cizalla en el caso de los niveles de NIR e IR son muy peque\~{n}os, como cabría esperar debido a que la magnitud de los vientos en similar. El perfil para una separación vertical de 10 km es el que se ajusta mejor a los valores de cizalla medidos por la sonda Day de Pioneer Venus en 30$^{\circ}$S \cite{Gierasch1997}. En el caso de la cizalla entre los niveles UV y NIR, no aparece un máximo local en 45$^{\circ}$S (con Galileo aparecía para 45$^{\circ}$N) éste no es tan significativo como el del caso de las imágenes de Galileo, si bien el valor medio del cuadrado de la cizalla es similar y por debajo de $40\cdot10^{-6}~s^{-2}$.\\

Por otro lado, el \emph{número de Richardson}\index{Richardson!número de}, ($Ri$), nos aporta información sobre la estabilidad de la atmósfera ante diferentes tipos de inestabilidades y se define como:
\begin{equation}
	Ri(z)=\frac{N^{2}(z)}{\left(\partial u/\partial z\right)^{2}},
	\label{Richardson}
\end{equation}
donde $\partial u/\partial z$ es la cizalla vertical del viento y $N$ es la frecuencia de Brunt-Väisällä\footnote{Para estimar la frecuencia de Brunt-Väisällä usamos los valores medidos por las sondas Pioneer Venus y Venera \citep{Gierasch1997}.}. Para el intervalo de altura de interés en nuestro trabajo (55-65 km) podemos considerar que es razonable suponer que el perfil $N(z)$ apenas varía con el tiempo ni con la latitud \citep{Gierasch1997}, por lo que será la cizalla vertical (ver Figura 3.6) la que determine la variación del número de Richardson con la latitud. Tenemos que cuando $Ri<0$ (N es imaginario) existe una situación de inestabilidad convectiva, cuando $0<Ri<0.25$ te\-ne\-mos inestabilidad dinámica o de cizalla y para $Ri>1$ tenemos situaciones de estabilidad \citep{Wallace2006}. De esta manera, la atmosfera será más inestable en las latitudes en las que tengamos valores elevados de cizalla vertical (número de Richardson peque\~{n}o).\\

Si bien los datos de viento no permiten por sí solos estimar la estabilidad de forma precisa, el fuerte incremento que se produce en la cizalla vertical a 45$^{\circ}$N (ver Figura 3.6A) está próximo al límite externo del \textit{collar polar} y es una región donde son probables las inestabilidades verticales, tanto convectivas como dinámicas. Además, la banda de latitud 40$^{\circ}$-50$^{\circ}$N es también la región donde se produce el mayor cambio de magnitud en el perfil meridional del viento zonal (ver Figura 3.3), con cambios en la cizalla meridional que van desde $\frac{\partial<u>}{\partial y}\sim0~m\cdot s^{-1}\cdot km^{-1}$ (45$^{\circ}$-50$^{\circ}$N) hasta $\frac{\partial<u>}{\partial y}\sim0.03~m\cdot s^{-1}\cdot km^{-1}$ (50$^{\circ}$-90$^{\circ}$N).\\

\subsection{Comparación con el viento ciclostrófico}\label{subsection-ciclostrophic-wind}\indent

La región no ecuatorial de la atmósfera de Venus entre $\sim10$ km y la cima de las nubes está, de forma aproximada, en \textit{equilibrio ciclostrófico}\index{Equilibrio!ciclostrófico} \citep{Schubert1983}. Este consiste en un estado dinámico en el que la componente me\-ri\-dio\-nal de la fuerza centrífuga que actúa sobre una parcela de aire que rota zonalmente es equilibrada por la componente meridional de la fuerza debida al gradiente de presión. La existencia de dicho equilibrio en la atmósfera superrotante de Venus fue sugerida por ver primera por \citet{Leovy1973}, y ha sido explorada a través de medidas de vientos y gradientes horizontales de presión y temperatura tanto durante la misión Pioneer Venus \citep{Seiff1980,Schubert1983} como en la misión Venus Express \citep{Piccialli2008}. Si la atmósfera está en equibrio ciclostrófico entonces tenemos que:
\begin{equation}
	\frac{u^{2}\cdot\tan\phi}{a}=-\frac{1}{\rho}\cdot\frac{\partial P}{\partial y}.
	\label{Cyclostrophic-balance}
\end{equation}
donde $u$ es la componente zonal de la velocidad del viento, $a$ es el radio de giro y $\phi$ es la latitud. El término de la izquierda representa la componente meridional (hacia el ecuador) de la fuerza centrífuga y el término de la derecha es la componente meridional (hacia el polo) de la fuerza debida al gradiente de presión (ver Figura 3.7).\\

\begin{figure}[h!]
	\centering
		\includegraphics[width=1.0\textwidth]{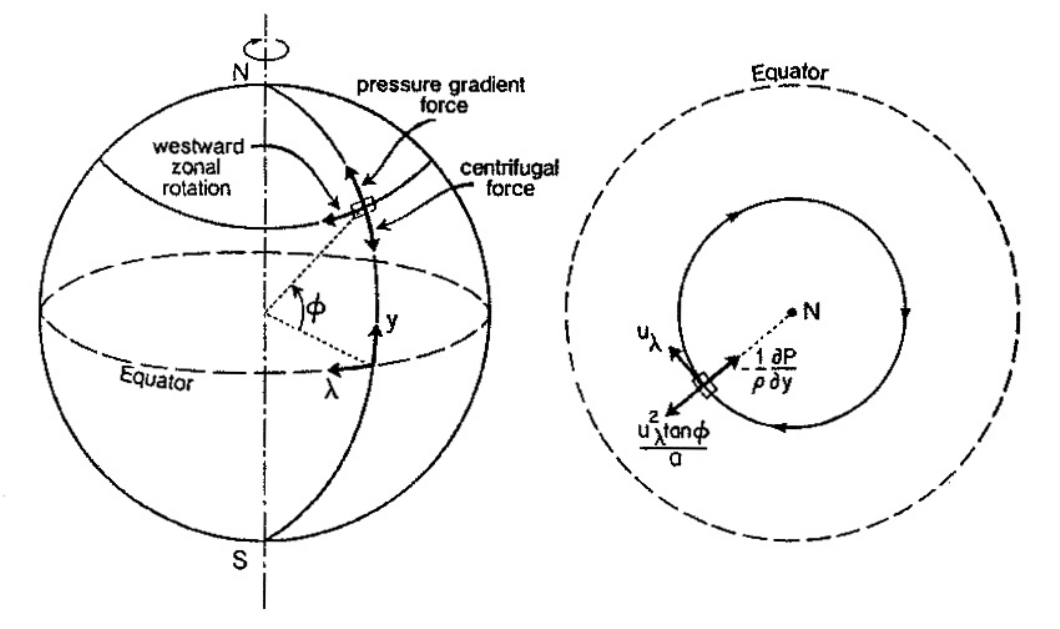}
	\label{fig:Cyclostrophic-Balance}
	\caption[Esquema ilustrativo del equilibrio Ciclostrófico.]{\scriptsize{Esquema ilustrativo de las fuerzas que actúan sobre el viento zonal de Venus en el caso de equilibrio ciclostrófico (ver Figura 14 en \citealt*{Schubert1983}).}}
\end{figure}

Si consideramos también que la atmósfera se comporta como un gas ideal y que está en equibrio hidrostático\index{Equilibrio!hidrostático}, entonces se puede llegar a la siguiente expresión (ver \citealt*{Schubert1983}):
\begin{equation}
	\frac{\partial u^{2}}{\partial\zeta}=-\frac{R}{\tan\phi}\cdot\frac{\partial T}{\partial\phi}.
	\label{Thermal-wind-eq}
\end{equation}
donde $\zeta=-\ln\frac{P}{P_{ref}}$ es la \textit{coordenada de presión normalizada} ($P_{ref}$ es una presión de referencia), $R$ es la constante de los gases en Venus y $T$ es la temperatura. Esta relación es la que conocemos como \textit{ecuación del viento térmico para el equilibrio ciclostrófico}\index{Viento Térmico!ecuación del} \citep{Schubert1983,Holton1992}, y nos dice que la derivada vertical de la energía cinética por unidad de masa es directamente proporcional al gradiente meridional de temperatura sobre una superficie de presión constante. Así, si la temperatura decrece hacia el polo, entonces el viento zonal aumenta con la altura, y viceversa.\\

Conocida la distribución vertical y horizontal de temperatura (y considerando un perfil de vientos correspondiente a un nivel dado) podemos integrar verticalmente la ecuación (\ref{Thermal-wind-eq}) para obtener la componente zonal del viento en diferentes alturas. Asímismo, los vientos obtenidos con este método pueden ser comparados con las medidas de viento a través de otras técnicas, examinando el rango de validez de la aproximación ciclostrófica en Venus. Diferentes perfiles de viento han sido obtenidos de esta manera u\-san\-do datos de presión y temperatura de la misión Pioneer Venus \citep{Schubert1983,Newman1984}, Galileo \citep{Roos-Serote1995} y Venus Express \citep{Piccialli2008}, y obteniendo en general un buen acuerdo en latitudes medias con otras medidas de vientos. Con este procedimiento se deduce que por encima de la cima de las nubes el viento zonal decrece con la altura \citep{Newman1984,Roos-Serote1995,Piccialli2008}.\\

En la Figura 3.8 comparamos nuestro perfil meridional de viento zonal en la cima de las nubes a partir de imágenes de Galileo, con el perfil de viento obtenido por \citet{Newman1984} usando datos térmicos de Pioneer Venus y suponiendo válida la aproximación de equilibrio ciclostrófico. Resulta destacable la presencia de una fuerte corriente en chorro a $\sim$50$^{\circ}$ de latitud, ausente en las medidas de viento usando seguimiento de estructuras nubosas. Se ha encontrado que dicha corriente en chorro ciclostrófica varía tanto en magnitud ($100-160~m\cdot s^{-1}$), como en altura ($z\sim65-70$ km) y latitud (45$^{\circ}$-65$^{\circ}$). Asímismo, la hora local parece influir en su posición \citep{Piccialli2008}, y el intervalo temporal sobre el que se promedian los datos térmicos parece afectar a la intensidad de la corriente en chorro, siendo ésta menor cuanto mayor sea el intervalo temporal \citep{Newman1984}. La existencia de esta corriente en chorro es atribuida al collar polar\index{Collar polar}, una zona de aire frío que rodea a la región polar y que da lugar a una inversión térmica vertical a la altura de las nubes \citep{Newman1984,Taylor1985,Roos-Serote1995,Piccialli2008}.\\

\begin{figure}[h!]
	\centering
		\includegraphics[width=0.8\textwidth]{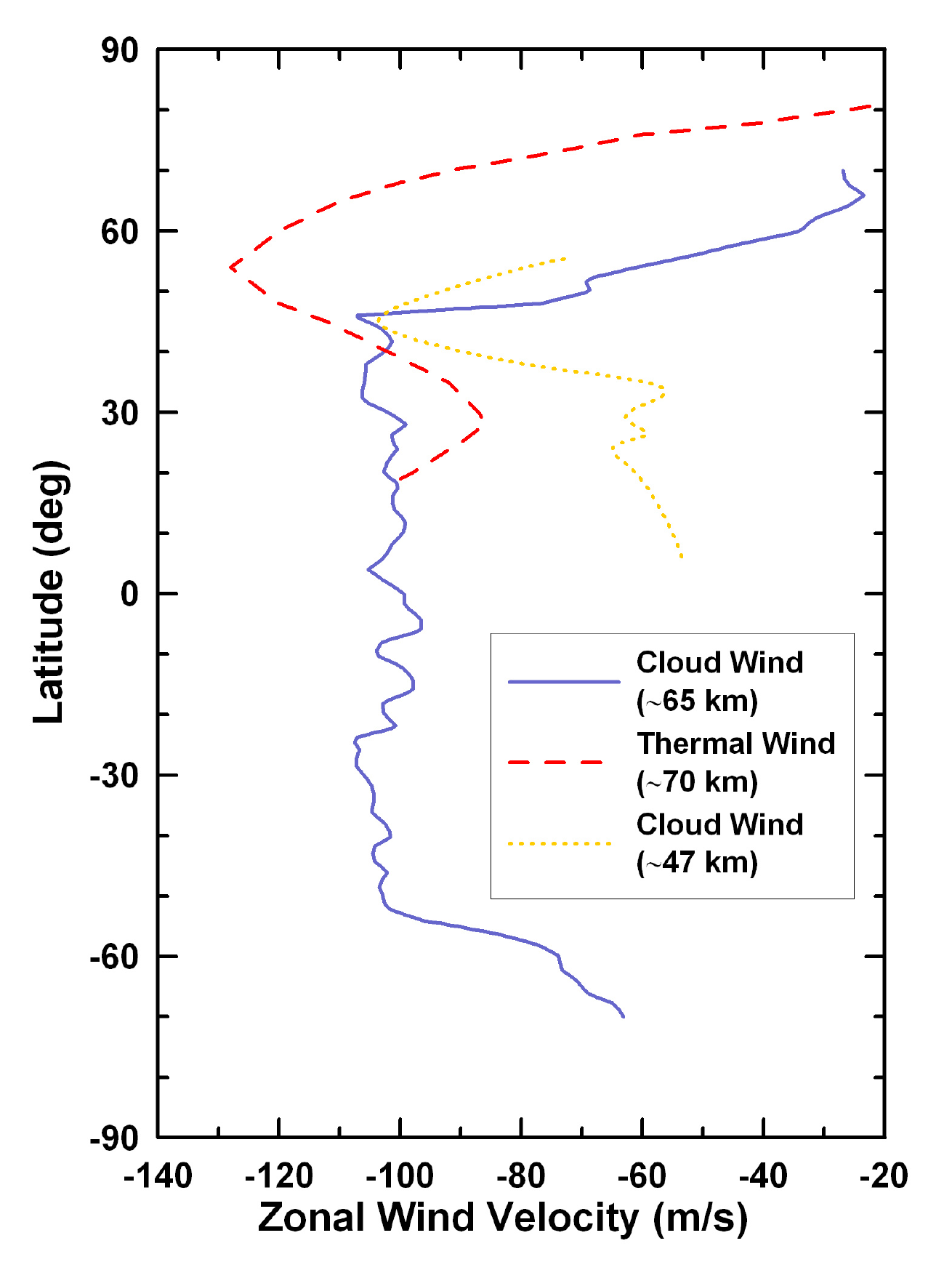}
	\label{fig:Cyclostrophic-Balance}
	\caption[Comparación con el viento zonal ciclostrófico.]{\scriptsize{Comparación entre vientos ciclostróficos y vientos obtenidos con seguimiento de trazadores nubosos. La línea continua violeta representa medidas de viento a partir de trazadores nubosos en la cima de las nubes durante el día Galileo-SSI), la línea en puntos suspensivos amarilla representa el mismo tipo de medidas pero en las nubes inferiores durante la noche (Galileo-NIMS), y la línea a trazos roja el perfil de viento en el lado nocturno a partir de datos térmicos de radio ocultación (Pioneer Venus), usando la ecuación del viento térmico y suponiendo válido el equilibrio ciclostrófico.}}
\end{figure}

Dado que los datos térmicos necesarios para calcular el viento ciclostrófico se han obtenido siempre en el lado nocturno del planeta y los vientos en las nubes se determinan analizando imágenes del lado diurno, es posible obtener cierta discrepancia entre los resultados obtenidos usando ambos métodos. Por ejemplo, es posible que la inversión térmica del collar polar sea menos acusada durante el día (la atmósfera en la cima de las nubes puede enfriarse hasta 15$^{\circ}$K durante la noche, \citealt*{Piccialli2008}), lo que en principio podría explicar la inexistencia de la corriente en chorro en nuestro perfil de Galileo correspondiente a observaciones diurnas en la cima de las nubes. Dicha discrepancia también podría identificarse con las regiones donde se incumple el equilibrio ciclostrófico. Durante la misión Pioneer Venus se dedujo a partir de datos térmicos promediados que el equilibrio ciclostrófico se rompe en la región ecuatorial y a partir de los 70-75 km de altura para latitudes menores que 40$^{\circ}$ y mayores que 70$^{\circ}$ (ver Figura 1 en \citealt*{Taylor1985}). No obstante, \citet{Roos-Serote1995} siembran la duda sobre las regiones de validez de dicho equilibrio, apoyándose en sus propios resultados y en que las condiciones de presión atmosférica parecen variar de a\~{n}o en a\~{n}o de manera importante \citep{Clancy1991}. En todo caso, el uso de una ecuación del viento térmico más general \citep{Li2008} para extraer los vientos zonales a partir del campo de temperaturas podría en un futuro eludir las limitaciones de aplicabilidad del equilibrio ciclostrófico descrito por la ecuación (\ref{Thermal-wind-eq}).\\

\section{Vientos Meridionales en Venus}\label{chapter-winds-meridwind}\indent

La Figura 3.9 muestra el perfil promediado de la componente meridional de la velocidad del viento a partir de imágenes tomadas por la cámara SSI de Galileo (Figura 3.9A) y VIRTIS de Venus Express (Figura 3.9B).\\

En el caso del sobrevuelo de la nave Galileo, observamos que la componente meridional es hacia el polo en ambos hemisferios, con el perfil me\-ri\-dio\-nal simétrico respecto del ecuador e incrementándose de 0 a $10~m\cdot s^{-1}$ desde el ecuador hasta los 45$^{\circ}$ donde se localiza el \textit{collar polar}. Las medidas de los vientos meridionales en las imágenes NIR tienen un error asociado del orden de su máximo valor ($10~m\cdot s^{-1}$), lo que impide evaluar con exactitud la circulación meridional. Se representan además las velocidades de las imágenes en violeta y el perfil que extrajo \citet{Belton1991}, comprobándose que existe buen acuerdo entre sus resultados y los nuestros.\\

\begin{figure}[h!]
	\centering
		\includegraphics[width=1.0\textwidth]{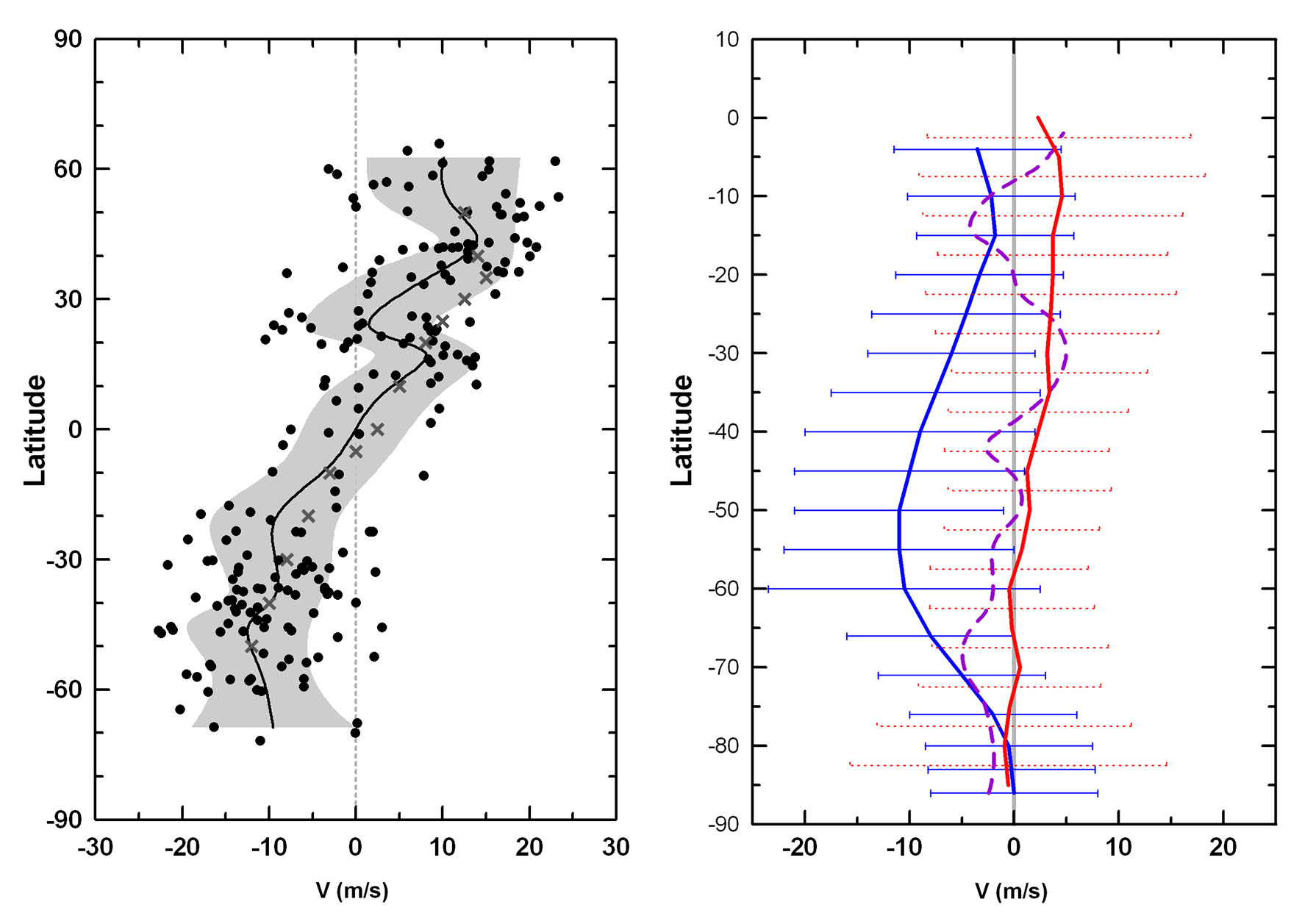}
	\label{fig:Venus-MeridionalWind}
	\caption[Viento meridional a distintas alturas con Galileo y VIRTIS-M.]{\scriptsize{Perfil promediado de los Vientos Meridionales. \textbf{(A)} nos muestra los resultados obtenidos con imágenes de Galileo de la cima de las nubes, representando nuestras medidas individuales (puntos), el promedio de éstas (línea continua) y las de \citet{Belton1991} (cruces). \textbf{(B)} nos ense\~{n}a el perfil promediado de la velocidad zonal a partir de las imágenes de VIRTIS en la cima de las nubes (línea azul, UV), base de la cima de las nubes (línea púrpura, NIR) y nubes profundas (línea roja, IR).}}
\end{figure}

En la Figura 3.9B podemos ver el perfil meridional deducido de las imáge\-nes de VIRTIS para los tres niveles de altura que estamos estudiando. Si bien seguimos teniendo errores del mismo orden que nuestras medidas ($9~m\cdot s^{-1}$ para todas las longitudes de onda), en el caso de las imágenes UV ($z\sim66$ km) podemos discernir que la componente meridional aumenta de $0~m\cdot s^{-1}$ en el ecuador a unos $10~m\cdot s^{-1}$ en 55$^{\circ}$S, disminuyendo a partir de entonces hasta anularse en el polo sur. Este patrón podría indicar la presencia de la rama superior de una \textit{célula de Hadley}\index{Hadley!célula de} \citep{Schubert1983,Gierasch1997} y el acusado descenso hacia el polo tras el máximo de velocidad a 55$^{\circ}$S es más probable que se deba a la acción del vórtice polar. Para las imágenes NIR ($z\sim66$ km) e IR ($z\sim47$ km) el valor medio de la componente me\-ri\-dio\-nal está por debajo de los $5~m\cdot s^{-1}$, por lo que no es posible extraer ningún patrón claro de tendencia debido al error de medida ($\sim9~m\cdot s^{-1}$). Todos estos resultados están en concordancia con los perfiles de componente me\-ri\-dio\-nal de misiones extraídos en la época de Pioneer Venus \citep{Limaye2007} y Galileo \citep{Belton1991,Peralta2007b}.\\

\section{Variabilidad de los vientos}

\subsection{Evolución Temporal de la superrotación}\indent

La Figura 3.10 muestra la evolución temporal de los vientos zonales en Venus a largo plazo, comparando para ello los perfiles de velocidad (promediados en longitud y temporalmente) obtenidos en las misiones espaciales Pioneer Venus (perfil promediado durante meses; \citealt*{Limaye2007}), Galileo (perfil promediado durante días; \citealt*{Peralta2007b,Carlson1991}) y Venus Express (perfil promediado durante meses; \citealt*{Sanchez-Lavega2008}). Para todos los niveles verticales estudiados observamos que el patrón general del perfil de los vientos (velocidad aproximadamente constante para latitudes medias y ecuatoriales, y caída de los vientos hacia los polos, ver sección \ref{chapter-winds-zonalwind}) se mantiene presente en todas las épocas.\\

\begin{figure}[h!]
	\centering
		\includegraphics[width=1.0\textwidth]{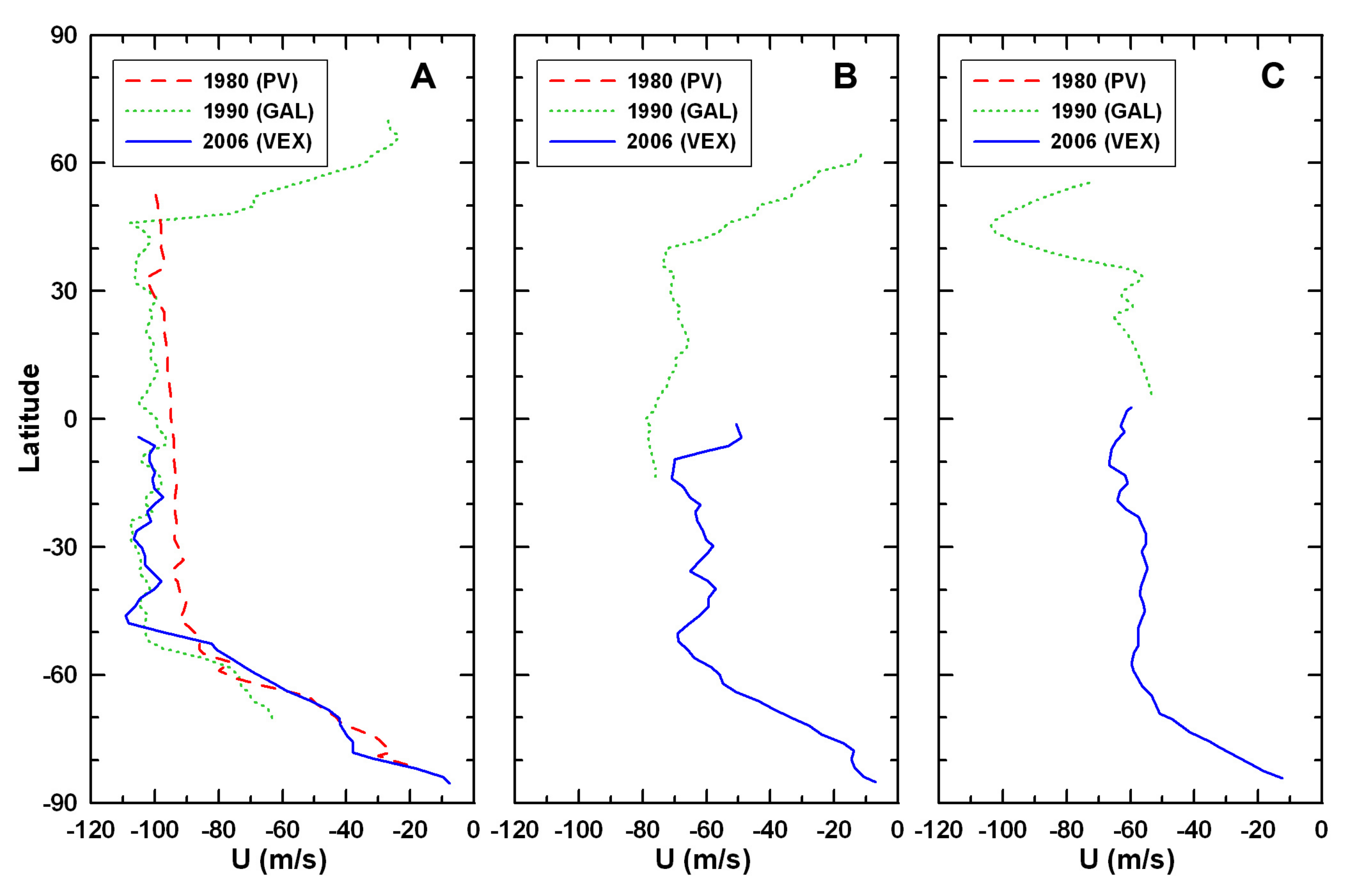}
	\label{fig:LongTermEvol-ZonalWind}
	\caption[Evolución temporal a largo plazo del viento zonal.]{\scriptsize{Evolución a largo plazo del perfil meridional del viento zonal en la atmósfera de Venus. La gráfica \textbf{A} nos muestra la evolución en la cima de las nubes ($\sim66$ km, UV), \textbf{B} los vientos en la base de la cima de las nubes ($\sim61$ km, NIR) y \textbf{C} los vientos en la base de las nubes inferiores ($\sim47$ km, IR). En rojo los datos de la cámara OCPP de Pioneer Venus \citep{Limaye2007}, en verde los de las cámaras SSI (UV y NIR) y NIMS (IR) de Galileo \citep{Peralta2007b}, y en azul los de VIRTIS en Venus Express \citep{Sanchez-Lavega2008}.}}
\end{figure}

El perfil correspondiente a la cima de las nubes parece ser el más estable en el tiempo, tanto en la región de caída como en la de velocidad constante. No obstante, la magnitud del viento zonal durante 1980 parece ser ligeramente inferior (10\%). Las medidas mostradas también parecen rebatir la existencia de las corrientes en chorro detectadas en latitudes medias con análisis anteriores de las imágenes de Mariner 10 y Pioneer Venus \citep{Rossow1980,Limaye1981,Schubert1983}. Por otro lado, los perfiles de velocidad zonal para la base de la cima de las nubes (NIR) y las nubes inferiores (IR) (Figuras 3.10B y 3.10C) no son fácilmente comparables en el tiempo ya que las imágenes de la misión Galileo se centraron en el hemisferio norte mientras que Venus Express cubre esencialmente el hemisferio sur de Venus. Sin embargo, en el caso de las imágenes de NIR existe un de\-cre\-ci\-miento aparente de la magnitud del viento ($\sim10~m\cdot s^{-1}$) entre 1990 y 2006, hecho que podría deberse a una alteración del perfil vertical del viento zonal o a otras causas que veremos a continuación. Las mayores discrepancias las tenemos para el caso de los perfiles en la parte baja de las nubes, donde \citet{Carlson1991} obtuvieron con el instrumento NIMS\index{NIMS} de Galileo una distribución latitudinal de vientos más acorde con la que se obtiene de forma indirecta con la ecuación del viento térmico y bajo la hipótesis de equilibrio ciclostrófico (ver discusión en la subsección \ref{subsection-ciclostrophic-wind}).\\

\begin{figure}[h!]
	\centering
		\includegraphics[width=0.9\textwidth]{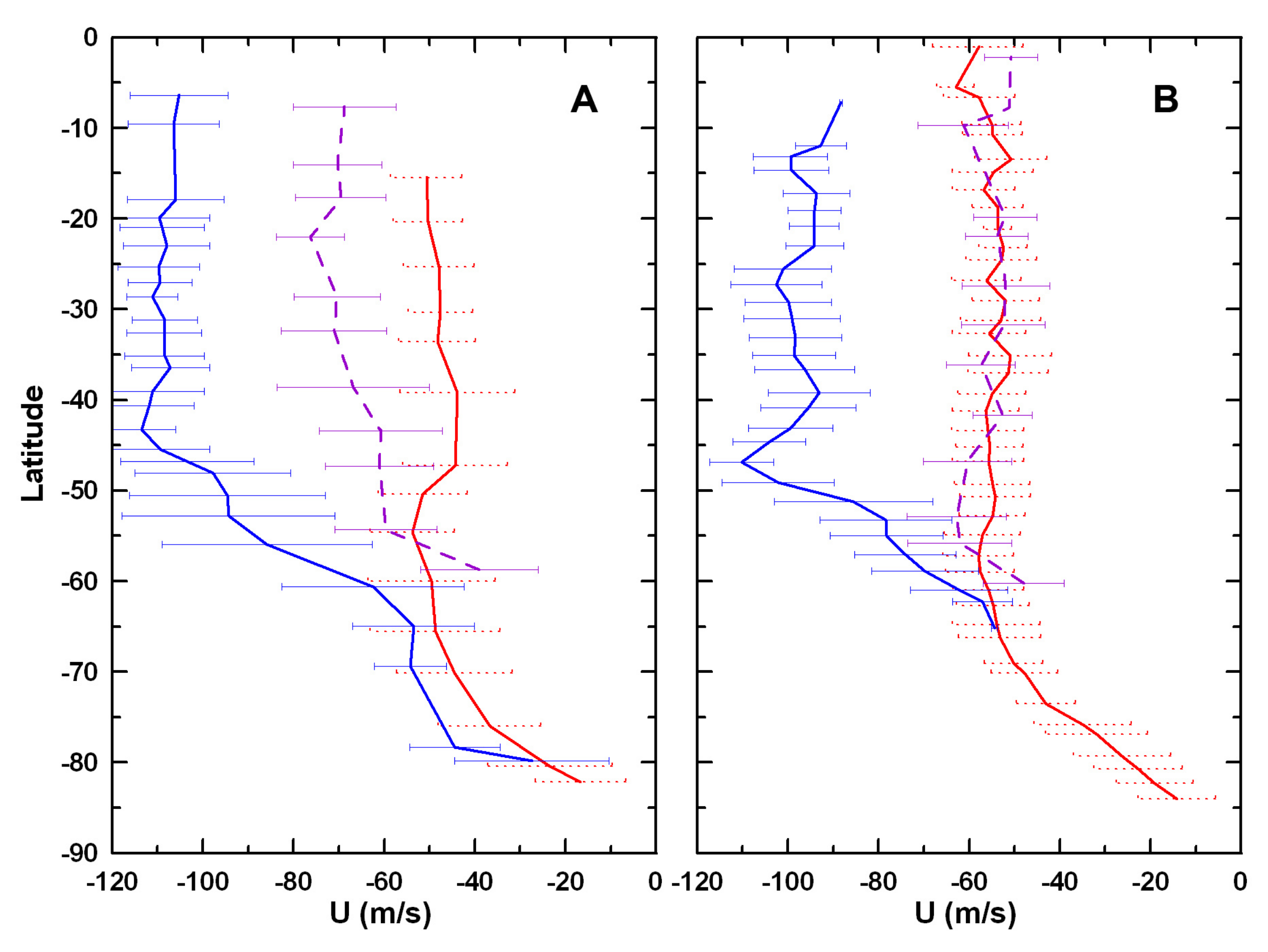}
	\label{fig:ShortTermEvol-ZonalWind}
	\caption[Evolución temporal a corto plazo del viento zonal.]{\scriptsize{Evolución a corto plazo del perfil meridional del viento zonal en la atmósfera de Venus \citep{Sanchez-Lavega2008}. La gráfica \textbf{A} muestra el viento durante la fase de inserción orbital (VOI) de Venus Express, realizada entre en abril de 2006. La gráfica \textbf{B} nos ense\~{n}a el viento durante los 3 meses siguientes (MTP001,MTP002 y MTP003). La línea azul hace alusión a los vientos en la cima de las nubes ($\sim66$ km, UV), la púrpura a la base de la cima de las nubes ($\sim61$ km, NIR) y la roja a las nubes inferiores.}}
\end{figure}

La variabilidad de la intensidad del viento no está únicamente presente al comparar datos de misiones espaciales separadas por décadas sino también al analizar los datos de VIRTIS-M en diferentes periodos de tiempo. En la Figura 3.11 se comparan los perfiles de viento zonal en los tres niveles de altura obtenidos con las imágenes de VIRTIS durante la fase de inserción orbital (VOI) y durante las órbitas efectuadas en los tres meses siguientes (MTP001, MTP002 y MTP003). Lo más llamativo es el cambio de la magnitud del viento en las imágenes NIR, con vientos superiores a los de las nubes inferiores durante la inserción orbital, y coincidentes posteriormente. Esta disminución de los vientos es del mismo orden que la que observamos entre la época de Galileo (1990) y Venus Express (2006), lo que podría estar indicando que la transición entre regímenes estables en el tiempo\footnote{\citet{delGenio1990} concluyeron que a estos niveles verticales existe un ciclo en la dinámica de 5-10 a\~{n}os relacionado con la aparición y desaparición de la onda Kelvin ecuatorial y con disminuciones del viento de $5-10~m\cdot s^{-1}$.} (del orden de a\~{n}os) es bastante rápida (menos de un mes). Frente a un cambio real en la dinámica, la brevedad con la que se produce este cambio indica, como hipótesis más probable, una variación del nivel de altura que se vi\-sua\-liza en las imágenes de longitudes de onda de NIR debido a cambios en la composición atmosférica y en el espesor óptico. No obstante, esta cuestión queda abierta y requiere el análisis de un mayor conjunto de datos.\\

\subsection{Influencia de la Marea Térmica}\indent

La \textit{marea térmica solar}\index{Marea térmica solar} constituye un fenómeno a tener en cuenta a la hora de estudiar la dinámica de la atmósfera de Venus. Las mareas térmicas atmosféricas son ondas de escala planetaria forzadas por el gradiente de absorción solar que existe entre la partes diurna y nocturna del planeta \citep{Pechmann1984}.\\

En la Figura 3.12 mostramos el efecto que tiene la marea térmica solar sobre la componente zonal del viento en la cima de las nubes ($\sim66$ km). Su influencia sobre el viento es evidente en el rango de latitudes comprendido entre 50$^{\circ}$S y 75$^{\circ}$S, donde lo acelera a un ritmo de $-2.5\pm0.5~m\cdot s^{-1}\cdot hr^{-1}$ de la ma\~{n}ana (9 horas en hora local) a la tarde (15 horas). No observamos esta aceleración en el resto de latitudes\footnote{Los resultados preliminares con la cámara VMC de Venus Express tampoco detectan efectos de marea térmica fuera del rango de latitudes mencionado \citep{Markiewicz2007b}.}. Esto entra en contradicción con la detección positiva en latitudes ecuatoriales con los datos de la misión Pioneer Venus \citep{delGenio1990}, diferencia que podría explicarse si te\-nemos en cuenta la alta dispersión y escaso número de medidas de velocidad de los datos de VIRTIS-M y VMC en bajas latitudes. También merece la pena mencionar que la amplitud de onda determinada ($\sim10$ m/s) coincide con la estimada durante la misión Pioneer Venus \citep{delGenio1990,Limaye2007} y Galileo \citep{Toigo1994}. Por otro lado, dentro del error de nuestras medidas no encontramos prueba alguna de la marea térmica ni en la base de la cima de las nubes ($\sim61$ km), ni en las nubes inferiores ($\sim47$ km) de la parte nocturna.\\

\begin{figure}[h!]
	\centering
		\includegraphics[width=0.6\textwidth]{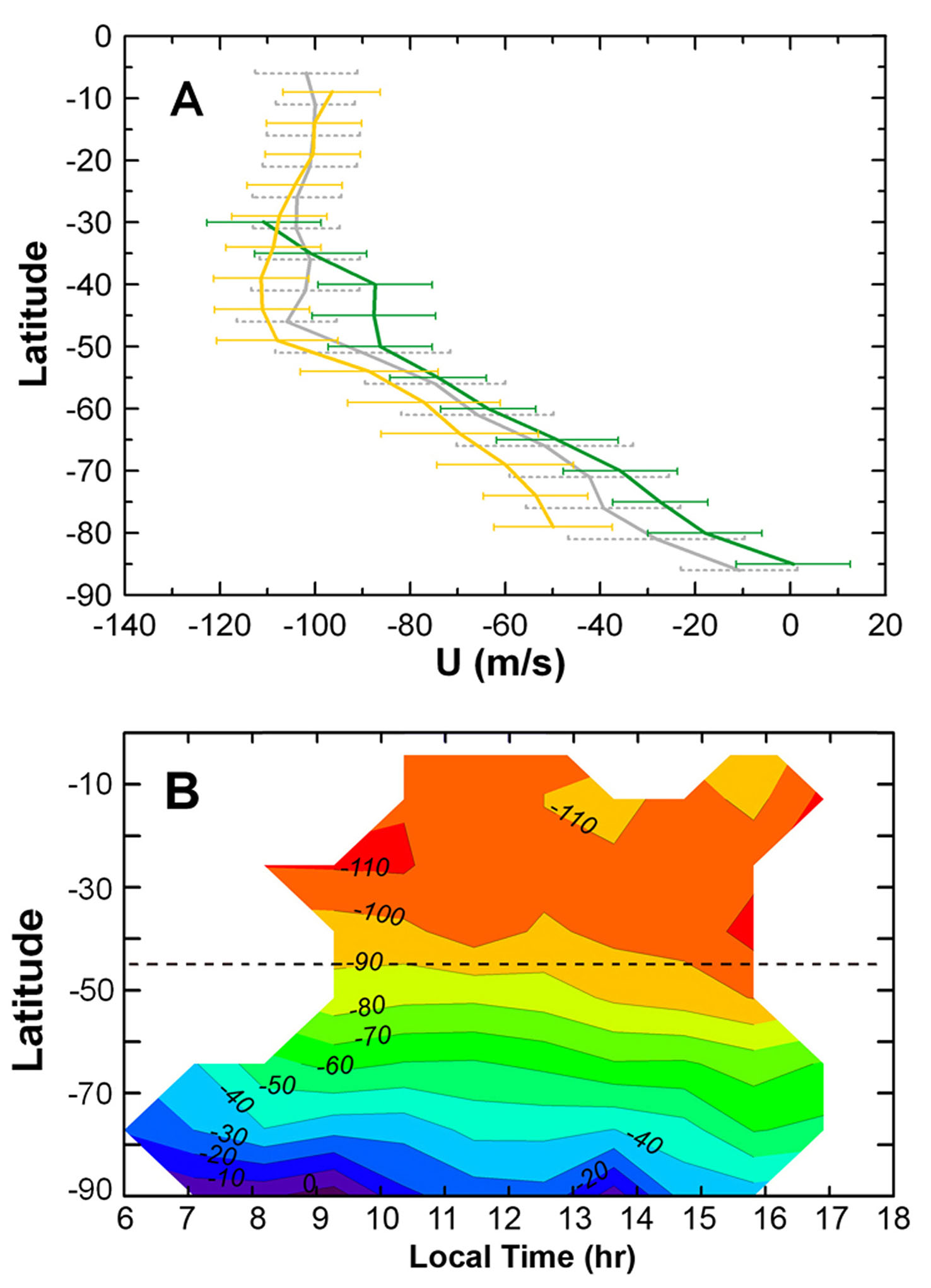}
	\label{fig:Venus-SolarTide}
	\caption[Efectos de la Marea Térmica sobre el viento zonal.]{\scriptsize{Efectos de la Marea Térmica sobre el viento zonal en la cima de las nubes ($380~nm$). En \textbf{A} tenemos cómo varían los vientos con la hora local. La curva verde representa el perfil de viento zonal para las $9\pm0.8$ horas, la curva amarilla el viento para las $15\pm0.5$ horas y la curva gris el viento para horas locales intermedias. En \textbf{B} visualizamos los efectos de la marea térmica representando contornos de velocidad constante en función de la latitud y la hora local.}}
\end{figure}

Es probable que las mareas térmicas jueguen un papel importante en el transporte de momento angular en la estratosfera de Venus, razón por la que se han llevado a cabo numerosos esfuerzos para caracterizar este fenómeno tanto teórica y numéricamente \citep{Fels1986,Leovy1987,Hou1990,Newman1992,Takagi2005,Takagi2006}. Así, en caso de que la marea semidiurna llegara a la superficie de Venus, la amplitud y la fase de las oscilaciones provocadas en la presión nos permitirían estimar el momento que ésta ejerce sobre la atmósfera \citep{Dobrovolskis1980}.\\

\subsection{Oscilación global de 5 días}\indent

A partir de los datos del viento es posible extraer información sobre los distintos modos de ondas de escala planetaria presentes en el nivel de las nubes. En nuestro caso, usando las imágenes de VIRTIS-M de la cima de las nubes ($380~nm$, $z\sim65$ km) hemos detectado una oscilación de escala global con un periodo de aproximado de 5 días y que afecta a las latitudes subpolares de Venus. En la Figura 3.13A se dispone el periodograma del viento promediado para cada día en el periodo de tiempo para el que hemos realizado medidas en UV (desde el 19 de febrero de 2006 hasta el 28 de enero de 2007, ver Tabla 3.2) y para los rangos de latitud 50$^{\circ}$-60$^{\circ}$S y 60$^{\circ}$-70$^{\circ}$S, ya que en otras latitudes no hay suficientes datos para llevar a cabo un análisis comparativo. En ésta se aprecia en ambos rangos de latitud un pico para un periodo de 4.9 días.\\

Supongamos ahora que tomamos todo el periodo de observación y a cada intervalo de 5 días asignamos unas coordenadas de fase entre 0$^{\circ}$ y 360$^{\circ}$ (la fase para una oscilación de 5 días). Así, para cada día que tengamos mediciones de viento asociamos un valor de fase dependiendo del día y la hora corres\-pon\-dien\-tes dentro del ciclo. De esta manera obtenemos un diagrama como el de la Figura 3.13B, donde podemos observar con claridad que nuestros datos se ajustan aceptablemente a una oscilación de 5 días. La amplitud de la oscilación es de $\sim12~m\cdot s^{-1}$ y esta onda podría estar relacionada con la onda ecuatorial de 4 días detectada ocasionalmente durante la misión Pioneer Venus a partir de datos de velocidad \cite{Rossow1990}, del estudio de las periodicidades de brillo en imágenes UV \citep{delGenio1982,delGenio1990} y en análisis detallados de los vientos zonales durante el sobrevuelo de Galileo \citep{Smith1996}.\\

\begin{figure}[h!]
	\centering
		\includegraphics[width=0.65\textwidth]{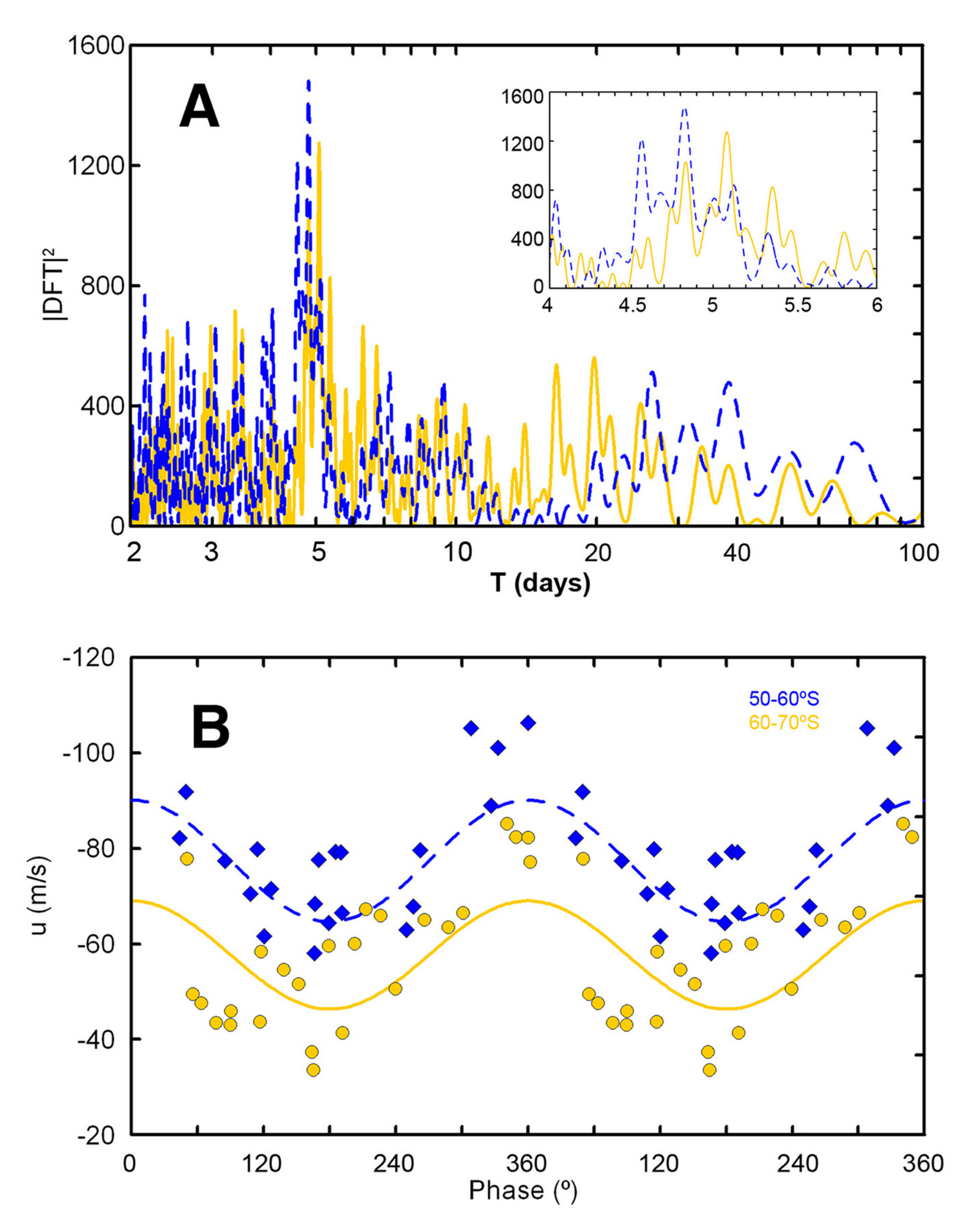}
	\label{fig:VEXVenus-5day}
	\caption[Oscilación de 5 días en el viento zonal.]{\scriptsize{Variaciones presentes en los datos de velocidad obtenidos con las imágenes de $380~nm$ de VIRTIS en Venus Express. En la fig.\textbf{A} se representa el periodograma de los vientos medios para 50-60$^{\circ}$S (línea cursiva azul) y 60-70$^{\circ}$S (línea continua amarilla). El periodograma muestra picos para un periodo de 4.9 días. En la fig.\textbf{(B)} se muestra la variación de los vientos medios para 50-60$^{\circ}$S (diamantes azules) y 60-70$^{\circ}$S (círculos amarillos) en términos de la fase para un periodo de oscilación de 5 días.}}
\end{figure}

\chapter{Turbulencia en las nubes}\label{chapter-turbulence}

\section{Introducción}\label{chapter-turbulence-intro}\indent

La atmósfera de Venus está permanentemente cubierta por nubes. En el capítulo \ref{chapter-winds} estudiamos los movimientos de estas nubes para determinar la dinámica general de la atmósfera. En este capítulo vamos a estudiar la distribución de brillo de las nubes para abordar una posible evaluación de las características de la turbulencia atmosférica en Venus.\\

En el caso de la Tierra, la turbulencia atmosférica es de tipo cuasigeostrófico y constituye un elemento clave en los procesos de circulación general ya que está ligada a la formación del flujo zonal del viento y de las corrientes en chorro, así como al modo en que la energía de origen baroclínico se distribuye en las distintas escalas espaciales de la atmósfera \citep{Vallis2006}. En el caso de la atmósfera de Venus (un planeta lentamente rotante), la turbulencia es de carácter distinto (interaccionan entre sí ondas, inestabilidades y convección que producen fluctuaciones en el viento rápidamente variables pero organizadas) aunque puede jugar un papel esencial a la hora de distribuir la energía cinética atmosférica de unas escalas a otras, pudiendo contribuir a la superrotación global de la atmósfera \citep{Rossow1979,Zhu2006}. En efecto, los fenómenos turbillonarios producto de la turbulencia bidimensional son teóricamente capaces de transportar momento del flujo zonal de latitudes medias hacia latitudes ecuatoriales \citep{Gierasch1997}. Antes de abordar la caracterización de la turbulencia en la atmósfera de Venus, presentaré brevemente algunos aspectos generales de la turbulencia y el estado de los estudios sobre turbulencia atmosférica aplicadas al caso terrestre y al de la atmósfera de Venus.\\

\subsection{Aspectos Generales de la Turbulencia}\indent

La turbulencia está presente en la mayor parte de los fluidos en movimiento que encontramos en la naturaleza, desde las grandes corrientes oceánicas hasta la capa límite planetaria terrestre. Paradójicamente, no existe una forma sa\-tisfactoria de definir la turbulencia. Para su descripción se suele a recurrir a un parámetro adimensional denominado \textit{número de Reynolds}\index{Reynolds!número de} y que se define como $\left(\frac{U\cdot L}{\nu}\right)$, donde $L$ es una escala representativa de longitud, $U$ es la velocidad promedio y $\nu$ es la viscosidad dinámica. Es decir, el \textit{número de Reynolds} no es otra cosa que la razón entre las fuerzas inerciales y las fuerzas viscosas, lo que ayuda a caracterizar los diferentes regímenes en un fluido. Así, para números de Reynold peque\~{n}os las fuerzas viscosas son las que do\-minan y tendremos un \textit{flujo laminar}\index{Flujo!laminar}, caracterizado por un flujo ``suave'' y cons\-tante, mientras que para números de Reynold elevados tendremos un \textit{flujo turbulento}\index{Flujo!turbulento}, con un sistema dominado por las fuerzas iner\-ciales, que tienden a causar de forma aleatoria fenómenos turbillona\-rios, vórtices y otras fluctuaciones. Se suele definir \textit{turbulencia}\index{Turbulencia} como la situación en la que un fluido que posee un \textit{número de Reynolds} elevado, altamente no lineal y ca\-rac\-te\-rizado por ser aparentemente impredecible y caótico \citep{Vallis2006}. Estas propiedades favorecen la difusión tanto de calor como de momento en los flui\-dos turbulentos tridimensionales. Asímismo, los fenómenos turbillonarios se presentan en un amplio rango de escalas espaciales, viniendo determinadas las más grandes por el fenómeno que esté perturbando la atmósfera y las más peque\~{n}as por la viscosidad.\\

Las características generales de un fluido turbulento se suelen estudiar a través del denominado \textit{espectro de potencias de energía cinética}\index{Espectro de potencias!de energía cinética}, que cons\-tituye una representación de la distribución de energía cinética $E(k)$ del sistema en las diferentes escalas espaciales (dadas por un número de onda $k$). En el caso de un régimen turbulento tridimensional cuasi-isótropo y estático con unas características energéticas específicas\footnote{Inyección de energía en las escalas grandes y transporte de ésta hacia las escalas peque\~{n}as donde finalmente se disipe por viscosidad.}, \citet{Kolmogorov1941} demostró en base a criterios dimensionales e hipótesis físicas razonables que el espectro de potencias de energía cinética se ajusta a la siguiente ley de potencias:
\begin{equation}
	E(k) = \zeta\cdot\epsilon^{2/3}\cdot k^{-5/3},
	\label{Kolmogorov-Law}
\end{equation}
llamada también \emph{ley de Kolmogorov}\index{Kolmogorov!ley de}. En ella $E(k)$ es la densidad espectral de energía, es decir la energía por unidad de masa y número de onda ($m^{3}\cdot s^{-2}$), $\epsilon$ es el flujo de energía de unas escalas a otras, con unidades de energía por unidad de masa y unidad de tiempo ($m^{2}\cdot s{-3}$) y $\zeta$ es una constante adimensional cuyo valor ($\zeta\approx1.5$) se determina experimentalmente \citep{Vallis2006}. Como puede verse en la Figura 4.1, dentro del espectro de potencias podemos distinguir varias regiones que se corresponden con diferentes escalas de interés \citep{Kundu2008}. Debido a que la energía se inyecta en escalas grandes ($L$) para $k\simeq L^{-1}$ tenemos que $E(k)$ se maximiza. La escala espacial $\eta$ a partir de la cual comienza a ser importante la disipación viscosa de energía se denomina \textit{microescala de Kolmogorov}\index{Kolmogorov!microescala de}. Para escalas comparables o inferiores a $\eta$ se produce una fuerte caída en $E(k)$. Si la escala a la que se inyecta energía al sistema es lo suficientemente grande, existe un rango de escalas que es intermedio entre la escala grande y las escalas de disipación y donde ni la inyección de energía ni la disipación son factores importantes en la dinámica. Dicho rango es conocido como el \emph{rango inercial}\index{Rango!inercial} porque en él los términos dominantes son los inerciales y sólo tenemos transferencia de energía de forma inercial (principalmente por estiramiento de vórtices). \citet{Kolmogorov1941} demostró que, si bien en el \textit{subrango inercial} tanto la disipación como la inyección de energía son despreciables, la transferencia de energía de unas escalas a otras depende del ritmo en que la energía es disipada por las escalas más peque\~{n}as ($\epsilon$), que coincide con el ritmo con el que se inyecta energía al sistema en las escalas grandes.\\

\begin{figure}[h!]
	\centering
		\includegraphics[width=0.7\textwidth]{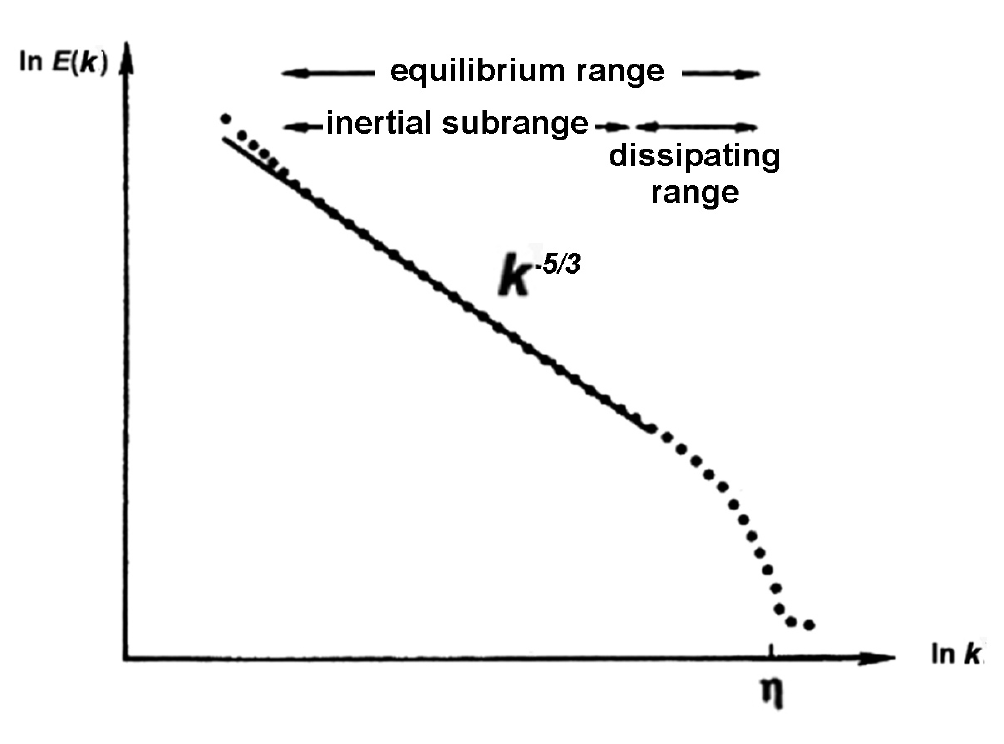}
	\label{fig:3D-Turbulence}
	\caption[Espectro de potencias clásico para turbulencia 3D.]{\scriptsize{Espectro de potencias clásico predicho por Kolmogorov para la turbulencia isótropa en 3D.}}
\end{figure}

Trasladémonos ahora al caso de fluidos atmosféricos, en los que la turbulencia de gran escala es esen\-cialmente bidimensional y donde además de la energía juega un papel fundamental otra variable dinámica: la enstrofía (esencialmente el cuadrado de la vorticidad, $s^{-2}$). La enstrofía\index{Enstrofía} es una magnitud que está directamente relacionada con la energía cinética asociada a los efectos de disipación dentro de un fluido \citep{Weiss1991}. En un sistema no disipativo tanto la energía total como la enstrofía deben conservarse. \citet{Kraichnan1967} de\-sarro\-lló una teoría para el caso de turbulencia en dos dimensiones para la que una fuerza cualquiera concentra energía en un rango determinado de números de onda, dentro del cual inyectará al sistema tanto energía como enstrofía (ver Figura 4.2). En dicha teoría se demuestra que, en ausencia de estiramiento de vórtices, la energía inyectada en una escala determinada se transporta de escalas peque\~{n}as a grandes\footnote{\citet{Fjortoft1953} ya había demostrado que en flujos bidimensionales la energía cinética no puede ser transportada de las escalas grandes a las peque\~{n}as.} (cascada inversa de energía) acompa\~{n}ada de un transporte de enstrofía en sentido inverso (cascada directa de enstrofía). La teoría de Kraichnan predice dos rangos inerciales caracterizados por las leyes \citep{Kundu2008}:
\begin{align}
  E(k) &= \zeta_{\epsilon}\cdot\epsilon^{2/3}\cdot k^{-5/3}, \hspace{1cm} &L^{-1}\ll &k < k_0  \label{Kraichnan-Law-1.6}\\
  E(k) &= \zeta_{\alpha}\cdot\alpha^{2/3}\cdot k^{-3},   \hspace{1cm} &k_0   <   &k \ll \eta^{-1}  \label{Kraichnan-Law-3.0}
\end{align}
donde $E(k)$ es la densidad espectral de energía ($m^{3}~s^{-2}$), $\epsilon$ es el ritmo al cual la energía se transmite hacia las escalas grandes ($m^{3}~s^{-3}$), $\zeta_{\epsilon}$ es la \textit{constante de Kolmogorov-Kraichnan}\index{Kolmogorov-Kraichnan!constante de}, $\alpha$ el flujo directo de enstrofía hacia las escalas peque\~{n}as ($s^{-3}$) y $k=2\pi/\lambda$ es el número de onda espacial ($m^{-1}$), $\epsilon$ es el flujo de energía que se transporta en el sistema ($m^{3}~s^{-3}$), y $\zeta_{\alpha}$ es una constante universal análoga a la constante adimensional $\zeta$ en la ley de Kolmogorov (\ref{Kolmogorov-Law}).\\
\begin{figure}[h!]
	\centering
		\includegraphics[width=0.7\textwidth]{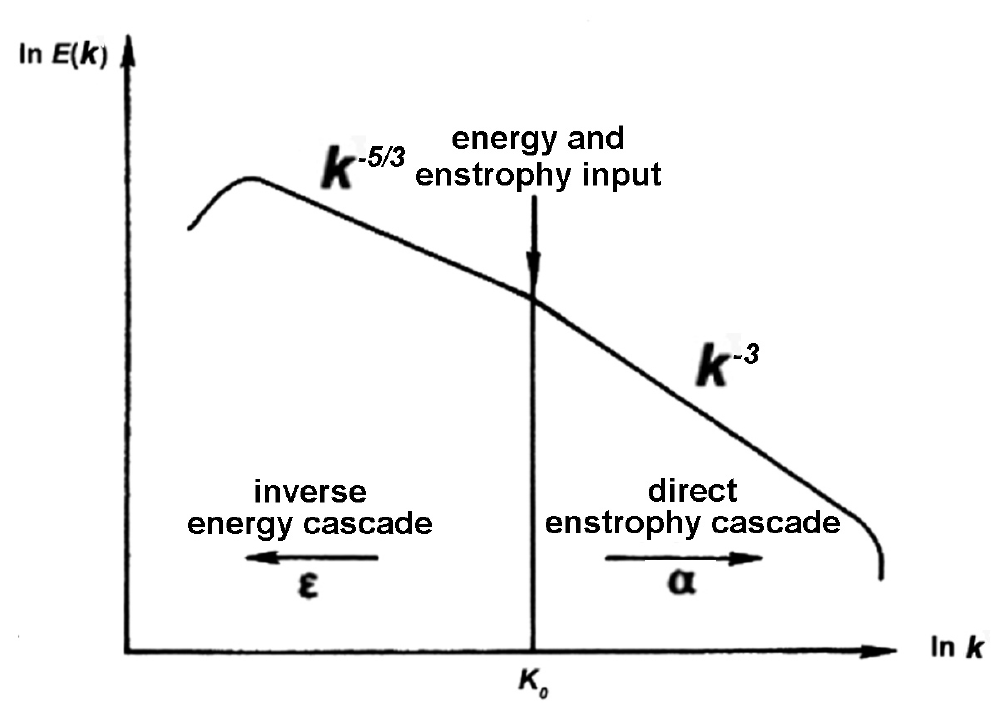}
	\label{fig:2D-Turbulence}
	\caption[Espectro de potencias clásico para turbulencia 2D.]{\scriptsize{Espectro de potencias clásico predicho por Kraichnan para la turbulencia en 2D debido a un forzamiento determinado en una escala $k_{0}$.}}
\end{figure}

\subsection{Turbulencia en las Atmósferas Planetarias}\indent

\citet*{Nastrom1984} obtuvieron por vez primera el espectro de potencias de energía cinética turbulenta de la atmósfera terrestre a partir de medidas de viento tomadas en unos 6,900 vuelos comerciales entre 1975 y 1979. En la Figura 4.3A podemos observar que en el espectro de potencias aparecen dos regiones con pendientes claramente diferenciadas: para las escalas grandes (escalas planetaria y sinóptica, $k<50$) tenemos una ley de potencias $k^{-3}$, similar a la que define la ecuación (\ref{Kraichnan-Law-3.0}). Para escalas más peque\~{n}as (sub-sinóptica y mesoescala, $k\sim50-10^{4}$) la atmósfera exhibe un comportamiento que se ajusta a la ley $k^{-5/3}$. Algunos autores encuentran ``paradójico'' \citep{Frisch1995,Lindborg1999} que el espectro que se obtiene para la atmósfera terrestre muestre los rangos en posiciones ``invertidas'' a las que predice Kraichnan, si bien \citet*{Gage1986} sugirieron que dicha discrepancia quedaría resuelta si existiera un sumidero de energía y enstrofía en la región intermedia. Sin embargo, esto parece poco probable si tenemos en cuenta que la región de transición varía de forma gradual de un comportamiento a otro, sin reflejar ningún cambio disipativo llamativo.\\

\begin{figure}[h!]
	\centering
		\includegraphics[width=1.0\textwidth]{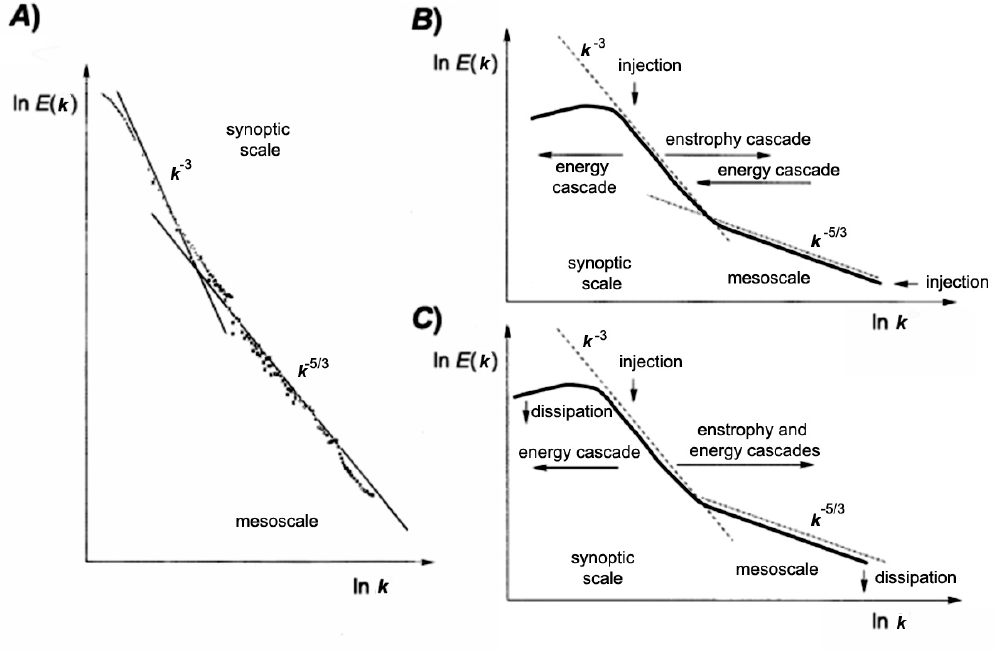}
	\label{fig:Earth-Turbulence}
	\caption[Espectro de potencias para la atmósfera terrestre.]{\scriptsize{Espectro de potencias de la energía cinética turbulenta medido a partir de la velocidad del viento zonal en la atmósfera terrestre (A) \citep{Nastrom1984}, interpretación teórica de Lilly (B) \citep{Lilly1989} y Tung (C) \citep{Tung2003}.}}
\end{figure}

Teniendo en cuenta que \citet{Kraichnan1967} sólo consideró el caso de una fuerza que actuaba sobre escalas intermedias, \citet{Lilly1989} investigó el caso de dos fuerzas, una actuando sobre las escalas más grandes y otra sobre las más peque\~{n}as, sugiriendo que era posible la coexistencia de los dos rangos inerciales (energía y enstrofía) en la región intermedia y descartando la necesidad de ningún sumidero tal como se muestra en la Figura 4.3B. No obstante, el modelo de Lilly no está exento de críticas. Por un lado \citet*{Maltrud1991} también desarrollaron un modelo con forzamiento en ambos extremos del espectro y obtuvieron resultados que se alejaban de los resultados experimentales (pendiente más pronunciada que $k^{-3}$ y una región de transición mucho más abrupta). Además, resulta difícil justificar la ``universalidad'' observada en los espectros de potencias\footnote{Se obtiene el mismo tipo de espectro de potencias tanto para medidas de velocidad como para medidas de temperatura \citep{Gage1986}.} y que ello sea resultado de dos forzamientos sin relación alguna y aplicados a ambos extremos del espectro de escalas espaciales \citep{Tung2003}. También se ha cuestionado seriamente la hipótesis de que el rango $k^{-5/3}$ sea el equivalente al de cascada inversa de energía predicho por Kraichnan ya que existen evidencias observacionales de la existencia de cascadas directas de energía en la región de mesoescala \citep{Lambert1981,Straus1999}. Modelos cuasi-geostróficos más recientes \citep{Tung2003} han logrado reproducir de manera satisfactoria el espectro de potencias de Nastrom-Gage introduciendo forzamiento únicamente a escalas sinópticas y disipación en escalas peque\~{n}as (viscosidad), tal como puede observarse en la Figura 4.3C.\\

\subsection{Turbulencia en la Atmósfera de Venus}\indent

En el caso de Venus no ha sido posible hasta ahora la obtención de un espectro de potencias de energía cinética ya que los errores de medida a la hora de determinar la velocidad del viento resultan ser del mismo orden que las fluctuaciones debidas a la turbulencia \citep{Gierasch1997}. Podemos tratar de resolver el problema de forma indirecta si consideramos que la distribución espacial del brillo de las nubes está relacionada con las escalas espaciales características de la dinámica atmosférica. De este modo el espectro de potencias del brillo\index{Espectro de potencias!de brillo} de las nubes puede intentar utilizarse para caracterizar el espectro de energía cinética. Este procedimiento fue llevado a cabo usando imágenes ultravioleta de las nubes de Venus y la Tierra por Mariner 10 \citep{Travis1978} y por Pioneer Venus \citep{Rossow1980,delGenio1982}.\\

Conviene, sin embargo, hacer referencia a algunas cuestiones respecto las limitaciones de esta metodología. \textbf{(1)} estos estudios están limitados al nivel de altura de las estructuras nubosas que aparezcan en nuestras imágenes. \textbf{(2)} las escalas espaciales más peque\~{n}as que podemos analizar vienen determinadas por la resolución de las imágenes. \textbf{(3)} resta determinar de forma estricta, a partir del análisis del movimiento, si las escalas espaciales de la distribución de brillo de las nubes se corresponden con las del espectro de e\-ner\-gía cinética. Las escalas espaciales que se derivan de la distribución de brillo de las nubes son producidas por la variabilidad espacial del espesor óptico, a su vez relacionada con cambios en la concentración de los compuestos absorbentes, cambios en la composición (que afecta al coeficiente de extinción) y modificaciones del grosor vertical de las nubes (esto último puede estar controlado por procesos químicos, microfísicos y dinámicos, \citealt{Esposito1997}). Actualmente sigue sin comprenderse bien el papel jugado por todos estos factores ni la manera en que pueden generar los contrastes de albedo observados o afectar a su distribución espacial y temporal. No obstante, existen muchos procesos dinámicos que se manifiestan en los campos de nubes con escalas espaciales y que contribuyen al espectro de potencias de brillo: procesos de misoescala determinados por la extensión vertical de las nubes (escala de altura de $\sim5-10$ km), procesos de mesoescala (``manchas redondeadas'', producto de la agrupación de células de convección\index{Convección!celulas de}, $\sim100-500$ km), de escala sinóptica (``estrías y estructuras lineales'', $\sim1000$ km de longitud) y planetaria (la estructura ``Y'').\\

A pesar de las limitaciones anteriores, el análisis del espectro de potencias de la distribución de brillo en las nubes es de por sí interesante pues nos aporta información sobre los procesos implicados en diferentes escalas espaciales de las nubes (dinámica, química y microfísica). Asímismo, la distribución espacial de brillo en las nubes resulta ser una herramienta válida para comparar distintas atmósferas planetarias, independientemente de las diferentes propiedades dinámicas, térmicas y químicas que pudieran existir entre ellas. A efectos de simplificar la comparación entre los resultados de distintos planetas independientemente del tama\~{n}o de éstos, es costumbre que en el estudio de espectro de potencias de brillo de las nubes se sustituya el número de onda $k=2\pi/\lambda$ ($m^{-1}$) por el \textit{número de onda zonal}\index{Número de onda!zonal} (adimensional) que describe las escalas espaciales en términos del número de escalas que caben dentro de un paralelo planetario. Dicho número de onda zonal viene dado por la expresión $\hat{k}=2\pi R_{p} \cos\varphi/\lambda,$ siendo $\lambda$ la longitud de onda característica, $R_p$ el radio planetario y $\varphi$ la latitud.\\

El primer análisis de espectros de potencias del brillo de las nubes de Venus fue llevado a cabo por \citet*{Travis1978} usando imágenes ultravioleta tomadas durante el sobrevuelo de Mariner 10 en los días 8-9 de febrero de 1974. Travis obtuvo espectros de potencias que cumplían la ley $\hat{k}^{-n}$, con $\hat{k}=3-30$ como rango de números de onda\footnote{El rango de números de onda era limitado debido a la longitud máxima cubierta por cada imagen (unos 150$^{\circ}$) y al suavizado previo que se aplicó a las imágenes originales.} y con valores de $n$ comprendidos entre -1.7 y -2.7. La nave Pioneer Venus obtuvo imágenes durante una década y los datos obtenidos durante 1979 y 1980 fueron utilizados para realizar un análisis de espectro de potencias más detallado \citep{Rossow1980,delGenio1982}. Se cubrió un rango de números de onda $\hat{k}=3-50$, con valores de $n$ entre -2.5 y -4, dependiendo de la latitud y del rango de números de onda en el que se realizara el ajuste. Los espectros de potencias obtenidos a partir de las observaciones de Mariner 10 y Pioneer Venus muestran diferencias que, si bien pueden atribuirse a que la instrumentación, calidad de los datos, condiciones observacionales y metodología son distintas, también podrían deberse a cambios temporales reales de la atmósfera de Venus entre ambas épocas. Más de 10 a\~{n}os después de la Pioneer Venus, la nave Galileo sobrevoló Venus en febrero de 1990 y fue la última misión que obtuvo imágenes de alta resolución hasta la llegada de Venus Express en abril de 2006. En este capítulo hacemos uso de las imágenes violeta obtenidas por Galileo para estudiar la distribución espacial de brillo de las nubes de Venus, comparando nuestros resultados con los obtenidos por Mariner 10 y Pioneer Venus así como con los resultados teóricos anteriores.\\

\section{Observaciones}\label{chapter-turbulence-observs}\indent

De las 77 imágenes obtenidas por la cámara SSI de la sonda Galileo en su sobrevuelo de febrero de 1990 escogimos 20 imágenes en el violeta ($418~nm$) cuya amplia cobertura espacial y calidad visual permitían cons\-truir varias composiciones de gran extensión sobre el planeta. El listado de estas imágenes y sus características aparecen resumidas en la Tabla \ref{tab:tabla-ImgsGalileoTurb}. El nivel de ruido de las imágenes obtenidas en el infrarrojo cercano ($986~nm$) y el pobre contraste de las nubes impidieron realizar un estudio similar en el nivel de las nubes sondeado por estas imágenes.\\

\begin{table}[h!]
  \caption{Imágenes de Galileo usadas en este estudio.}
	\label{tab:tabla-ImgsGalileoTurb}
	\centering
  \begin{spacing}{0.6}

		\begin{tabular}{*{7}{>{\scriptsize}c}}
			& & & & & & \\
			\hline\hline
			& & & & & & \\
			\textit{Número} & \textit{Fecha} & \textit{Hora} & \textit{Rango de} & \textit{Resolución} & \textit{Distancia} & \multirow{2}{*}{\textit{Composiciones}} \\
      \textit{de Imagen} & \textit{(dd/mm/aa)} & \textit{(hh:mm:ss)} & \textit{Latitudes} & \textit{Espacial (km/pix)} & \textit{(km)}  \\
			& & & & & & \\
			\hline
			& & & & & & \\
			18229900 & 11/02/1990 & 09:23:53 & 04$^{\circ}$N-78$^{\circ}$N &  6.36 &   632,468 & B         \\
			18222800 & 11/02/1990 & 09:53:12 & 04$^{\circ}$N-80$^{\circ}$N &  6.47 &   643,447 & C         \\
			18220100 & 11/02/1990 & 21:13:41 & 02$^{\circ}$N-85$^{\circ}$N &  9.06 &   897,811 & B         \\
			18220145 & 11/02/1990 & 21:14:11 & 01$^{\circ}$N-85$^{\circ}$N &  9.06 &   898,000 & B         \\
			18223100 & 11/02/1990 & 21:44:01 & 01$^{\circ}$N-85$^{\circ}$N &  9.16 &   909,135 & C         \\
			18223145 & 11/02/1990 & 21:44:31 & 01$^{\circ}$N-85$^{\circ}$N &  9.17 &   909,323 & C         \\
			18227600 & 12/02/1990 & 11:58:24 & 04$^{\circ}$S-86$^{\circ}$N & 12.40 & 1,227,765 & B         \\
			18229400 & 12/02/1990 & 13:57:43 & 06$^{\circ}$S-86$^{\circ}$N & 12.85 & 1,272,224 & A y C     \\
			18224400 & 13/02/1990 & 05:58:16 & 26$^{\circ}$S-86$^{\circ}$N & 16.49 & 1,629,971 & B         \\
			18226300 & 13/02/1990 & 07:58:36 & 28$^{\circ}$S-85$^{\circ}$N & 17.00 & 1,674,768 & C         \\
			18228400 & 13/02/1990 & 10:00:56 & 29$^{\circ}$S-85$^{\circ}$N & 17.40 & 1,720,316 & A y C     \\
			18220200 & 13/02/1990 & 12:00:15 & 29$^{\circ}$S-86$^{\circ}$N & 17.85 & 1,764,734 & B         \\
			18223900 & 14/02/1990 & 05:28:46 & 59$^{\circ}$S-85$^{\circ}$N & 21.82 & 2,155,032 & A,B,C,D,E \\
			18225700 & 14/02/1990 & 07:28:05 & 63$^{\circ}$S-84$^{\circ}$N & 22.27 & 2,199,442 & A         \\
			18223300 & 15/02/1990 & 04:58:16 & 90$^{\circ}$S-84$^{\circ}$N & 27.14 & 2,679,705 & D         \\
			18225200 & 15/02/1990 & 06:58:35 & 88$^{\circ}$S-86$^{\circ}$N & 27.60 & 2,724,499 & A y E     \\
			18222800 & 16/02/1990 & 04:28:46 & 87$^{\circ}$S-86$^{\circ}$N & 32.47 & 3,204,862 & A y D     \\
			18224600 & 16/02/1990 & 06:28:04 & 88$^{\circ}$S-85$^{\circ}$N & 32.92 & 3,249,288 & E         \\
			18222200 & 17/02/1990 & 03:58:15 & 87$^{\circ}$S-85$^{\circ}$N & 37.80 & 3,729,724 & D         \\
			18224100 & 17/02/1990 & 05:58:34 & 88$^{\circ}$S-84$^{\circ}$N & 38.26 & 3,774,530 & E         \\
			& & & & & & \\
			\hline
			& & & & & & \\
			
		\end{tabular}
    \begin{flushleft}
\scriptsize{\textit{Nota:} Construimos cinco composiciones de imágenes con la mayor cobertura posible sobre el planeta (A, B, C, D y E). La composición \textbf{A} es la de más alta resolución, con una extensión de 360$^{\circ}$ de longitud y latitudes entre 6$^{\circ}$N y 54$^{\circ}$N. Usamos 7 imágenes para las composiciones \textbf{B} y \textbf{C}, que cubren latitudes del hemisferio norte hasta 60$^{\circ}$N. Cada una de las composiciones \textbf{D} y \textbf{E} cubren latitudes entre 50$^{\circ}$S y 50$^{\circ}$N, cada una construida con 4 imágenes.}
    \end{flushleft}

  \end{spacing}
  
\end{table}

La Figura 4.4 muestra dos de las cinco composiciones de planisferios usa\-das en este trabajo, cubriendo una de ellas paralelos completos del hemisferio norte hasta 70$^{\circ}$N (Fig. 4.4, composición A) mientras que la otra muestra paralelos casi completos de ambos hemisferios y se extiende en latitud hasta 60$^{\circ}$S (Fig. 4.4, composición E). En vista de la buena superposición de las distintas imágenes, queda patente la validez de aproximar la atmósfera a un sólido rígido con un periodo de rotación de 4.4 días. La resolución espacial de las imágenes originales empleadas varían entre $\sim$0.15$^{\circ}$/pix hasta 0.50$^{\circ}$/pix para altas latitudes. Además de las brillantes estructuras alargadas que se extienden desde la banda polar brillante hasta bajas latitudes también podemos ver claramente la oscura ``Y'' en la región ecuatorial, así como muchas estructuras de menor tama\~{n}o, y células convectivas a 220$^{\circ}$ de longitud y 20$^{\circ}$N \citep{Belton1991,Toigo1994,Peralta2007a}.\\

\begin{figure}[h!]
	\centering
		\includegraphics[width=1.0\textwidth]{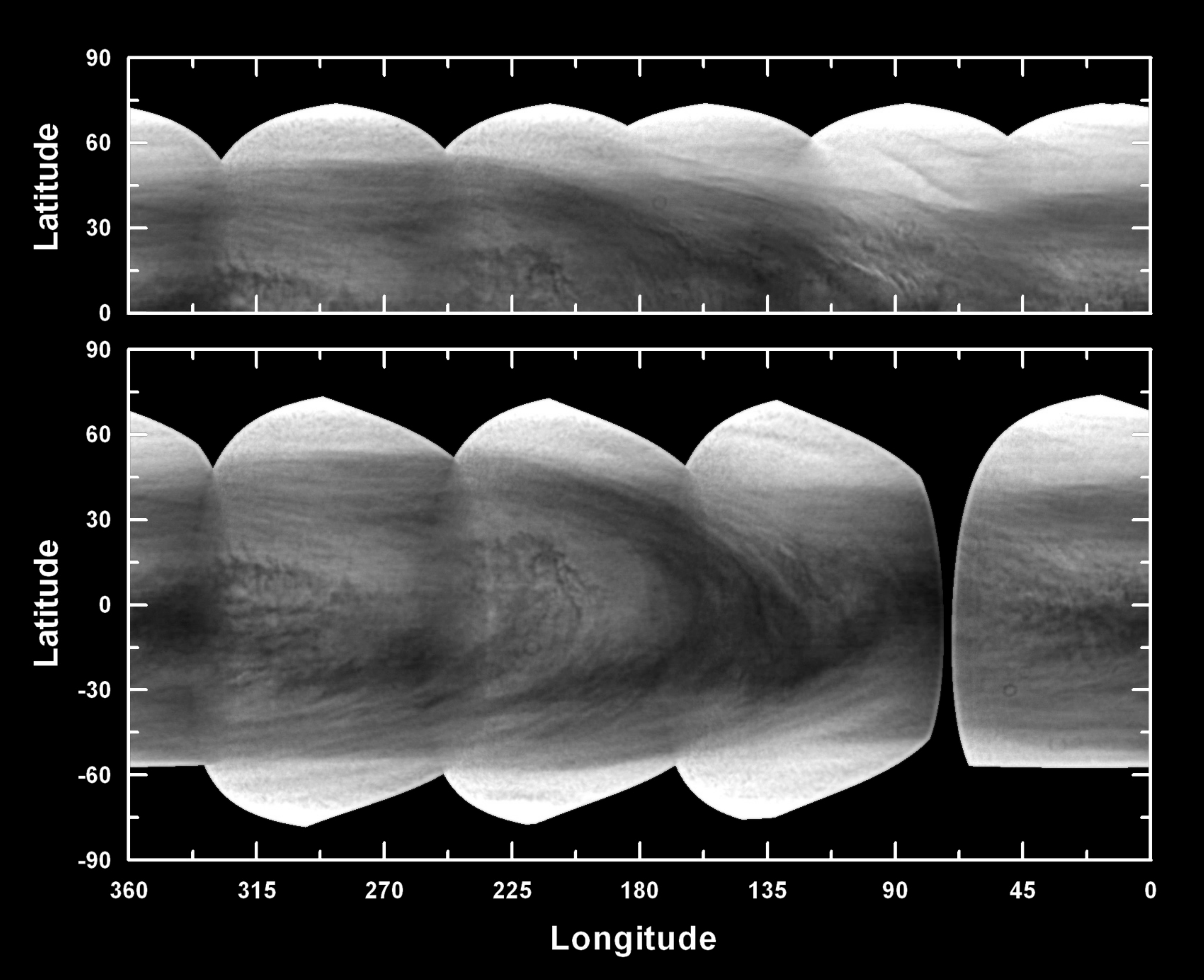}
	\label{fig:GalileoVenus-cylindrical-projection}
	\caption[Proyecciones cilíndricas de la atmósfera de Venus.]{\scriptsize{Composiciones de Venus a partir de proyecciones cilíndricas de imágenes de Galileo: A (arriba) y E(abajo). La composición A se extiende 360$^{\circ}$ en longitud y abarca de 0$^{\circ}$ a 50$^{\circ}$N. La composición E es muy similar a las composiciones B, C y D, e incluye el hemisferio sur aunque abarcando menos de 360$^{\circ}$. Estas imágenes han sido procesadas para un mejor contraste.}}
\end{figure}

\section{Análisis}\label{chapter-turbulence-analysis}\indent

Utilizamos el software PLIA \citep{Hueso2008b} para navegar las imágenes, procesarlas y llevar a cabo sus proyecciones cilíndricas. Previamente a la navegación y proyección de las imágenes corregimos los defectos de éstas (ver Capítulo \ref{chapter-observs}). Un tipo de defectos que presentan las imágenes SSI son ``manchas'' debidas a la ocasional deposición de partículas de polvo sobre la cámara. Aunque estos defectos suelen corregirse calibrando las imágenes con un ``flat-field'', en nuestro caso no pudimos realizar dicha calibración ya que no se tomaron ``flat-fields'' adecuados durante el sobrevuelo de Venus, tal y como se vio en el capítulo \ref{chapter-observs}.\\

Una vez eliminados los defectos, navegadas las imágenes y proyectadas de forma cilíndrica (``planisferios'') con una resolución de 0.25$^{\circ}$/pix, del mismo modo que \citet*{Rossow1980} corregimos los efectos que la geometría esférica del planeta tiene sobre la dispersión de la luz aplicando la corrección de Minnaert\index{Minnaert!corrección de} (\ref{Minnaert-Law}) en la aproximación Lambertiana (ver capítulo \ref{chapter-observs}). Finalmente eliminamos los píxeles pertenecientes a lugares donde la resolución original de la imagen sea peor que 0.5$^{\circ}$/pix, lo que automáticamente excluye los píxeles cercanos al limbo del planeta y, por ende, las regiones de mayor oscurecimiento por el efecto centro-limbo. Las proyecciones cilíndricas de cada imagen individual presentan una extensión típica de 120$^{\circ}$ en longitud para latitudes ecuatoriales, y 50$^{\circ}$ para latitudes más elevadas, consecuencia del elevado ángulo de fase que presenta Venus en nuestras observaciones ($\sim$45$^{\circ}$).\\

Para construir imágenes globales del planeta proyectamos las imágenes en planisferios con intervalos de longitud y latitud constante utilizando el software PLIA, superponiendo las imágenes adyacentes y obtenido cinco composiciones globales con una cobertura prácticamente global de bandas amplias de latitud. Para realizar estas composiciones es necesario conocer el periodo de rotación de las nubes superiores. Nuestras propias medidas sobre estas imágenes (ver Capítulo \ref{chapter-winds} y \citealt{Peralta2007b}) y análisis anteriores \citep{Belton1991} muestran un perfil meridional de vientos zonales prácticamente constante con un valor medio de $<u>=-101~m/s$ y un periodo de rotación de 4.4 días en la región ecuatorial-tropical. Si bien este periodo de rotación varía con la latitud (disminuyendo hacia las latitudes altas) y el viento llega a experimentar variaciones del orden de los $10~m/s$ (ver Capítulo \ref{chapter-winds} y \citealt{Sanchez-Lavega2008}) se ha demostrado que no se mejora sustancialmente la apariencia de las composiciones por el hecho de aplicar una rotación dependiente de la latitud \citep{delGenio1982} por lo que en nuestro trabajo aplicamos una rotación de sólido rígido de 4.4 días para elaborar las composiciones. Con objeto de obtener información sobre los números de onda más peque\~{n}os se realizaron composiciones del hemisferio norte (0$^{\circ}$-70$^{\circ}$N) cubriendo 360$^{\circ}$ de longitud, así como 4 composiciones más que cubrían un rango de longitud ligeramente inferior (330$^{\circ}$) pero en ambos hemisferios (70$^{\circ}$N-50$^{\circ}$S). Sin embargo, para que la superposición sea buena no sólo debería ser aceptable considerar la atmósfera como un sólido rígido en rotación, sino que las estructuras nubosas no deberían sufrir cambios drásticos durante una rotación completa de la nubes. Este último hecho no es preocupante ya que, si bien las formaciones nubosas de misoescala ($<100$ km) sí que pueden experimentar transformaciones significativas en este pe\-rio\-do, no ocurre lo mismo para las formaciones de mayores escalas. Dado un par de planisferios consecutivos, la región de solapamiento se determina de la siguiente manera:
\begin{equation}
	B(i,j)=\frac{B_{Left}(i,j)\cdot (i_{Right}-i)+B_{Right}(i,j)\cdot (i-i_{Left})}{i_{Right}-i_{Left}},
	\label{overlapping}
\end{equation}
donde $i$ y $j$ son los índices horizontal y vertical del planisferio respectivamente, $B(i,j)$ es el brillo en el punto ($i$,$j$), $B_{Left}(i,j)$ es el valor de brillo del planisferio a la izquierda de la zona de solapamiento, $B_{Right}(i,j)$ es el valor de brillo del planisferio a la izquierda de la zona de solapamiento, e $i_{Left}$ e $i_{Right}$ son los límites izquierdo y derecho de la región de solapamiento para el paralelo de latitud $j$. El uso de la ecuación (\ref{overlapping}) permite obtener una transición suave de un planisferio a otro, si bien dicha transición sigue siendo algo abrupta para las zonas de solapamiento con pocos píxeles (ver Fig. 4.4).\\

\begin{figure}[h!]
	\centering
		\includegraphics[width=0.75\textwidth]{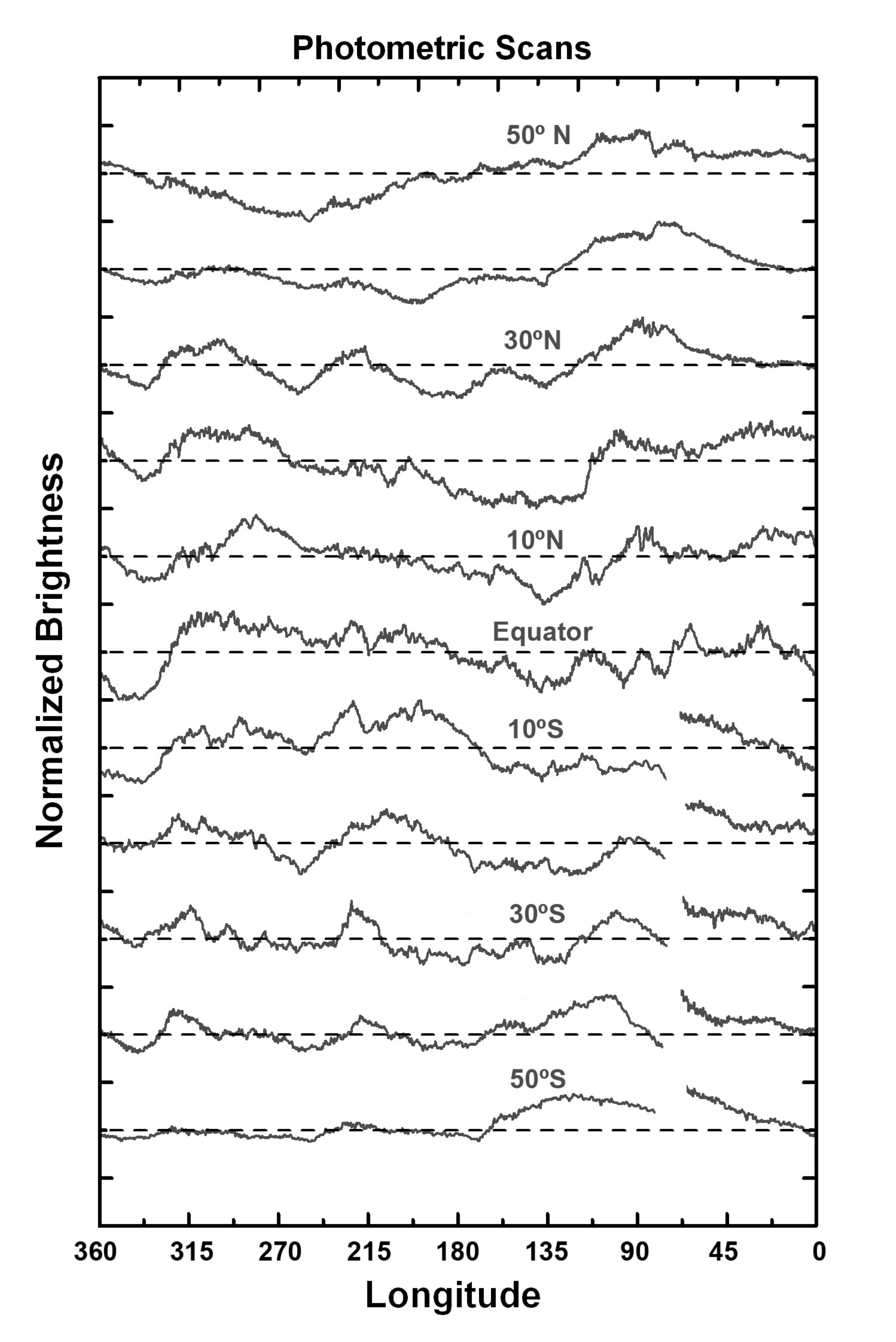}
	\label{fig:GalileoVenus-PhotoScans}
	\caption[Ejemplos de Cortes Fotométricos.]{\scriptsize{Ejemplos de Cortes Fotométricos. Los cortes desde el ecuador hasta 50$^{\circ}$N provienen de la composición A, mientras que los del hemisferio sur proceden de la composición E (ver Fig.4.4). Los cortes fotométricos están representados como desviaciones respecto a un valor promedio cero y representan variaciones de brillo de en torno a un 25\% con respecto al valor medio.}}
\end{figure}

Se define un \textit{corte fotométrico}\index{Fotométrico!corte} como la secuencia de valores de brillo co\-rres\-pon\-dien\-te a los pixeles del planisferio contenidos en un paralelo determinado. Dada la resolución de los planisferios, se extraen cortes fotométricos cada 0.25$^{\circ}$ de latitud. Se pueden ver ejemplos de estos cortes en la Figura 4.5. Usamos estos cortes fotométricos para calcular los espectros de potencias correspondientes a cada latitud. El espectro de potencias\index{Espectro de potencias!de brillo} se obtiene elevando al cuadrado la amplitud de la transformada de Fourier correspondiente a cada corte fotométrico. Debido a que el orden de los valores de potencia varía de unas regiones a otras de la atmósfera de Venus (el brillo promedio es distinto), para facilitar la comparación se normalizan todos los espectros de potencia para $\hat{k}=10$. Si bien el análisis de Fourier de imágenes individuales sólo podría resolver escalas de hasta $\hat{k}=3$ (debido a que en el mejor de los casos podríamos ver como máximo algo menos de 180$^{\circ}$ de longitud), una de nuestras composiciones cubre paralelos completos por lo que podremos llegar a estimar potencias de hasta $\hat{k}=1$ (360$^{\circ}$). Sin embargo, el resto de composiciones cubren tan solo 330$^{\circ}$ de longitud y para calcular el espectro de potencias de estos paralelos incompletos se procede generalmente a forzar la periodicidad en los cortes fotométricos multiplicándolos por una función que los anule en los extremos. En estudios anteriores se trabajó con cortes fotométricos de $\sim$120$^{\circ}$ \citep{Travis1978,Rossow1980,delGenio1982} a los que se aplicaba una función de Hanning\index{Hanning!función de} \citep{Blackman1958}, que se define como:
\begin{equation}
	H(i)=\alpha-(1-\alpha)\cdot \cos\left(\frac{2\pi i}{N}\right),
	\label{hanning}
\end{equation}
donde $N$ es el número de píxeles que compone el corte fotométrico, $i$ es el índice para los píxeles del corte (con valores de 0 a $N-1$), y $\alpha$ es un parámetro variable que toma normalmente el valor de 0.5. No obstante, la aplicación de la función de Hanning no resulta aconsejable para cortes fotométricos con extensiones próximas a 360$^{\circ}$. Tras realizar diversas pruebas con cortes fotométricos del campo de nubes de Venus se comprobó que este procedimiento elimina parte de la información del espectro y que para cortes fotométricos de más de 315$^{\circ}$ es conveniente no aplicar dicha función. Otro inconveniente que se deriva de aplicar la función de Hanning es el ruido que genera para números de onda $\hat{k}\geq10$, si bien éste es despreciable en comparación con otras fuentes de ruido.\\

Además de todo lo mencionado anteriormente, es conveniente tener en cuenta otras fuentes de valor espurio para la potencia:
\begin{enumerate}
	\item El \textit{aliasing}\index{Aliasing} es el efecto por el cual dos se\~{n}ales continuas diferentes se tornan indistinguibles cuando se digitalizan. Cuando esto sucede la se\~{n}al original no puede ser reconstruida de forma unívoca a partir de la se\~{n}al digital. Al estar nuestras imágenes digitalizadas con una resolución espacial determinada, el \textit{aliasing} contribuye con una potencia falsa para todos los números de onda. Dicho error se vuelve más importante a medida que tenemos números de onda cercanos a la re\-solución espacial, hasta llegar al número de onda de \textit{Nyquist}\index{Número de onda!de Nyquist}, el mayor número de onda que puede resolver nuestra imagen (en nuestro caso $\hat{k}_{Nyquist}=360$ ya que la resolución de las imágenes originales es de 0.50$^{\circ}$/pixel). Esta condición impone un límite superior a los valores de $\hat{k}$ que podemos estudiar. Nosotros lo establecemos con un cierto margen en $\hat{k}\approx 200$.
  \item El rango dinámico de brillo\index{Imágenes,rango dinámico de} también afecta a los valores de potencia. Las imágenes de Venus tomadas por la cámara SSI de Galileo tienen un rango dinámico de 6-7 bits, mientras que las variaciones de brillo de las nubes son de $\sim 25\%$ \citep{Belton1991}, lo que nos deja unos 30 niveles de gris ($\sim 5$~bits) para cuantificar dichas variaciones\footnote{Esto ha sido verificado para todas las latitudes mediante un examen directo de cada corte fotométrico.}. Para determinar de qué modo afecta esta cuantización del rango dinámico a los espectros de potencia experimentamos con funciones periódicas sintéticas cuyo espectro de potencias cumpliera una ley exponencial simple de la forma:
    \begin{equation}
	    P(\hat{k})=C_{\hat{k}}\cdot \hat{k}^{-n},
	    \label{ley-potencias}
    \end{equation}
donde $P(\hat{k})$ es la intensidad del espectro de potencias en el número de ondas $\hat{k}$, $C_{\hat{k}}$ es una constante de normalización y $n$ es la pendiente que exhibe el espectro cuando es representado en escala logarítimica. El resultado de comparar los espectros de potencias que derivan de la función sintética pura y de la función sintética digitalizada a 5 bits para dos valores de $n$ predeterminados puede observarse en la Figura 4.6. Como vemos, para $n=5/3$ la se\~{n}al digitalizada con 5 bits puede reproducir fielmente las características del espectro sintético hasta $\hat{k}\sim100$, mientras que para $n=3$ sólo tenemos buenos resultados hasta $\hat{k}\sim50$.
    \begin{figure}[h!]
	    \centering
		    \includegraphics[width=1.0\textwidth]{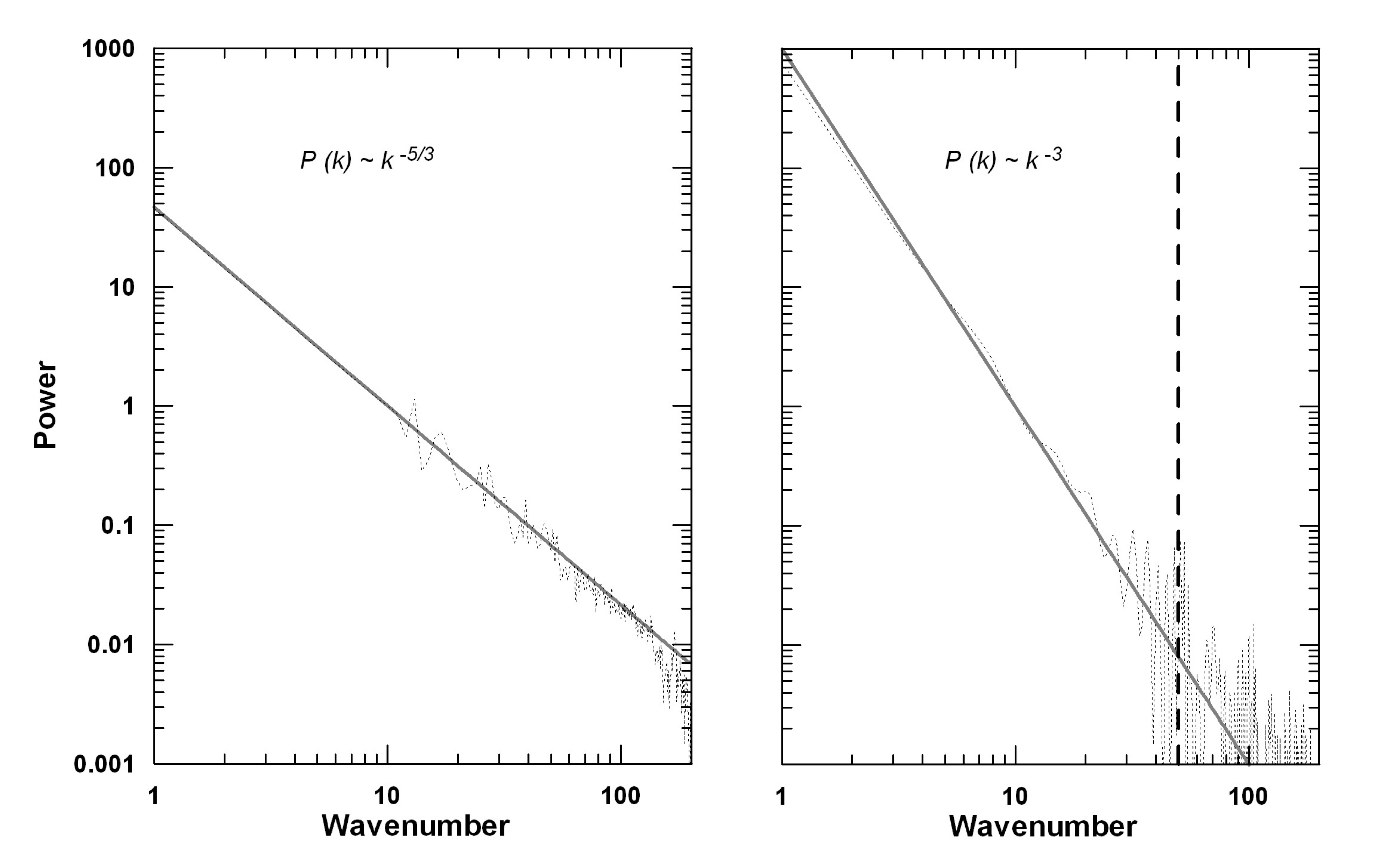}
	    \label{fig:Digitization-effect}
	    \caption[Efectos de la digitalización sobre los espectros.]{\scriptsize{Espectros de potencias generados a partir de cortes fotométricos sintéticos (línea continua) y su versión digitalizada usando 5 bits (línea en puntos suspensivos). Se analizan los casos de pendiente con valores clásicos de la turbulencia: -5/3 y -3. Para espectros de potencias variando con pendiente -3, la línea discontinua marca el número de onda ($\hat{k}=50$) a partir del cual el ruido debido a la digitalización comienza a ser importante.}}
    \end{figure}
    
  \item La composición de los planisferios en sí misma puede introducir ruido por diversos motivos: (a) existencia de discontinuidades entre las imá\-ge\-nes debido a cambios en la morfología de las nubes durante el espacio de tiempo transcurrido entre ambas, (b) fallos a la hora de corregir el efecto de oscurecimiento hacia el limbo, y (c) que el periodo de rotación de las formaciones nubosas en la zona de superposición sea ligeramente distinto al periodo de rotación asumido. Este ruido del que estamos hablando se manifiesta en forma de máximos de potencia para números de onda iguales al número de planisferios empleados para la composición o múltiplos de éste. No obstante hemos podido comprobar que el ruido consecuencia de variar el periodo de 4.4 días a por ejemplo 4.0 días no afecta sensiblemente a las características ge\-ne\-ra\-les del espectro ni al valor del exponente $n$. Se estima que el error introducido debido a este efecto no es superior al 20\% en el valor de $n$.
  
\end{enumerate}

Por tanto, en base a los puntos 1 y 2 restringimos nuestro análisis de los espectros de potencias a números de onda $\hat{k}\leq50$, y en base al punto 3 establecemos un error mínimo de 0.2 para los valores de $n$ que caracterizan los espectros de potencias de la distribución de brillo en las nubes de Venus.\\

\section{Resultados}\label{chapter-turbulence-results}\indent

Teniendo en cuenta la morfología de las formaciones nubosas observadas en longitudes de onda del violeta, en la cima de las nubes se suelen distinguir tres regiones \citep{Rossow1980,delGenio1982}: (1) la \textit{región polar}\index{Región!polar} con latitudes entre 50$^{\circ}$ y el polo, (2) la \textit{región de latitudes medias}\index{Región!latitudes medias} entre 20$^{\circ}$ y 50$^{\circ}$, y (3) la \textit{región ecuatorial}\index{Región!ecuatorial} entre 20$^{\circ}$S y 20$^{\circ}$N. En las regiones polares no suelen observarse estructuras de peque\~{n}a escala, mostrando una textura suave y más brillante en comparación con las otras regiones de la atmósfera. Por el contrario, las regiones de latitudes medias y ecua\-to\-ria\-les exhiben una apariencia moteada por la continua presencia de una gran variedad de formaciones nubosas oscuras, estructuras en forma de células, estrías y la típica estructura ``Y''. Teniendo en cuenta la morfología de las estructuras de gran escala observadas con las imágenes de Galileo, hemos seleccionado en nuestro estudio con espectros de potencias un total de seis bandas de latitud: 70$^{\circ}$-50$^{\circ}$N, 50$^{\circ}$-30$^{\circ}$N, 30$^{\circ}$-10$^{\circ}$N, 10$^{\circ}$N-10$^{\circ}$S, 10$^{\circ}$-30$^{\circ}$S, y 30$^{\circ}$-50$^{\circ}$S. Con esto pretendemos no sólo estudiar el ecuador y latitudes medias, sino también las regiones de transición entre ellas. Con vistas a una comparación temporal más fácil, la mayor parte de las bandas seleccionadas en este trabajo coinciden con las que fueron analizadas durante la misión Pioneer Venus \citep{delGenio1982}.\\

Con objeto de obtener un espectro de potencias de ruido reducido promediamos todos los espectros de potencias correspondientes a los cortes fotométricos de todas las composiciones y contenidos en cada banda de latitud de 20$^{\circ}$. Este promedio se justifica en base a que durante la época del sobrevuelo de Galileo Venus mostraba una atmósfera bastante homogénea en la dirección meridional, con ausencia de estrías de latitudes medias, células anulares, o cinturones circumecuatoriales (formaciones que, sin embargo, sí estaban presentes en la época de Pioneer Venus) \citep{Toigo1994}. A fin de comparar de forma cuantitativa los distintos espectros de potencias en diferentes bandas de latitud y rangos de números de onda, los espectros de potencias ya promediados son ajustados a una ley de potencias $P(\hat{k})$ del tipo dado por la ecuación (\ref{ley-potencias}). Así, para cada espectro de potencias se calcula la pendiente $n$ usando un ajuste lineal de mínimos cuadrados, con lo que podemos obtener el valor medio de $n$ así como su desviación estándar en cada banda de latitud.\\

\begin{figure}[h!]
	\centering
		\includegraphics[width=1.0\textwidth]{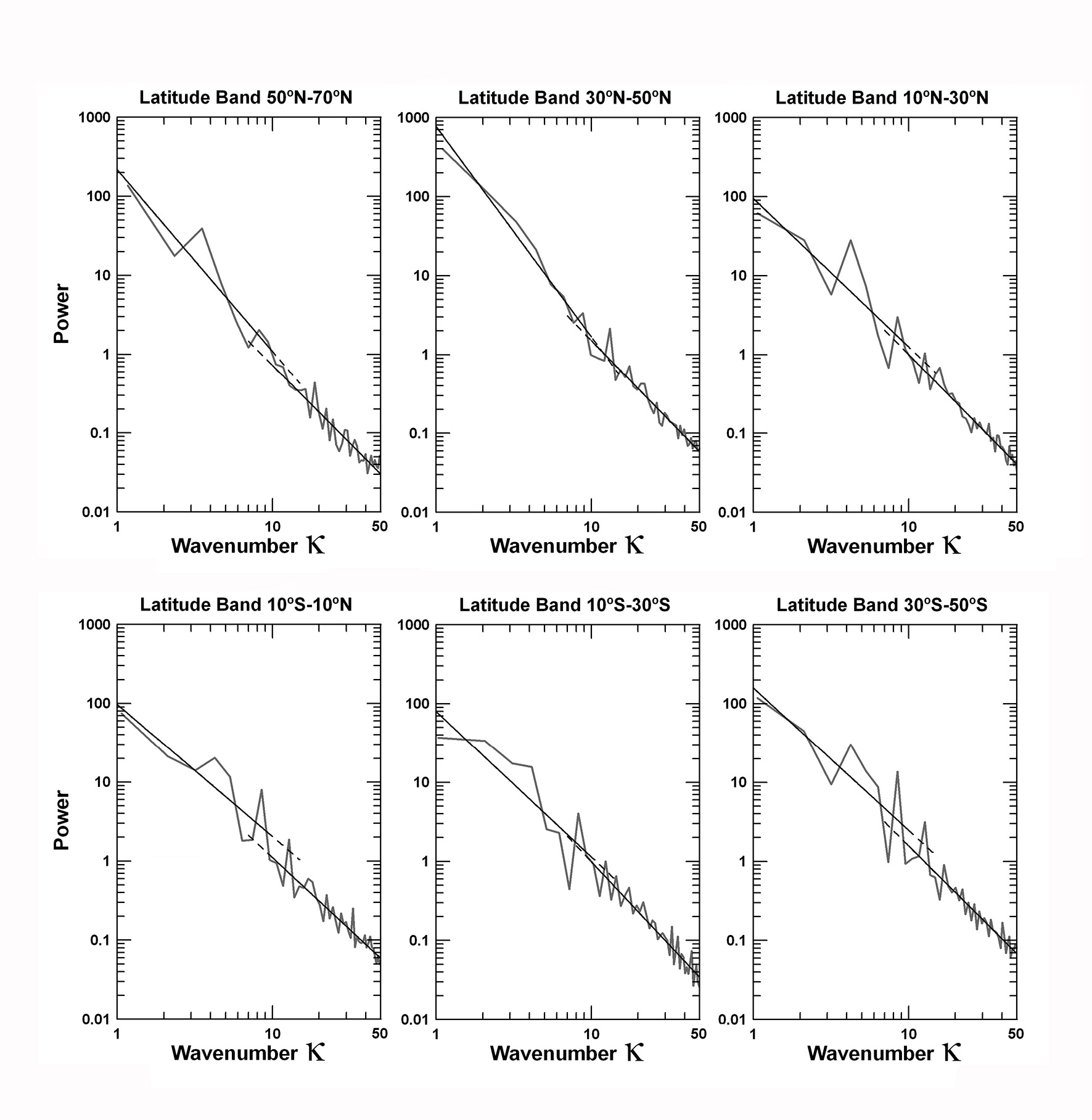}
	\label{fig:GalileoVenus-BrightSpectra}
	\caption[Espectros de potencias de brillo en Venus.]{\scriptsize{Ejemplos de espectros de potencias promediados en diferentes bandas de latitud. Los espectros son promediados de forma logarítmica para tres composiciones y para bandas de latitud de 20$^{\circ}$ (líneas grises). Todos los espectros están normalizados, con valor 1 para la potencia $\hat{k}=10$. Las líneas negras continuas designan los ajustes logarítmicos para los rangos de número de onda 1-10 y 10-50, y se representan ligeramente extendidas más allá del rango de ajuste (líneas negras discontinuas).}}
\end{figure}

En la Figura 4.7 mostramos los espectros de potencias así como sus ajustes a la ley de potencias (\ref{ley-potencias}) para dos rangos de números de onda: $\hat{k}=1-10$ y $\hat{k}=10-50$. Las pendientes $n$ para estos rangos y para cada banda de latitud vienen dadas en la Tabla \ref{tab:tabla-PendientesEspectros}, con las barras de error calculadas a partir de la desviación estándar de las pendientes de todos los cortes fotométricos dentro de una banda de latitud. La Tabla \ref{tab:tabla-PendientesEspectros} también permite estudiar cómo afecta a los resultados el tener cortes fotométricos de paralelos completos o de 330$^{\circ}$, y observamos que aunque los resultados son muy similares, en el caso de usar cortes fotométricos incompletos las barras de error son mayores.\\

\begin{table}[h!]
	\caption{Pendientes de los espectros de potencias.}
	\label{tab:tabla-PendientesEspectros}
	\centering
  \begin{spacing}{0.6}

		\begin{tabular}{*{7}{>{\scriptsize}c}}
			& & & & & & \\
			\hline\hline
			& & & & & & \\
			\multirow{3}{*}{\textit{Rango $\hat{k}$}} & \multicolumn{6}{c}{\scriptsize\textit{Pendiente n en cada banda de latitud}} \\
			& & & & & & \\
                                        & \textit{70$^{\circ}$N-50$^{\circ}$N} & \textit{50$^{\circ}$N-30$^{\circ}$N} & \textit{30$^{\circ}$N-10$^{\circ}$N} & \textit{10$^{\circ}$N-10$^{\circ}$S} & \textit{10$^{\circ}$S-30$^{\circ}$S} & \textit{30$^{\circ}$S-50$^{\circ}$S}  \\
			& & & & & & \\
			\hline
			& & & & & & \\
      \multicolumn{7}{l}{\scriptsize{(1) \textit{Composición A (360$^{\circ}$ de longitud)}}} \\
			& & & & & & \\
			 1-10 &  \textemdash  & $-2.5\pm 0.3$ & $-2.1\pm 0.4$ & $-2.0\pm 0.2$ &  \textemdash  &  \textemdash    \\
			10-50 &  \textemdash  & $-2.1\pm 0.5$ & $-1.9\pm 0.4$ & $-2.1\pm 0.3$ &  \textemdash  &  \textemdash    \\
			& & & & & & \\
      \multicolumn{7}{l}{\scriptsize{(2) \textit{Promedio de todas las Composiciones (A, B, C, D y E)}}} \\
			& & & & & & \\
			 1-10 & $-2.3\pm 0.3$ & $-2.7\pm 0.7$ & $-1.9\pm 0.4$ & $-1.7\pm 0.4$ & $-1.9\pm 0.2$ & $-1.8\pm 0.5$   \\
			10-50 & $-1.9\pm 0.5$ & $-2.0\pm 0.5$ & $-2.0\pm 0.4$ & $-1.8\pm 0.4$ & $-2.1\pm 0.4$ & $-1.9\pm 0.4$   \\
			& & & & & & \\
			\hline
			& & & & & & \\

		\end{tabular}
  \end{spacing}
\end{table}

Todos los espectros de potencias decrecen monótonamente y muestran un cambio de pendiente significativo para números de onda elevados. El hecho de que los números de onda $\hat{k}=1-3$ dominen el espectro indica que las longitudes de onda de escala planetaria son las que dominan la distribución de brillo de las nubes en todas las latitudes. Además, la presencia de picos para $\hat{k}\sim 4$ y sus armónicos constituye ruido a\~{n}adido como consecuencia de usar cuatro planisferios para realizar las composiciones. Para cuantificar dicho ruido nos basta con comparar el espectro de potencias de la composición con el espectro de un planisferio individual (éste último restringido a $\hat{k}\geq3$), del que se obtiene una fluctuación de $\sim0.2$ en los valores de las pendientes. Dicho fluctuación es del mismo orden que el error mínimo que se deduce de fuentes de error anteriormente discutidas, y no se ve afectado por el periodo de rotación de las formaciones nubosas.\\

Por otro lado, ya que muchas de las estructuras celulares ubicadas en la región ecuatorial y latitudes medias poseen extensiones meridionales de $\sim200$ km, parece razonable estudiar también cómo varía la pendiente de los espectros con la latitud promediando en bandas mucho más estrechas. La Figura 4.8 nos muestra esta variación latitudinal ``fina'' de las pendientes para los dos rangos de números de onda tenidos en cuenta hasta ahora: $\hat{k}=1-10$ (correspondiente a escalas de $L\sim 30,000-3,000$ km) y $\hat{k}=10-50$ (escalas de $L\sim 3,000-500$ km). Una vez más hemos promediado para todas las composiciones, aunque esta vez usando bandas de latitud de 2$^{\circ}$ de ancho. Observamos que para los dos rangos de números de onda estudiados, la pendiente $n$ muestra un valor aproximado a -5/3 entre 45$^{\circ}$S y 30$^{\circ}$N. Fuera de esas latitudes, para el rango $\hat{k}=1-10$ ($L\sim 30,000-3,000$ km) se aprecia un crecimiento de la pendiente hasta -3, mientras que en el rango de números de onda elevados ($\hat{k}=10-50$, $L\sim 3,000-500$ km) no parece haber variaciones significativas de la pendiente $n$ con la latitud. La discrepancia observada para los comportamientos a altas latitudes de ambos rangos de escalas espaciales refleja la diferente naturaleza de las nubes que se encuentran en la zona subpolar (con un menor predominio de las escalas espaciales más peque\~{n}as) con respecto a las ubicadas en la zona ecuatorial y tropical.\\

\begin{figure}[h!]
	\centering
		\includegraphics[width=1.0\textwidth]{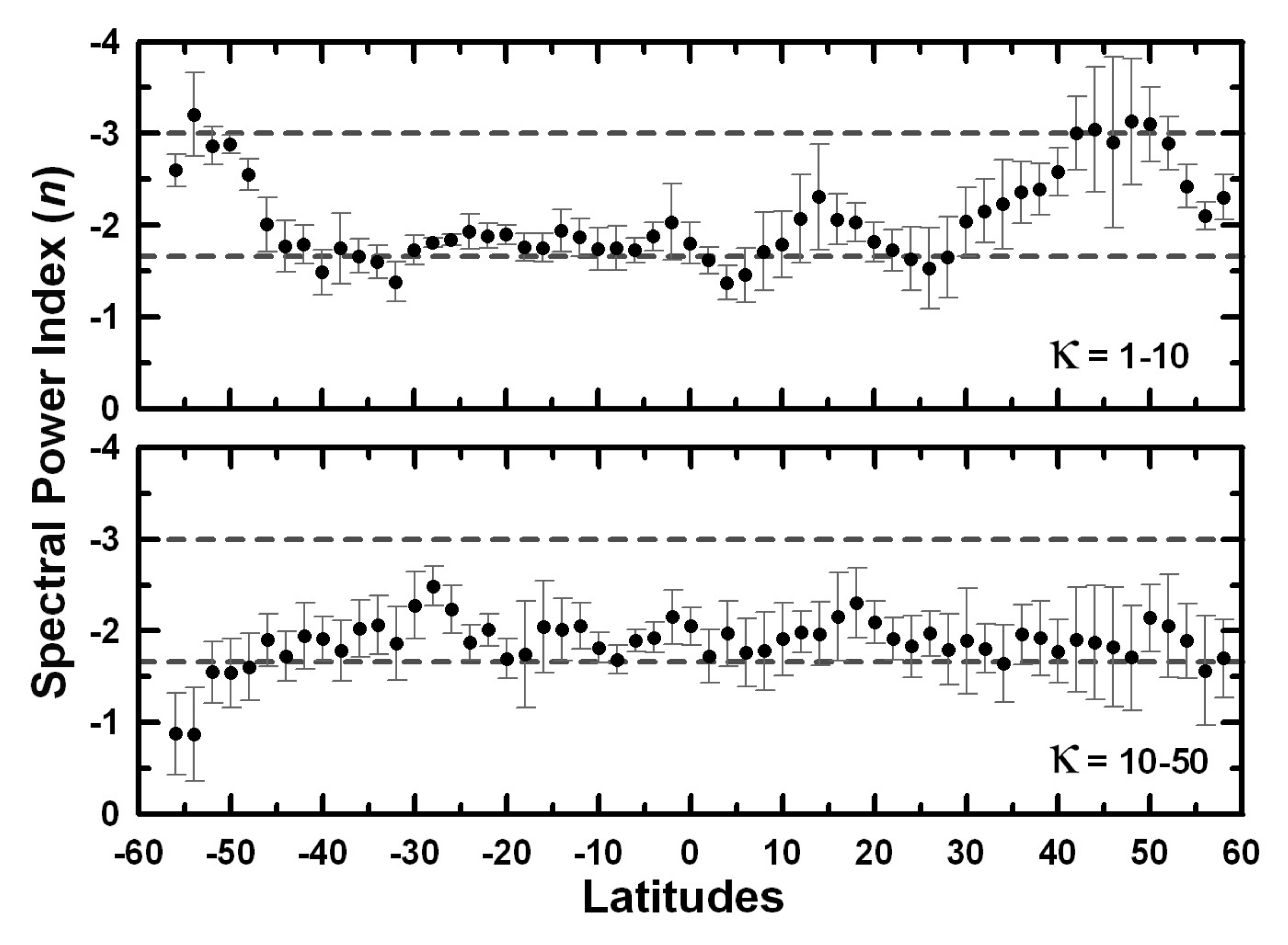}
	\label{fig:GalileoVenus-SpectraSlopes}
	\caption[Dependencia con la latitud de la pendiente del espectro.]{\scriptsize{Dependencia con la latitud de la pendiente $n$ del espectro de potencias a partir de las composiciones de imágenes de Galileo. Arriba están representadas las pendientes para el intervalo $\hat{k}=1-10$ y abajo para el intervalo $\hat{k}=10-50$. Las líneas a trazos marcan el valor de las pendientes $n=-5/3$ y $n=-3$.}}
\end{figure}

\section{Discusión}\label{chapter-turbulence-discuss}\indent

¿Hasta qué punto puede la distribución de brillo de las nubes reflejar las ca\-racterísticas de la turbulencia en Venus? ¿De qué manera están relacionados los espectros de potencias de la distribución de brillo con los espectros de potencias de energía cinética? \citet{Travis1978} encontró evidencias de esta relación al comparar para las nubes de la Tierra los espectros de energía cinética medidos unos a\~{n}os antes \citep{Julian1970,Julian1974} con los de distribución de brillo obtenidos con imágenes de la nave Mariner 10. También se han llevado a cabo modelos numéricos aplicables a Venus para estudiar la turbulencia en dos dimensiones sobre esferas débilmente rotantes \citep{Rossow1979,Yoden1993}, si bien estos modelos obvian muchas de las condiciones reales de la atmósfera de Venus y obtienen espectros de energía cinética con pendientes mucho más pronunciadas que las que se observan en los espectros de distribución de brillo.\\

En nuestros análisis de la distribución de brillo de las nubes en Venus a diferentes latitudes (ver Tabla \ref{tab:tabla-PendientesEspectros}) se observan los siguientes comportamientos:
\begin{enumerate}
	\item Entre 45$^{\circ}$S y 30$^{\circ}$N, el espectro de potencias exhibe un valor de pen\-dien\-te muy próximo a -5/3 para todos los números de onda ($\hat{k}=1-50$). Si asumimos que la distribución observada de escalas en las nubes se debe a la turbulencia entonces el régimen turbulento parece seguir, o bien la ley de potencias para turbulencia tridimensional descrita por la ecuación (\ref{Kolmogorov-Law}), o bien la ley de potencias para la turbulencia bidimensional dada por la ecuación (\ref{Kraichnan-Law-1.6}).
  \item Fuera del rango de latitudes anteriormente comentado (45$^{\circ}$S a 30$^{\circ}$N), el espectro de potencias guarda una mayor similitud con las ca\-rac\-te\-rís\-ti\-cas que muestra el espectro de energía cinética de la Tierra: una pendiente con $n\approx-3$ para números de onda peque\~{n}os (1-10), y una pendiente con $n\cong-5/3$) para números de onda mayores (10-50). Sin embargo, el número de onda para el cual se produce el cambio de régimen parece ser distinto en ambos planetas, con $\hat{k}=10$ ($L\sim3000$ km) para el espectro de brillo en Venus, y $\hat{k}=50-60$ ($L\sim500$ km) para el espectro de energía cinética en la Tierra. Este cambio de régimen del espectro de potencias para altas latitudes parece guardar relación con la influencia del vórtice polar tanto sobre el régimen turbulento dominante como sobre la morfología de las nubes y por ende la distribución de brillo mostrada.
\end{enumerate}

En la Tabla \ref{tab:tabla-EvolucionPendientes} examinamos la evolución temporal de las pendientes en espectros del brillo de las nubes para diferentes bandas de latitud, tomando para ello los valores medidos en 1974 durante las misión Mariner 10 \citep{Travis1978} y entre 1979-1980 durante la misión Pioneer Venus \citep{Rossow1980,delGenio1982}. Cada autor seleccionó el rango de números de onda en función de comportamiento del espectro de potencias obtenido a partir de las diferentes imágenes tomadas en cada misión. Si a partir de los datos de la Tabla \ref{tab:tabla-EvolucionPendientes} realizamos un promedio de los valores de pendiente para todas las bandas de latitud y rangos de números de onda, obtenemos para cada época de observación: $\left\langle n\right\rangle=-2.2\pm0.4$ (1974, Mariner 10), $\left\langle n\right\rangle=-2.6\pm0.3$ (1979-1980, Pioneer Venus), $\left\langle n\right\rangle=-2.1\pm0.3$ (1990, Galileo). Nótese que los números de onda estudiados en todas las épocas entran dentro del rango $\hat{k}=1-50$, escalas zonales situadas entre la escala planetaria ($L\sim30,000$ km) y la sub-sinóptica ($L\sim500$ km), y en ninguno de los casos se llega a la misoescala ($L\sim20-200$ km).\\

\begin{table}[h!]
	\caption{Evolución Temporal de las pendientes.}
	\label{tab:tabla-EvolucionPendientes}
	\centering
  \begin{spacing}{0.6}

		\begin{tabular}{*{6}{>{\scriptsize}c}}
			& & & & & \\
			\hline\hline
			& & & & & \\
			\multicolumn{2}{c}{\scriptsize\textit{Mariner 10 (1974)}} & \multicolumn{2}{c}{\scriptsize\textit{Pioneer Venus (1979-80)}} & \multicolumn{2}{c}{\scriptsize\textit{Galileo (1990)}} \\
			\multicolumn{2}{c}{\scriptsize\textit{$\hat{k}=3-30$}} & \multicolumn{2}{c}{\scriptsize\textit{$\hat{k}=5-50$}} & \multicolumn{2}{c}{\scriptsize\textit{$\hat{k}=1-50$}} \\
			& & & & & \\
      \textit{Rango Latitud} & \textit{n} & \textit{Rango Latitud} & \textit{n} & \textit{Rango Latitud} & \textit{n}  \\
			& & & & & \\
			\hline
			& & & & & \\
			  \textemdash  &  \textemdash  &  \textemdash  &  \textemdash  &  70$^{\circ}$N - 50$^{\circ}$N  &   $-2.2\pm0.2$   \\
			  \textemdash  &  \textemdash  &  50$^{\circ}$N - 30$^{\circ}$N  &      -3.0     &  50$^{\circ}$N - 30$^{\circ}$N  &   $-2.3\pm0.3$   \\
			  \textemdash  &  \textemdash  &  30$^{\circ}$N - 10$^{\circ}$N  &      -2.3     &  30$^{\circ}$N - 10$^{\circ}$N  &   $-2.0\pm0.2$   \\
			  15$^{\circ}$N - 15$^{\circ}$S  &      -2.7     &  10$^{\circ}$N - 10$^{\circ}$S  &      -2.4     &  10$^{\circ}$N - 10$^{\circ}$S  &   $-2.0\pm0.2$     \\
			  15$^{\circ}$S - 35$^{\circ}$S  &      -2.3     &  10$^{\circ}$S - 30$^{\circ}$S  &      -2.5     &  10$^{\circ}$S - 30$^{\circ}$S  &   $-2.0\pm0.2$   \\
			  35$^{\circ}$S - 55$^{\circ}$S  &      -1.7     &  30$^{\circ}$S - 50$^{\circ}$S  &      -2.7     &  30$^{\circ}$S - 50$^{\circ}$S  &   $-2.1\pm0.2$   \\
			  \textemdash  &  \textemdash  &  50$^{\circ}$S - 70$^{\circ}$S  &      -2.9     &  \textemdash  &  \textemdash    \\
			& & & & & \\
			\hline
			& & & & & \\

		\end{tabular}
  \end{spacing}
\end{table}

Observando la evolución temporal descrita en la Tabla \ref{tab:tabla-EvolucionPendientes} y teniendo en cuenta que el error a la hora de determinar la pendiente promedio es del orden del 10\% parece evidente que se han dado cambios temporales en algunas bandas de latitudes. Las imágenes tomadas en la misión Pioneer Venus \citep{Rossow1980,delGenio1982} muestran espectros más inclinados que los nuestros, lo que guarda relación con las diferencias en la morfología general de las nubes observada en ambas épocas \citep{Toigo1994}.\\

A continuación compararé las variaciones latitudinales experimentadas por las pendientes, la cizalla vertical y la componente zonal del viento. En la Figura 4.9 es posible visualizar tanto las pendientes de los espectros de brillo como las velocidades zonales y cizallas verticales durante las épocas de Pioneer Venus y Galileo\footnote{Excluimos de este estudio los escasos datos de Mariner 10 ya que tienen asociados errores de medida demasiado grandes.}. En primer lugar vemos que en ambas épocas las pendientes crecen de manera acusada en las latitudes más altas por encima de (40$^{\circ}$N y por debajo de 50$^{\circ}$S). Este crecimiento coincide en ambos hemisferios con la fuerte caída de los vientos hacia el polo (que es al mismo tiempo la zona de mayor cizalla meridional $\left(\partial \left\langle u\right\rangle/\partial y\right)$), y durante el sobrevuelo de Galileo el inicio del crecimiento en las pendientes se corresponde en latitud con la zona de mayor intensificación de la cizalla vertical del viento. En segundo lugar, ambas gráficas parecen indicar en latitudes medias y ecua\-to\-ria\-les una anti-correlación entre la intensidad del viento zonal y las pendientes de los espectros de potencias: el ligero aumento en la velocidad del viento parece corresponderse con un decrecimiento en el valor de las pendientes. Estos resultados podrían indicar una relación entre la intensidad del viento zonal y la morfología de las nubes, hecho que parece corroborarse cuando se comparan las imágenes en ultravioleta de la cima de las nubes en Venus a lo largo de las tres misiones y se comprueba que la evolución a largo plazo de la morfología de las nubes siempre viene acompa\~{n}ada de cambios notables en el perfil de los vientos.\\

\begin{figure}[h!]
	\centering
		\includegraphics[width=1.0\textwidth]{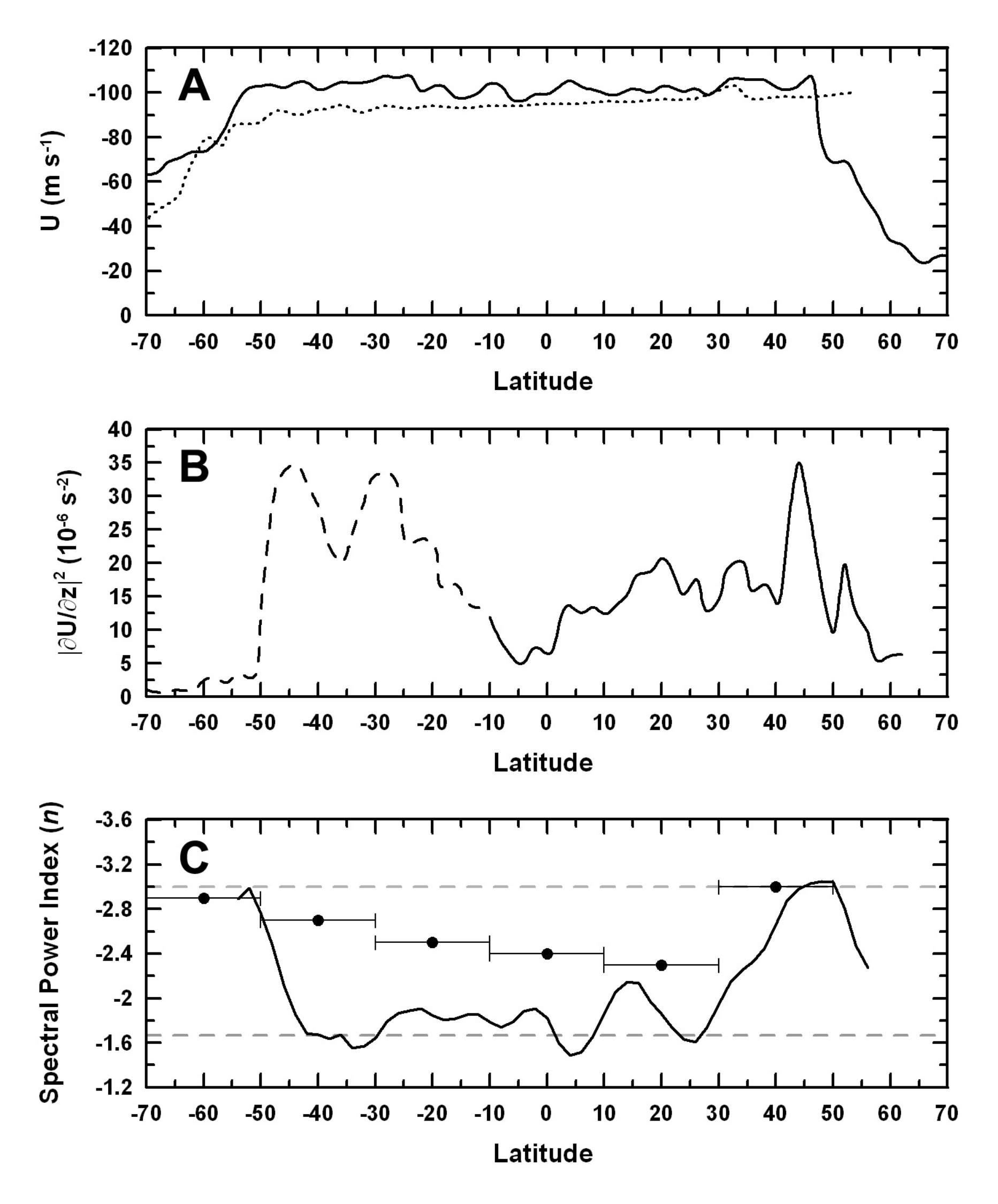}
	\label{fig:Venus-Winds-SpectraSlopes}
	\caption[Pendientes de los espectros en las diferentes misiones espaciales.]{\scriptsize{Comparación entre los perfiles meridionales del viento zonal (\textbf{A}) durante las misiones Pioneer Venus (línea en puntos suspensivos) y Galileo (línea continua), la cizalla vertical del viento (\textbf{B}) durante las misiones Galileo (línea continua) y Venus Express (línea a trazos), y la pendiente para $\hat{k}=1-10$ de los espectros de brillo (\textbf{C}) durante las misiones Pioneer Venus (puntos) y Galileo (línea continua). Aunque no disponemos datos de espectros de brillo durante la misión Venus Express, se a\~{n}ade la cizalla vertical en (\textbf{B}) para suplir la falta de datos de cizalla para el hemisferio en la misión Galileo. En la gráfica (\textbf{C}) la líneas grises a trazos designan los mismos valores de pendiente que en la Figura 4.8. y reflejan los valores de pendiente característicos de los regímenes turbulentos estudiados.}}
\end{figure}

Finalmente, dejaremos para trabajos posteriores la determinación directa de la distribución espectral de energía cinética (técnicamente mucho más difícil de obtener con los datos actuales) y su comparación con la distribución espectral del brillo de las nubes. Este avance fundamental podría venir de la mano de mediciones realizadas con los instrumentos VIRTIS y VMC a bordo de Venus Express, que podrían permitir de forma simultánea el estudio de la turbulencia tanto con espectros de energía cinética (a partir de mediciones de velocidad) como con espectros de la distribución de brillo. Asímismo harían posible extender este análisis a escalas más peque\~{n}as y a diferentes niveles verticales de la atmósfera. De igual manera, un análisis de la evolución temporal de ambos enfoques así como la aplicación de éstos a otros planetas podría ayudar a comprender con mayor profundidad los mecanismos que ligan la morfología de las nubes a la dinámica en las atmósferas planetarias.

\chapter{Ondas de gravedad}\label{chapter-gravitywaves}

\section{Introducción}\label{chapter-gravwaves-intro}\indent

Las atmósferas planetarias son capaces de desarrollar diferentes tipos de perturbaciones ondulatorias. Las fuerzas restauradoras que permiten el desarrollo de ondas en una atmósfera pueden deberse a la compresibilidad (ondas acústicas\index{Ondas!acústicas}), a la rotación de un sistema (ondas de Rossby) o a la flotación (ondas de gravedad\index{Ondas!de gravedad}). Como veremos más adelante, en la atmósfera de Venus adquieren un particular interés un tipo específico de ondas denominadas \textit{ondas internas de gravedad} en las que la fuerza restauradora es la fuerza de flotación debido a desviaciones internas de la densidad atmosférica \citep{Holton1992}. Estas ondas se desarrollan únicamente en condiciones de estabilidad atmosférica frente a movimientos verticales. En las estratosferas (regiones de alta estabilidad) de la mayoría de los planetas se suelen detectar gracias a las oscilaciones que provocan en la distribución vertical de temperaturas, aunque también podemos contemplar sus efectos en los campos de nubes de la Tierra \citep{Houze1993}, Marte \citep{Read2004}, Júpiter \citep{Hunt1979,Flasar1986,Reuter2007} y Venus \citep{Gierasch1997,Markiewicz2007b}.\\

Las ondas de gravedad resultan de gran interés para el estudio de la dinámica atmosférica ya que su existencia nos alerta de la presencia de condiciones de estabilidad vertical y sus características nos permiten cuantificar este parámetro fundamental que es en general difícil de obtener. Por otro lado, también intervienen en un gran número de fenómenos dinámicos (incluyendo tanto la agitación turbulenta como el transporte vertical de energía y momento). En el caso de la atmósfera de Venus, estas ondas podrían jugar un papel importante en la circulación general de la atmósfera transportando momento desde la superficie, si bien no se ha llegado a ningún consenso respecto al grado en que éstas pudieran alimentar y mantener la superrotación atmosférica \citep{Hou1987,Gierasch1987,Leroy1995}.\\

Gracias al descenso de varias sondas espaciales (9 sondas Venera y 5 sondas Pioneer Venus) a diferentes latitudes y horas locales, se ha determinado la existencia de al menos dos capas estables \citep{Kliore1980,Kliore1985,Gierasch1987} (ver Figura 5.1): una capa entre 30 y 40 km, y otra por encima de los 55 km. Debido a que dichos niveles albergan nubes densas \citep{Esposito1997}, es posible observar en ellas ondas de gravedad manifestándose como patrones regulares en el albedo de las nubes \citep{Markiewicz2007b}. Las regiones por debajo de los 30 km de altura, y la situada entre 48 y 55 km son básicamente inestables y por tanto no reúnen las condiciones favorables para el desarrollo y propagación de ondas internas de gravedad, si bien podrían tener un papel en la generación de las perturbaciones que dan lugar a éstas \citep{Baker1998,Baker2000a,Baker2000b}.\\

\begin{figure}[h!]
	\centering
		\includegraphics[height=0.6\textheight]{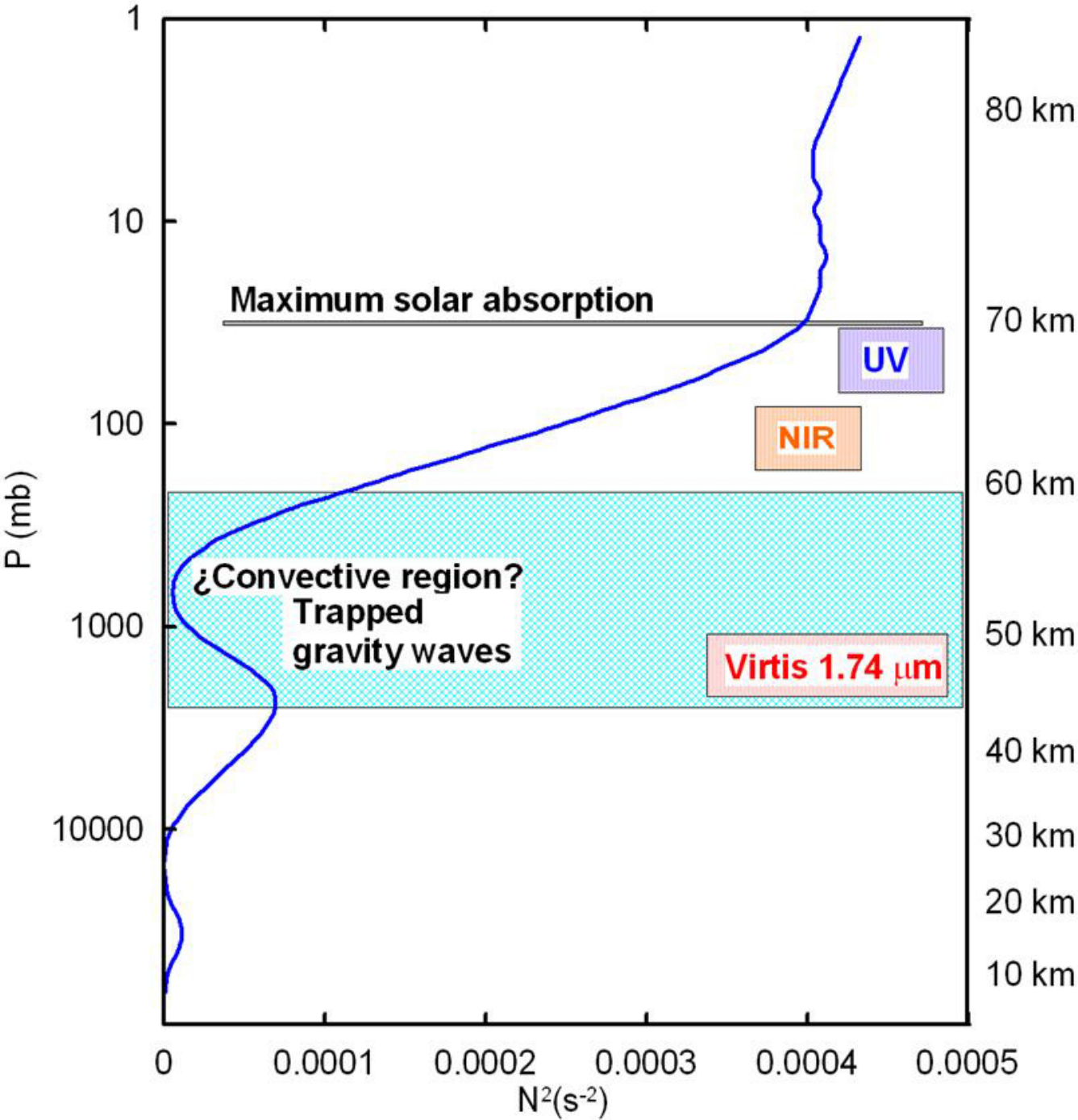}
	\label{fig:Venus-BruntVaisalla}
	\caption[Perfil vertical de la frecuencia de Brunt-Väisälä en Venus.]{\scriptsize{Perfil vertical esquemático simplificado de la frecuencia de Brunt-Väisälä\index{Brunt-Väisälä!frecuencia de} en la atmósfera de Venus. Existen dos regiones importantes de estabilidad: entre 30 y 48 km y por encima de los 55 km. Las regiones por debajo de los 30 km de altura, y la situada entre 48 y 55 km son básicamente inestables. También se indican los niveles de altura que se corresponden con las longitudes de onda de las imágenes de nuestro estudio: $380~nm$ (la cima de las nubes), $980~nm$ (base de la cima de las nubes) y $1.74~\mu m$ (nubes inferiores).}}
\end{figure}

Las imágenes del instrumento VIRTIS-M han mostrado la presencia de ondas de mesoescala en las nubes de Venus a diferentes alturas. Estas ondas han sido identificadas como ondas de gravedad y en este capítulo presentaré el trabajo que llevamos a cabo para su identificación y caracterización en dos niveles distintos de la atmósfera de Venus, siendo esta la primera vez que se estudian las ondas de gravedad en la parte baja de las nubes. Comparando dichas ondas de gravedad con un modelo lineal clásico y de acuerdo con anteriores estudios \citep{Gierasch1987,Leroy1995}, comprobamos que en Venus existe una región estable entre 30-48 km, ampliamente extendida (cubriendo latitudes entre 40$^{\circ}$ y 80$^{\circ}$S) y persistente en el tiempo.\\

\section{Estudios previos de ondas de gravedad en Venus}\indent

En Venus, las ondas de gravedad al nivel de las nubes fueron descubiertas exami\-nando imágenes en ultravioleta correspondientes a la cima de las nubes ($\sim65$ km) tomadas por las naves Mariner 10 \citep{Belton1976a} y Pioneer Venus \citep{Rossow1980}. Las imágenes de Mariner 10 mostraron en el ecuador de Venus los denominados \textit{cinturones circumecuatoriales}\index{Cinturón circumecuatorial}: unas estructuras alargadas de unos 5,000 km, orientadas meridionalmente, con un patrón periódico que se repetía cada 500 km y que se desplazaban hacia el sur con una velocidad de fase de $\sim20$ m/s \citep{Belton1976a}. Dichos cin\-tu\-ro\-nes circumecuatoriales también fueron observados durante la misión Pioneer Venus, además de un nuevo tipo de trenes de onda con una disposición más zonal y compuesto por estrías oscuras de $\sim2,000$ km, se\-paradas entre sí $\sim200$ km \citep{Rossow1980}. En 1990, la nave Galileo volvió a tomar imágenes de Venus, aunque esta vez no se encontró evidencia de ondas de gravedad de escala peque\~{n}a ni de cinturones circumecuatoriales \citep{Belton1991,Toigo1994,Peralta2007a}. Esta ausencia de ondas de gravedad estaba aparentemente relacionada con una menor abundancia de nubes de aspecto celular probablemente convectivas. Las imágenes de alta resolución obtenidas por la cámara VMC (Venus Monitoring Camera)\index{Cámara!VMC} de Venus Express han mostrado recientemente en diferentes latitudes patrones nubosos claramente identificables con ondas de gravedad de longitudes de onda de tan solo unas decenas de kilómetros \citep{Markiewicz2007b}.\\

La principal evidencia de la existencia de ondas de gravedad en Venus proviene, sin embargo, de medidas ``in situ'' efectuadas por las cuatro sondas de la misión espacial Pioneer Venus. Estas sondas se adentraron en la atmósfera de Venus proporcionando perfiles verticales tanto de temperatura (por encima de los 110 km de altura) como de velocidad del viento (zona baja de las nubes) que evidenciaron variaciones ondulatorias asociadas a longitudes de onda verticales entre 5 y 10 km \citep{Seiff1980,Counselman1980}. También se han encontrado variaciones verticales similares en sondeos de temperatura a partir de datos de emisión térmica \citep{Taylor1980} entre la cima de las nubes ($\sim65$ km) y la base de la termosfera ($\sim100$ km) en experimentos de radio ocultación efectuados por Pioneer Venus \citep{Kliore1980}, Venera 9 \citep{Kolosov1980} y Magallanes \citep{Hinson1995}. Gracias al estudio de las fluctuaciones respecto al valor medio de la densidad atmosférica medidas con el espectrómetro de masas en Pioneer Venus \citep{Niemann1980,Kasprzak1988} y en las emisiones de $CO_{2}$ de la alta atmósfera utilizando datos de VIRTIS \citep{Garcia2008}, se han detectado también ondas de gravedad con longitudes de onda verticales entre 100-600 km en niveles superiores de la termosfera ($z\gtrsim100$ km).\\

\section{Observaciones de ondas en las nubes}\label{chapter-gravwaves-observs}\indent

Las imágenes usadas en este trabajo cubren los periodos de tiempo comprendidos entre el 12 de abril de 2006 (inserción orbital) y el 9 de marzo de 2007 para el lado nocturno de Venus (rango infrarrojo), y desde la inserción orbital hasta el 28 de julio de 2007 para las imágenes del lado diurno (rango del visible), un mayor periodo necesario para poder compensar el escaso número de órbitas que contienen imágenes de lado diurno con alta resolución y buen contraste. De esta manera, teniendo en cuenta tanto la calidad de imagen como la resolución espacial, seleccionamos un total de 112 órbitas en el visible y 116 en el infrarrojo.\\

En la Figura 5.2 mostramos ejemplos de algunas de las ondas observadas en el campo de nubes con imágenes tomadas en $380~nm$ (UV), $980~nm$ (NIR) y $1.74~\mu m$ (IR). En todos ellos se observa un patrón alternante de estrías bri\-llan\-tes y oscuras dispuestas espacialmente de forma regular. Dichas estrías nos muestran los efectos de las ondas de gravedad sobre el campo de nubes, lo que se traduce en cambios de la reflectividad para las nubes de imágenes en UV y NIR (observaciones del lado diurno) y cambios de la opacidad para imágenes en IR (observaciones del lado nocturno). Las observaciones nos muestran paquetes de ondas con valles y crestas distribuidos prácticamente en la dirección zonal, y extendiéndose de manera perpendicular al paquete y no más allá de unos cuantos grados de latitud. En muchos casos los valles son más anchos que las crestas, o simplemente desaparecen. Muchos de los paquetes son muy regulares, mientras que en otros casos las crestas no siempre son equidistantes, o bien algunas son más extensas en latitud que otras, o bien muestran una ligera curvatura. Todas estas irregularidades no son exclusivas de las ondas observadas en campos de nubes de Venus, sino que han podido también visualizarse en las nubes de Júpiter \citep{Flasar1986}.\\

\begin{figure}[h!]
	\centering
		\includegraphics[height=0.6\textheight]{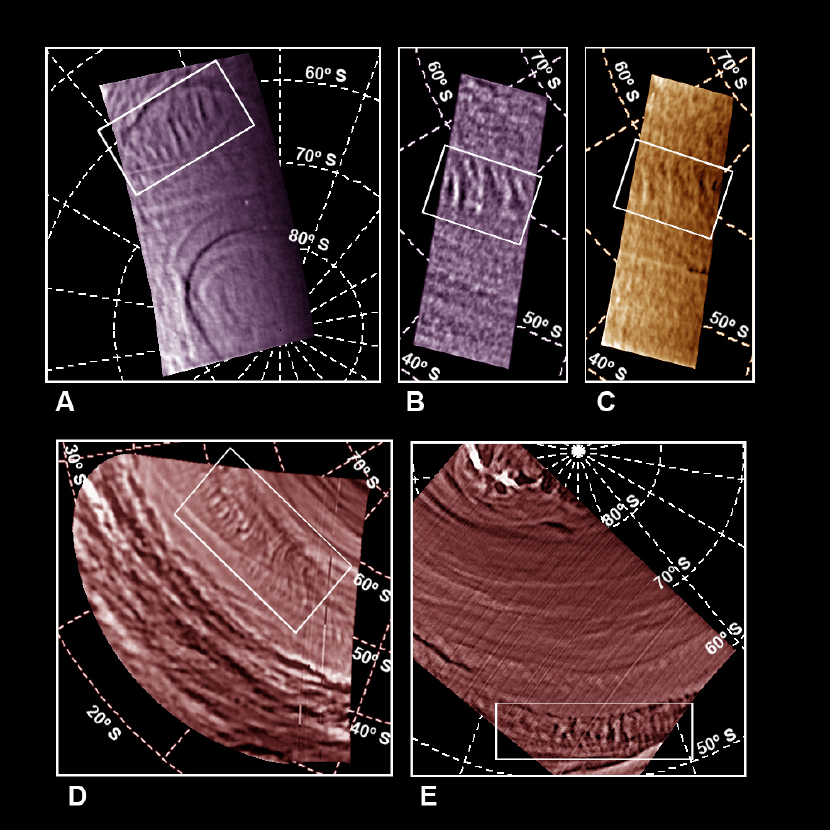}
	\label{fig:VEXVenus-WavesExamples}
	\caption[Ejemplos de ondas en las nubes de Venus.]{\scriptsize{Ejemplos de ondas en las nubes de Venus a partir de imágenes de VIRTIS-M. \textbf{(A)} y \textbf{(B)} nos muestran ondas observadas a $380~nm$ en la cima de las nubes. La onda que aparece en \textbf{(C)} se corresponde con una imagen de la base de la cima de las nubes a $980~nm$ y se correlaciona con la estructura observada en $380~nm$ \textbf{(B)}, lo que podría corresponderse con una onda que se extiende verticalmente o a que los niveles de ambas longitudes de onda estén más cerca para ciertas latitudes. En \textbf{(D)} y \textbf{(E)} tenemos paquetes de ondas más grandes y pertenece a imágenes de las nubes inferiores a $1.74~\mu m$.}}
\end{figure}

En esta tesis llevamos a cabo una búsqueda sistemática de ondas similares a las que pueden verse en la Figura 5.2, lo que llevó a identificar hasta 6 paquetes de onda en UV, uno solo\footnote{Este paquete de ondas se observó simultáneamente en la región subpolar de Venus en UV y NIR, por lo que podríamos tener el caso de una onda que se extiende verticalmente entre ambos niveles de nubes, o más probablemente una región en la que ambos niveles verticales se encuentran muy próximos} en NIR y un total de 30 paquetes de onda en las imágenes de IR. En algunos casos se observan varios paquetes de onda en la misma imagen, cada uno localizado en una latitud diferente. La mayor abundancia de ondas observada en las imágenes de IR se traduce también en una mayor diversidad en las propiedades que muestran. Respecto al contraste promedio que exhiben las imágenes, éste varía con la longitud de onda usada. En las imágenes de IR tenemos contrastes que varía desde 15~\% (donde las ondas son claramente observables) hasta valores del orden del 1~\% (ondas difíciles de ver). Las imágenes de UV en general muestran contrastes débiles (1~\%), por lo que las ondas no eran visibles hasta después de procesar mucho las imágenes. Es importante anotar que este estudio no descarta la existencia de ondas más sutiles dentro del campo de nubes, ya sea por un contraste insuficiente en las imágenes de estudio, o por tener longitudes de onda peque\~{n}as que no puedan ser resueltas por VIRTIS, como las que han sido encontradas en la nube superior por la cámara de más alta resolución VMC, con longitudes de onda de pocos kilómetros a unas decenas \citep{Markiewicz2007b}.\\

En la Figura 5.3 se muestran mapas en términos de hora local y de la latitud del número de imágenes analizadas tanto en el lado diurno como en el nocturno con una resolución adecuada para el estudio de estas ondas (mejor que 30 km/pixel), lo que se corresponde en ambos casos con latitudes entre el polo sur y 30$^{\circ}$S. La Figura 5.3 muestra también la localización de estas ondas que está correlacionada con el número de observaciones de alta calidad de cada región del planeta.\\

\begin{figure}[h!]
	\centering
		\includegraphics[width=0.8\textwidth]{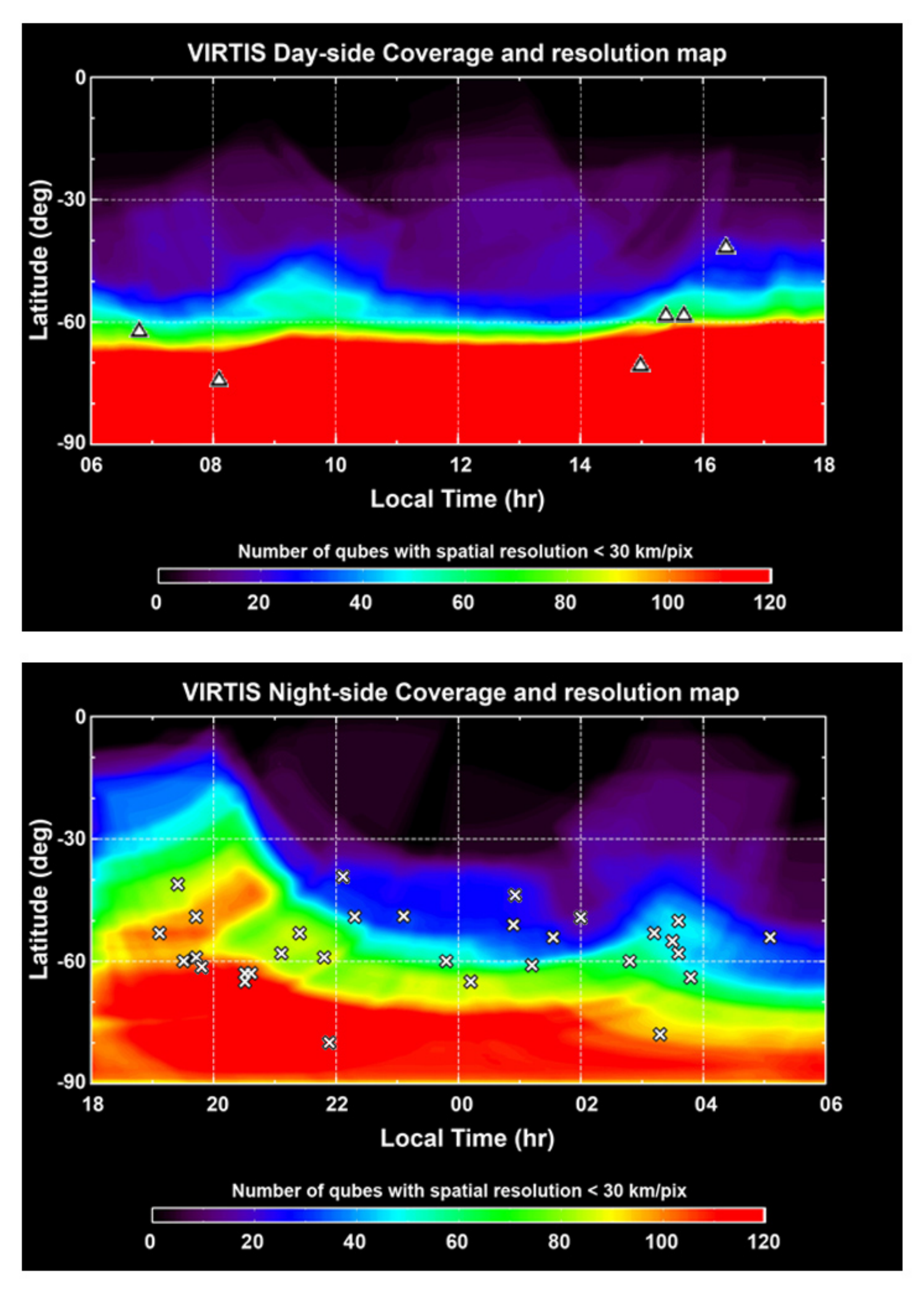}
	\label{fig:Waves-ResolutionMaps}
	\caption[Mapas globales del número de imágenes de alta resolución.]{\scriptsize{Mapas globales del número de imágenes de alta resolución espacial (mejor que 30 km/pix) en función de la latitud y la hora local de Venus. Arriba tenemos el mapa de las observaciones del lado diurno y abajo las tomadas del lado nocturno. Un valor de 30 km/pix es una resolución suficiente para poder observar ondas con longitudes de onda mayores que $\sim100$ km. Aunque en número escaso, también existen imágenes con resoluciones del orden de 15-20 km/pix. Adicionalmente se muestra que las posiciones de las ondas encontradas (triángulos para las ondas en la cima de las nubes y cruces para las de las nubes inferiores) se correlacionan con las regiones con un mayor número de observaciones de alta resolución.}}
\end{figure}

\section{Resultados}\indent

Para todos estos paquetes de onda\index{Ondas!paquetes de} se realizó una medición de las distintas propiedades observables tales como su localización tanto geográfica (latitud y longitud) como en términos de insolación (latitud y hora local), la longitud del paquete de ondas\index{Ondas!paquetes de!longitud} (es decir, su extensión en la dirección perpendicular a la de los frentes de onda), su anchura\index{Ondas!paquetes de!ancho} (es decir, su extensión en la dirección de los frentes de onda) y la orientación\index{Ondas!paquetes de!orientación} respecto de los paralelos (las líneas de latitud constante). Debido a las muy variadas características de las imágenes y patrones de ondas observados, se procedió a un procesado individualizado combinando aumento del contraste, máscara de enfoque y filtros de suavizado. Los resultados se muestran en la Tabla \ref{tab:table-PropertiesWavePackets} (ver página \pageref{tab:table-PropertiesWavePackets}).\\

\begin{table}[h!]
	\caption{Propiedades de los paquetes de ondas.}
	\label{tab:table-PropertiesWavePackets}
	\centering
  \begin{spacing}{0.6}

		\begin{tabular}{*{7}{>{\scriptsize}c}}
		  \multicolumn{7}{l}{Tabla \ref{tab:table-PropertiesWavePackets}A: Propiedades de los paquetes de ondas en imágenes UV.} \\
			& & & & & & \\
			\hline\hline
			& & & & & & \\
			\multirow{2}{*}{\textit{órbita}} & \textit{Fecha} & \textit{Latitud} & \textit{Hora} & \textit{Longitud del} & \textit{Ancho del} & \textit{Orientación} \\
                                   & \textit{(dd/mm/aa)} & \textit{(grad)} & \textit{Local} & \textit{Paquete (km)} & \textit{Paquete (km)} & \textit{(grad)} \\
			& & & & & & \\
			\hline
			& & & & & & \\
			 59 & 18/06/2006 & -41 & 16.4 & $>860$ & 170 & 25 \\
		 	 59 & 18/06/2006 & -58 & 15.7 & $>865$ & 340 & 30 \\
			170 & 07/10/2006 & -62 & 06.8 &  1275  & 340 & 25 \\
			255 & 31/12/2006 & -70 & 15.0 &   980  & 420 & 40 \\
			388 & 13/05/2007 & -74 & 08.1 &   640  & 335 &  2 \\
			& & & & & & \\
			\hline
			& & & & & & \\
			& & & & & & \\
			& & & & & & \\
			& & & & & & \\

		  \multicolumn{7}{l}{Tabla \ref{tab:table-PropertiesWavePackets}B: Propiedades de los paquetes de ondas en imágenes NIR} \\
			& & & & & & \\
			\hline\hline
			& & & & & & \\
			\multirow{2}{*}{\textit{órbita}} & \textit{Fecha} & \textit{Latitud} & \textit{Hora} & \textit{Longitud del} & \textit{Ancho del} & \textit{Orientación} \\
                                   & \textit{(dd/mm/aa)} & \textit{(grad)} & \textit{Local} & \textit{Paquete (km)} & \textit{Paquete (km)} & \textit{(grad)} \\
			& & & & & & \\
			\hline
			& & & & & & \\
		 	 59 & 18/06/2006 & -58 & 15.4 & $>860$ & 350 & 35 \\
			& & & & & & \\
			\hline
			& & & & & & \\
			& & & & & & \\
			& & & & & & \\
			& & & & & & \\

		  \multicolumn{7}{l}{Tabla \ref{tab:table-PropertiesWavePackets}C: Propiedades de los paquetes de ondas en imágenes IR} \\
			& & & & & & \\
			\hline\hline
			& & & & & & \\
			\multirow{2}{*}{\textit{órbita}} & \textit{Fecha} & \textit{Latitud} & \textit{Hora} & \textit{Longitud del} & \textit{Ancho del} & \textit{Orientación} \\
                                   & \textit{(dd/mm/aa)} & \textit{(grad)} & \textit{Local} & \textit{Paquete (km)} & \textit{Paquete (km)} & \textit{(grad)} \\
			& & & & & & \\
			\hline
			& & & & & & \\
			 84 & 13/07/2006 & -80 & 21.9 &  $>600$ & 225 & 50 \\
		 	 96 & 25/07/2006 & -48 & 02.0 &   1750  & 520 & 15 \\
			 97 & 26/07/2006 & -39 & 22.1 &    660  & 160 & 35 \\
		 	 97 & 26/06/2006 & -61 & 19.8 &    570  & 160 & 40 \\
			100 & 29/07/2006 & -55 & 03.5 & $>1525$ & 270 &  2 \\
			112 & 10/08/2006 & -60 & 23.8 &    610  & 105 & 20 \\
			112 & 10/08/2006 & -49 & 23.1 &    410  & 170 & 40 \\
			112 & 10/08/2006 & -49 & 22.3 &    730  & 320 & 25 \\
			112 & 10/08/2006 & -53 & 21.4 &    785  & 200 &  8 \\
			112 & 10/08/2006 & -58 & 21.1 &    575  & 205 &  7 \\
			113 & 11/08/2006 & -54 & 01.5 & $>1140$ & 180 & 17 \\
			114 & 12/08/2006 & -54 & 05.1 &  $>645$ & 205 & 14 \\
			118 & 16/08/2006 & -65 & 00.2 &    635  & 265 &  2 \\
			141 & 08/09/2006 & -65 & 20.5 &    520  & 370 & 11 \\
			142 & 09/09/2006 & -64 & 03.8 & $>1415$ & 160 & 13 \\
			161 & 28/09/2006 & -58 & 03.6 &   1050  & 310 & 17 \\
			164 & 01/10/2006 & -50 & 03.6 &    335  & 380 & 20 \\
			166 & 03/10/2006 & -53 & 03.2 &  $>790$ & 225 & 12 \\
			228 & 04/12/2006 & -51 & 00.9 &    690  & 370 & 10 \\
			231 & 07/12/2006 & -78 & 03.3 &    280  & 130 & 30 \\
			261 & 06/01/2007 & -59 & 21.8 &    620  & 240 & 13 \\
			290 & 04/02/2007 & -63 & 20.6 &  $>440$ & 250 &  4 \\
			313 & 27/02/2007 & -60 & 02.8 &  $>860$ & 315 & 11 \\
			315 & 01/03/2007 & -61 & 01.2 &    760  & 335 &  0 \\
			317 & 04/03/2007 & -53 & 19.1 & $>1230$ & 170 & 10 \\
			317 & 04/03/2007 & -59 & 19.7 &    685  & 200 & 15 \\
			317 & 04/03/2007 & -63 & 20.5 &    450  & 145 &  7 \\
			318 & 05/03/2007 & -60 & 19.5 &    640  & 150 &  0 \\
			321 & 07/03/2007 & -49 & 19.7 & $>1300$ & 210 & 12 \\
			323 & 10/03/2007 & -41 & 19.4 &    615  & 200 &  0 \\
			& & & & & & \\
			\hline
			& & & & & & \\
		\end{tabular}

  \end{spacing}
\end{table}

\subsection{Relación entre las ondas y la topografía}\indent

¿Existe alguna relación entre las ondas observadas y la orografía de la superficie? En la Tierra \citep{Holton1992,Salby1996} y Marte \citep{Read2004} se suelen observar un tipo de ondas de gravedad llamadas \textit{ondas de sotavento}\index{Ondas!de sotavento}. Éstas se forman sobre las cima de las monta\~{n}as (o sobre cráteres) cuando sopla el viento, curvándose las líneas de corriente verticalmente siguiendo la estructura orográfica (ver Figura 14.13 en \citealt{Holton1992}). Si la humedad ambiente en la cima es alta, se produce condensación en las crestas (gotitas de agua en la Tierra, cristales de $CO_{2}$ en Marte) y las bandas de nubes se hacen visibles. En Venus, las ondas de este estudio están localizadas muy por encima de la superficie, y no parecen ser producto de condensación adicional de $SO_{4}H_{2}$ ya que se forman sobre la nube ya existente, siendo quizás resultado de la acumulación o incremento de la densidad en las crestas de las ondas. Las ondas de gravedad que estamos observando se localizan principalmente en los regiones de estabilidad (ver Figura 5.1): la situada entre 30 y 48 km de altura, y la localizada a partir de los 55 km. También se ha detectado una peque\~{n}a región de estabilidad entre 10 y 20 km de altura. El resto de intervalos son zonas de inestabilidad donde $N^{2}\sim0$ y donde las ondas de gravedad no pueden propagarse. El que las ondas detectadas entre 30 y 48 km sean producidas por la orografía parece improbable, al menos en el hemisferio sur del planeta donde la superficie es bastante llana. A esto hay que a\~{n}adir que por debajo de los 10 km de altitud los vientos en Venus son insignificantes, con velocidades que no exceden los $1.5~m\cdot s^{-1}$ cerca de la superficie \citep{Kerzhanovich1980}, y que las regiones inestables intermedias constituyen una barrera natural a la propagación vertical de las ondas de gravedad (si bien, en caso de que exista en ellas convección podrían constituir una fuente de ondas como discutiremos en la sección \ref{chapter-gravwaves-discuss}.\\

La Figura 5.4A nos muestra la localización de las ondas en términos de latitud y topografía de la superficie. Para la nube inferior (imágenes en IR) la Figura 5.4A parece evidenciar una ausencia de relación entre las ondas y las elevaciones de superficie. Este resultado está en contraposición con los resultados de las observaciones en el hemisferio norte, donde los globos VEGA registraron oscilaciones en la velocidad vertical del viento en la nube inferior, claramente influenciadas por elevaciones monta\~{n}osas \citep{Sagdeev1986}. Dichas oscilaciones fueron identificadas con ondas de gravedad, originadas por la interacción del viento con cadenas monta\~{n}osas del área de Afrodita Terra y que rondan alturas de entre 2 y 5 km de altura \citep{Young1987,Young1994}. Sin embargo esta dualidad de resultados puede explicarse si se tiene en cuenta que el hemisferio sur de Venus es prácticamente plano, y que las únicas elevaciones de interés se encuentran en latitudes tropicales difíciles de muestrear con VIRTIS. Así pues, estas ondas parecen estar presentes por doquier en la nube inferior, excepto quizás en la región interna del vórtice polar sur. Esto último podría ser consecuencia de un efecto observacional más que de una ausencia real ya que la parte interior del dipolo es particularmente oscura en la longitud de onda IR utilizada.\\

\begin{figure}[h!]
	\centering
		\includegraphics[width=1.0\textwidth]{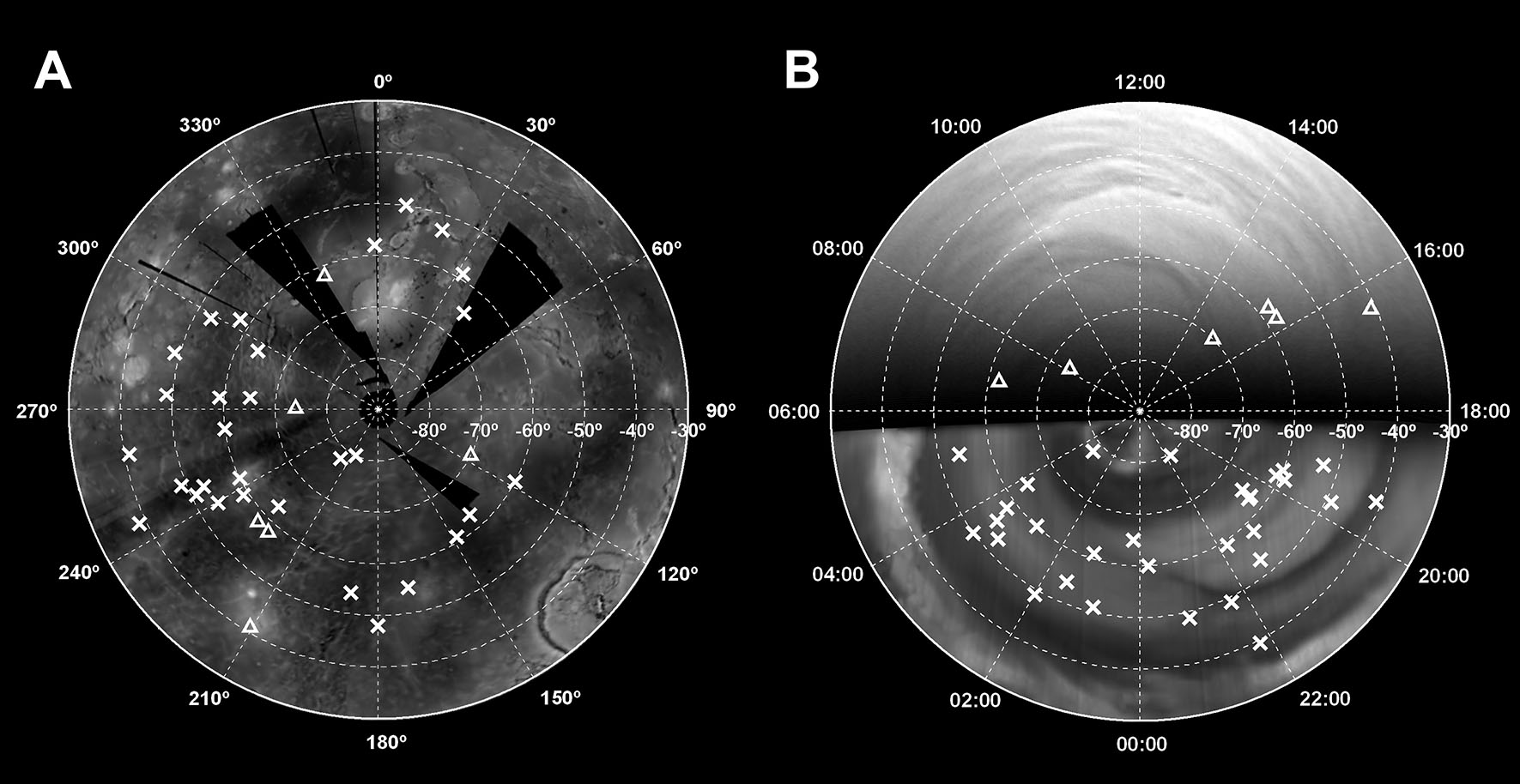}
	\label{fig:Waves-locations}
	\caption[Relación de las ondas con la topografía y la hora local.]{\scriptsize{Mapa de la localización de las ondas en función de la longitud geográfica, hora local, morfología de las nubes y topografía de la superficie de Venus. Las ondas de la cima de las nubes se representan con triángulos y las de las nubes inferiores con cruces. \textbf{(A)} Posición de los paquetes de ondas sobre una proyección de la superficie del hemisferio sur obtenida por el instrumento IDRS de la nave Magallanes \citep{Saunders1991}. La elevación de la superficie varía entre -1.3 km (llanuras oscuras) hasta +2.5 km (monta\~{n}as brillantes). Debido al sampleo de las observaciones, tenemos una mayor cantidad de ondas entre las longitudes 210$^{\circ}$-300$^{\circ}$. \textbf{(B)} Posición de los paquetes de ondas en función de la latitud y la hora local, al mismo tiempo que observamos las estructuras nubosas típicas de cada nivel con una proyección polar del hemisferio sur a partir de una imagen de Venus tomada durante la inserción orbital de Venus Express.}}
\end{figure}

\subsection{Relación entre las ondas y la insolación}\indent

La Figura 5.4B nos muestra la posición de las ondas en función de la latitud y la hora local. La escasez de ondas detectadas en las imágenes co\-rres\-pon\-dien\-tes a la cima de las nubes (UV y NIR) impide cualquier conclusión clara sobre su distribución, pero parecen agruparse en torno a las primeras horas de la ma\~{n}ana y bien avanzada la tarde. Por otro lado, ni en la cima de las nubes ni en las nubes inferiores encontramos ondas en latitudes más ecuatoriales que 40$^{\circ}$S. Aunque esto podría deberse a la escasez de imágenes de resolución aceptable para esta región, resulta interesante comentar que en ambos niveles existe una región de transición hacia estructuras nubosas distintas, de bandas oscuras en la cima de las nubes (UV) y de aspecto turbulento en la nube inferior (IR). Esta última podría tanto dificultar la propia visualización de las ondas de gravedad como impedir que éstas se generen en caso de re\-presentar una región activa de convección celular.\\

En el lado diurno podría existir una relación entre la presencia de las ondas de gravedad en la cima de las nubes y la convección. Por un lado, es posible que la radiación solar penetre hasta capas situadas por debajo del nivel superior de las nubes\footnote{De hecho, justo por debajo de la cima de las nubes tenemos que la constante de enfriamiento radiativo $\tau_{rad}\sim100$ días ($\tau_{dia}\sim117$ días), por lo que el aire tendría tiempo para enfriarse durante la noche y se calentaría durante el día (ver Figura 20 en \citealt*{Schubert1983}).} y que caldee dicha región, dispararando la convección y excitando de esta manera ondas de gravedad hacia arriba. Por otro lado, la fuerte deposición de radiación solar que se produce en la parte superior de las nubes modifica el perfil vertical de temperatura de tal ma\-ne\-ra que reduce el ritmo con el que la temperatura disminuye con la altura. Consecuentemente esto produce estabilidad atmosférica (ya que entonces el gradiente térmico vertical es menor que el gradiente adiabático seco, $\Gamma<\Gamma_{d}$), y esta estabilidad, a su vez, favorece la existencia y propagación de ondas de gravedad en estos niveles de la parte diurna del planeta.\\

\subsection{Propiedades de las ondas}\indent

Los paquetes de onda tienen longitudes comprendidas entre 640 y 1,200 km en la cima de las nubes (UV), con un valor medio de 920 km, mientras que los de la nube inferior rondan el rango comprendido entre 280 y 1,700 km, con un valor medio de unos 760 km. Por otro lado el ancho promedio de los paquetes es de 320 y 230 km para la cima y la parte baja de las nubes respectivamente. Además de las magnitudes previamente mencionadas, también se midió la longitud de onda, el número de crestas y la velocidad de fase\index{Velocidad!de fase}\footnote{ésta última siempre que podíamos observar al menos 1 hora de evolución temporal de un mismo campo de nubes} (la velocidad con la que se mueven los valles y las crestas). Se consideró importante estimar el viento promedio en que el que se movían los paquetes de onda a partir de trazadores cercanos a éstas en vez de usar datos de perfiles de viento procedentes de promedios temporales, ya que, como se vio en el capítulo \ref{chapter-winds}, el viento a la altura de las nubes está sujeto a importantes variaciones temporales \citep{Sanchez-Lavega2008}. Dentro de las escalas de tiempo estudiadas los paquetes de onda apenas mostraban cambios de morfología y variaciones de albedo, lo que indica que en estas escalas dichas ondas se comportan como no dispersivas. Todas estas propiedades son des\-cri\-tas en detalle en la Tabla \ref{tab:table-PropertiesWaves} (ver página \pageref{tab:table-PropertiesWaves}).\\

\begin{table}[h!]
	\caption{Propiedades de las ondas.}
	\label{tab:table-PropertiesWaves}
	\centering
  \begin{spacing}{0.6}

		\begin{tabular}{*{7}{>{\scriptsize}c}}
		  \multicolumn{7}{l}{Tabla \ref{tab:table-PropertiesWaves}A: Propiedades de las ondas en imágenes UV} \\
			& & & & & & \\
			\hline\hline
			& & & & & & \\
			\multirow{2}{*}{\textit{órbita}} & \textit{Latitud} & \textit{Número de} & \textit{Longitud de} & \textit{Ancho} & $c_{x}$ & $|\stackrel{\rightarrow}{c}| - \overline{u}$ \\
                                   & \textit{(grad)} & \textit{Crestas} & \textit{Onda (km)} & \textit{Cresta (km)} & \textit{(m/s)} & \textit{(m/s)} \\
			& & & & & & \\
			\hline
			& & & & & & \\
			 59 & -41 &  8 &  95 &  45 &        -113 &         -12 \\
		 	 59 & -58 &  7 & 115 &  60 &        -108 &         -40 \\
			170 & -62 &  5 & 210 & 125 &         -62 &         0.5 \\
			255 & -70 & 11 &  90 &  45 &         -29 &          14 \\
			388 & -74 &  4 & 170 & 100 &         -76 &         -36 \\
			& & & & & & \\
			\hline
			& & & & & & \\
			& & & & & & \\
			& & & & & & \\
			& & & & & & \\

		  \multicolumn{7}{l}{Tabla \ref{tab:table-PropertiesWaves}B: Propiedades de las ondas en imágenes NIR} \\
			& & & & & & \\
			\hline\hline
			& & & & & & \\
			\multirow{2}{*}{\textit{órbita}} & \textit{Latitud} & \textit{Número de} & \textit{Longitud de} & \textit{Ancho} & $c_{x}$ & $|\stackrel{\rightarrow}{c}| - \overline{u}$ \\
                                   & \textit{(grad)} & \textit{Crestas} & \textit{Onda (km)} & \textit{Cresta (km)} & \textit{(m/s)} & \textit{(m/s)} \\
			& & & & & & \\
			\hline
			& & & & & & \\
			 59 & -58 &  7 & 125 &  75 & \textemdash & \textemdash \\
			& & & & & & \\
			\hline
			& & & & & & \\
			& & & & & & \\
			& & & & & & \\
			& & & & & & \\

		  \multicolumn{7}{l}{Tabla \ref{tab:table-PropertiesWaves}C: Propiedades de las ondas en imágenes IR} \\
			& & & & & & \\
			\hline\hline
			& & & & & & \\
			\multirow{2}{*}{\textit{órbita}} & \textit{Latitud} & \textit{Número de} & \textit{Longitud de} & \textit{Ancho} & $c_{x}$ & $|\stackrel{\rightarrow}{c}| - \overline{u}$ \\
                                   & \textit{(grad)} & \textit{Crestas} & \textit{Onda (km)} & \textit{Cresta (km)} & \textit{(m/s)} & \textit{(m/s)} \\
			& & & & & & \\
			\hline
			& & & & & & \\
			 84 & -80 &  6 &  63 &  40 &         -26 &        -1.0 \\
		 	 96 & -48 & 12 & 155 &  70 &         -64 &        -9.0 \\
			 97 & -39 &  8 & 100 &  35 &         -59 &        -3.1 \\
		 	 97 & -61 &  8 &  80 &  30 &         -58 &        -2.7 \\
			100 & -55 & 12 & 130 &  65 & \textemdash & \textemdash \\
			112 & -60 &  8 &  60 &  30 & \textemdash & \textemdash \\
			112 & -49 &  5 & 103 &  50 & \textemdash & \textemdash \\
			112 & -49 &  7 &  90 &  20 & \textemdash & \textemdash \\
			112 & -53 &  6 & 125 &  70 & \textemdash & \textemdash \\
			112 & -58 &  4 & 105 &  45 & \textemdash & \textemdash \\
			113 & -54 &  6 & 140 &  90 & \textemdash & \textemdash \\
			114 & -54 &  5 & 100 &  40 & \textemdash & \textemdash \\
			118 & -65 &  7 &  90 &  50 & \textemdash & \textemdash \\
			141 & -65 &  4 & 180 &  75 &         -44 &        -0.2 \\
			142 & -64 & 18 &  75 &  44 & \textemdash & \textemdash \\
			161 & -58 & 14 &  75 &  40 &         -61 &        -5.1 \\
			164 & -50 &  4 & 100 &  50 &         -64 &        -1.8 \\
			166 & -53 &  6 & 130 &  60 & \textemdash & \textemdash \\
			228 & -51 &  5 & 130 &  65 &         -45 &       -29.7 \\
			231 & -78 &  5 & 105 &  65 & \textemdash & \textemdash \\
			261 & -59 &  4 & 145 &  65 &         -42 &         9.0 \\
			290 & -63 &  4 & 120 &  66 & \textemdash & \textemdash \\
			313 & -60 &  7 & 135 &  65 &         -60 &         5.0 \\
			315 & -61 &  6 & 100 &  55 &         -64 &         3.1 \\
			317 & -53 &  9 & 100 &  55 &         -59 &         2.4 \\
			317 & -59 &  9 &  90 &  50 &         -60 &         1.1 \\
			317 & -63 &  6 &  90 &  45 &         -66 &        -6.1 \\
			318 & -60 &  5 & 100 &  50 &         -54 &        -7.5 \\
			321 & -49 & 12 & 105 &  50 & \textemdash & \textemdash \\
			323 & -41 &  5 & 110 &  55 &         -50 &         3.3 \\
			& & & & & & \\
			\hline
			& & & & & & \\
		\end{tabular}

  \end{spacing}
\end{table}

Vemos que aunque las longitudes de onda medidas están entre los 90 y 170 km para los paquetes en la cima de las nubes (UV) y entre 60 y 180 km para la parte baja de las nubes (IR), no podemos descartar ondas de menor longitud de onda (menos de 50 km) que escapen a la resolución que permiten nuestras mejores imágenes. Observamos también que las velocidades de fase respecto al viento promedio son peque\~{n}as en ambos niveles, lo que es compatible con la idea de interpretar estas ondas como ondas de gravedad, tal y como discutiremos más adelante.\\

También es de destacar que las propiedades de estos paquetes de ondas no parecen depender de la latitud, tal como viene mostrado en la Figura 5.5: ni la longitud de onda (Figura 5.5A), ni la longitud del paquete (Figura 5.5B), ni el ancho del paquete (Figura 5.5C) parecen tener relación alguna con la latitud. Sólo la orientación de los paquetes (Figura 5.5D) parece crecer li\-ge\-ra\-men\-te en magnitud conforme nos desplazamos a latitudes mayores. El caso más extremo es el de un paquete de ondas en el límite exterior del dipolo (80$^{\circ}$S, en la órbita 84), que forma el mayor ángulo con los paralelos.\\

\begin{figure}[h!]
	\centering
		\includegraphics[height=0.7\textheight]{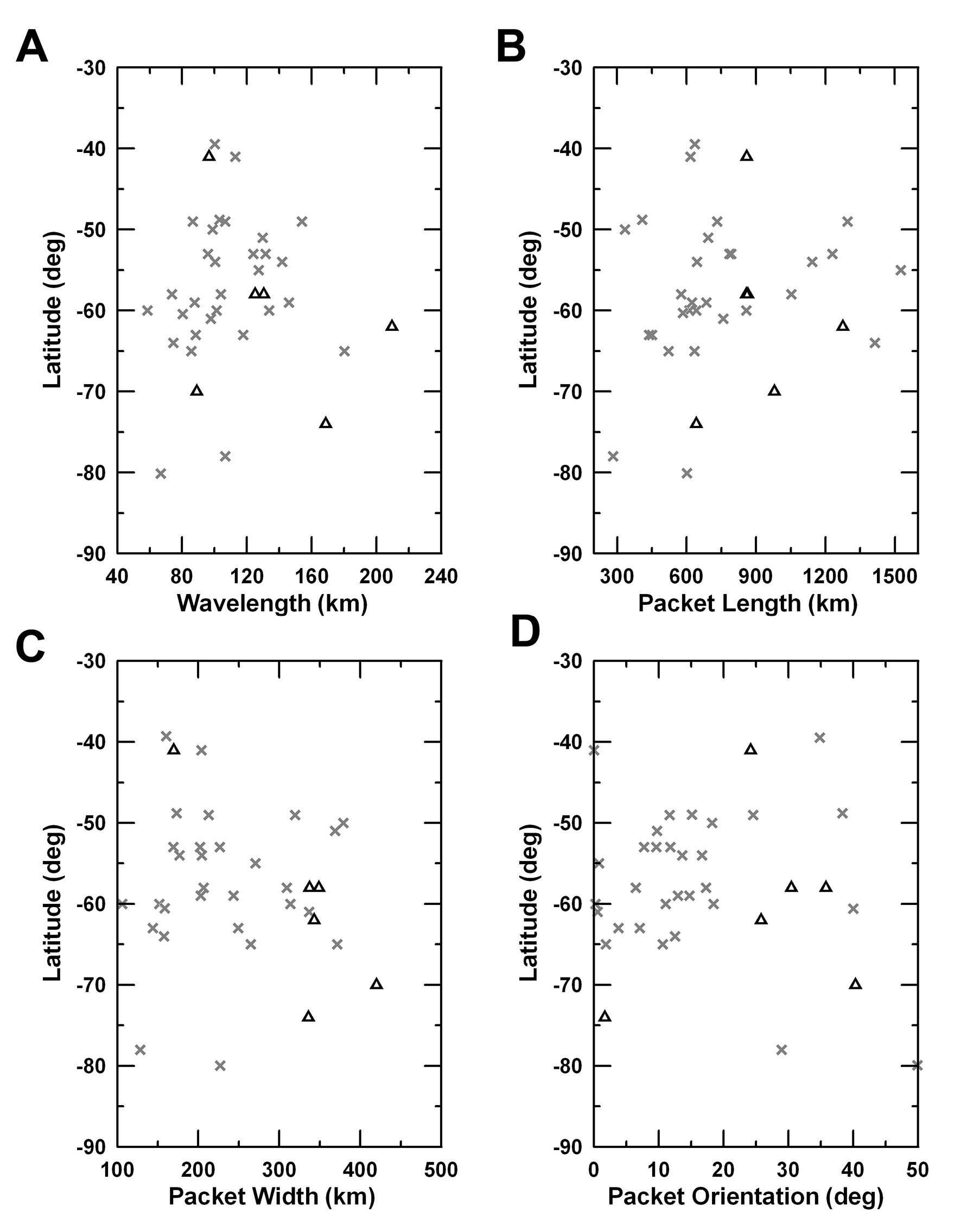}
	\label{fig:Waves-dependence-lat}
	\caption[Propiedades de las ondas en función de la latitud.]{\scriptsize{Propiedades de las ondas en función de la latitud. Las gráficas \textbf{A}, \textbf{B}, \textbf{C} y \textbf{D} muestran respectivamente la longitud de onda, longitud del paquete, ancho del paquete y orientación del paquete respecto a los paralelos. Los triángulos y las cruces denotan respectivamente ondas de gravedad en la cima de las nubes y en las nubes inferiores.}}
\end{figure}

La calidad de las imágenes resulta ser un factor determinante a la hora de extraer las propiedades arriba mencionadas, a lo que hay que a\~{n}adir también la habilidad del observador para hacer un seguimiento preciso de los detalles, tanto de los frentes de onda (en el caso de querer estimar longitudes de onda y velocidades de fase) como de detalles fuera de los paquetes de ondas (en el caso de querer estimar la velocidad promedio del viento). Aunque el error de navegación en el caso de las imágenes de VIRTIS es inferior a 1 pixel, las mediciones fueron hechas de forma manual por lo que el factor humano podría implicar errores de medida mayores. Estimamos que las longitudes de onda son precisas hasta un 10 \% de su valor, que la longitud y ancho de los paquetes tienen un error de unas decenas de kilómetros, y que la incertidumbre en las velocidades de fase es de unos $8~m\cdot s^{-1}$.\\

\subsection{Relación con el viento zonal y la cizalla vertical}\indent

En la Figura 5.6 se analiza la posible influencia del perfil de vientos sobre la localización de las ondas. En ella se muestran los perfiles horizontales temporalmente promediados de la componente zonal del viento para la cima de las nubes y las nubes inferiores, obtenidos a partir de las imágenes de VIRTIS \citep{Sanchez-Lavega2008}. Superpuestos a estos perfiles, se aprecian las velocidades zonales de los paquetes de onda estudiados en este trabajo. Aparentemente la posición latitudinal de las ondas no parece estar correlacionada con cambios en el perfil zonal. Teniendo en cuenta que el ancho de los paquetes de ondas tampoco parece estar correlacionado con la cizalla meridional del viento, es razo\-na\-ble pensar que el ancho de los paquetes podría guardar relación con posibles variaciones meridionales en la estabilidad atmosférica. Esto parece confirmado por análisis preliminares de perfiles de temperatura obtenidos en experimentos de radio-ocultación efectuados por Venus Express \citep{Bird2008}.\\

\begin{figure}[h!]
	\centering
		\includegraphics[height=0.5\textheight]{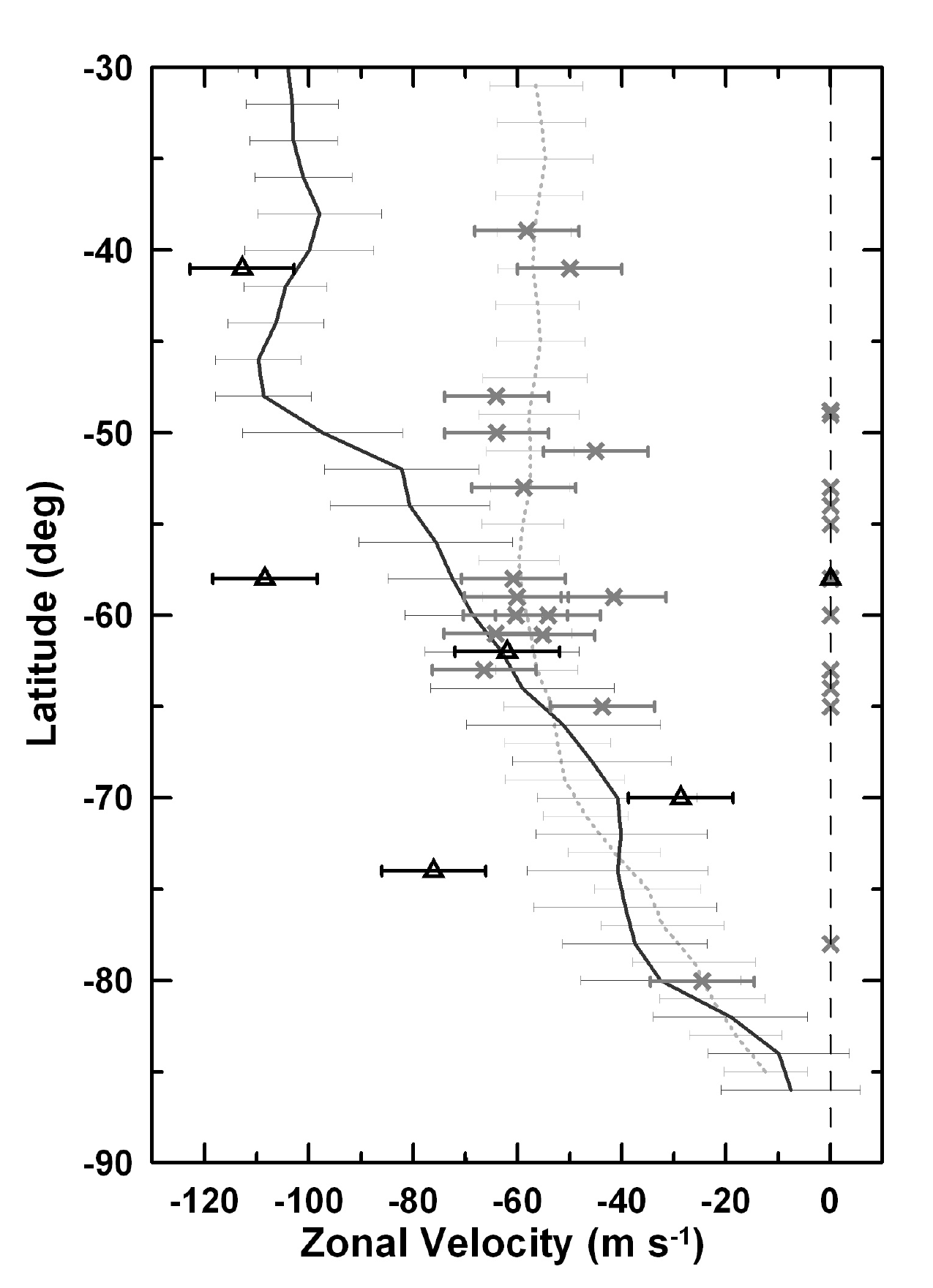}
	\label{fig:Waves-vs-Winds}
	\caption[Velocidades de fase y viento zonal medio.]{\scriptsize{Velocidades de fase de las ondas y perfil del viento zonal en los dos niveles de nubes. Los triángulos negros y la línea continua representan respectivamente velocidades de fase de las ondas y viento zonal en la cima de las nubes. De manera equivalente, las cruces grises y la línea discontinua hacen alusión a velocidades de fase y viento zonal en las nubes inferiores. Las ondas para las que no tenemos información dinámica se muestran en la línea discontinua centrada en velocidad zonal cero. Las barras de error del los prefiles de viento contienen el error de medida y/o estadístico en cada latitud \citep{Sanchez-Lavega2008}, mientras que las barras de error de las velocidades de fase de las ondas representan el error de medida para cada paquete de onda.}}
\end{figure}

Respecto a los paquetes de ondas encontrados en la cima de las nubes, su mayor velocidad de fase es probablemente consecuencia de una más alta estabilidad en este nivel nuboso \citep{Kliore1985,Gierasch1997}. Al contrario que las ondas de las nubes inferiores, estas ondas no están confinadas verticalmente, por lo que podrían propagarse verticalmente hasta niveles críticos donde serían absorbidas por efectos de la cizalla vertical del viento, o disipadas \citep{Schubert1983}.\\

En la Figura 5.7 se representa la cizalla vertical del viento en función de la latitud y el número de ondas detectados en intervalos de 10 a 20$^{\circ}$, comparando las longitudes de onda y del paquete promediadas en dichos intervalos de latitud. En este caso tampoco se aprecia relación alguna de estos parámetros de onda con la cizalla vertical del viento, existiendo abundantes ondas tanto en regiones de alta cizalla vertical como en regiones de baja cizalla y siendo las características de las ondas similares en ambos casos.\\

\begin{figure}[h!]
	\centering
		\includegraphics[height=0.7\textheight]{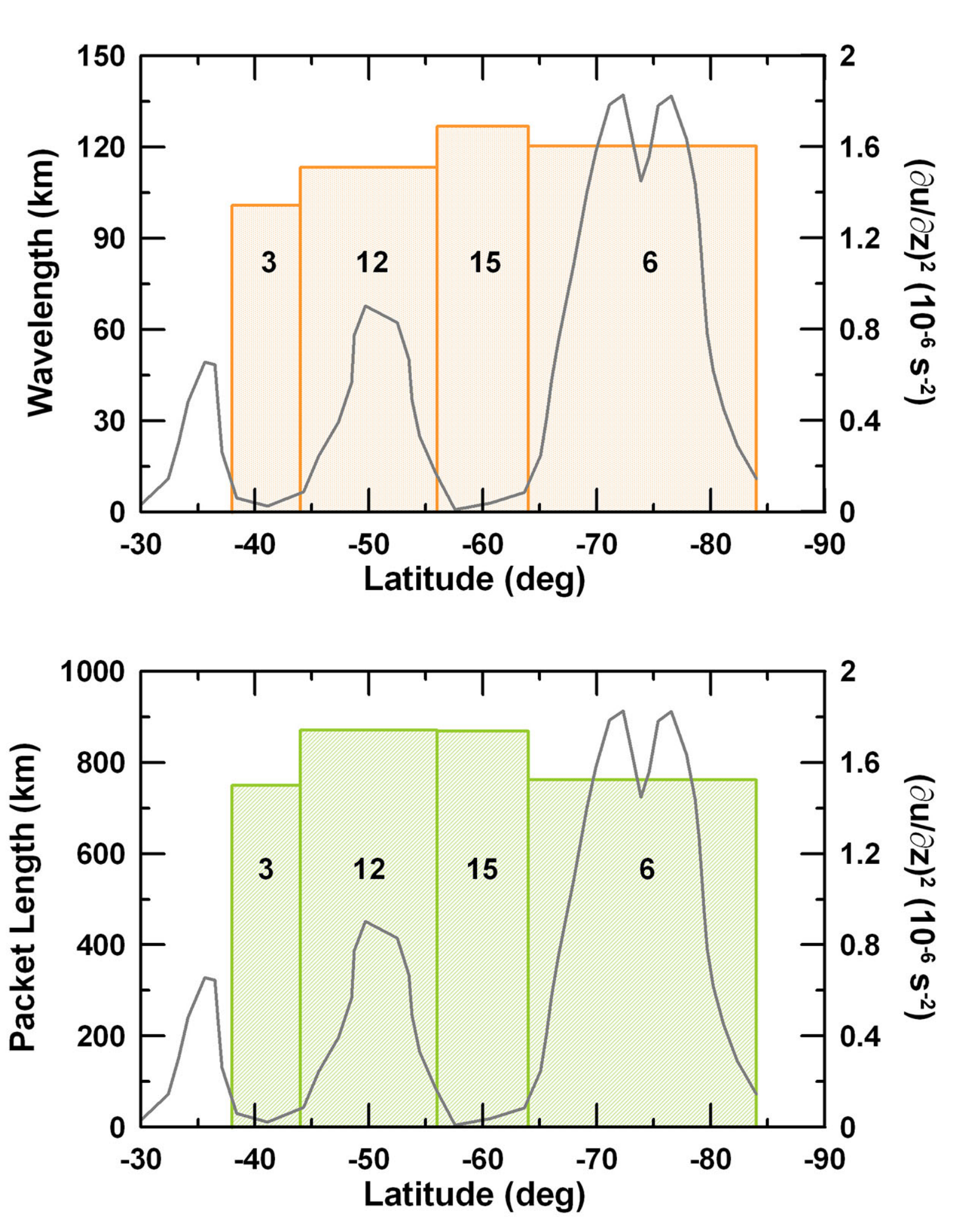}
	\label{fig:WavLeng-PacketLeng-Shear-vs-Lat}
	\caption[Relación entre las ondas y la cizalla vertical.]{\scriptsize{Cizalla vertical del viento entre en función de la latitud. Los datos están divididos en diferentes rangos de latitud, dependiendo de si tenemos valores peque\~{n}os o elevados de cizalla. Los números representan el número de paquetes de ondas encontrados en cada región de interés, y en cada una de ellas superponemos el valor medio de la longitud de onda (gráfica superior) y la longitud del paquete (gráfica inferior).}}
\end{figure}

\section{Interpretación de los resultados}\label{chapter-gravwaves-interpretation}\indent

En base a las características descritas anteriormente, interpretamos los paquetes de ondas observados como una manifestación en los campos de nubes de ondas internas de gravedad. Por un lado, descartamos que sean ondas de Rossby causadas por el efecto Coriolis debido a que las escalas espaciales de nuestras ondas son muy peque\~{n}as comparadas con la escala sinóptica de escala planetaria característica de las ondas de Rossby\footnote{A esto hay que a\~{n}adir que sus frecuencias temporales son muy grandes comparadas con la lentísima rotación de Venus.}. Por otro lado, sus frecuencias temporales son mucho más peque\~{n}as que la corres\-pon\-diente frecuencia acústica de corte, por lo que también podemos desechar las ondas acústicas. Tampoco pueden tratarse de ondas Kelvin ya que las latitudes en que se propagan no poseen apenas cizalla vertical del viento y las ondas poseen una extensión meridional demasiado grande como para deberse a la cizalla meridional. En el caso de la Tierra se puede observar en ocasiones estructuras nubosas alineadas de manera similar a las que tenemos en nuestras imágenes y conocidas como \textit{convección de rodillo}\index{Convección!de rodillo}. Sin embargo, este fenómeno es característico de la capa límite planetaria y depende de un delicado equilibrio entre condiciones de baja estabilidad vertical e ines\-ta\-bi\-li\-dad dinámica debido a la cizalla vertical del viento (valores peque\~{n}os del número de Richardson) \citep{Lemone1973}. Nuestros paquetes de ondas, sin embargo, están localizados a varias decenas de kilómetros por encima de la superficie y en regiones donde el número de Richardson es elevado ($Ri>10$) \citep{Gierasch1997}.\\

Con objeto de profundizar en la naturaleza de estas ondas y determinar sus características procederemos a continuación a plantear su relación de dispersión\index{Dispersión!relación de} a partir de un modelo lineal clásico para las ondas de gravedad \citep{Holton1992}. Si despreciamos el efecto Coriolis y asumimos que el mo\-vi\-mien\-to es bidimensional y que es válida la aproximación de Boussinesq\index{Boussinesq!aproximación de}, entonces al aplicar la teoría de perturbaciones a un modelo lineal con viento promedio y densidad constantes se determina la siguiente relación de dispersión (ver Apéndice \ref{appendix-waves}):
\begin{equation}
	\bar{\omega} = \omega-\bar{u}\cdot k = \pm\frac{k}{\sqrt{k^{2}+m^{2}}}\cdot N
	\label{rel-dispersion}
\end{equation}
donde $\omega$ es la frecuencia angular de la perturbación de onda, $\bar{u}$ es el viento promedio proyectado en la dirección de la propagación de la fase de la onda, $\bar{\omega}$ es la \textit{frecuencia angular intrínseca}\index{Intrínseca!frecuencia} (la que mediría un observador que se mueva conforme al viento), $N$ es la frecuencia de Brunt-Väisälä\index{Brunt-Väisälä!frecuencia de}, y $k$ y $m$ son los números de onda horizontal y vertical correspondientes a las longitudes de onda horizontal y vertical $\lambda_{x}$ y $\lambda_{z}$ mediante las fórmulas $k=2\pi/\lambda_{x}$ y $m=2\pi/\lambda_{z}$. De la ecuación (\ref{rel-dispersion}) resulta sencillo obtener las siguientes expresiones para las componentes horizontal y vertical de la velocidad de fase intrínseca\index{Velocidad!de fase|textbf}\index{Intrínseca!velocidad de fase} (respecto del viento promedio):
\begin{align}
	\hat{c}_{x}= \frac{\bar{\omega}}{k} &= \pm\frac{N}{\sqrt{k^{2}+m^{2}}},                  \label{vel-fase-horz}\\
	\hat{c}_{z}= \frac{\bar{\omega}}{m} &= \pm\frac{N}{\sqrt{k^{2}+m^{2}}}\cdot\frac{k}{m}.  \label{vel-fase-vert}
\end{align}
\\

A continuación examinaremos las propiedades de las ondas para las que se tienen medidas de velocidad de fase de acuerdo con este modelo simple de ondas internas de gravedad. Las nubes que se observan en las imágenes de UV están localizadas a $\sim66$ km de altura, nivel próximo a la transición de la troposfera a la mesosfera de Venus y donde la estabilidad atmosférica aumenta de forma brusca. En base a los datos proporcionados por di\-fe\-ren\-tes sondas podemos considerar que un valor razonable para la frecuencia de Brunt-Väisälä en este nivel es $N^{2}\approx 260\cdot 10^{-6}~s^{-2}$ \citep{Seiff1985,delGenio1990,Gierasch1997}. Por encima de estos niveles la estabilidad de la atmósfera aumenta hasta alcanzar valores muy superiores típicos de la mesosfera y por debajo la estabilidad decrece hasta prácticamente anularse entre 48 y 55 km, zona donde la convección puede desarrollarse. Las nubes inferiores (las que actúan como fuentes de opacidad a la radiación emitida) están localizadas justo debajo, en un rango de altitud (43-48 km) caracterizado por altos valores de estabilidad con un valor re\-pre\-sen\-ta\-ti\-vo de la frecuencia de Brunt-Väisälä de $N^{2}\approx 65\cdot 10^{-6}~s^{-2}$. Dicho valor es muy semejante al obtenido por la sonda Pioneer Venus Night a la altura de 45 km \citep{Kliore1985,Gierasch1997} y está dentro del rango de $500-100\cdot 10^{-6}~s^{-2}$ obtenido en estos niveles por el conjunto completo de sondas Pioneer Venus. Esta capa estable podría actuar como un conducto donde quedarían atrapadas las ondas excitadas por convección desde los inestables niveles inferiores \citep{Baker1998}.\\

Dadas las frecuencias de Brunt-Väisälä anteriores para ambos niveles de nubes (cima de las nubes y nubes inferiores), la Figura 5.8 nos muestra la relación de dispersión que se obtiene para diferentes valores de la longitud de onda vertical $\lambda_{z}$. La cima de las nubes (Figura 5.8A) parece agrupar las ondas en longitudes de onda peque\~{n}as (5 km) y grandes\footnote{El paquete de ondas detectado simultáneamente en UV y NIR corresponde a uno de los ejemplos de longitud de onda vertical grande.} (15 km), o alternativamente en regiones con valores peque\~{n}os y elevados de estabilidad atmosférica. Las ondas de las nubes inferiores (Figura 5.8B) son más abundantes y se distribuyen homogéneamente en regiones con valor peque\~{n}o de longitud de onda vertical (5-10 km), lo que es razonable en vista de la extensión vertical del conducto de estabilidad estática en el que pueden propagarse ($\sim18$ km). Esto es, además, consistente con el hecho de que longitudes de onda más extensas verticalmente resultarían dispersadas por efecto de la cizalla vertical del viento \citep{Leroy1995}.\\

\begin{figure}[h!]
	\centering
		\includegraphics[width=1.0\textwidth]{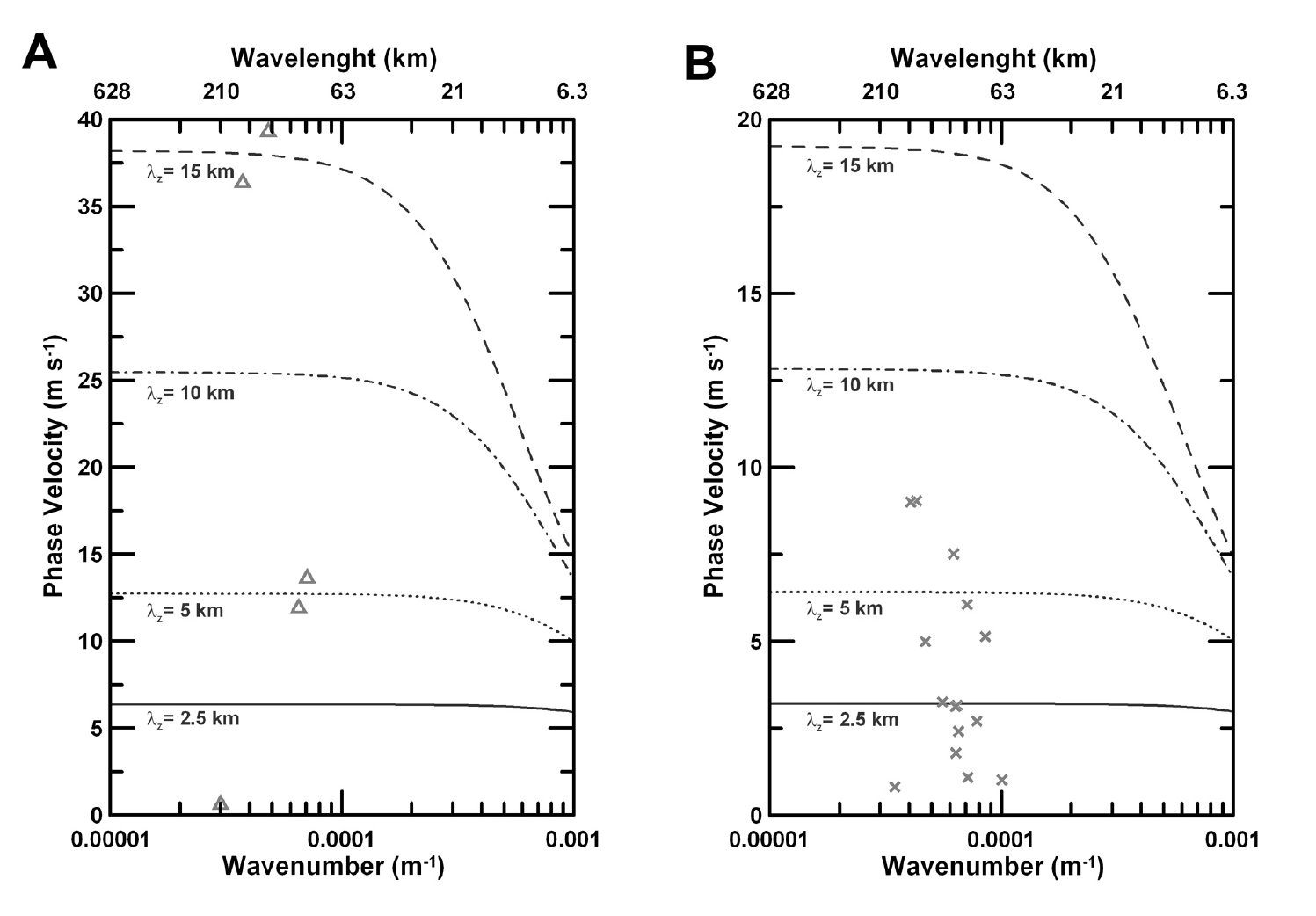}
	\label{fig:Wave-DispersionRelation}
	\caption[Relación de dispersión teórica de las ondas.]{\scriptsize{Comparativa de las ondas encontradas y la relación de dispersión que se deduce para una onda interna de gravedad a partir de la ecuación (\ref{vel-fase-horz}). \textbf{(A)} muestra el caso de la cima de las nubes y \textbf{(B)} el caso de las nubes inferiores. Las velocidades de fase están expresadas respecto al viento zonal promedio, y la relación de dispersión se representa para diferentes longitudes de onda verticales dentro del rango $\lambda_{z}=2-15$ km. Fijamos el valor de la frecuencia de Brunt-Väisällä a una cantidad característica de cada nivel de altura ($N^{2}=260\cdot10^{-6}~s^{-2}$ para la cima de las nubes y $N^{2}=65\cdot10^{-6}~s^{-2}$ para las nubes inferiores).}}
\end{figure}

Se puede apreciar también que las velocidades de fase respecto al viento zonal medio tienen valores peque\~{n}os en ambos casos, que es lo que se espera para ondas inducidas por convección que poseen velocidad de fase relativa peque\~{n}a en la región donde se generen. La Figura 5.8 evidencia claramente un comportamiento escasamente dispersivo de las ondas, implicando que tanto la longitud de onda vertical como la estabilidad atmosférica apenas deberían mostrar variaciones a lo largo de los cientos de kilómetros que se extiende horizontalmente el paquete de ondas.\\

En el caso de las nubes inferiores es posible profundizar en el estudio de la doble dependencia de la velocidad de fase con los parámetros $N^{2}$ (cuyos valores están acotados por las medidas ``in situ'' de las sondas Pioneer Venus) y $\lambda_{z}$ (a su vez con valores acotados mediante argumentos físicos razonables)\footnote{Esto no es viable para la cima de las nubes debido a las escasas observaciones y a las aparentes grandes variaciones que muestran tanto $\lambda_{z}$ como $N^{2}$.}, estudiando la función de desviación del conjunto de nuestras medidas con respecto a los parámetros libres, $\chi^{2}(N,m)$. Esta función representa, para cada valor de los parámetros libres, una medida de cuánto se desvían las velocidades de fase medidas de los valores que predice la relación de dispersión (\ref{rel-dispersion}). Se define esta función\index{Test de Chi-cuadrado} de la siguiente manera \citep{NumRecipes1992}:
\begin{equation}	\chi^{2}(N^{2},m)=\sum^{N}_{i=1}\left(\frac{c_{x,i}-c_{x}(N^{2},m)}{\sigma_{c}}\right)^{2},
	\label{chi2}
\end{equation}
donde $N$ es el número de velocidades de fase medidas, $c_{x,i}$ es la velocidad de fase de la onda $i$, $c_{x}$ es la velocidad de fase teórica de la ecuación (\ref{rel-dispersion}) y $\sigma_{c}$ es la variancia de las velocidades de fase medidas (e introduce los errores de medida).\\

\begin{figure}[h!]
	\centering
		\includegraphics[height=0.6\textheight]{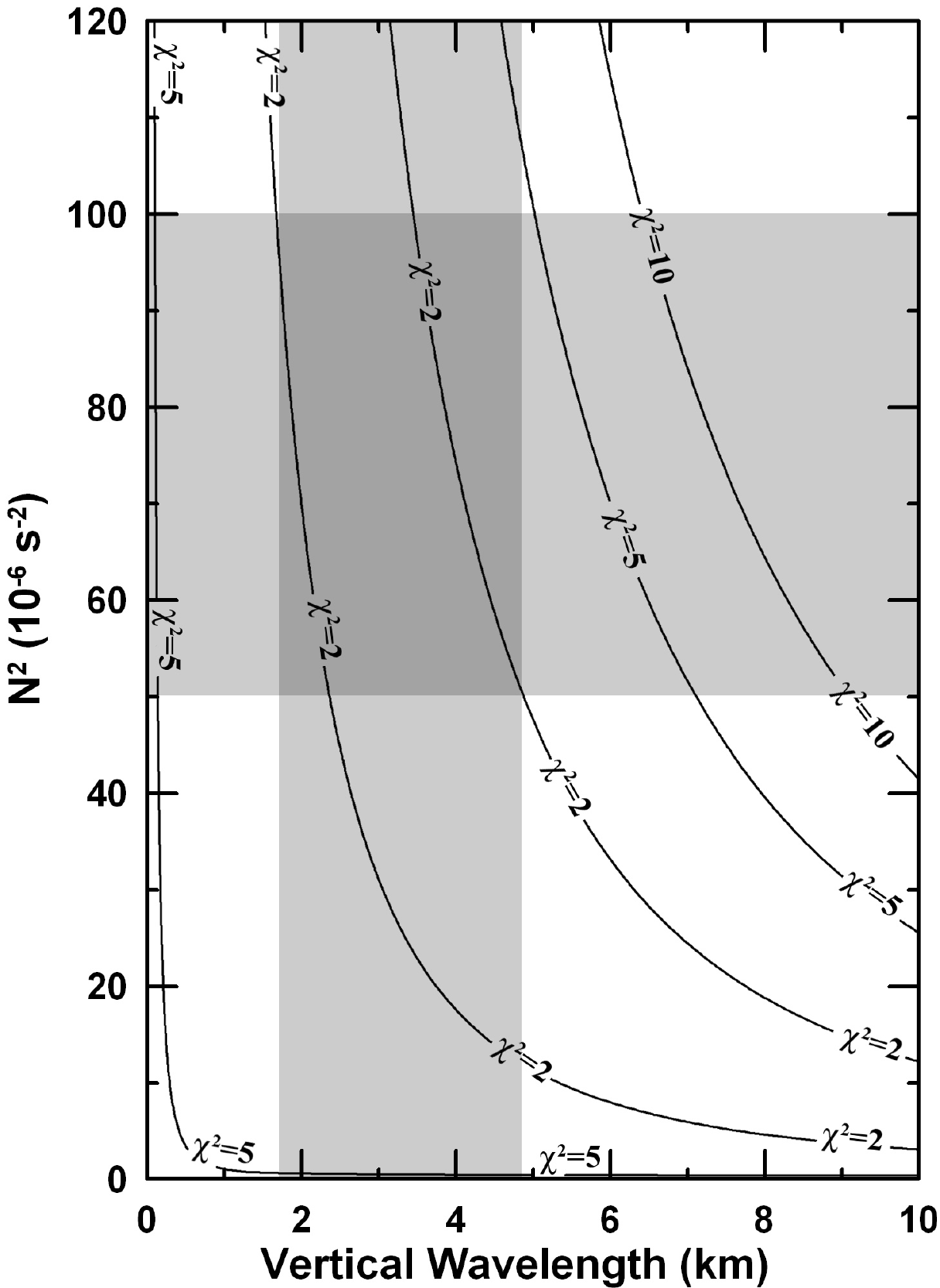}
	\label{fig:Chi2-test}
	\caption[Test de Chi-cuadrado para el modelo teórico de las ondas.]{\scriptsize{Test de Chi-cuadrado aplicado al espacio de parámetros definido por la ec. (\ref{vel-fase-horz}) para las ondas de las nubes inferiores. Las regiones de confianza correspondientes a valores de $\chi^{2}$ mayores que un valor dado se representan como líneas de contorno. El margen gris horizontal indica el rango de frecuencias de Brunt-Väisällä posibles en la región de las ondas, mientras que el margen gris vertical nos indica el rango de longitudes de onda verticales compatibles con la región de confianza de mejor ajuste (la contenida entre las líneas $\chi^{2}=2$).}}
\end{figure}

La Figura 5.9 muestra los valores de la función $\chi^{2}$ variando los parámetros libres. Los contornos con $\Delta\chi^{2}$ constante en torno al valor mínimo pueden usarse para determinar las regiones de confianza de $N^{2}$ y $\lambda_{z}$. Una \textit{región de confianza}\index{Confianza!región de} es una región del espacio de parámetros donde tenemos una alta probabilidad de encontrar los valores reales de $N^{2}$ y $\lambda_{z}$. Así, en el supuesto de que todos los paquetes de ondas medidos tuvieran el mismo valor de $N^{2}$ y $\lambda_{z}$, y que los errores de medida obedecieran el comportamiento de una distribución normal, la región contenida entre los contornos con $\chi^{2}=2$ se correspondería con la región donde existe un 68 \% de probabilidad de encontrar los valores reales de $N^{2}$ y $\lambda_{z}$. De igual manera, las regiones más extensas contenidas entre los contornos $\chi^{2}=5$ y $\chi^{2}=10$ tendrían un 90 y 95 \% respectivamente de contener los valores reales \citep{NumRecipes1992}. Asímismo, si en la Figura 5.9 fijamos el valor de $N^{2}$ (por ejemplo el usado previamente, $N^{2}=65\cdot 10^{-6}~s^{-2}$) entonces la región de confianza donde tenemos un mejor ajuste entre experimento y teoría queda limitada al rango de longitudes de onda verticales $\lambda_{z}$ entre 2 y 5 km (consistente con la Figura 5.8B). Este rango de longitudes de onda verticales encaja también con el valor de 2.5 km encontrado en las fluctuaciones de temperatura medidas por la nave Magallanes por encima de la base de la nube intermedia \citep{Hinson1995}. Como estas ondas tienen un número de onda vertical $m$ mucho mayor que el número de onda horizontal $k$, la ecuación (\ref{vel-fase-horz}) se puede simplificar:
\begin{equation}
	\hat{c}_{x}\approx\pm\frac{N}{m}.
\end{equation}
lo que corresponde a ondas viajeras horizontalmente no dispersivas, estando de acuerdo con nuestras observaciones.\\

Si bien un tratamiento no lineal de estas ondas escapa a las pretensiones de esta tesis, los modelos generales más avanzados de este tipo de ondas en la Tierra muestran que en el límite $\left| c-\overline{u}\right|\rightarrow 0$ las ondas de gravedad alcanzan una situación crítica en la que se produce su ruptura y donde ya no es válida la aproximación lineal. Los peque\~{n}os valores de la velocidad de fase relativa de las ondas de Venus las sitúan, por tanto, relativamente cerca de dicho régimen crítico, si bien somos capaces de verlas con una morfología coherente durante secuencias de imágenes separadas por 1-2 horas. Bajo la aproximación WKB, a medida que nos aproximamos al régimen crítico, la longitud de onda vertical disminuye gradualmente hasta que la onda finalmente colapsa. Sin embargo, la onda se ``estanca'' en este proceso de colapso, tardándose un tiempo infinito en llegar al régimen crítico \citep{Salby1996}, proceso que es acompa\~{n}ado por un aumento de la amplitud de la onda. Las ondas de IR que mejor se visualizan son aquellas con menores velocidades de fase relativas, sugiriendo la posibilidad de ondas de gran amplitud moviéndose cerca del régimen crítico.\\

\section{Discusión}\label{chapter-gravwaves-discuss}\indent

El principal candidato para excitar la formación de ondas de gravedad en la baja atmósfera de Venus es sin duda la \textit{convección}\footnote{Otras fuentes alternativas podrían ser la interacción del viento con cadenas monta\~{n}osas o la inestabilidad de Kelvin-Helmholtz en las zonas de alta cizalla del viento.} \citep{Gierasch1997}. No en vano numerosos trabajos teóricos y numéricos referidos tanto a la Tierra \citep{Stull1976} como a Venus \citep{Leroy1995,Baker1998,Baker2000a,Baker2000b,Yamamoto2003b,McGouldrick2008} han trabajado con la hipótesis de la convección como fuente de ondas de gravedad. En el caso de Venus dichos trabajos han demostrado que la convección de tipo compresible puede penetrar de forma significativa (llamaremos a este material convectivo inyectado plumas convectivas\index{Convección!plumas de}) en la capa estable entre 30-48 km, ya sea por movimientos convectivos hacia arriba desde la capa inestable entre 20-30 km, como por movimientos convectivos hacia abajo desde la capa inestable entre 48-55 km de altura\footnote{También se ha apuntado la posibilidad de que la capa inestable superior pudiera afectar de igual manera a la región estable por encima de los 55 km.}. \citet{Baker2000b} mostraron que la convección es más penetrante a través de las capas estables cuando existen valores elevados de cizalla vertical del viento, pudiendo interactuar entre sí dentro de la capa estable entre 30-48 km las plumas convectivas de la capa inestable superior (material convectivo moviéndose hacia abajo) y las plumas convectivas de la capa inestable inferior (material convectivo moviéndose hacia arriba), generando ondas de gravedad de mayores longitudes de onda. En casos como este podría darse el caso de que plumas convectivas penetren simultáneamente en ambas capas estables por lo que podría haber cierto grado de correlación entre las ondas de gravedad de la cima de las nubes y de la nube inferior. No obstante, el escaso número de ondas localizadas en las imágenes UV, y el hecho de que VIRTIS no pueda observar ambos niveles verticales simultáneamente, obstaculizan esta tarea.\\

La presencia de estas dos capas estables donde las nubes se asientan son un fenómeno estable en el tiempo, tal y como evidencian nuestras observaciones y las de la mayoría de las misiones espaciales anteriores (Pioneer Venus, Venera y los globos VEGA; ver \citealt{Gierasch1997}). Las ondas de gravedad detectadas en el hemisferio sur entre 40$^{\circ}$S y 75$^{\circ}$S sin duda conllevan la existencia de alta estabilidad en el nivel muestreado por las imágenes IR de VIRTIS. La zona entre 40$^{\circ}$S y el ecuador no está suficientemente observada como para extraer la misma conclusión y podría estar dominada por movimientos convectivos más fuertes. Por otro lado, la presencia del dipolo en la región situada entre 80$^{\circ}$S y el polo podría enmascarar la presencia de ondas adicionales. Si bien las ondas observadas en la nube inferior son frecuentes y abundantes, no son un fenómeno tan persistente como la propia estabilidad de esta región, por lo que parece más conveniente achacar los periodos de ausencia de ondas a una variabilidad en los mecanismos responsables de la propia excitación de éstas.\\

La mayor parte de los modelos numéricos para las ondas de gravedad en Venus se centran en ondas de longitudes de onda peque\~{n}as (no mayores que 30 km, ver refs. \citealt*{Baker1998,Baker2000a,Baker2000b}), mientras que las que observamos en nuestras imágenes rondan longitudes de onda superiores\footnote{Las imágenes de VIRTIS no permiten distinguir longitudes de onda tan peque\~{n}as pues la mejor resolución ronda los 15-20 km/pixel.}, de unos 100 km. \citet*{McGouldrick2008} sugieren que tanto una mayor cizalla vertical del viento zonal como la penetración de plumas de convección más gruesas podrían generar ondas de gravedad de mayores longitudes de onda horizontales. En nuestro caso no logramos observar indicios de fuerte actividad convectiva en ninguno de los niveles, y nuestros datos de VIRTIS indican que los valores más elevados de cizalla vertical del viento se dan para la región entre 61 y 66 km de altura, con un valor de $\partial \overline{u}/\partial z=8\pm 2$ m/s por km entre el ecuador y latitudes medias, y menos de 2 m/s por km en latitudes cercanas al polo \citep{Sanchez-Lavega2008}.\\

Si bien resulta complicado evaluar el papel que pueden jugar las ondas de gravedad en el transporte global de energía y momento en la atmósfera de Venus, podemos llevar a cabo una estimación si consideramos que nuestras ondas se comportan según el modelo lineal anteriormente descrito. Así, des\-pe\-jan\-do la frecuencia angular $\omega$ de la relación de dispersión (\ref{rel-dispersion}) podemos obtener una expresión para la \textit{velocidad de grupo}\index{Velocidad!de grupo} de las ondas de gravedad:
\begin{align}
	c_{gx}= \frac{\partial\omega}{\partial k} &= \bar{u} \pm \frac{m^{2}}{\left(k^{2}+m^{2}\right)^{3/2}}\cdot N,     \label{vel-group-horz}\\
	c_{gz}= \frac{\partial\omega}{\partial m} &= \mp \frac{k \cdot m}{\left(k^{2}+m^{2}\right)^{3/2}}\cdot N.  \label{vel-group-vert}
\end{align}
De acuerdo con estas expresiones, los paquetes de onda de la parte baja de las nubes tendrían una velocidad de grupo prácticamente horizontal, con una componente vertical de apenas $c_{gz}\sim 1~m\cdot s^{-1}$ y una componente horizontal $c_{gx}\sim 12~m\cdot s^{-1}$.\\

Las ondas de gravedad transmiten energía y momento en la dirección perpendicular a la dirección de la propagación de la fase. Aunque la mayor parte de los paquetes de ondas son zonales, algunos de ellos muestran una orientación distinta y podrían ser capaces de transportar cierta cantidad de momento meridionalmente. Sin embargo, al igual que podemos encontrar ondas de gravedad que zonalmente se mueven más deprisa que el viento zonal promedio, también hay ondas que se mueven más despacio que éste. Respecto a la dirección meridional, encontramos ondas con un ligero movimiento hacia el ecuador, y otras moviéndose levemente hacia el polo. A todo esto hemos de a\~{n}adir el inconveniente infranqueable de carecer de información sobre la componente vertical del movimiento ondulatorio. De todos modos, si bien las ondas son capaces de transportar energía y momento, no parece muy pro\-ba\-ble que éstas jueguen un papel importante en mantener la superrotación atmosférica ni que sean responsables del viento global en la atmósfera de Venus \citep{Leroy1995}. Lo variable de su presencia así como su escasa ocupación en área sugieren que esto es así.\\

\section{Conclusiones}\label{chapter-gravwaves-concluss}\indent

En este capítulo hemos realizado un estudio de las ondas de mesoescala presentes en la atmósfera de Venus a diferentes niveles verticales en las nubes, usando para ello imágenes en tres longitudes de onda ($380~nm$, $980~nm$ y $1.74~\mu m$) captadas por la cámara espectral VIRTIS a bordo de la nave Venus Express. Hemos identificado dichas ondas como ondas internas de gravedad, siendo la primera vez que en Venus se caracterizan ondas simultáneamente a tres niveles distintos de la atmósfera, y habiendo observado también por vez primera la rica actividad ondulatoria presente en la nube inferior.\\

La nube inferior presenta abundantes paquetes de ondas, de características diversas tanto en morfología como en propiedades, y cubriendo un amplio rango de latitudes (40$^{\circ}$S-80$^{\circ}$S). No parece haber una relación clara entre las propiedades de dichas ondas y la latitud o el perfil del viento zonal, aunque en nuestro conjunto de imágenes ha sido imposible encontrar ondas en latitudes tropicales y ecuatoriales. Tampoco parecen estar correlacionadas con accidentes topográficos ni con el ciclo de insolación, lo que podría estar indicando el papel que juega la convección profunda como fuente de esta actividad ondulatoria.\\

También hemos presentado una primera interpretación de las propiedades de estas ondas haciendo uso de un modelo lineal clásico para las ondas de gravedad \citep{Holton1992} que confirma la naturaleza de estas ondas. Haciendo uso de un análisis estadístico sobre la dependencia de estas ondas con la estabilidad y la longitud de onda vertical (Figuras 5.8B y 5.9) concluimos que las longitudes de onda verticales son del orden de 2-5 km en las nubes in\-fe\-rio\-res, y entre 5-15 km en la cima de las nubes (Figura 5.8A). Estos resultados concuerdan con las longitudes de onda verticales determinadas a partir de las fluctuaciones de diversas magnitudes físicas medidas en la nube intermedia por las sondas Pioneer Venus \citep{Counselman1980} o la Magallanes \citep{Hinson1995}.\\

Este estudio deja abiertos importantes temas de interés, tales como la identidad específica del fenómeno que excita estas ondas, la distribución la\-titudinal y temporal tanto de la estabilidad atmosférica como de la convección, y una profundización sobre el papel que juegan estas ondas en la dinámica global de la atmósfera de Venus.

\chapter{Conclusions}

In this thesis I have exposed the results of my research on three important topics related to the atmospheric dynamics of Venus: the winds, the turbulence and the mesoscale gravity waves. For this purpose, I used imagery data obtained in 1990 by the camera SSI onboard Galileo spacecraft, and in 2006-2007 by the imaging spectrometer VIRTIS onboard Venus Express. As distinct vertical levels of the cloud regions can be sensed depending on the wavelength we are observing with, three different ranges were selected: UV range ($418~nm$ with SSI; $380~nm$ with VIRTIS) for sensing the top of the clouds at $\sim66$ km, NIR range ($986~nm$ with SSI; $980~nm$ with VIRTIS) for sensing the base of the top of the clouds at $\sim61$ km, and IR range ($1.74~\mu m$ with VIRTIS) for sensing the lower clouds at $\sim47$ km. The UV and NIR observations are made using the reflected radiation from the dayside, while the IR observations are useful when imaging the thermal radiation of the nightside.\\

\section{Results on winds at the cloud level}\indent

As clouds have proved to be valid tracers for measuring the wind in the atmosphere of Venus, we used images from the two missions. Due to observations' constraints, the Venus Express dataset only allowed the measurement of winds in the southern hemisphere, while Galileo let us infer winds from both hemispheres. Images were examined in two wavelengths in Galileo and in three wavelengths in the Venus Express case, enabling us to sense the wind at three vertical levels within the cloud layers.\\

Results on the \textit{zonal} component of the wind are displayed in Figures 3.3 (Galileo, see page \pageref{fig:GalileoVenus-ZonalWind}) and 3.4 (Venus Express, see page \pageref{fig:VirtisVenus-ZonalWind}). Although with different magnitude, the meridional profiles of the zonal wind show a common behaviour with a constant magnitude in the equator and midlatitudes, and an important decrease poleward of midlatitudes until they become zero at the pole. At the top of the clouds (UV) the winds are on the order of $<u>=-102\pm 10 m\cdot s^{-1}$ and this behaviour doesn't seem strictly symmetrical relative to the equator, starting to diminish at 45$^{\circ}$N in the northern hemisphere and at 50$^{\circ}$S in the southern hemisphere. The decrease in the north is more marked than in the south. At the base of the top of the clouds (NIR) winds have a magnitude of $<u>=-60\pm 10 m\cdot s^{-1}$ and the drop of the winds occurs at 40$^{\circ}$N (Galileo) and 55$^{\circ}$S (Venus Express). In the lower clouds (IR) winds are as speedy as at the NIR level ($<u>=-60\pm 10 m\cdot s^{-1}$) with the decrease beginning at 65$^{\circ}$S.\\

The variation with latitude for the \textit{meridional} component of the wind from Galileo and Venus Express images is displayed in Figures 3.9A and 3.9B respectively (see page \pageref{fig:Venus-MeridionalWind})). In the case of the top of the clouds, the profile is symmetrical relative to the equator, the meridional component is polewards in both hemispheres and increasing from zero to a maximum of $\sim10~m\cdot s^{-1}$ when reaching 45$^{\circ}$N and 55$^{\circ}$S. This behaviour is indicative of a Hadley cell circulation at the top of the clouds. Meridional winds at the base of the top of the clouds (NIR) and lower clouds (IR) showed slower magnitudes ($<v>=5~m\cdot s^{-1}$), lower than the error of measurement what inhibits the possibility of getting confident conclussions about the profile's trend.\\

A study about the temporal variability of the winds is also presented in this thesis. Figure 3.10 (see page \pageref{fig:LongTermEvol-ZonalWind}) displays the \textit{long-term variability} of the zonal winds through profiles obtained by previous missions along 26 years. The pattern exhibited in our zonal winds profiles is present throughout the different missions, settling down as a stable feature for the wind dynamics in Venus. An apparent decrease of the winds ($\sim10~m\cdot s^{-1}$) is detected between 1990 and 2006 at the top of the clouds (UV). In addition, a strong decrease of the wind speeds at the base of the top of the clouds ($\sim20~m\cdot s^{-1}$) was also detected from the Venus Express images (NIR) between the spacecraft orbit insertion and the following months (see Figure 3.11 in page \pageref{fig:ShortTermEvol-ZonalWind}). Against a real change in the dynamics, the possibility of a variation the height level that is visualized at NIR wavelength due to changes in the atmospheric composition seems more plausible. Both \textit{thermal tides} and \textit{5-day global scale waves} have been detected too within subpolar and mid-latitudes at the top of the clouds (UV), the former accelerating the winds ($-2.5\pm0.5~m\cdot s^{-1}$ per hour) between the morning and the afternoon, an the latter causing perturbations with an amplitude of $12~m\cdot s^{-1}$ (see pages \pageref{fig:Venus-SolarTide} and \pageref{fig:VEXVenus-5day}).\\

\section{Results on turbulence at the top of the clouds}\indent

As the error associated to wind speed measurements inhibits studying the turbulence in the atmosphere of Venus through the kinetic energy power spectra, an alternative approach is used as a proxy. The brightness power spectra at the top of the clouds serve a diagnostic tool for the atmospheric turbulence if we consider cloud brightness distribution is related to the spatial scales inherent in the atmospheric dynamics. Cylindrical projections from Galileo violet images of the top of the clouds were elaborated in order to infer the brightness power spectra of the photometric scans. Our study covers a range of latitudes between 70$^{\circ}$S and 70$^{\circ}$N.\\

Both temporal and spatial changes are detected through this analysis at the top of the clouds. On the one hand, the cloud brightness power spectra display notable alterations in the slope values when comparing results belonging to the periods of Mariner 10, Pioneer Venus and Galileo flyby (see Table \ref{tab:tabla-EvolucionPendientes} on page \pageref{tab:tabla-EvolucionPendientes}). On the other hand, the equatorial and mid-latitudes regions at the top of the clouds exhibit brightness power spectra with a slope $\approx-5/3$ (similar to Kolmogorov's law and what's expected for the three-dimensional turbulence), whereas in the subpolar regions the brightness power spectra display a slope $\approx-3$ for the highest spatial scales and $\approx-5/3$ for the lowest ones (see Figure 4.8 on page \pageref{fig:GalileoVenus-SpectraSlopes}), a result close to what has been obtained for kinetic energy power spectrum in the case of the Earth. It is also shown that this behaviour seems highly related to the meridional distribution of the zonal winds, with nearly constant values for the equator and midlatitudes, and a steep decrease of the wind from the subpolar regions (see Figure 4.9 on page \pageref{fig:Venus-Winds-SpectraSlopes}).\\

\section{Results on gravity waves}\indent

Venus Express data corroborates that mesoscale gravity waves are an abundant feature within the cloud region. Images from the VIRTIS-M instrument have allowed us to study mesoscale waves at the top of the clouds, for the first time in the lower clouds, confirming the presence of ducted waves within the region of stability sensed at $1.74~\mu m$ in the nightside. We have characterized these waves both morphological and dinamically, identifying them as internal gravity waves.\\

These waves are nearly zonal, have wavelengths of 60-150 km, propagate westward with low phase velocities relative to the zonal flow and are confined in wave packets of 400 to 1,800 km in length. Packet waves in the lower clouds seem more abundant and morphologically varied than at the top of the clouds, covering latitudes between 40$^{\circ}$S and 75$^{\circ}$S. None of the wave properties seem to vary systematically with latitude (see Figure 5.5 on page \pageref{fig:Waves-dependence-lat}) or the zonal wind profile (see Figure 5.6 on page \pageref{fig:Waves-vs-Winds}), and show no apparent relation to solar heating or surface elevations sources (see Figure 5.4 on page \pageref{fig:Waves-locations}).\\

And examination of their properties in terms of a linear model was also carried out, resulting in highly non-dispersive gravity waves with vertical wavelengths on the order of 2 to 5 km (see Figures 5.8 and 5.9 on pages \pageref{fig:Wave-DispersionRelation} and \pageref{fig:Chi2-test}), in accordance temperature scintillations carried out during Magallanes mission. This work opens questions about the specific source of the wave activity observed, the latitudinal and temporal distributions of atmospheric stability and convection, and the role of these waves in the overall dynamics of the Venus atmosphere.

\appendix

\chapter{Estimación del error de velocidad}\indent\label{appendix-errors}

La calidad de las imágenes juega un papel primordial a la hora de estimar la incertidumbre de nuestras medidas de velocidad del viento. Si bien la resolución espacial de las imágenes es a priori el factor más decisivo en la calidad de éstas, también contribuyen al error de nuestras medidas la se\-pa\-ra\-ción de tiempo entre las imágenes usadas y su rango dinámico de brillo (que afectan a la visualización e identificación de los trazadores nubosos a pesar de disponer de una buena reso\-lución espacial, como es el caso de las imágenes de la cámara SSI de Galileo). Las ecuaciones que hemos usado para calcular la velocidad del viento son las siguientes:
\begin{align}
	u&=(R+H)\cdot\cos\phi\cdot\frac{\Delta\lambda}{\Delta t}\cdot\frac{\pi}{180},  \label{vel-zonal-appendix}\\
	v&=(R+H)\cdot\frac{\Delta\phi}{\Delta t}\cdot\frac{\pi}{180}.                  \label{vel-merid-appendix}
\end{align}
donde $\Delta\lambda$ y $\Delta\phi$ son los desplazamientos del detalle nuboso medidos en grados de longitud y latitud respectivamente, $\Delta t$ es el intervalo de tiempo entre el par de imágenes, $\phi$ es la latitud promedio del detalle nuboso, $R$ es el radio planetario (en este caso el radio de Venus) y $H$ es la altura correspondiente al nivel de nubes que estemos observando.\\

A efectos de facilitar la identificación de los detalles, todos los pares de imá\-ge\-nes usados para medir velocidades se proyectaron geométricamente (de forma cilídrica o polar) con la misma resolución espacial. Esto permite comparar más fácilmente imágenes tomadas con diferentes ángulos de visión y/o distancia entre la nave y el planeta (lo que evidentemente afecta a la resolución espacial de la imagen resultante). Para evitar la pérdida de información, en esta tesis se ha procurado proyectar las imágenes a una malla de mejor resolución que la imagen original, con lo que usar el error asociado a las coordenadas de latitud y longitud de nuestras proyecciones necesariamente subestima el error real de nuestras medidas. Alternativamente, y con objeto de simplificar en lo posible la estimación de nuestro error de velocidad, trabajaremos con el error asociado a la resolución espacial (medida en \textit{metros}) de las imágenes originales. Las ecuaciones (\ref{vel-zonal-appendix}) y (\ref{vel-merid-appendix}) pueden, así, expresarse:
\begin{align}
	u&=\frac{\Delta X}{\Delta t},  \label{vel-zonal-new}\\
	v&=\frac{\Delta Y}{\Delta t}.  \label{vel-merid-new}
\end{align}
donde $\Delta X$ y $\Delta Y$ son los desplazamientos longitudinal y latitudinal del detalle nuboso (expresados en metros) y $\Delta t$ es el intervalo de tiempo entre las imágenes (medido en segundos).\\

Ya que la velocidad se mide de manera indirecta (a través del desplazamiento espacial y el intervalo temporal transcurrido) necesitamos saber cómo se propaga el error de nuestras medidas. De forma genérica, supongamos que tenemos una magnitud $f=f(A,B,C,\ldots)$ donde $A,B,C,\ldots$ son variables no correlacionadas entre sí y directamente mensurables, el error absoluto $\delta f$ viene dado por la siguiente expresión \citep{Bevington1992}:
\begin{align}
	\delta f= \left|\frac{\partial f}{\partial A}\right|\cdot\delta A + \left|\frac{\partial f}{\partial B}\right|\cdot\delta B + \left|\frac{\partial f}{\partial C}\right|\cdot\delta C + \ldots \label{err-propag-general}
\end{align}
donde $\delta A,\delta B,\delta C,\ldots$ son los errores de medida de los parámetros directamente mensurables. Si ahora suponemos que $f$ es de la forma $f=\frac{A}{B}$, entonces tenemos que aplicando (\ref{err-propag-general}), elevando al cuadrado y operando llegaremos al siguiente resultado:
\begin{align}
  \delta f = \left|\frac{1}{B}\right|\cdot\delta A + \left|\frac{-A}{B^{2}}\right|\cdot\delta B   \label{error-propag}
\end{align}
Si ahora consideramos que el error cometido en la variable $B$ tiene una contribución despreciable frente al error de la variable $A$, obtendremos la si\-guien\-te expresión:
\begin{align}
  \delta f \approx \frac{\delta A}{B}   \label{error-propag-approx}
\end{align}
Esta aproximación es razonable en el caso de $A=\Delta X$ y $B=\Delta t$, ya que el tiempo de las imágenes de VIRTIS-M y de la cámara SSI se proporcionan con una precisión de la milésima de segundo, por lo que $\delta(\Delta t)$ no afecta especialmente al error final.\\

Aplicando (\ref{error-propag-approx}) a las ecuaciones (\ref{vel-zonal-new})-(\ref{vel-merid-new}) tenemos que el error de medida para las componentes zonal y meridional de la velocidad es:
\begin{align}
	\delta u &\approx \pm\frac{\delta(\Delta X)}{\Delta t},   \label{vel-zonal-error}\\
	\delta v &\approx \pm\frac{\delta(\Delta Y)}{\Delta t}.   \label{vel-merid-error}
\end{align}
es decir, que el error en la velocidad viene determinado por la resolución espacial y el intervalo de tiempo entre las imágenes utilizadas.\\

En el caso de las imágenes de la cámara SSI de Galileo la mayor parte de las imágenes analizadas tenían una resolución espacial entre 11 y 40 km/pixel (ver Tabla \ref{tab:tabla-imagsGalileo} en página \pageref{tab:tabla-imagsGalileo}), por lo que en el peor de los casos podemos considerar que el error a la hora de estimar el desplazamiento es como mucho de 2 píxeles ($\delta(\Delta X)=\delta(\Delta Y)\sim80~km$). Los pares de imágenes analizadas estaban separados por lo general 2 horas entre sí, por lo que $\Delta t\cong7200~s$. De esta manera, tomando el mejor y el peor valor de resolución espacial obtenemos tanto para la componente zonal como la meridional errores comprendidos entre $\delta u\sim \pm~3-11~m\cdot s^{-1}$.\\

De manera similar podemos proceder con las imágenes de VIRTIS-M. En este caso tenemos que la resolución espacial de las imágenes utilizadas varía entre 15 y 30 km/pixel \citep{Sanchez-Lavega2008}, siendo el intervalo temporal entre pares de imágenes de aproximadamente 1 hora ($\Delta t\cong3600~s$). Así, el error en las velocidades a partir de las imágenes de VIRTIS-M varía entre $\delta u\sim \pm~8-16~m\cdot s^{-1}$.

\chapter{Deducción de la Relación de Dispersión}\indent\label{appendix-waves}

En este apéndice utilizaré el método de perturbaciones para derivar la relación de dispersión de las ondas internas de gravedad. Por simplicidad asumiré que el movimiento es bidimensional, despreciaré el efecto Coriolis y la disipación viscosa. También asumiré que la atmósfera es incompresible excepto en la dirección vertical (aproximación de Boussinesq\index{Boussinesq!aproximación de}). Con estas consideraciones, las ecuaciones que describen el fluido son:\\
\begin{align}
  \frac{\partial u}{\partial t}+u\frac{\partial u}{\partial x}+w\frac{\partial u}{\partial z}&=-\frac{1}{\rho}\frac{\partial p}{\partial x},      \label{x-mov-eq}\\
  \frac{\partial w}{\partial t}+u\frac{\partial w}{\partial x}+w\frac{\partial w}{\partial z}&=-\frac{1}{\rho}\frac{\partial p}{\partial z}-g,    \label{z-mov-eq}\\
  \frac{\partial u}{\partial x}+\frac{\partial w}{\partial z}&=0,    \label{cont-eq}\\
  \frac{\partial\theta}{\partial t}+u\frac{\partial\theta}{\partial x}+w\frac{\partial\theta}{\partial z}&=0.    \label{thermal-eq}
\end{align}
\\
donde $u$ y $w$ son las velocidades horizontal y vertical respectivamente, $p$ es la presión, $g$ la aceleración de la gravedad y $\theta$ es la temperatura potencial. En caso de que los movimientos atmosféricos sean adiabáticos y que la ecuación de estado de las parcelas de aire vengan dadas por la de un gas ideal, tenemos que la temperatura potencial se relaciona con la presión y la densidad a través de la siguiente relación:\\
\begin{align}
  \theta = T\cdot\left(\frac{p_{s}}{p}\right)^{\kappa} = \frac{p}{\rho R}\cdot\left(\frac{p_{s}}{p}\right)^{\kappa}    \label{adiab-eq}
\end{align}
\\
donde $R$ es la constante de los gases en Venus, $p_{s}$ es una presión de referencia (por ejemplo la de la superficie) y el exponente $\kappa=R/c_{p}$, siendo $c_{p}$ el calor específico a presión constante.\\

Si expresamos las variables que definen el fluido como la suma de un estado básico independiente del tiempo y una componente menor que representa las peque\~{n}as variaciones temporales (perturbaciones) con respecto al fluido promedio\index{Perturbaciones!teoría de}, tenemos entonces que las variables del sistema se pueden expresar de la siguiente manera:\\
\begin{align*}
  u &= \bar{u} + u'(x,z,t),\\
  w &= w'(x,z,t),\\
  \rho &= \bar{\rho} + \rho'(x,z,t),\\
  p &= \bar{p}(z) + p'(x,z,t),\\
  \theta &= \bar{\theta}(z) + \theta'(x,z,t).
\end{align*}
\\
donde presuponemos que $\bar{w}=0$ y que tanto el flujo zonal $\bar{u}$ como la densidad $\bar{\rho}$ son constantes. Además, el promedio temporal de todas las variables perturbadas es cero.\\

A continuación se sustituye esta definición de las variables del fluido en el sistema de ecuaciones (\ref{x-mov-eq})-(\ref{thermal-eq}) y se desprecian los términos que contengan productos de perturbaciones. En el caso de la ecuación de movimiento vertical (\ref{z-mov-eq}), desarrollamos los términos del lado derecho de la igualdad de la siguiente manera:\\
\begin{equation}
\begin{split}
  \frac{-1}{\rho}\cdot\frac{\partial p}{\partial z}-g &= \frac{-1}{\bar{\rho}+\rho'}\cdot\left(\frac{\partial\bar{p}}{\partial z}+\frac{\partial p'}{\partial z}\right)-g=\\ &=\frac{-1}{\bar{\rho}}\cdot\left(1+\frac{\rho'}{\bar{\rho}}\right)^{-1}\cdot \left(\frac{\partial\bar{p}}{\partial z}+\frac{\partial p'}{\partial z}\right)-g=\\ &\approx\frac{-1}{\bar{\rho}}\cdot\left(1-\frac{\rho'}{\bar{\rho}}\right)\cdot \left(\frac{\partial\bar{p}}{\partial z}+\frac{\partial p'}{\partial z}\right)-g=\\ &=\frac{-1}{\bar{\rho}}\cdot\frac{\partial p'}{\partial z}-\frac{\rho'}{\bar{\rho}}\cdot g
\end{split}
\end{equation}
\\
donde se ha empleado la aproximación $\left(1+x\right)^{-1}\approx 1-x$ (válida cuando $|x|<<1$), hemos despreciado los términos que contenían productos de perturbaciones y se asume que la presión promedio cumple la ecuación de equilibrio hidrostático:\\
\begin{align}\label{hydro-eq}
  \frac{\partial\bar{p}}{\partial z} = -\bar{\rho}\cdot g
\end{align}
\\
De esta manera, llegamos a que las ecuaciones (\ref{x-mov-eq})-(\ref{thermal-eq}) pueden escribirse como:
\begin{align}
  \left(\frac{\partial}{\partial t}+\bar{u}\frac{\partial}{\partial x}\right)\cdot u' + \frac{1}{\bar{\rho}}\cdot\frac{\partial p'}{\partial x}&= 0,      \label{pert-x-mov-eq}\\
  \left(\frac{\partial}{\partial t}+\bar{u}\frac{\partial}{\partial x}\right)\cdot w' + \frac{1}{\bar{\rho}}\cdot\frac{\partial p'}{\partial z}+\frac{\rho'}{\bar{\rho}}\cdot g &= 0,    \label{pert-z-mov-eq}\\
  \frac{\partial u'}{\partial x}+\frac{\partial w'}{\partial z}&=0,    \label{pert-cont-eq}\\
  \left(\frac{\partial}{\partial t}+\bar{u}\frac{\partial}{\partial x}\right)\cdot \theta' + w'\cdot\frac{\partial\bar{\theta}}{\partial z} &= 0.    \label{pert-thermal-eq}
\end{align}
\\
Es posible eliminar la densidad $\rho'$ expresándola en términos de la temperatura potencial. Para ello, nos basta con operar sobre la ecuación (\ref{adiab-eq}) de la siguiente manera:\\
\begin{align*}
  \theta &= \frac{p}{\rho R}\cdot\left(\frac{p_{s}}{p}\right)^{\kappa} = const\cdot\frac{p^{1-\kappa}}{\rho} ,\\
  \ln\theta &= \gamma^{-1}\cdot\ln p - \ln\rho + const.
\end{align*}
\\
y hemos tenido en cuenta que para un gas ideal $1-\kappa=1-R/c_{p}=c_{v}/c_{p}=\gamma^{-1}$. Esta relación debe cumplirse también para las variables promediadas temporalmente, por lo que:\\
\begin{align}
  \ln\bar{\theta} &= \gamma^{-1}\cdot\ln\bar{p} - \ln\bar{\rho} + const,  \label{aver-adiab-eq} \\
  \ln\left[\bar{\theta}\cdot\left(1+\frac{\theta'}{\bar{\theta}}\right)\right] &= \gamma^{-1}\cdot\ln\left[\bar{p}\cdot\left(1+\frac{p'}{\bar{p}}\right)\right] - \ln\left[\bar{\rho}\cdot\left(1+\frac{\rho'}{\bar{\rho}}\right)\right] + const,\\
  \ln\left(1+\frac{\theta'}{\bar{\theta}}\right) &= \gamma^{-1}\cdot\ln\left(1+\frac{p'}{\bar{p}}\right) - \ln\left(1+\frac{\rho'}{\bar{\rho}}\right) + const.  \label{adiab-expression1}
\end{align}
\\
donde para obtener (\ref{adiab-expression1}) hemos usado $\ln(a\cdot b)=\ln(a)+\ln(b)$ y la relación (\ref{aver-adiab-eq}). Teniendo en cuenta que $\ln(1+x)\approx x$ para $x<<1$ podemos obtener la siguiente expresión para $\rho'$:\\
\begin{align}
  \frac{\theta'}{\bar{\theta}}\approx \frac{1}{\gamma}\cdot\frac{p'}{\bar{p}} - \frac{\rho'}{\bar{\rho}}, \\
  \rho' \approx -\bar{\rho}\cdot\frac{\theta'}{\bar{\theta}} + \frac{p'}{c_{s}^{2}}.  \label{adiab-expression2}
\end{align}
\\
donde $c_{s}^{2}=\bar{p}\gamma/\bar{\rho}$ es el cuadrado de la velocidad del sonido. En la ecuación (\ref{adiab-expression2}) vemos que a las fluctuaciones en la densidad contribuyen tanto los cambios de temperatura como los cambios de presión. En el caso de las ondas de gravedad, podemos despreciar la contribución de los cambios de presión frente a la de los cambios de temperatura \citep{Holton1992} por lo que:\\
\begin{align}
  \frac{\theta'}{\bar{\theta}} \approx -\frac{\rho'}{\bar{\rho}}  \label{adiab-expression3}
\end{align}
\\
lo que nos permite finalmente eliminar la variable $\rho'$ y reescribir el sistema como:
\begin{align}
  \left(\frac{\partial}{\partial t}+\bar{u}\frac{\partial}{\partial x}\right)\cdot u' + \frac{1}{\bar{\rho}}\cdot\frac{\partial p'}{\partial x}&= 0,      \label{pert-x-mov-eq}\\
  \left(\frac{\partial}{\partial t}+\bar{u}\frac{\partial}{\partial x}\right)\cdot w' + \frac{1}{\bar{\rho}}\cdot\frac{\partial p'}{\partial z}-\frac{\theta'}{\bar{\theta}}\cdot g &= 0,    \label{pert-z-mov-eq}\\
  \frac{\partial u'}{\partial x}+\frac{\partial w'}{\partial z}&=0,    \label{pert-cont-eq}\\
  \left(\frac{\partial}{\partial t}+\bar{u}\frac{\partial}{\partial x}\right)\cdot \theta' + w'\cdot\frac{\partial\bar{\theta}}{\partial z} &= 0.    \label{pert-thermal-eq}
\end{align}
\\

Combinando convenientemente las ecuaciones (\ref{pert-x-mov-eq})-(\ref{pert-thermal-eq}) podemos llegar a una ecuación de una sola incógnita. Así, restando $\partial(\ref{pert-x-mov-eq})/\partial z$ a $\partial(\ref{pert-z-mov-eq})/\partial x$ podemos eliminar $p'$,\\
\begin{align}
  \left(\frac{\partial}{\partial t}+\bar{u}\frac{\partial}{\partial x}\right) \cdot \left(\frac{\partial w'}{\partial x}-\frac{\partial u'}{\partial z}\right) - \frac{g}{\bar{\theta}}\cdot \frac{\partial\theta'}{\partial x} = 0,  \label{y-vort-eq} \\
  \left(\frac{\partial}{\partial t}+\bar{u}\frac{\partial}{\partial x}\right)^{2} \cdot \left(\frac{\partial^{2} w'}{\partial x^{2}}+\frac{\partial^{2} w'}{\partial z^{2}}\right) + N^{2}\cdot\frac{\partial^{2} w'}{\partial x^{2}} = 0.  \label{w-eq}
\end{align}
\\
donde la ecuación (\ref{w-eq}) ha sido obtenida aplicando (\ref{pert-cont-eq}) y (\ref{pert-thermal-eq}) a (\ref{y-vort-eq}), y donde $N^{2}=g\cdot\frac{\partial\ln\bar{\theta}}{\partial z}$ es el cuadrado de la frecuencia de Brunt-Väisällä\index{Brunt-Väisälä!frecuencia de} (que asumimos constante con la altura, ver \citealt{Holton1992}).\\

La ecuación (\ref{w-eq}) tiene soluciones armónicas del tipo:\\
\begin{align}
  w' = \Re\left[\hat{w}\cdot e^{i\phi}\right] = w_{r} \cos\phi-w_{i} \sin\phi \label{sol-w-eq}
\end{align}
\\
donde $\Re[\ldots]$ designa la parte real de una variable o expresión, $\hat{w}=w_{r}+i w_{i}$ es la amplitud de oscilación, $\phi=k x+m z-\omega t$ es la fase (que se supone que varía linealmente con $x$, $z$ y $t$), $k$ y $m$ son los números de onda horizontal y vertical respectivamente, y $\omega$ es la frecuencia angular. Bajo la hipótesis de que $\hat{w}\neq f(x,z,t)$ y sustituyendo esta solución en la ecuación (\ref{w-eq}) llegamos a la relación de dispersión\index{Dispersión!relación de|textbf}:\\
\begin{align}
  (\omega-\bar{u} k)^{2}\cdot(k^{2}+m^{2})-N^{2} k^{2}   \label{disp-rel1}
\end{align}
\\
de tal manera que,\\
\begin{align}
  \bar{\omega} \equiv \omega-\bar{u} k = \pm\frac{N k}{\sqrt{k^{2}+m^{2}}}   \label{disp-rel2}
\end{align}
\\
donde $\bar{\omega}$ es la \textit{frecuencia intrínseca}\index{Intrínseca!frecuencia}, es decir, la frecuencia que mediría un observador que se mueve conforme al viento promedio $\bar{u}$, y el signo positivo (negativo) corresponde a ondas cuya fase se desplaza hacia el este (oeste) respecto del viento promedio.

\backmatter

\bibliographystyle{apalike}
\addcontentsline{toc}{chapter}{Referencias}

\begin{thebibliography}{}

\bibitem[{Akima}, 1978]{Akima1978}
{Akima}, N. (1978).
\newblock {A method of bivariate interpolation and smooth surface fitting for
  irregularly distributed data points}.
\newblock {\em ACM Transactions on Mathematical Software}, 4:148--159.

\bibitem[{Allen} and {Crawford}, 1984]{Allen1984}
{Allen}, D.~A. and {Crawford}, J.~W. (1984).
\newblock {Cloud structure on the dark side of Venus}.
\newblock {\em Nature}, 307:222--224.

\bibitem[{Baker} et~al., 1998]{Baker1998}
{Baker}, R.~D., {Schubert}, G., and {Jones}, P.~W. (1998).
\newblock {Cloud-level penetrative compressible convection in the Venus
  atmosphere}.
\newblock {\em Journal of Atmospheric Sciences}, 55:3--18.

\bibitem[{Baker} et~al., 2000a]{Baker2000a}
{Baker}, R.~D., {Schubert}, G., and {Jones}, P.~W. (2000a).
\newblock {Convectively generated internal gravity waves in the lower
  atmosphere of Venus. I. No wind shear.}
\newblock {\em Journal of Atmospheric Sciences}, 57:184--199.

\bibitem[{Baker} et~al., 2000b]{Baker2000b}
{Baker}, R.~D., {Schubert}, G., and {Jones}, P.~W. (2000b).
\newblock {Convectively generated internal gravity waves in the lower
  atmosphere of Venus. II. Mean wind shear and wave-mean flow interaction.}
\newblock {\em Journal of Atmospheric Sciences}, 57:200--215.

\bibitem[{Belton} et~al., 1991]{Belton1991}
{Belton}, M.~J.~S., {Gierasch}, P.~J., {Smith}, M.~D., {Helfenstein}, P.,
  {Schinder}, P.~J., {Pollack}, J.~B., {Rages}, K.~A., {Morrison}, D.,
  {Klaasen}, K.~P., and {Pilcher}, C.~B. (1991).
\newblock {Images from Galileo of the Venus cloud deck}.
\newblock {\em Science}, 253:1531--1536.

\bibitem[{Belton} et~al., 1992]{Belton1992}
{Belton}, M.~J.~S., {Klaasen}, K.~P., {Clary}, M.~C., {Anderson}, J.~L.,
  {Anger}, C.~D., {Carr}, M.~H., {Chapman}, C.~R., {Davies}, M.~E., {Greeley},
  R., {Anderson}, D., {Bolef}, L.~K., {Townsend}, T.~E., {Greenberg}, R., {Head
  III}, J.~W., {Neukum}, G., {Pilcher}, C.~B., {Veverka}, J., {Gierasch},
  P.~J., {Fanale}, F.~P., {Ingersoll}, A.~P., {Masursky}, H., {Morrison}, D.,
  and {Pollack}, J.~B. (1992).
\newblock {The Galileo Solid-State Imaging Experiment}.
\newblock {\em Space Science Reviews}, 60:413--455.

\bibitem[{Belton} et~al., 1976]{Belton1976a}
{Belton}, M.~J.~S., {Smith}, G.~R., {Schubert}, G., and {del Genio}, A.~D.
  (1976).
\newblock {Cloud patterns, waves and convection in the Venus atmosphere}.
\newblock {\em Journal of Atmospheric Sciences}, 33:1394--1417.

\bibitem[{Bevington} and {Robinson}, 1992]{Bevington1992}
{Bevington}, P.~R. and {Robinson}, D.~K. (1992).
\newblock {\em {Data reduction and error analysis for the physical sciences}}.
\newblock New York: McGraw-Hill, |c1992, 2nd ed.

\bibitem[{Bird} et~al., 2007]{Bird2008}
{Bird}, M.~K., {P{\"a}tzold}, M., {Tellmann}, S., {H{\"a}usler}, B., and
  {Tyler}, G.~L. (2007).
\newblock {Structure of the Venus Neutral Atmosphere and Ionosphere from the
  Venus Express Radio Science Investigation - VeRa.}
\newblock In {\em Bulletin of the American Astronomical Society}, volume~40 of
  {\em Bulletin of the American Astronomical Society}, pages 468--+.

\bibitem[{Blackman} and {Tukey}, 1958]{Blackman1958}
{Blackman}, J.~W. and {Tukey}, R.~B. (1958).
\newblock {\em {The Measurement of Power Spectra}}.
\newblock International Geophysics Series. Dover Publication.

\bibitem[{Bougher} et~al., 1997]{Bougher1997}
{Bougher}, S.~W., {Alexander}, M.~J., and {Mayr}, H.~G. (1997).
\newblock {Upper atmosphere dynamics: global circulation and gravity waves}.
\newblock In {\em VENUS II: Geology, Geophysics, Atmosphere, and Solar Wind
  Environment}, pages 259--291. University of Arizona Press.

\bibitem[{Carlson} et~al., 1991]{Carlson1991}
{Carlson}, R.~W., {Baines}, K.~H., {Kamp}, L.~W., {Weissman}, P.~R., {Smythe},
  W.~D., {Ocampo}, A.~C., {Johnson}, T.~V., {Matson}, D.~L., {Pollack}, J.~B.,
  and {Grinspoon}, D. (1991).
\newblock {Galileo infrared imaging spectroscopy measurements at Venus}.
\newblock {\em Science}, 253:1541--1548.

\bibitem[{Clancy} and {Muhleman}, 1991]{Clancy1991}
{Clancy}, R.~T. and {Muhleman}, D.~O. (1991).
\newblock {Long-term (1979-1990) changes in the thermal, dynamical, and
  compositional structure of the Venus mesosphere as inferred from microwave
  spectral line observations of C-12O, C-13O, and CO-18}.
\newblock {\em Icarus}, 89:129--146.

\bibitem[{Colin}, 1979]{Colin1979}
{Colin}, L. (1979).
\newblock {Encounter with Venus}.
\newblock {\em Science}, 203:743--745.

\bibitem[{Counselman} et~al., 1980]{Counselman1980}
{Counselman}, C.~C., {Gourevitch}, S.~A., {King}, R.~W., {Loriot}, G.~B., and
  {Ginsberg}, E.~S. (1980).
\newblock {Zonal and meridional circulation of the lower atmosphere of Venus
  determined by radio interferometry}.
\newblock {\em Journal of Geophysical Research}, 85:8026--8030.

\bibitem[{Crisp} et~al., 1991]{Crisp1991}
{Crisp}, D., {McMuldroch}, S., {Stephens}, S.~K., {Sinton}, W.~M., {Ragent},
  B., {Hodapp}, K.-W., {Probst}, R.~G., {Doyle}, L.~R., {Allen}, D.~A., and
  {Elias}, J. (1991).
\newblock {Ground-based near-infrared imaging observations of Venus during the
  Galileo encounter}.
\newblock {\em Science}, 253:1538--1541.

\bibitem[{de Bergh} et~al., 2006]{deBergh2006}
{de Bergh}, C., {Moroz}, V.~I., {Taylor}, F.~W., {Crisp}, D., {B{\'e}zard}, B.,
  and {Zasova}, L.~V. (2006).
\newblock {The composition of the atmosphere of Venus below 100 km altitude: An
  overview}.
\newblock {\em Planetary and Space Science}, 54:1389--1397.

\bibitem[{del Genio} and {Rossow}, 1982]{delGenio1982}
{del Genio}, A.~D. and {Rossow}, W.~B. (1982).
\newblock {Temporal variability of ultraviolet cloud features in the Venus
  stratosphere}.
\newblock {\em Icarus}, 51:391--415.

\bibitem[{del Genio} and {Rossow}, 1990]{delGenio1990}
{del Genio}, A.~D. and {Rossow}, W.~B. (1990).
\newblock {Planetary-scale waves and the cyclic nature of cloud top dynamics on
  Venus}.
\newblock {\em Journal of Atmospheric Sciences}, 47:293--318.

\bibitem[{Dobrovolskis} and {Ingersoll}, 1980]{Dobrovolskis1980}
{Dobrovolskis}, A.~R. and {Ingersoll}, A.~P. (1980).
\newblock {Atmospheric tides and the rotation of Venus. I - Tidal theory and
  the balance of torques}.
\newblock {\em Icarus}, 41:1--17.

\bibitem[{Drossart} et~al., 2007]{Drossart2007}
{Drossart}, P., {Piccioni}, G., {Adriani}, A., {Angrilli}, F., {Arnold}, G.,
  {Baines}, K.~H., {Bellucci}, G., {Benkhoff}, J., {B{\'e}zard}, B., {Bibring},
  J.-P., {Blanco}, A., {Blecka}, M.~I., {Carlson}, R.~W., {Coradini}, A., {di
  Lellis}, A., {Encrenaz}, T., {Erard}, S., {Fonti}, S., {Formisano}, V.,
  {Fouchet}, T., {Garcia}, R., {Haus}, R., {Helbert}, J., {Ignatiev}, N.~I.,
  {Irwin}, P.~G.~J., {Langevin}, Y., {Lebonnois}, S., {L{\'o}pez-Valverde},
  M.~A., {Luz}, D., {Marinangeli}, L., {Orofino}, V., {Rodin}, A.~V.,
  {Roos-Serote}, M.~C., {Saggin}, B., {S{\'a}nchez-Lavega}, A., {Stam}, D.~M.,
  {Taylor}, F.~W., {Titov}, D., {Visconti}, G., {Zambelli}, M., {Hueso}, R.,
  {Tsang}, C.~C.~C., {Wilson}, C.~F., and {Afanasenko}, T.~Z. (2007).
\newblock {Scientific goals for the observation of Venus by VIRTIS on ESA/Venus
  express mission}.
\newblock {\em Planetary and Space Science}, 55:1653--1672.

\bibitem[{Esposito} et~al., 1997]{Esposito1997}
{Esposito}, L.~W., {Bertaux}, J.~L., {Krasnopolsky}, V., {Moroz}, V.~I., and
  {Zasova}, L.~V. (1997).
\newblock {Chemistry of Lower Atmosphere and Clouds}.
\newblock In {\em VENUS II: Geology, Geophysics, Atmosphere, and Solar Wind
  Environment}, pages 415--458. University of Arizona Press.

\bibitem[{Fels}, 1986]{Fels1986}
{Fels}, S.~B. (1986).
\newblock {An approximate analytical method for calculating tides in the
  atmosphere of Venus}.
\newblock {\em Journal of Atmospheric Sciences}, 43:2757--2772.

\bibitem[{Fj\o rtoft}, 1953]{Fjortoft1953}
{Fj\o rtoft}, R. (1953).
\newblock {On the changes in the spectral distribution of kinetic energy for
  2-dimensional, non-divergent flow}.
\newblock {\em Tellus}, 5:225--230.

\bibitem[{Flasar} and {Gierasch}, 1986]{Flasar1986}
{Flasar}, F.~M. and {Gierasch}, P.~J. (1986).
\newblock {Mesoscale waves as a probe of Jupiter's deep atmosphere}.
\newblock {\em Journal of Atmospheric Sciences}, 43:2683--2707.

\bibitem[{Frisch}, 1995]{Frisch1995}
{Frisch}, U. (1995).
\newblock {\em {Turbulence}}.
\newblock Cambridge University Press.

\bibitem[{Gage} and {Nastrom}, 1986]{Gage1986}
{Gage}, K.~S. and {Nastrom}, G.~D. (1986).
\newblock {Theoretical Interpretation of Atmospheric Wavenumber Spectra of Wind
  and Temperature Observed by Commercial Aircraft During GASP.}
\newblock {\em Journal of Atmospheric Sciences}, 43:729--740.

\bibitem[{Garc{\'i}a} et~al., 2008]{Garcia2008}
{Garc{\'i}a}, R., {L{\'o}pez-Valverde}, M.~A., {Drossart}, P., and {Piccioni},
  G. (2008).
\newblock {Gravity waves in Venus upper atmosphere revealed by $CO_{2}$ Non LTE
  emissions}.
\newblock \textit{Submitted to Journal of Geophysical Research: Planets}.

\bibitem[{Gaulme} et~al., 2008]{Gaulme2008}
{Gaulme}, P., {Schmider}, F.~X., {Grec}, C., {L{\'o}pez-Ariste}, A.,
  {Widemann}, T., and {Gelly}, G. (2008).
\newblock {Venus wind map at cloud top level with the MTR/THEMIS visible
  spectrometer, I: Instrumental performance and first results}.
\newblock {\em Planetary and Space Science}, 56:1335--1343.

\bibitem[{Gierasch}, 1987]{Gierasch1987}
{Gierasch}, P.~J. (1987).
\newblock {Waves in the atmosphere of Venus}.
\newblock {\em Nature}, 328:510--512.

\bibitem[{Gierasch} et~al., 1997]{Gierasch1997}
{Gierasch}, P.~J., {Goody}, R.~M., {Young}, R.~E., {Crisp}, D., and {et al.}
  (1997).
\newblock {The General Circulation of the Venus Atmosphere}.
\newblock In {\em VENUS II: Geology, Geophysics, Atmosphere, and Solar Wind
  Environment}, pages 459--500. University of Arizona Press.

\bibitem[{Gonzalez} and {Woods}, 1992]{Gonzalez1992}
{Gonzalez}, R.~C. and {Woods}, R.~E. (1992).
\newblock {\em Digital Image Processing}.
\newblock Addison-Wesley Longman Publishing Co., Inc., Boston, MA, USA.

\bibitem[{Hinson} and {Jenkins}, 1995]{Hinson1995}
{Hinson}, D.~P. and {Jenkins}, J.~M. (1995).
\newblock {Magellan radio occultation measurements of atmospheric waves on
  Venus}.
\newblock {\em Icarus}, 114:310--327.

\bibitem[{Holton}, 1992]{Holton1992}
{Holton}, J.~R. (1992).
\newblock {\em {An introduction to dynamic meteorology}}.
\newblock International geophysics series, San Diego, New York: Academic Press,
  |c1992, 3rd ed.

\bibitem[{Hou} and {Farrell}, 1987]{Hou1987}
{Hou}, A.~Y. and {Farrell}, B.~F. (1987).
\newblock {Superrotation induced by critical-level absorption of gravity waves
  on Venus - an assessment}.
\newblock {\em Journal of Atmospheric Sciences}, 44:1049--1061.

\bibitem[{Hou} et~al., 1990]{Hou1990}
{Hou}, A.~Y., {Fels}, S.~B., and {Goody}, R.~M. (1990).
\newblock {Zonal superrotation above Venus' cloud base induced by the
  semidiurnal tide and the mean meridional circulation}.
\newblock {\em Journal of Atmospheric Sciences}, 47:1894--1901.

\bibitem[{Houze}, 1993]{Houze1993}
{Houze}, R.~A. (1993).
\newblock {\em {Cloud Dynamics}}.
\newblock Academic Press.

\bibitem[{Hueso} et~al., 2008]{Hueso2008b}
{Hueso}, R., {Legarreta}, J., {Rojas}, J.~F., {Peralta}, J., {P{\'e}rez-Hoyos},
  S., {del R{\'i}o-Gaztelurrutia}, T., and {S{\'a}nchez-Lavega}, A. (2008).
\newblock {Introducing PLIA: The Planetary Laboratory for Image Analysis}.
\newblock \textit{Submitted to Planetary \& Space Science}.

\bibitem[{Hunt} and {Muller}, 1979]{Hunt1979}
{Hunt}, G.~E. and {Muller}, J.~P. (1979).
\newblock {Voyager observations of small-scale waves in the equatorial region
  of the jovian atmosphere}.
\newblock {\em Nature}, 280:778--780.

\bibitem[{Julian} and {Cline}, 1974]{Julian1974}
{Julian}, P.~R. and {Cline}, A.~K. (1974).
\newblock {The Direct Estimation of Spatial Wavenumber Spectra of Atmospheric
  Variables.}
\newblock {\em Journal of Atmospheric Sciences}, 31:1526--1539.

\bibitem[{Julian} et~al., 1970]{Julian1970}
{Julian}, P.~R., {Washington}, W.~M., {Hembree}, L., and {Ridley}, C. (1970).
\newblock {On the Spectral Distribution of Large-Scale Atmospheric Kinetic
  Energy.}
\newblock {\em Journal of Atmospheric Sciences}, 27:376--387.

\bibitem[{Kasprzak} et~al., 1988]{Kasprzak1988}
{Kasprzak}, W.~T., {Hedin}, A.~E., {Mayr}, H.~G., and {Niemann}, H.~B. (1988).
\newblock {Wavelike perturbations observed in the neutral thermosphere of
  Venus}.
\newblock {\em Journal of Geophysical Research}, 93:11237--11245.

\bibitem[{Kerzhanovich} et~al., 1980]{Kerzhanovich1980}
{Kerzhanovich}, V.~V., {Mararov}, Y.~F., {Marov}, M.~Y., {Rozhdestvenskii},
  M.~K., and {Sorokin}, V.~P. (1980).
\newblock {Venera 11 and Venera 12 - Preliminary evaluations of wind velocity
  and turbulence in the atmosphere of Venus}.
\newblock {\em Moon and Planets}, 23:261--270.

\bibitem[{Klaasen} et~al., 1997]{Klaasen1997}
{Klaasen}, K.~P., {Belton}, M.~J.~S., {Breneman}, H.~H., {McEwen}, A.~S.,
  {Davies}, M.~E., {Sullivan}, R.~J., {Chapman}, C.~R., {Neukum}, G.,
  {Heffernan}, C.~M., {Harch}, A.~P., {Kaufman}, J.~M., {Merline}, W.~J.,
  {Gaddis}, L.~R., {Cunningham}, W.~F., {Helfenstein}, P., and {Colvin}, T.~R.
  (1997).
\newblock {Inflight performance characteristics, calibration, and utilization
  of the Galileo SSI camera}.
\newblock {\em Optical Engineering}, 36:3001--3027.

\bibitem[{Kliore} et~al., 1992]{Kliore1985}
{Kliore}, A.~J., {Keating}, G.~M., and {Moroz}, V.~I. (1992).
\newblock {Venus international reference atmosphere (1985)}.
\newblock {\em Planetary and Space Science}, 40:573--573.

\bibitem[{Kliore} and {Patel}, 1980]{Kliore1980}
{Kliore}, A.~J. and {Patel}, I.~R. (1980).
\newblock {Vertical structure of the atmosphere of Venus from Pioneer Venus
  orbiter radio occultations}.
\newblock {\em Journal of Geophysical Research}, 85:7957--7962.

\bibitem[{Kolmogorov}, 1941]{Kolmogorov1941}
{Kolmogorov}, A. (1941).
\newblock {The Local Structure of Turbulence in Incompressible Viscous Fluid
  for Very Large Reynolds' Numbers}.
\newblock {\em Akademiia Nauk SSSR Doklady}, 30:301--305.

\bibitem[{Kolosov} et~al., 1980]{Kolosov1980}
{Kolosov}, M.~A., {Yakovlev}, O.~I., {Efimov}, A.~I., {Matyugov}, S.~S.,
  {Timofeeva}, T.~S., {Chub}, E.~V., {Pavelyev}, A.~G., {Kucheryavenkov},
  A.~I., {Kalashnikov}, I.~E., and {Milekhin}, O.~E. (1980).
\newblock {Investigation of the Venus Atmosphere and Surface by the Method of
  Radiosounding Using VENERA-9 and 10 Satellites}.
\newblock {\em Acta Astronautica}, 7:219--234.

\bibitem[{Kraichnan}, 1967]{Kraichnan1967}
{Kraichnan}, R.~H. (1967).
\newblock {Inertial Ranges in Two-Dimensional Turbulence}.
\newblock {\em Physics of Fluids}, 10:1417--1423.

\bibitem[{Kundu} and {Cohen}, 2008]{Kundu2008}
{Kundu}, P.~K. and {Cohen}, I.~M. (2008).
\newblock {\em {Fluid Mechanics: Fourth Edition}}.
\newblock Fluid Mechanics: Fourth Edition.~Edited by Pijush K.~Kundu and Ira
  M.~Cohen with contributions by P.~S.~Ayyaswamy and H.~H.~Hu.~ISBN
  978-0-12-373735-9.~Published by Academic Press, Elsevier, Inc., London,
  England, 2008.

\bibitem[{Lambert}, 1981]{Lambert1981}
{Lambert}, S.~J. (1981).
\newblock {A diagnostic study of global energy and enstrophy fluxes and
  spectra}.
\newblock {\em Tellus}, 33:411--414.

\bibitem[{Lellouch} et~al., 2008]{Lellouch2008}
{Lellouch}, E., {Paubert}, G., {Moreno}, R., and {Moullet}, A. (2008).
\newblock {Monitoring Venus' mesospheric winds in support of Venus Express:
  IRAM 30-m and APEX observations}.
\newblock {\em Planetary and Space Science}, 56:1355--1367.

\bibitem[{Lemone}, 1973]{Lemone1973}
{Lemone}, M.~A. (1973).
\newblock {The Structure and Dynamics of Horizontal Roll Vortices in the
  Planetary Boundary Layer.}
\newblock {\em Journal of Atmospheric Sciences}, 30:1077--1091.

\bibitem[{Leovy}, 1973]{Leovy1973}
{Leovy}, C.~B. (1973).
\newblock {Rotation of the upper atmosphere of Venus.}
\newblock {\em Journal of Atmospheric Sciences}, 30:1218--1220.

\bibitem[{Leovy}, 1987]{Leovy1987}
{Leovy}, C.~B. (1987).
\newblock {Zonal winds near Venus' cloud top level - an analytic model of the
  equatorial wind speed.}
\newblock {\em Icarus}, 69:193--201.

\bibitem[{Leroy} and {Ingersoll}, 1995]{Leroy1995}
{Leroy}, S.~S. and {Ingersoll}, A.~P. (1995).
\newblock {Convective Generation of Gravity Waves in Venus's Atmosphere:
  Gravity Wave Spectrum and Momentum Transport.}
\newblock {\em Journal of Atmospheric Sciences}, 52:3717--3737.

\bibitem[{Li} et~al., 2008]{Li2008}
{Li}, L., {Conrath}, B.~J., {Flasar}, F.~M., and {Gierasch}, P.~J. (2008).
\newblock {Revisit of the thermal wind equation: application to planetary
  atmospheres at low latitudes}.
\newblock \textit{Submitted to Journal of Atmospheric Sciences}.

\bibitem[{Lilly}, 1989]{Lilly1989}
{Lilly}, D.~K. (1989).
\newblock {Two-Dimensional Turbulence Generated by Energy Sources at Two
  Scales.}
\newblock {\em Journal of Atmospheric Sciences}, 46:2026--2030.

\bibitem[{Limaye}, 1977]{Limaye1977}
{Limaye}, S.~S. (1977).
\newblock {\em {Venus Stratospheric Circulation: a Diagnostic Study.}}
\newblock PhD thesis, THE UNIVERSITY OF WISCONSIN - MADISON.

\bibitem[{Limaye}, 2007]{Limaye2007}
{Limaye}, S.~S. (2007).
\newblock {Venus atmospheric circulation: Known and unknown}.
\newblock {\em Journal of Geophysical Research (Planets)}, 112:4--+.

\bibitem[{Limaye} et~al., 1988]{Limaye1988a}
{Limaye}, S.~S., {Grassotti}, C., and {Kuetemeyer}, M.~J. (1988).
\newblock {Venus: Cloud level circulation during 1982 as determined from
  Pioneer cloud photopolarimeter images. I - Time and zonally averaged
  circulation}.
\newblock {\em Icarus}, 73:193--211.

\bibitem[{Limaye} et~al., 1982]{Limaye1982}
{Limaye}, S.~S., {Grund}, C.~J., and {Burre}, S.~P. (1982).
\newblock {Zonal mean circulation at the cloud level on Venus - Spring and fall
  1979 OCPP observations}.
\newblock {\em Icarus}, 51:416--439.

\bibitem[{Limaye} and {Suomi}, 1981]{Limaye1981}
{Limaye}, S.~S. and {Suomi}, V.~E. (1981).
\newblock {Cloud motions on Venus - Global structure and organization}.
\newblock {\em Journal of Atmospheric Sciences}, 38:1220--1235.

\bibitem[{Lindborg}, 1999]{Lindborg1999}
{Lindborg}, E. (1999).
\newblock {Can the atmospheric kinetic energy spectrum be explained by
  two-dimensional turbulence?}
\newblock {\em Journal of Fluid Mechanics}, 388:259--288.

\bibitem[{Linkin} et~al., 1986]{Linkin1986}
{Linkin}, V.~M., {Kerzhanovich}, V.~V., {Lipatov}, A.~N., {Pichkadze}, K.~M.,
  {Shurupov}, A.~A., {Terterashvili}, A.~V., {Ingersoll}, A.~P., {Crisp}, D.,
  {Grossman}, A.~W., {Young}, R.~E., {Seiff}, A., {Ragent}, B., {Blamont},
  J.~E., {Elson}, L.~S., and {Preston}, R.~A. (1986).
\newblock {VEGA balloon dynamics and vertical winds in the Venus middle cloud
  region}.
\newblock {\em Science}, 231:1417--1419.

\bibitem[{Maltrud} and {Vallis}, 1991]{Maltrud1991}
{Maltrud}, M.~E. and {Vallis}, G.~K. (1991).
\newblock {Energy spectra and coherent structures in forced two-dimensional and
  beta-plane turbulence}.
\newblock {\em Journal of Fluid Mechanics}, 228:321--342.

\bibitem[{Markiewicz} et~al., 2007a]{Markiewicz2007a}
{Markiewicz}, W.~J., {Titov}, D.~V., {Ignatiev}, N., {Keller}, H.~U., {Crisp},
  D., {Limaye}, S.~S., {Jaumann}, R., {Moissl}, R., {Thomas}, N., {Esposito},
  L., {Watanabe}, S., {Fiethe}, B., {Behnke}, T., {Szemerey}, I., {Michalik},
  H., {Perplies}, H., {Wedemeier}, M., {Sebastian}, I., {Boogaerts}, W.,
  {Hviid}, S.~F., {Dierker}, C., {Osterloh}, B., {B{\"o}ker}, W., {Koch}, M.,
  {Michaelis}, H., {Belyaev}, D., {Dannenberg}, A., {Tschimmel}, M., {Russo},
  P., {Roatsch}, T., and {Matz}, K.~D. (2007a).
\newblock {Venus Monitoring Camera for Venus Express}.
\newblock {\em Planetary and Space Science}, 55:1701--1711.

\bibitem[{Markiewicz} et~al., 2007b]{Markiewicz2007b}
{Markiewicz}, W.~J., {Titov}, D.~V., {Limaye}, S.~S., {Keller}, H.~U.,
  {Ignatiev}, N., {Jaumann}, R., {Thomas}, N., {Michalik}, H., {Moissl}, R.,
  and {Russo}, P. (2007b).
\newblock {Morphology and dynamics of the upper cloud layer of Venus}.
\newblock {\em Nature}, 450:633--636.

\bibitem[{Marov}, 1978]{Marov1978}
{Marov}, M.~Y. (1978).
\newblock {Results of Venus missions}.
\newblock {\em Annual Review of astronomy and astrophysics}, 16:141--169.

\bibitem[{McGouldrick} and {Toon}, 2008]{McGouldrick2008}
{McGouldrick}, K. and {Toon}, O.~B. (2008).
\newblock Observable effects of convection and gravity waves on the venus
  condensational cloud.
\newblock {\em Planetary and Space Science}, 56:1112--1131.

\bibitem[{McMahon}, 1996]{McMahon1996}
{McMahon}, S.~K. (1996).
\newblock {Overview of the Planetary Data System}.
\newblock {\em Planetary and Space Science}, 44:3--12.

\bibitem[{Minnaert}, 1941]{Minnaert1941}
{Minnaert}, M. (1941).
\newblock {The reciprocity principle in lunar photometry}.
\newblock {\em The Astrophysical Journal}, 93:403--410.

\bibitem[{Nastrom} et~al., 1984]{Nastrom1984}
{Nastrom}, G.~D., {Jasperson}, W.~H., and {Gage}, K.~S. (1984).
\newblock {Kinetic energy spectrum of large- and mesoscale atmospheric
  processes}.
\newblock {\em Nature}, 310:36--38.

\bibitem[{Newman} and {Leovy}, 1992]{Newman1992}
{Newman}, M. and {Leovy}, C. (1992).
\newblock {Maintenance of strong rotational winds in Venus' middle atmosphere
  by thermal tides}.
\newblock {\em Science}, 257:647--650.

\bibitem[{Newman} et~al., 1984]{Newman1984}
{Newman}, M., {Schubert}, G., {Kliore}, A.~J., and {Patel}, I.~R. (1984).
\newblock {Zonal winds in the middle atmosphere of Venus from Pioneer Venus
  radio occultation data}.
\newblock {\em Journal of Atmospheric Sciences}, 41:1901--1913.

\bibitem[{Niemann} et~al., 1980]{Niemann1980}
{Niemann}, H.~B., {Kasprzak}, W.~T., {Hedin}, A.~E., {Hunten}, D.~M., and
  {Spencer}, N.~W. (1980).
\newblock {Mass spectrometric measurements of the neutral gas composition of
  the thermosphere and exosphere of Venus}.
\newblock {\em Journal of Geophysical Research}, 85:7817--7827.

\bibitem[{Pajares} and {de la Cruz}, 2001]{Pajares2001}
{Pajares}, G. and {de la Cruz}, J.~M. (2001).
\newblock {\em {Visi{\'o}n por computador: Im{\'a}genes digitales y
  aplicaciones}}.
\newblock Ra-Ma.

\bibitem[{Pechmann} and {Ingersoll}, 1984]{Pechmann1984}
{Pechmann}, J.~B. and {Ingersoll}, A.~P. (1984).
\newblock {Thermal tides in the atmosphere of Venus: Comparison of model
  results with observations}.
\newblock {\em Journal of Atmospheric Sciences}, 41:3290--3313.

\bibitem[{Peralta} et~al., 2005]{Peralta2005}
{Peralta}, J., {Hueso}, R., {Barrado}, N., and {S{\'a}nchez-Lavega}, A. (2005).
\newblock {Introducing PLIA: Planetary Laboratory for Image Analysis}.
\newblock In {\em Bulletin of the American Astronomical Society}, volume~37 of
  {\em Bulletin of the American Astronomical Society}, pages 653--+.

\bibitem[{Peralta} et~al., 2007a]{Peralta2007b}
{Peralta}, J., {Hueso}, R., and {S{\'a}nchez-Lavega}, A. (2007a).
\newblock {A reanalysis of Venus winds at two cloud levels from Galileo SSI
  images}.
\newblock {\em Icarus}, 190:469--477.

\bibitem[{Peralta} et~al., 2007b]{Peralta2007a}
{Peralta}, J., {Hueso}, R., and {S{\'a}nchez-Lavega}, A. (2007b).
\newblock {Cloud brightness distribution and turbulence in Venus using Galileo
  violet images}.
\newblock {\em Icarus}, 188:305--314.

\bibitem[{Peralta} et~al., 2008]{Peralta2008}
{Peralta}, J., {Hueso}, R., {S{\'a}nchez-Lavega}, A., {Piccioni}, G.,
  {Lanciano}, O., and {Drossart}, P. (2008).
\newblock {Characterization of mesoscale gravity waves in the upper and lower
  clouds of Venus from VEX-VIRTIS images}.
\newblock {\em Journal of Geophysical Research}, 113:E00B18--+.

\bibitem[{Piccialli} et~al., 2008]{Piccialli2008}
{Piccialli}, A., {Titov}, D., {Grassi}, D., {Khatuntsev}, I., {Drossart}, P.,
  {Piccioni}, G., and {Migliorini}, A. (2008).
\newblock {Retrieval of the cyclostrophic wind in the Venus mesosphere from the
  VIRTIS/Venus Express temperature sounding.}
\newblock In {\em 37th COSPAR Scientific Assembly. Held 13-20 July 2008, in
  Montr{\'e}al, Canada., p.2429}, volume~37 of {\em COSPAR, Plenary Meeting},
  pages 2429--+.

\bibitem[{Piccioni} et~al., 2007]{Piccioni2007}
{Piccioni}, G., {Drossart}, P., {S{\'a}nchez-Lavega}, A., {Hueso}, R.,
  {Taylor}, F.~W., {Wilson}, C.~F., {Grassi}, D., {Zasova}, L., {Moriconi}, M.,
  {Adriani}, A., {Lebonnois}, S., {Coradini}, A., {B{\'e}zard}, B., {Angrilli},
  F., {Arnold}, G., {Baines}, K.~H., {Bellucci}, G., {Benkhoff}, J., {Bibring},
  J.~P., {Blanco}, A., {Blecka}, M.~I., W., C.~R., {di Lellis}, A., {Encrenaz},
  T., {Erard}, S., {Fonti}, S., {Formisano}, V., {Fouchet}, T., {Garc{\'i}a},
  R., {Haus}, R., {Helbert}, J., {Ignatiev}, N.~I., {Irwin}, P.~G.~J.,
  {Langevin}, Y., {L{\'o}pez-Valverde}, M.~A., {Luz}, D., {Marinangeli}, L.,
  {Orofino}, V., {Rodin}, A.~V., {Roos-Serote}, M.~C., {Saggin}, B., {Stam},
  D.~M., {Titov}, D., {Visconti}, G., {Zambelli}, M., and {the VIRTIS-Venus
  Express Technical Team} (2007).
\newblock {South-polar features on Venus similar to those near the north pole}.
\newblock {\em Nature}, 450:637--640.

\bibitem[{Piccioni} et~al., 2006]{Piccioni2006}
{Piccioni}, G., {Drossart}, P., and {VIRTIS/Venus Express Team} (2006).
\newblock {First Results from VIRTIS on Venus Express1. From Surface to Cloud
  Level}.
\newblock In {\em Bulletin of the American Astronomical Society}, volume~38 of
  {\em Bulletin of the American Astronomical Society}, pages 510--+.

\bibitem[{Press} et~al., 1992]{NumRecipes1992}
{Press}, W.~H., {Teukolsky}, S.~A., {Vetterling}, W.~T., and {Flannery}, B.~P.
  (1992).
\newblock {\em {Numerical Recipes in Fortran 77: The art of Scientific
  Computing}}.
\newblock Cambridge University Press.

\bibitem[{Preston} et~al., 1986]{Preston1986}
{Preston}, R.~A., {Hildebrand}, C.~E., {Purcell}, G.~H., {Ellis}, J.,
  {Stelzried}, C.~T., {Finley}, S.~G., {Sagdeev}, R.~Z., {Linkin}, V.~M.,
  {Kerzhanovich}, V.~V., {Altunin}, V.~I., {Kogan}, L.~R., {Kostenko}, V.~I.,
  {Matveenko}, L.~I., {Pogrebenko}, S.~V., {Strukov}, I.~A., {Akim}, E.~L.,
  {Alexandrov}, Y.~N., {Armand}, N.~A., {Bakitko}, R.~N., {Vyshlov}, A.~S.,
  {Bogomolov}, A.~F., {Gorchankov}, Y.~N., {Selivanov}, A.~S., {Ivanov}, N.~M.,
  {Tichonov}, V.~F., {Blamont}, J.~E., {Boloh}, L., {Laurans}, G., {Boischot},
  A., {Biraud}, F., {Ortega-Molina}, A., {Rosolen}, C., and {Petit}, G. (1986).
\newblock {Determination of Venus winds by ground-based radio tracking of the
  VEGA balloons}.
\newblock {\em Science}, 231:1414--1416.

\bibitem[{Prinn}, 1985]{Prinn1985}
{Prinn}, R.~G. (1985).
\newblock {The Photochemistry of the Atmosphere of Venus}.
\newblock In {\em The Photochemistry of Atmospheres: Earth, the Other Planets,
  and Comets}, pages 281--336. Academic Press.

\bibitem[{Ragent} et~al., 1985]{Ragent1985}
{Ragent}, B., {Esposito}, L.~W., {Tomasko}, M.~G., {Marov}, M.~I., and {Shari},
  V.~P. (1985).
\newblock {Particulate matter in the Venus atmosphere}.
\newblock {\em Advances in Space Research}, 5:85--115.

\bibitem[{Read} and {Lewis}, 2004]{Read2004}
{Read}, P.~L. and {Lewis}, S.~R. (2004).
\newblock {\em {The Martian Climate Revisited: Atmosphere and Environment of a
  Desert Planet}}.
\newblock Springer.

\bibitem[{Reuter} et~al., 2007]{Reuter2007}
{Reuter}, D.~C., {Simon-Miller}, A.~A., {Lunsford}, A., {Baines}, K.~H.,
  {Cheng}, A.~F., {Jennings}, D.~E., {Olkin}, C.~B., {Spencer}, J.~R., {Stern},
  S.~A., {Weaver}, H.~A., and {Young}, L.~A. (2007).
\newblock {Jupiter Cloud Composition, Stratification, Convection, and Wave
  Motion: A View from New Horizons}.
\newblock {\em Science}, 318:223--.

\bibitem[{Roos-Serote} et~al., 1995]{Roos-Serote1995}
{Roos-Serote}, M., {Drossart}, P., {Encrenaz}, T., {Lellouch}, E., {Carlson},
  R.~W., {Baines}, K.~H., {Taylor}, F.~W., and {Calcutt}, S.~B. (1995).
\newblock {The thermal structure and dynamics of the atmosphere of Venus
  between 70 and 90 KM from the Galileo-NIMS spectra}.
\newblock {\em Icarus}, 114:300--309.

\bibitem[{Rossow} et~al., 1990]{Rossow1990}
{Rossow}, W.~B., {del Genio}, A.~D., and {Eichler}, T. (1990).
\newblock {Cloud-tracked winds from Pioneer Venus OCPP images}.
\newblock {\em Journal of Atmospheric Sciences}, 47:2053--2084.

\bibitem[{Rossow} et~al., 1980]{Rossow1980}
{Rossow}, W.~B., {del Genio}, A.~D., {Limaye}, S.~S., and {Travis}, L.~D.
  (1980).
\newblock {Cloud morphology and motions from Pioneer Venus images}.
\newblock {\em Journal of Geophysical Research}, 85:8107--8128.

\bibitem[{Rossow} and {Williams}, 1979]{Rossow1979}
{Rossow}, W.~B. and {Williams}, G.~P. (1979).
\newblock {Large-scale motion in the Venus stratosphere}.
\newblock {\em Journal of Atmospheric Sciences}, 36:377--389.

\bibitem[{Sagdeev} et~al., 1986]{Sagdeev1986}
{Sagdeev}, R.~Z., {Linkin}, V.~M., {Blamont}, J.~E., and {Preston}, R.~A.
  (1986).
\newblock {The VEGA Venus Balloon Experiment}.
\newblock {\em Science}, 231:1407--1408.

\bibitem[{Salby}, 1996]{Salby1996}
{Salby}, M.~L. (1996).
\newblock {\em {Fundamentals of Atmospheric Physics}}, volume~61 of {\em
  International Geophysics Series}.
\newblock Academic Press, San Diego, U.S.A.

\bibitem[{S{\'a}nchez-Lavega} et~al., 2008]{Sanchez-Lavega2008}
{S{\'a}nchez-Lavega}, A., {Hueso}, R., {Piccioni}, G., {Drossart}, P.,
  {Peralta}, J., {P{\'e}rez-Hoyos}, S., {Wilson}, C.~F., {Taylor}, F.~W.,
  {Baines}, K.~H., {Luz}, D., {Erard}, S., and {Lebonnois}, S. (2008).
\newblock {Variable winds on Venus mapped in three dimensions}.
\newblock {\em Geophysical Research Letters}, 35:13204--+.

\bibitem[{Saunders} et~al., 1991]{Saunders1991}
{Saunders}, R.~S., {Arvidson}, R.~E., {Head}, J.~W., {Schaber}, G.~G.,
  {Stofan}, E.~R., and {Solomon}, S.~C. (1991).
\newblock {An overview of Venus geology}.
\newblock {\em Science}, 252:249--252.

\bibitem[{Schubert}, 1983]{Schubert1983}
{Schubert}, G. (1983).
\newblock {General circulation and the dynamical state of the Venus}.
\newblock In {\em VENUS}, pages 681--765. University of Arizona Press.

\bibitem[{Schubert} et~al., 1977]{Schubert1977}
{Schubert}, G., {Counselman}, III, C.~C., {Pettengill}, G., {Shapiro}, I.~I.,
  {Hansen}, J., {Travis}, L., {Limaye}, S.~S., {Suomi}, V.~E., {Seiff}, A., and
  {Taylor}, F. (1977).
\newblock {Dynamics, winds, circulation and turbulence in the atmosphere of
  Venus}.
\newblock {\em Space Science Reviews}, 20:357--387.

\bibitem[{Seiff}, 1983]{Seiff1983}
{Seiff}, A. (1983).
\newblock {Temperature structure of the Venus atmosphere}.
\newblock In {\em VENUS}, pages 215--279. University of Arizona Press.

\bibitem[{Seiff} et~al., 1980]{Seiff1980}
{Seiff}, A., {Kirk}, D.~B., {Young}, R.~E., {Blanchard}, R.~C., {Findlay},
  J.~T., {Kelly}, G.~M., and {Sommer}, S.~C. (1980).
\newblock {Measurements of thermal structure and thermal contrasts in the
  atmosphere of Venus and related dynamical observations - Results from the
  four Pioneer Venus probes}.
\newblock {\em Journal of Geophysical Research}, 85:7903--7933.

\bibitem[{Seiff} et~al., 1985]{Seiff1985}
{Seiff}, A., {Schofield}, J.~T., {Kliore}, A.~J., {Taylor}, F.~W., and
  {Limaye}, S.~S. (1985).
\newblock {Models of the structure of the atmosphere of Venus from the surface
  to 100 kilometers altitude}.
\newblock {\em Advances in Space Research}, 5:3--58.

\bibitem[{Smith} and {Gierasch}, 1996]{Smith1996}
{Smith}, M.~D. and {Gierasch}, P.~J. (1996).
\newblock {Global-Scale Winds at the Venus Cloud-Top Inferred from Cloud Streak
  Orientations}.
\newblock {\em Icarus}, 123:313--323.

\bibitem[{Straus} and {Ditlevsen}, 1999]{Straus1999}
{Straus}, D.~M. and {Ditlevsen}, P. (1999).
\newblock {Two-dimensional turbulence properties of the ECMWF reanalyses}.
\newblock {\em Tellus Series A}, 51:749--772.

\bibitem[{Stull}, 1976]{Stull1976}
{Stull}, R.~B. (1976).
\newblock {Internal Gravity Waves Generated by Penetrative Convection}.
\newblock {\em Journal of Atmospheric Sciences}, 33:1279--1286.

\bibitem[{Suomi} and {Limaye}, 1978]{Suomi1978}
{Suomi}, V.~E. and {Limaye}, S.~S. (1978).
\newblock {Venus - Further evidence of vortex circulation}.
\newblock {\em Science}, 201:1009--1011.

\bibitem[{Svedhem} et~al., 2007]{Svedhem2007}
{Svedhem}, H., {Titov}, D.~V., {McCoy}, D., {Lebreton}, J.-P., {Barabash}, S.,
  {Bertaux}, J.-L., {Drossart}, P., {Formisano}, V., {H{\"a}usler}, B.,
  {Korablev}, O., {Markiewicz}, W.~J., {Nevejans}, D., {P{\"a}tzold}, M.,
  {Piccioni}, G., {Zhang}, T.~L., {Taylor}, F.~W., {Lellouch}, E., {Koschny},
  D., {Witasse}, O., {Eggel}, H., {Warhaut}, M., {Accomazzo}, A.,
  {Rodriguez-Canabal}, J., {Fabrega}, J., {Schirmann}, T., {Clochet}, A., and
  {Coradini}, M. (2007).
\newblock {Venus Express: The first European mission to Venus}.
\newblock {\em Planetary and Space Science}, 55:1636--1652.

\bibitem[{Takagi} and {Matsuda}, 2005]{Takagi2005}
{Takagi}, M. and {Matsuda}, Y. (2005).
\newblock {Sensitivity of thermal tides in the Venus atmosphere to basic zonal
  flow and Newtonian cooling}.
\newblock {\em Geophysical Research Letters}, 32:2203--+.

\bibitem[{Takagi} and {Matsuda}, 2006]{Takagi2006}
{Takagi}, M. and {Matsuda}, Y. (2006).
\newblock {Dynamical effect of thermal tides in the lower Venus atmosphere}.
\newblock {\em Geophysical Research Letters}, 33:13102--+.

\bibitem[{Taylor} et~al., 1980]{Taylor1980}
{Taylor}, F.~W., {Beer}, R., {Chahine}, M.~T., {Diner}, D.~J., {Elson}, L.~S.,
  {Haskins}, R.~D., {McCleese}, D.~J., {Martonchik}, J.~V., {Reichley}, P.~E.,
  {Bradley}, S.~P., {Delderfield}, J., {Schofield}, J.~T., {Farmer}, C.~B.,
  {Froidevaux}, L., {Leung}, J., {Coffey}, M.~T., and {Gille}, J.~C. (1980).
\newblock {Structure and meteorology of the middle atmosphere of Venus -
  Infrared remote sensing from the Pioneer orbiter}.
\newblock {\em Journal of Geophysical Research}, 85:7963--8006.

\bibitem[{Taylor} et~al., 1985]{Taylor1985}
{Taylor}, F.~W., {Schofield}, J.~T., and {Valdes}, P.~J. (1985).
\newblock {Temperature structure and dynamics of the middle atmosphere of
  Venus}.
\newblock {\em Advances in Space Research}, 5:5--23.

\bibitem[{Titov} et~al., 2006]{Titov2006}
{Titov}, D.~V., {Svedhem}, H., {Koschny}, D., {Hoofs}, R., {Barabash}, S.,
  {Bertaux}, J.-L., {Drossart}, P., {Formisano}, V., {H{\"a}usler}, B.,
  {Korablev}, O., {Markiewicz}, W.~J., {Nevejans}, D., {P{\"a}tzold}, M.,
  {Piccioni}, G., {Zhang}, T.~L., {Merritt}, D., {Witasse}, O., {Zender}, J.,
  {Accomazzo}, A., {Sweeney}, M., {Trillard}, D., {Janvier}, M., and {Clochet},
  A. (2006).
\newblock {Venus Express science planning}.
\newblock {\em Planetary and Space Science}, 54:1279--1297.

\bibitem[{Toigo} et~al., 1994]{Toigo1994}
{Toigo}, A., {Gierasch}, P.~J., and {Smith}, M.~D. (1994).
\newblock {High resolution cloud feature tracking on Venus by Galileo}.
\newblock {\em Icarus}, 109:318--336.

\bibitem[{Travis}, 1978]{Travis1978}
{Travis}, L.~D. (1978).
\newblock {Nature of the atmospheric dynamics on Venus from power spectrum
  analysis of Mariner 10 images}.
\newblock {\em Journal of Atmospheric Sciences}, 35:1584--1595.

\bibitem[{Tung} and {Orlando}, 2003]{Tung2003}
{Tung}, K.~K. and {Orlando}, W.~W. (2003).
\newblock {The $k^{-3}$ and $k^{-5/3}$ Energy Spectrum of Atmospheric
  Turbulence: Quasigeostrophic Two-Level Model Simulation}.
\newblock {\em Journal of Atmospheric Sciences}, 60:824--835.

\bibitem[{Vallis}, 2006]{Vallis2006}
{Vallis}, G.~K. (2006).
\newblock {\em {Atmospheric and Oceanic Fluid Dynamics: Fundamentals and
  Large-Scale Circulation}}.
\newblock Cambridge University Press, Cambridge, U.K.

\bibitem[{Wallace} and {Hobbs}, 2006]{Wallace2006}
{Wallace}, J.~M. and {Hobbs}, P.~V. (2006).
\newblock {\em {Atmospheric Science}}.
\newblock Academic Press.

\bibitem[{Weiss}, 1991]{Weiss1991}
{Weiss}, J. (1991).
\newblock {The dynamics of enstrophy transfer in two-dimensional
  hydrodynamics}.
\newblock {\em Physica D Nonlinear Phenomena}, 48:273--294.

\bibitem[{Widemann} et~al., 2007]{Widemann2007}
{Widemann}, T., {Lellouch}, E., and {Campargue}, A. (2007).
\newblock {New wind measurements in Venus' lower mesosphere from visible
  spectroscopy}.
\newblock {\em Planetary and Space Science}, 55:1741--1756.

\bibitem[{Yamamoto}, 2003]{Yamamoto2003b}
{Yamamoto}, M. (2003).
\newblock {Gravity Waves and Convection Cells Resulting from Feedback Heating
  of Venus' Lower Clouds}.
\newblock {\em Journal of the Meteorological Society of Japan}, 81:885--892.

\bibitem[{Yoden} and {Yamada}, 1993]{Yoden1993}
{Yoden}, S. and {Yamada}, M. (1993).
\newblock {A numerical experiment on two-dimensional decaying turbulence on a
  rotating sphere.}
\newblock {\em Journal of Atmospheric Sciences}, 50:631--643.

\bibitem[{Young} et~al., 1994]{Young1994}
{Young}, R.~E., {Walterscheid}, R.~L., {Schubert}, G., {Pfister}, L., {Houben},
  H., and {Bindschadler}, D.~L. (1994).
\newblock {Characteristics of Finite Amplitude Stationary Gravity Waves in the
  Atmosphere of Venus.}
\newblock {\em Journal of Atmospheric Sciences}, 51:1857--1875.

\bibitem[{Young} et~al., 1987]{Young1987}
{Young}, R.~E., {Walterscheid}, R.~L., {Schubert}, G., {Seiff}, A., {Linkin},
  V.~M., and {Lipatov}, A.~N. (1987).
\newblock {Characteristics of gravity waves generated by surface topography on
  Venus - Comparison with the VEGA Balloon results}.
\newblock {\em Journal of Atmospheric Sciences}, 44:2628--2639.

\bibitem[{Zasova} et~al., 1999]{Zasova1999}
{Zasova}, L.~V., {Khatountsev}, I.~A., {Moroz}, V.~I., and {Ignatiev}, N.~I.
  (1999).
\newblock {Structure of the Venus middle atmosphere: Venera 15 fourier
  spectrometry data revisited}.
\newblock {\em Advances in Space Research}, 23:1559--1568.

\bibitem[{Zhu}, 2006]{Zhu2006}
{Zhu}, X. (2006).
\newblock {Maintenance of equatorial superrotation in the atmospheres of Venus
  and Titan}.
\newblock {\em Planetary and Space Science}, 54:761--773.

\end{thebibliography}

\newpage{\cleardoublepage}

\addcontentsline{toc}{chapter}{\'{i}ndice alfab\'{e}tico}
\printindex

\end{document}